\RequirePackage{heppennames2}
\RequirePackage{lhchiggs}

\documentclass{SciPost}
\hypersetup{
    colorlinks,
    linkcolor={red!50!black},
    citecolor={blue!50!black},
    urlcolor={blue!80!black}
}

\usepackage[bitstream-charter]{mathdesign}
\DeclareSymbolFont{usualmathcal}{OMS}{cmsy}{m}{n}
\DeclareSymbolFontAlphabet{\mathcal}{usualmathcal}

\fancypagestyle{SPstyle}{
\fancyhf{}
\lhead{\colorbox{scipostblue}{\bf \color{white} ~SciPost Physics Community Reports }}
\rhead{{\bf \color{scipostdeepblue} ~Submission }}

\fancyfoot[C]{\textbf{\thepage}}
}

\usepackage{graphicx}
\usepackage{tabularx}
\usepackage{subcaption}
\usepackage{multirow}
\usepackage{graphbox}
\usepackage{amsmath}
\usepackage{hyperref}
\DeclareRobustCommand{\ttus}{\_\allowbreak{}}
\usepackage{color}
\usepackage{cite}
\usepackage{todonotes}
\usepackage{float}
\usepackage{booktabs}
\usepackage{multirow}
\usepackage{authblk}
\usepackage[section]{placeins}
\newcommand{\imineq}[2]{\vcenter{\hbox{\includegraphics[height=#2ex]{#1}}}}

\usepackage{xspace}

\usepackage{tikz}
\usepackage{array}
\usepackage{slashed,mathtools}
\usepackage{psfrag,epsf,color}
\usepackage{xcolor}
\definecolor{cb}{HTML}{648FFF}
\definecolor{cc}{HTML}{FE6100}
\definecolor{s}{HTML}{DC267F}
\definecolor{c}{HTML}{FE6100}
\definecolor{hc}{HTML}{A2E03A}
\definecolor{sc}{HTML}{FFCF00}

\newcommand\ResNTwo{{\rm N}_2}

\newcommand\ResNbTwo{\overline{{\rm N}}_2}
\newcommand\NkLONkLLtimesNLOmt{{\rm (N}^{k}{\rm LO+N}^k{\rm LL)\otimes NLO}_{m_t}}
\newcommand\NkLOtimesNLOmt{{\rm N}^{k}{\rm LO \otimes NLO}_{m_t}}
\newcommand\NkLONkLL{{\rm N}^{k}{\rm LO+N}^k{\rm LL}}
\newcommand\NkLO{{\rm N}^{k}{\rm LO}}
\newcommand\NLOmt{{\rm NLO}_{m_t}}
\newcommand\NNLOtimesNLOmt{{\rm NNLO\otimes NLO}_{m_t}}
\newcommand\NNLONNLLtimesNLOmt{{\rm (NNLO+NNLL)\otimes NLO}_{m_t}}
\newcommand\NtLOtimesNLOmt{{\rm N}^{3}{\rm LO \otimes NLO}_{m_t}}
\newcommand\NtLONtLLtimesNLOmt{{\rm (N}^{3}{\rm LO+N}^3{\rm LL)\otimes NLO}_{m_t}}

\newcommand{\ftapprox}{FT$_{\mathrm{approx}}$\xspace}
\newcommand{\geneva}{\textsc{Geneva}\xspace}
\newcommand{\hhgrid}{\textsc{HHgrid}\xspace}
\newcommand{\openloops}{\textsc{OpenLoops}}
\newcommand{\pythiaEight}{\texttt{Pythia 8}\xspace}

\newcommand{\preprintnumbers}{%
  \begin{flushright}
    \small
    LHCHWG-2026-010
  \end{flushright}
  \vspace{1em}
}

\begin{document}

\pagestyle{SPstyle}

\preprintnumbers

\begin{center}{\Large \textbf{\color{scipostdeepblue}{
 Higgs Boson Pair Production via Gluon Fusion: Higher-Order Corrections and Theoretical Uncertainties\\
}}}\end{center}

\begin{center}\textbf{
Ajjath A~H\textsuperscript{1,2},
Simone Alioli\textsuperscript{3},
Emanuele Bagnaschi\textsuperscript{4},
Arunima Bhattacharya\textsuperscript{5},
Huan-Yu Bi\textsuperscript{6},
Roberto Bonciani\textsuperscript{7},
Marco Bonetti\textsuperscript{8,9},
Francisco Campanario\textsuperscript{5},
Sauro Carlotti\textsuperscript{9},
Jamie Chang\textsuperscript{10,11},
Long-Bin Chen\textsuperscript{12},
Xuan Chen\textsuperscript{13},
Yuesheng Dai\textsuperscript{13},
Joshua Davies\textsuperscript{14},
Giuseppe Degrassi\textsuperscript{15},
Pier~Paolo Giardino\textsuperscript{16},
Massimiliano Grazzini\textsuperscript{17},
Ramona Gr\"ober\textsuperscript{18},
Gudrun Heinrich\textsuperscript{9},
Li-Hong Huang\textsuperscript{19},
Rui-Jun Huang\textsuperscript{19},
Sebastian Jaskiewicz\textsuperscript{20},
Stephen Jones\textsuperscript{1},
Stefan Kallweit\textsuperscript{17},
Matthias Kerner\textsuperscript{9},
Hai~Tao Li\textsuperscript{13},
Shi-Yuan Li\textsuperscript{13},
Jonas~M. Lindert\textsuperscript{21},
Yan-Qing Ma\textsuperscript{19},
Giulia Marinelli\textsuperscript{22},
Javier Mazzitelli\textsuperscript{10},
Margarete M\"uhlleitner\textsuperscript{9},
Davide Napoletano\textsuperscript{3},
Philipp Rendler\textsuperscript{9},
Jonathan Ronca\textsuperscript{18},
Johannes Schlenk\textsuperscript{23},
Kay Sch\"onwald\textsuperscript{17,28},
Hua-Sheng Shao\textsuperscript{24},
Michael Spira\textsuperscript{10},
Thomas Stone\textsuperscript{1,25},
Daniel Stremmer\textsuperscript{26},
Robert Szafron\textsuperscript{27},
William~J. Torres~Bobadilla\textsuperscript{14},
Yannick Ulrich\textsuperscript{14},
Augustin Vestner\textsuperscript{9},
Marco Vitti\textsuperscript{26},
Jian Wang\textsuperscript{13},
Huai-Min Yu\textsuperscript{3} and
\mbox{Hantian Zhang}\textsuperscript{28}
}\end{center}
{\textit{Editors: Fabio Monti\textsuperscript{29}, Arantxa Ruiz Martínez\textsuperscript{30}, Angela Taliercio\textsuperscript{31}, Andreas Papaefstathiou\textsuperscript{32} and Ludovic Scyboz\textsuperscript{33}}}

\begin{center}
{\bf 1} Institute for Particle Physics Phenomenology, Durham University, Durham, United Kingdom
\\
{\bf 2} Centre for High Energy Physics, Indian Institute of Science, C. V. Raman Avenue, Bengaluru-560012, India
\\
{\bf 3} Universit\`a degli Studi di Milano-Bicocca and INFN Sezione di Milano-Bicocca, Milano, Italy
\\
{\bf 4} INFN, Laboratori Nazionali di Frascati, Frascati, Italy
\\
{\bf 5} Theory Division, IFIC, University of Valencia-CSIC, Paterna, Valencia, Spain
\\
{\bf 6} School of Physics and Optoelectronic Engineering, Hainan University, Haikou, China
\\
{\bf 7} Dipartimento di Fisica e Astronomia, Universit\`a di Firenze, Sesto Fiorentino, Italy
\\
{\bf 8} Institute for Theoretical Physics, University of T\"ubingen, T\"ubingen, Germany, and Institute for Astroparticle Physics, Karlsruhe Institute of Technology, Eggenstein-Leopoldshafen, Germany
\\
{\bf 9} Institute for Theoretical Physics, Karlsruhe Institute of Technology, Karlsruhe, Germany
\\
{\bf 10} PSI Center for Neutron and Muon Sciences, Villigen PSI, Switzerland
\\
{\bf 11} Institute for Theoretical Physics, ETH Z\"urich, Z\"urich, Switzerland
\\
{\bf 12} School of Physics and Electronic Engineering, Guangzhou University, Guangzhou, China
\\
{\bf 13} School of Physics, Shandong University, Jinan, China
\\
{\bf 14} Department of Mathematical Sciences, University of Liverpool, Liverpool, United Kingdom
\\
{\bf 15} Dipartimento di Matematica e Fisica, Universit\`a di Roma Tre, and INFN Sezione di Roma Tre, Rome, Italy
\\
{\bf 16} Departamento de F\'isica Te\'orica and Instituto de F\'isica Te\'orica UAM/CSIC, Universidad Aut\'onoma de Madrid, Madrid, Spain
\\
{\bf 17} Department of Physics, University of Zurich, Zurich, Switzerland
\\
{\bf 18} Dipartimento di Fisica e Astronomia ``G.~Galilei'', Universit\`a di Padova, and INFN Sezione di Padova, Padova, Italy
\\
{\bf 19} School of Physics, Peking University, Beijing, China
\\
{\bf 20} Albert Einstein Center for Fundamental Physics, Institut f\"ur Theoretische Physik, Universit\"at Bern, Bern, Switzerland
\\
{\bf 21} Department of Physics and Astronomy, University of Sussex, Brighton, United Kingdom
\\
{\bf 22} Deutsches Elektronen-Synchrotron DESY, Hamburg, Germany
\\
{\bf 23} Department of Astrophysics, University of Zurich, Zurich, Switzerland
\\
{\bf 24} Laboratoire de Physique Th\'eorique et Hautes Energies (LPTHE), Sorbonne Universit\'e and CNRS, Paris, France
\\
{\bf 25} Physik Department T35, Technische Universit\"at M\"unchen, Garching, Germany
\\
{\bf 26} Institute for Theoretical Particle Physics, Karlsruhe Institute of Technology, Karlsruhe, Germany
\\
{\bf 27} Department of Physics, Brookhaven National Laboratory, Upton, New York, U.S.A.
\\
{\bf 28} Theoretical Physics Department, CERN, European Organization for Nuclear Research, Geneva, Switzerland
\\
{\bf 29} Experimental Physics Department, CERN, European Organization for Nuclear Research, Geneva, Switzerland
\\
{\bf 30} Instituto de Física Corpuscular (IFIC), Centro Mixto Universidad de Valencia - CSIC, Valencia, Spain
\\
{\bf 31} Northwestern University, Evanston, Illinois, USA
\\
{\bf 32} Kennesaw State University, Marietta, Georgia, USA
\\
{\bf 33} School of Physics and Astronomy, Monash University, Clayton, Australia

\end{center}

\section*{\color{scipostdeepblue}{Abstract}}
\textbf{\boldmath{%
In this contribution, the higher-order QCD and electroweak corrections to Standard Model Higgs boson pair production via the gluon-fusion mechanism, $gg\to hh$, are summarized and the different sources of theoretical uncertainty are assessed. The discussion includes finite top quark mass effects, matching to parton showers, approximate NNLO and N$^3$LO QCD corrections, NLO electroweak effects, and uncertainties associated with the top quark mass scheme and perturbative scale choices. In addition, we provide an updated state-of-the-art recommendation for the inclusive gluon-fusion Higgs boson pair production cross section and the corresponding Higgs boson pair invariant-mass distribution.
}}

\vspace{\baselineskip}

\noindent\textcolor{white!90!black}{%
\fbox{\parbox{\dimexpr\linewidth-2\fboxsep-2\fboxrule\relax}{%
\textcolor{white!40!black}{\begin{tabularx}{\linewidth}{@{}X>{\raggedleft\arraybackslash}p{0.3\linewidth}@{}}%
  {\small Copyright attribution to authors. \newline
  This work is a submission to SciPost Phys. Comm. Rep. \newline
  License information to appear upon publication. \newline
  Publication information to appear upon publication.}
  &
  {\small Received Date \newline Accepted Date \newline Published Date}%
\end{tabularx}}
}}
}

\tableofcontents




\section{Introduction}
The discovery of a scalar resonance at the Large Hadron Collider (LHC) \cite{Aad:2012tfa,Chatrchyan:2012xdj} with a mass of 125 GeV, which is compatible with the
Standard Model (SM) Higgs boson \cite{Higgs:1964ia, Higgs:1964pj,
Englert:1964et, Guralnik:1964eu, Higgs:1966ev, Kibble:1967sv}, completed
our description of strong and electroweak interactions. However, determining the coupling strengths of the boson's self-interactions
and interactions with the other SM particles is essential to establish its identity as the SM Higgs boson.
Specifically, Higgs boson production and decay processes at the LHC \cite{Khachatryan:2016vau,ATLAS:2019nkf,CMS:2020xwi}
are suitable channels to extract these couplings and will be pursued in future runs.

\begin{figure}[t]
\vspace*{0.3cm}
\begin{center}
\includegraphics[width=0.9\textwidth]{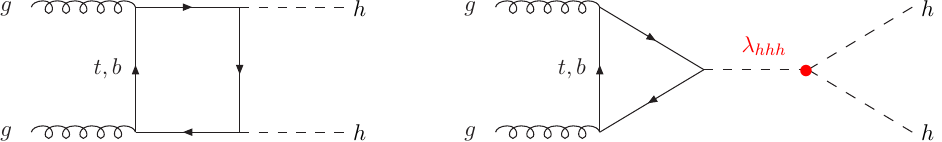}
\caption{Typical diagrams contributing to Higgs boson pair
production via gluon fusion. The contribution of the trilinear Higgs boson
self-coupling is marked in red.}
\label{fg:hhdia}
\end{center}
\end{figure}

Among the Higgs couplings, the trilinear self-coupling $\lambda_{hhh}$ --- a key parameter of the SM scalar sector
and a natural window into physics beyond the SM --- is still
relatively unconstrained and can be probed directly in Higgs boson pair production.
Higgs pair production via gluon fusion is mediated by triangle and box
diagrams involving closed top-quark loops at leading order (LO) \cite{Glover:1987nx,
Plehn:1996wb}, see Fig.~\ref{fg:hhdia}; bottom-quark loops contribute only at the sub-percent level.
%
The box diagrams (left) and the triangle diagrams (right) interfere destructively; the latter is the one sensitive to the trilinear
Higgs boson self-coupling $\lambda_{hhh}$, where the cross section $\Delta\sigma/\sigma \sim -\Delta\lambda_{hhh}/\lambda_{hhh}$
in the vicinity of the SM value of $\lambda_{hhh}$. Therefore, determining
the trilinear Higgs boson self-coupling from Higgs boson pair production requires
a reduction of the theoretical uncertainty on the corresponding cross section.

Relating the basic parameters of the Lagrangian
to the physical observables we measure is, in general, plagued by experimental and
theoretical uncertainties, which require a thorough assessment. On the theoretical side, the usual procedure
is to study the dependence of the prediction on the factorization and renormalization scales: this dependence
originates from the involvement of parton densities and the strong coupling in most production
processes, and is thus relevant for the QCD uncertainties. In some processes,
the renormalization scale dependence of the Yukawa coupling is included too, e.g.~in
$h\to q\bar q$ ($q=b,c$). In a high-precision context, the uncertainties due to unknown higher-order
electroweak corrections need to be considered to obtain a complete estimate of the
theoretical uncertainties.

In this report, we focus on theoretical predictions for Higgs pair production
in the SM: we review state-of-the-art
QCD and EW calculations, and provide updated
recommendations for inclusive cross sections, together with a detailed
assessment of the theoretical uncertainties --- a prerequisite for any extraction of
the self-coupling and subsequent interpretation in extended models.

The QCD corrections were first obtained at next-to-leading order (NLO) in the heavy-top limit (HTL) \cite{Dawson:1998py}, later including the full top quark mass dependence
up to NLO \cite{Borowka:2016ehy,Borowka:2016ypz, Baglio:2018lrj,Baglio:2020ini,Baglio:2020wgt} and at next-to-next-to-leading order (NNLO) in the HTL \cite{deFlorian:2013uza,
deFlorian:2013jea, Grigo:2014jma}. Meanwhile, the next-to-next-to-next-to-leading order (N$^3$LO) QCD
corrections have been computed in the HTL,
resulting in a small further increase of the cross section
\cite{Banerjee:2018lfq, Chen:2019lzz, Chen:2019fhs, Chen:2026zmi}, as well as the N$^3$LO+N$^3$LL corrections~\cite{Ajjath:2022kpv}, where the residual scale uncertainties are at the percent level. Consequently, other theory uncertainties now play a dominant role.

The QCD corrections
increase the total LO cross section by about a factor of two. The full NLO results have been matched to parton showers \cite{Heinrich:2017kxx, Jones:2017giv,Bagnaschi:2023rbx} and the full NNLO results in the
HTL have been merged with the NLO mass effects and
supplemented by additional top quark mass effects in the double-real
corrections \cite{Grazzini:2018bsd}.
Apart from the renormalization and factorization scale uncertainties, the following uncertainties are assessed in this report: uncertainties induced by the top quark mass scheme and scale dependence, uncertainties due to missing finite top quark mass effects and the impact of NLO electroweak corrections.

\subsection{General considerations}

The amplitude for $g(p_1) g(p_2) \to h(p_3) h(p_4)$ can be decomposed as~\cite{Glover:1987nx}\footnote{In Ref.~\cite{Bi:2023bnq}, the existence of an additional $\Delta_5$ tensor structure, containing a single Levi-Civita symbol and related to electroweak corrections, is considered. This new tensor structure plays no role at NLO since it vanishes upon contraction with the Born amplitude.}
\begin{align}
    \begin{aligned}
        \mathcal{M}_{ab}&=\delta_{ab}\,\epsilon^\mu (p_1,n_1)\epsilon^\nu (p_2,n_2)\,\mathcal{M}_{\mu\nu}   \,,\\
        \mathcal{M}^{\mu\nu}&=\frac{\alpha_s}{8\pi v^2}\left[
        F_1(\hat{s},\hat{t},m_h^2,m_t^2,D)\,T_1 ^{\mu\nu} +
        F_2(\hat{s},\hat{t},m_h^2,m_t^2,D)\,T_2 ^{\mu\nu}
        \right],
    \end{aligned}
    \label{eq:FFdeco}
\end{align}
where $m_h$ and $m_t$ are the Higgs boson and top quark masses, respectively, $D$ is the number of spacetime dimensions, $a$ and $b$ are colour indices, and the tensor structures $T_{1,2}^{\mu\nu}$ read
\begin{align}
    \begin{aligned}
        T_1^{\mu\nu} &= g^{\mu\nu} - \frac{1}{p_{12}}
        p_2^\mu p_1^\nu
        \,,\\
        T_2^{\mu\nu} &= g^{\mu\nu} + \frac{1}{p_{12} p_T^2}
    \left(
        m_h^2 p_2^\mu p_1^\nu - 2 p_{23} p_3^\mu p_1^\nu - 2 p_{13} p_2^\mu p_3^\nu + 2 p_{12} p_3^\mu p_3^\nu
    \right)
    \end{aligned}
\end{align}
with
\begin{align}
    p_T^2 = \frac{2 p_{13} p_{23}}{p_{12}} - m_h^2\quad\mbox{and}\quad p_{ij}=p_i\cdot p_j
    \,.
\end{align}
The form factors $F_{1,2}$ are related to helicity amplitudes by
\begin{align}
    \begin{aligned}
        \mathcal{M}^{++}    = \mathcal{M}^{--} &= -\frac{\alpha_s}{8\pi v^2} \,F_1(\hat{s},\hat{t},m_h^2,m_t^2,D) \,,\\
        \mathcal{M}^{+-}    = \mathcal{M}^{-+} &= -\frac{\alpha_s}{8\pi v^2} \,F_2(\hat{s},\hat{t},m_h^2,m_t^2,D)
        \,,
    \end{aligned}
\end{align}
and can be extracted by applying the projectors $P_{1,2}^{\mu\nu}$ to $\mathcal{M}^{\mu\nu}$ as
\begin{align}
    \label{eq:in:ffe}
    \begin{aligned}
        P_1^{\mu\nu} \mathcal{M}_{\mu\nu}   &= \frac{\alpha_s}{8\pi v^2}\,F_1(\hat{s},\hat{t},m_h^2,m_t^2,D)
        \,, \\
        P_2^{\mu\nu} \mathcal{M}_{\mu\nu}   &= \frac{\alpha_s}{8\pi v^2}\,F_2(\hat{s},\hat{t},m_h^2,m_t^2,D)\,,
    \end{aligned}
\end{align}
with
\begin{align}
    \begin{aligned}
        P_1^{\mu\nu}    &=
        \frac{1}{4}\frac{D-2}{D-3} T_1^{\mu\nu} - \frac{1}{4}\frac{D-4}{D-3} T_2^{\mu\nu}
        \,,\\
        P_2^{\mu\nu}    &=
        \frac{1}{4}\frac{D-2}{D-3} T_2^{\mu\nu} - \frac{1}{4}\frac{D-4}{D-3} T_1^{\mu\nu}
        \,.
    \end{aligned}
\end{align}
Higher-order corrections require the determination of both virtual and real contributions. In the latter case, extra partons have to be considered, producing new tensor structures. For QCD corrections, all possible non-zero combinations of coloured partons must be considered, while NLO EW corrections do not produce real corrections, thanks to Furry's theorem.

This report is organized as follows. Section~\ref{sec:nlo-qcd} reviews the
NLO QCD corrections with full top quark mass effects, including the matching
to parton showers and the treatment of finite top quark mass effects in the relevant
amplitudes. Section~\ref{sec:nnlo-qcd} discusses the approximate NNLO QCD
predictions, while Section~\ref{sec:n3lo-qcd} summarizes the N$^3$LO QCD
corrections and the impact of soft-gluon resummation.
Section~\ref{sec:scheme-scale-uncertainties} is devoted to the top quark mass
scheme and scale uncertainties, and Section~\ref{sec:ew-sm} reviews the
electroweak corrections in the Standard Model. The purpose of these sections
is to describe the methods and intermediate results, in some cases obtained
with choices of parameters and PDF sets that differ from those used in the
final recommendations; the definitive numbers are the recommendations collected
at the end of the report. The final cross section recommendations are presented
in Section~\ref{sec:recs}. In particular, Table~\ref{tab:final_xs_reco_125}
shows the combined final recommendation, including the dominant uncertainty
stemming from the top quark mass scheme dependence, and
Table~\ref{tab:final_xs_reco_125_mh} shows the variation of the recommended
cross section with the Higgs-boson mass. Section~\ref{subsec:mhh_distribution}
presents the recommended differential distributions for the Higgs boson pair invariant mass, $m_{hh}$.

\section{NLO QCD corrections with full top quark mass effects}
\label{sec:nlo-qcd}

\subsection{Full NLO QCD corrections and matching to parton showers}
\label{sec:fullPS}


\textbf{Sophia Borowka, Nicolas Greiner, Gudrun Heinrich, Stephen P. Jones, Matthias Kerner, Johannes Schlenk, Ulrich Schubert, Tom Zirke}~\cite{Borowka:2016ehy,Borowka:2016ypz,Heinrich:2017kxx,Jones:2017giv,Heinrich:2019bkc}.

\noindent In this section we briefly describe the calculation of the NLO QCD corrections to $gg\to hh$ including the full top quark mass dependence via a numerical method based on sector decomposition~\cite{Binoth:2000ps,Borowka:2015mxa}, explained in detail in Refs.~\cite{Borowka:2016ehy,Borowka:2016ypz}, and the matching of these fixed-order results to parton showers~\cite{Heinrich:2017kxx,Jones:2017giv,Heinrich:2019bkc}.

\paragraph{Method.}
\label{intro_ampli}

The virtual two-loop amplitude has been generated with an extension of the program \texttt{GoSam}~\cite{GoSam:2014yla,Braun:2025afl}, and the form factors extracted according to Eq.~\eqref{eq:in:ffe}.
The reduction of the integrals occurring in the amplitude to master integrals has been performed using \texttt{Reduze}~\cite{vonManteuffel:2012np}.
 A complete reduction had been achieved only for the planar sectors, while a partial reduction of the
non-planar sectors was sufficient for the numerical evaluation. This led to 145 planar master
integrals plus 70 non-planar integrals, and a further 112 integrals that differ by a crossing.
As these integrals contain four independent mass scales,
$\hat{s},\hat{t},m_t^2, m_h^2$, their analytical calculation is a hard task.
We calculated all integrals numerically using the program \texttt{SecDec}-3.0~\cite{Borowka:2015mxa}, which has since evolved into \texttt{pySecDec}~\cite{Borowka:2017idc,Heinrich:2023til}, using $\sim$16 dual \texttt{Nvidia Tesla K20X} GPU nodes.

The amplitudes for the real radiation have been generated with \texttt{GoSam}~\cite{GoSam:2014yla,Braun:2025afl}.
As the process is loop-induced, the NLO real corrections involve one-loop $2\to 3$ amplitudes where single-unresolved radiation can occur.
We used the Catani-Seymour dipole formalism~\cite{Catani:1996vz} to isolate the corresponding infrared poles.

The dependence on the trilinear Higgs boson self-coupling modifier
$\kappa_\lambda \equiv \lambda_{hhh}/\lambda_{hhh}^{\rm SM}$
was also assessed in Ref.~\cite{Borowka:2016ypz}, for the first time
at full NLO.
In the HTL at leading order, the $\kappa_\lambda$-dependence is
given by
\begin{align}
|{\cal M}|^2&\sim  \frac{2}{9} -  \frac{4}{3}\,m_h^2\,\frac{\kappa_\lambda}{\hat{s}-m_h^2} +
  2\,m_h^4\,\frac{\kappa_\lambda^2}{(\hat{s}-m_h^2)^2}\;.\label{eq:loheft}
\end{align}
For $\kappa_\lambda=1$, this expression vanishes at the Higgs boson pair production threshold
$\hat{s} \sim 4 m_h^2$. For larger values of $\kappa_\lambda$, the minimum is
shifted; the expression in Eq.~\eqref{eq:loheft}, viewed as a function of $\hat{s}$, has a zero of order $2$ at 
$\hat{s}=m_h^2(1+3\kappa_\lambda)$ due to the destructive interference between box- and
triangle-type contributions --- both $|\mathcal{M}|^2$ and its first derivative vanish.
In the full theory, the amplitude does not vanish completely at these points,
but nonetheless also gets small.

\paragraph{Results.}

\begin{figure}
\centering
\begin{subfigure}{0.49\textwidth}
\includegraphics[width=\textwidth]{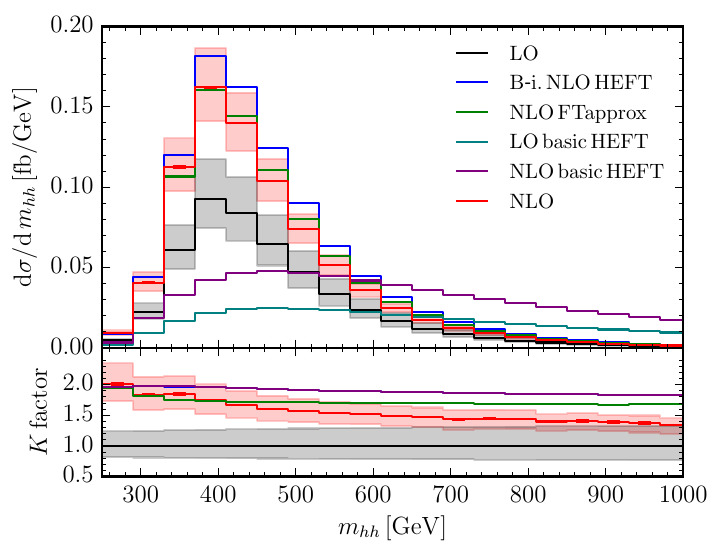}
\caption{Higgs boson pair invariant mass distribution.\label{subfig:mhh}}
\end{subfigure}
\begin{subfigure}{0.49\textwidth}
\includegraphics[width=\textwidth]{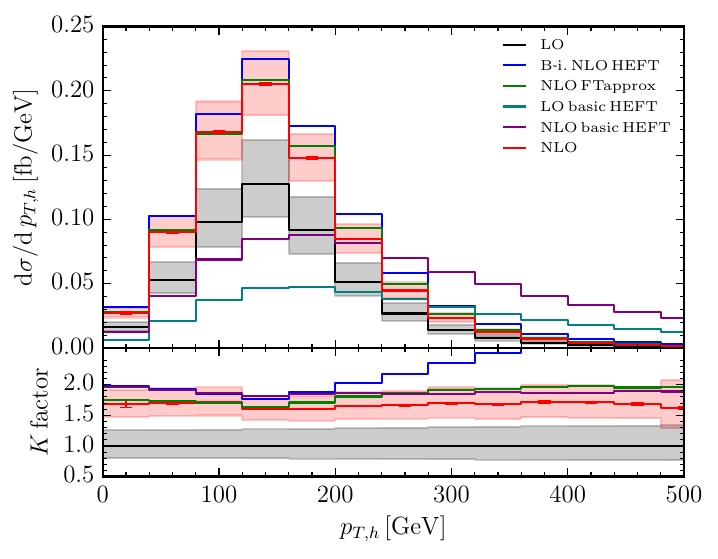}
\caption{$p_{T,h}$ distribution.\label{subfig:pth14}}
\end{subfigure}
\caption{(a) SM Higgs boson pair invariant mass distribution $m_{hh}$, (b) transverse momentum distribution of (any) Higgs boson, $p_{T,h}$, both at
  $\sqrt{s}=14$\,TeV. Predictions include the Born-improved (``B-i."), approximate full theory (FT$_{\rm{approx}}$), and full NLO corrections. In this figure, HEFT denotes the HTL, and ``basic'' means not rescaled by the full-$m_t$ Born amplitude. Figures from Ref.~\cite{Borowka:2016ypz}.\label{fig:mhh}}
\end{figure}

Differential results for the Higgs boson pair invariant mass distribution $m_{hh}$ and the transverse momentum distribution of a Higgs boson, $p_{T,h}$, are shown in Fig.~\ref{fig:mhh}.
Comparing the full NLO result with the HTL prediction (called HEFT in Fig.~\ref{fig:mhh}), one can clearly see that the high-energy region is not well described by this approximation, even if it is rescaled with the full Born (called ``Born-improved HEFT'', blue curve). Without rescaling by the full Born (called ``basic HEFT''), the prediction fails to describe the distribution except in the very low-$m_{hh}$ bins, as this approximation is only valid below the production threshold of a virtual top quark pair. The dark green curve denotes the so-called $\mathrm{NLO}_{\mathrm{FTapprox}}$ (``approximate Full Theory'')~\cite{Maltoni:2014eza}, where the real radiation part is calculated with full $m_t$ dependence, while the virtual corrections are calculated in the HTL. This approximation has been lifted to NNLO accuracy in Ref.~\cite{Grazzini:2018bsd}, described in Section~\ref{sec:NNLO_FTapprox}.

The total cross section at full NLO, at $\sqrt{s}=14$\,TeV, was found to be smaller than the Born-rescaled NLO cross section in the HTL by about 15\%, and smaller than the $\mathrm{NLO}_{\mathrm{FTapprox}}$ result by about 4\%.

\begin{figure}
\centering
\begin{subfigure}{0.49\textwidth}
\includegraphics[width=\textwidth]{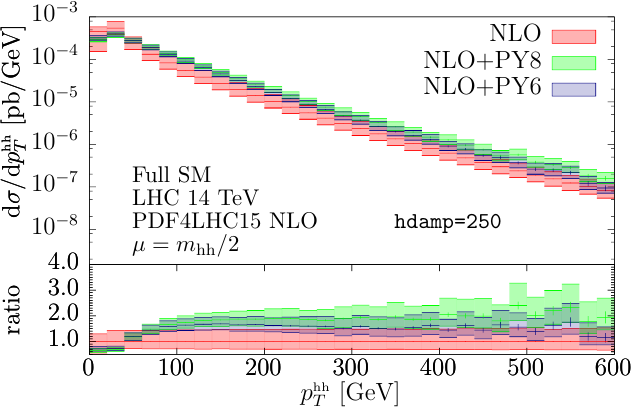}
\caption{Figure from Ref.~\cite{Heinrich:2017kxx}.\label{subfig:pthh_pythia_shower}}
\end{subfigure}
\begin{subfigure}{0.49\textwidth}
\includegraphics[width=\textwidth]{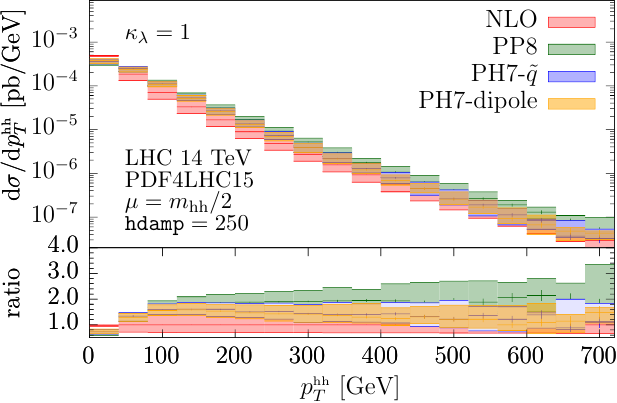}
\caption{Figure from Ref.~\cite{Heinrich:2019bkc}.\label{subfig:pthh_herwig_shower}}
\end{subfigure}
\caption{SM Higgs boson pair transverse momentum spectrum, comparing fixed-order NLO with different parton showers. All results are based on the {\tt ggHH} ({\tt Powheg-Box-V2}) code. (a) Comparison of fixed-order NLO with its matching to Pythia 6 (PY6) and Pythia 8 (PY8)~\cite{Sjostrand:2014zea}; (b) comparison of the Powheg results showered with Pythia 8 (PP8), with the Herwig 7~\cite{Bellm:2017bvx,Bellm:2025pcw} $\tilde{q}$ (PH7-$\tilde q$) and dipole (PH7-dipole) showers, respectively.\label{fig:gghh_nlo_shower}}
\end{figure}

\paragraph{Matching to parton showers.}
The NLO result calculated in
Refs.~\cite{Borowka:2016ehy,Borowka:2016ypz} has been matched to the {\tt Powheg-Box-V2}~\cite{Alioli:2010xd} and {\tt MG5\_aMC@NLO}~\cite{Alwall:2014hca} in Ref.~\cite{Heinrich:2017kxx} and to {\tt Sherpa}~\cite{Gleisberg:2008ta} in Ref.~\cite{Jones:2017giv}, where the {\tt Powheg-Box-V2} implementation, i.e. the public code {\tt ggHH}, was later extended to include variations of the trilinear Higgs boson self-coupling~\cite{Heinrich:2019bkc} and the leading five anomalous couplings within non-linear Higgs Effective Field Theory (HEFT)~\cite{Buchalla:2018yce,Heinrich:2020ckp}.
In Refs.~\cite{Heinrich:2017kxx,Heinrich:2020ckp} it has been demonstrated that the differences between different parton showers and matching schemes can be quite large for radiation-sensitive observables such as the net transverse momentum of the Higgs boson pair, while they are very moderate for the invariant mass of the Higgs boson pair, see Fig.~\ref{fig:gghh_nlo_shower}.
It also has been shown that the NLO results in the HTL can differ more from the full NLO result than some prominent EFT benchmark points, underlining again the importance of the NLO corrections with full top quark mass dependence.

The code {\tt ggHH\_SMEFT} contains NLO QCD results combined with leading and subleading operators in SMEFT~\cite{Heinrich:2022idm,Heinrich:2023rsd}, as well as running Wilson coefficients~\cite{Heinrich:2024rtg}, see also Ref.~\cite{Alasfar:2023xpc} for Effective Field Theory descriptions of Higgs boson pair production. This code can be interfaced to parton showers within the {\tt Powheg-Box-V2} in the same way as the {\tt ggHH} code.

\subsection[Full NLO calculation with free MH and mt]{Full NLO calculation with free \boldmath $m_h$ and $m_t$}
\label{sec:nlo2}


\textbf{Julien Baglio, Francisco Campanario, Seraina Glaus, Milada Margarete Mühl\-leitner, Jonathan Ronca, Michael Spira, Juraj Streicher~\cite{Baglio:2018lrj,Baglio:2020ini}.}

\begin{figure}[!hbtp]
\vspace*{-0cm}
\begin{center}
\includegraphics[width=0.5\textwidth]{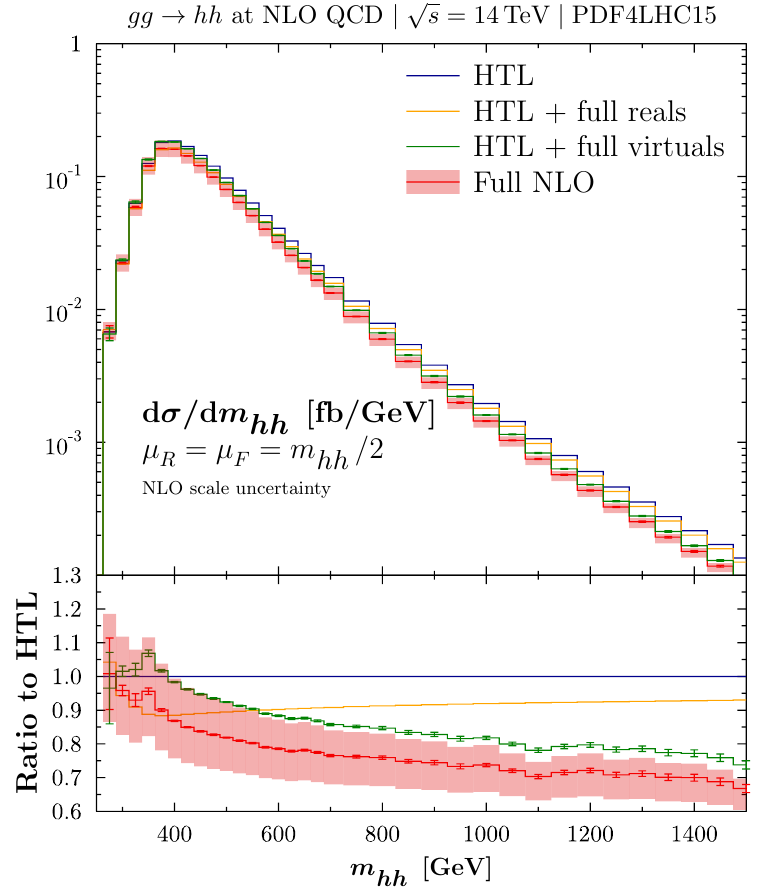}
\vspace*{-0cm}
\caption{Invariant-mass distributions for SM Higgs boson pair production via gluon fusion at the 14 TeV LHC as a function of $m_{hh}$: HTL results (in blue), HTL results including the full real corrections (in yellow), HTL results including the full virtual corrections (in green, including the numerical error), and full NLO QCD results (in red, including the numerical error). Results shown are from Refs.~\cite{Baglio:2018lrj,Baglio:2020ini}.}
\label{fg:dist_mhh}
\end{center}
\end{figure}
\noindent A second independent numerical calculation of the NLO QCD corrections, including the full top quark mass dependence, has been performed in Refs.~\cite{Baglio:2018lrj,Baglio:2020ini,Baglio:2020wgt}. In these works, the projection onto the two contributing form factors has been utilized in $D$ dimensions, with Feynman parametrization applied diagram by diagram, with no tensor reduction. This leads to compact expressions such that the free choice of Higgs boson and top quark masses could be retained. Since the bottom loops contribute at the percent level only, they have been neglected. The singularities were isolated by suitable end-point subtractions and dedicated infrared subtraction terms that built on the explicit structure of the Feynman-parameter integrals. The virtual $t\bar t$ thresholds were handled by using a complex virtual top quark mass $m_t^2 \to m_t^2 (1-i\bar\epsilon)$ and approaching the narrow-width limit $\bar\epsilon\to 0$ by decreasing the value of $\bar\epsilon$ to a minimal value of about 0.025 (or smaller) and performing a Richardson extrapolation \cite{Richardson} to obtain reliable numerical results for $\bar\epsilon\to 0$ as described in detail in Refs.~\cite{Baglio:2018lrj,Baglio:2020ini}. The real corrections have been obtained by computing the corresponding matrix elements with {\tt FeynArts} \cite{Hahn:2000kx} and {\tt FormCalc} \cite{Hahn:1998yk} while using {\tt Collier~1.2} \cite{Denner:2016kdg} for the scalar one-loop integrals. Choosing a Higgs boson mass of $m_h=125$ GeV and a top quark mass of $m_t=172.5$ GeV, we obtained a reduction of the total cross section from the top quark mass effects beyond the Born-improved HTL by about 15\% in agreement with \cite{Borowka:2016ypz,Borowka:2016ehy}. The small residual differences between the results of both calculations could be traced back to the different choices of the top quark mass, i.e.~172.5 GeV in \cite{Baglio:2018lrj,Baglio:2020ini} compared to 173 GeV in \cite{Borowka:2016ypz,Borowka:2016ehy}. The results for the distribution in the invariant Higgs boson pair mass $m_{hh}$ are shown in Fig.~\ref{fg:dist_mhh}. Whilst the top quark mass effects beyond the Born-improved HTL of the finite part of the real corrections amount to about $-10\%$, the top quark mass effects of the virtual corrections are greater for large values of the invariant Higgs boson pair mass. In this context, the explicit structure of the virtual $t\bar t$ threshold at $m_{hh} = 345$ GeV should be emphasized, which is in agreement with the ${\cal P}$-wave structure at LO in accordance with the single Higgs boson case \cite{Graudenz:1992pv,Spira:1995rr}. Results at NLO QCD with variable values of $\kappa_\lambda \equiv \lambda_{hhh}/\lambda_{hhh}^{\rm SM}$ are also available \cite{Baglio:2020ini,Baglio:2020wgt}.

Different scheme and scale choices of the virtual top quark mass can be considered for this calculation as a result of the freedom of choice of the top quark mass as an input parameter. This constitutes a (re)discovered additional source of theoretical uncertainties\footnote{The same treatment of these uncertainties has been applied to single-Higgs production, where these uncertainties turned out to be small due to the small scale defined by the Higgs mass \cite{Anastasiou:2016cez}.}. The explicit discussion of these uncertainties can be found in Section~\ref{sec:NLO-uncertainties}.

\subsection{Amplitude expansion around small transverse momentum and mass}
\label{sec:expGPL}


\textbf{Emanuele Bagnaschi, Roberto Bonciani, Giuseppe Degrassi, Pier Paolo Giardino, Ramona Gröber~\cite{Bonciani:2018omm,Bagnaschi:2023rbx}.}

\noindent An alternative approach to the fully numerical evaluation of the virtual corrections is provided by expansions. In particular, an expansion in small $p_T$ as originally proposed in \cite{Bonciani:2018omm} can cover more than 95\% of the relevant phase space for SM Higgs boson pair production. This expansion assumes that $p_T^2$ and $m_h^2$ are much smaller than $4 m_t^2$, allowing one to Taylor expand in these quantities while keeping the partonic centre-of-mass energy $\hat{s}$ arbitrary. Concretely, this means expanding in the forward region, $p_3\sim - p_1$ where $p_3$ is the incoming Higgs momentum and $p_1$ the momentum of a gluon. This expansion reduces the number of scales in the two-loop integrals to one, which allows for the analytical determination. In Ref.~\cite{Bonciani:2018omm} all the two-loop integrals were expressed analytically in terms of generalized harmonic polylogarithms, apart from two elliptic integrals that were determined in terms of series expansions around singular points \cite{Bonciani:2018uvv} following the approach of \cite{Pozzorini:2005ff, Aglietti:2007as}.

\begin{figure}[ht]
\centering
\includegraphics[width=.69\textwidth]{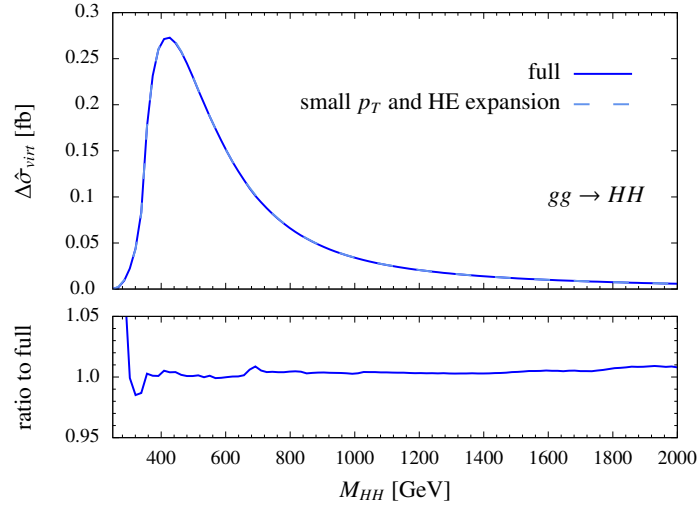}
\caption{Finite part of the NLO virtual contribution, for the full NLO result from Ref.~\cite{Davies:2019dfy} (dark blue), and for the combination of small-$p_T$ and high-energy (HE) expansions (dashed, light blue). The results are in excellent agreement over the whole range of invariant masses.
Figure from Ref.~\cite{Bellafronte:2022jmo}.}
\label{fig:pt-highE-exp}
\end{figure}

In the high-energy limit, the described expansion is no longer valid, as large $p_T$ values can be achieved. It must therefore be combined with a complementary expansion as provided in Ref.~\cite{Davies:2018qvx}. This reference expands the amplitudes in the high-energy limit, namely for $\hat{s}, \hat{t}, \hat{u} \gg m_t^2$. The expansion allows all integrals to be expressed analytically in terms of harmonic polylogarithms \cite{Davies:2018ood}.

The two expansions have been combined by means of Pad\'e approximations in Ref.~\cite{Bellafronte:2022jmo}, showing that a very reliable result, with a $< 1\%$ difference from the numerical grid of Ref.~\cite{Davies:2019dfy}, is obtained, as can be seen in Fig.~\ref{fig:pt-highE-exp}. Note
that in the first bin, the numerical grid entering the full result of Ref.~\cite{Davies:2019dfy} suffers from a large statistical uncertainty (because the grid is populated according to the SM cross section, which
is small at the production threshold).

This approach has been used for the POWHEG implementation of Higgs boson pair production at NLO QCD presented in Ref.~\cite{Bagnaschi:2023rbx}, for which the real corrections have been computed with \texttt{MadLoop} \cite{Hirschi:2011pa}. Due to the flexibility of the analytic approach, the top quark mass renormalization scheme uncertainty can be computed fully differentially, showing that, depending on the kinematic variable considered, it can be even larger in certain parts of the phase space than for the inclusive cross section. The code so far allows one to vary the top quark Yukawa coupling and the trilinear Higgs boson self-coupling, and further extensions to BSM scenarios can be achieved easily due to the flexibility of the approach.

A similar approach has been followed for the public code \texttt{ggxy} \cite{Davies:2025qjr} relying on a deeper expansion in small Mandelstam $\hat{t}$ combined with a high-energy expansion \cite{Davies:2023vmj}, see next section.


Other approaches, for instance, include a combination of large mass expansion with a threshold expansion combined via Pad\'e approximants \cite{Grober:2017uho}, or an expansion in the Higgs mass $m_h$ with numerical evaluation of the two-loop integrals \cite{Xu:2018eos}.

Finally, we want to emphasize that if one were to associate an uncertainty to the analytic approach with respect to the numeric one, it would be below 1\% and hence currently completely negligible with respect to other theory uncertainties.

\subsection{Full NLO QCD corrections with \textbf{\texttt{ggxy}}}


\textbf{Joshua Davies, Kay Schönwald, Matthias Steinhauser, Daniel Stremmer~\cite{Davies:2018qvx,Davies:2023vmj,Davies:2025qjr}.}

\noindent An alternative to the numerical approaches described in Sections~\ref{sec:fullPS} and~\ref{sec:nlo2} is to consider series
expansions of the two-loop virtual amplitude in complementary kinematic regions; here we consider in particular both the high-energy and forward-scattering regions.

The
``high-energy'' expansion has been considered in Ref.~\cite{Davies:2018qvx}, which assumes the
scale hierarchy $s,|t| > m_t^2 > m_h^2$ to produce a Taylor series in $m_h$ and a deep
asymptotic expansion in $m_t$. Pad\'e approximants are used to greatly increase
the region of validity of the expansion, ultimately producing a good description
of the amplitude for $p_T$ values above around 150 GeV. The high-energy expansion
also provides the input required by the resummation of \cite{Jaskiewicz:2024xkd}
(see Section~\ref{sec:scet}).
The small-$t$ expansion of \cite{Davies:2023vmj} also performs a Taylor expansion
in $m_h$, followed by an expansion in the forward limit, around $t=0$. This
produces a description for smaller $p_T$ values, below around 150 GeV, similar in
spirit to the small-$p_T$ expansion of \cite{Bonciani:2018omm}.

The combination of both regions
covers the whole phase space~\cite{Davies:2023vmj} and can
thus be used to compute the two-loop
virtual amplitude. The numerical evaluation
is
extremely fast and yet retains full dependence on the input parameters,
in particular the masses of the Higgs boson and top quark, and the ability to change
the top quark renormalization scheme and scale without expensive numerical integration.
The two-loop virtual amplitudes based on this approach are publicly available in the \texttt{C++}
library \texttt{ggxy}~\cite{Davies:2025qjr}, which is able to produce total cross sections and differential distributions in about 30 minutes on a laptop.

An example is shown in Table~\ref{tab::ggxy_kappa}, where we present the total cross sections at NLO QCD accuracy for various
values of $\kappa_\lambda \equiv \lambda_{hhh}/\lambda_{hhh}^{\rm SM}$. We show results for $\sqrt{s}=13$, $13.6$ and $14$~TeV,
and use both the on-shell and $\overline{\rm MS}$ top quark masses. Each value in Table~\ref{tab::ggxy_kappa} is obtained by averaging four seeds, each with a runtime of about $30$ minutes on a single core of an AMD Ryzen Threadripper PRO 3955WX processor, yielding a total runtime of $72$ hours on a single core for the complete table.\footnote{Using the fact that the dependence on $\kappa_\lambda$ is quadratic would of course cut the runtime by half.}
\begin{table}[t!]
    \centering
    \renewcommand{\arraystretch}{1.2}
    \begin{tabular}{cc@{\hskip 7mm}lll}
        \toprule
         \multirow{2}{*}{$\kappa_\lambda$}&\multirow{2}{*}{Top quark mass scheme}
         & \multicolumn{3}{c}{$\sigma^{\rm NLO}_{\rm \tt ggxy}$ [fb]} \\
         \cmidrule(lr){3-5}
         & & $\sqrt{s}=13$~TeV & $\sqrt{s}=13.6$~TeV & $\sqrt{s}=14$~TeV\\
        \midrule
        \multirow{2}{*}{$-5.0$} & on-shell & $ 511.7(4)^{+17.2\%}_{-14.4\%} $ & $ 564.0(4)^{+17.1\%}_{-14.3\%} $ & $ 599.3(5)^{+16.9\%}_{-14.1\%} $ \\
               & $\overline{\rm MS},\,\mu_t=m_{hh}/2$ & $ 521.8(4)^{+16.9\%}_{-14.2\%} $ & $ 574.3(4)^{+16.8\%}_{-14.0\%} $ & $ 609.8(5)^{+16.7\%}_{-13.9\%} $   \\ 
        \midrule
        \multirow{2}{*}{$-0.6$} & on-shell & $ 90.80(7)^{+15.8\%}_{-13.8\%} $ & $ 100.27(7)^{+15.7\%}_{-13.6\%} $ & $ 106.88(8)^{+15.6\%}_{-13.5\%} $ \\
               & $\overline{\rm MS},\,\mu_t=m_{hh}/2$ & $ 89.51(7)^{+16.2\%}_{-13.9\%} $ & $ 98.85(7)^{+16.0\%}_{-13.7\%} $ & $ 105.15(8)^{+15.9\%}_{-13.6\%} $ \\ 
        \midrule
        \multirow{2}{*}{$0.0$} & on-shell & $ 61.54(5)^{+15.3\%}_{-13.5\%} $ & $ 68.14(5)^{+15.2\%}_{-13.4\%} $ & $ 72.60(6)^{+15.0\%}_{-13.3\%} $  \\
              & $\overline{\rm MS},\,\mu_t=m_{hh}/2$ & $ 59.84(5)^{+15.9\%}_{-13.8\%} $ & $ 66.01(5)^{+15.8\%}_{-13.6\%} $ & $ 70.34(6)^{+15.7\%}_{-13.5\%} $ \\ 
        \midrule
        \multirow{2}{*}{$1.0$} & on-shell & $ 27.80(3)^{+13.9\%}_{-12.9\%} $ & $ 30.85(3)^{+13.7\%}_{-12.7\%} $ & $ 32.93(3)^{+13.6\%}_{-12.6\%} $  \\
              & $\overline{\rm MS},\,\mu_t=m_{hh}/2$ & $ 25.91(2)^{+15.4\%}_{-13.6\%} $ & $ 28.65(3)^{+15.2\%}_{-13.5\%} $ & $ 30.59(3)^{+15.2\%}_{-13.4\%} $ \\ 
        \midrule
        \multirow{2}{*}{$2.4$} & on-shell & $ 12.053(9)^{+14.8\%}_{-13.3\%} $ & $ 13.381(9)^{+14.7\%}_{-13.2\%} $ & $ 14.26(1)^{+14.6\%}_{-13.1\%} $  \\
              & $\overline{\rm MS},\,\mu_t=m_{hh}/2$ & $ 11.056(8)^{+17.6\%}_{-14.7\%} $ & $ 12.240(9)^{+17.4\%}_{-14.6\%} $ & $ 13.069(9)^{+17.3\%}_{-14.5\%} $  \\ 
        \midrule
        \multirow{2}{*}{$5.0$} & on-shell & $ 80.38(6)^{+19.5\%}_{-15.4\%} $ & $ 88.14(6)^{+19.3\%}_{-15.3\%} $ & $ 93.60(7)^{+19.2\%}_{-15.2\%} $ \\
              & $\overline{\rm MS},\,\mu_t=m_{hh}/2$ & $ 84.67(6)^{+18.7\%}_{-15.0\%} $ & $ 92.93(7)^{+18.5\%}_{-14.9\%} $ & $ 98.69(7)^{+18.5\%}_{-14.8\%} $ \\
        \bottomrule
    \end{tabular}
    \caption{Total cross sections for $\sqrt{s}=13, 13.6, 14~\textup{TeV}$ in the on-shell and $\overline{\rm MS}$ top quark mass schemes. Results are obtained with $m_t=172.5$~GeV, $m_h=125~\textup{GeV}$ and $\mu_R=\mu_F=m_{hh}/2$ with the NLO NNPDF3.1 \cite{NNPDF:2017mvq} PDF set.}   \label{tab::ggxy_kappa}
\end{table}

\begin{figure}[t]
  \begin{center}
  \begin{tabular}{cc}
     \includegraphics[width=0.45\textwidth]{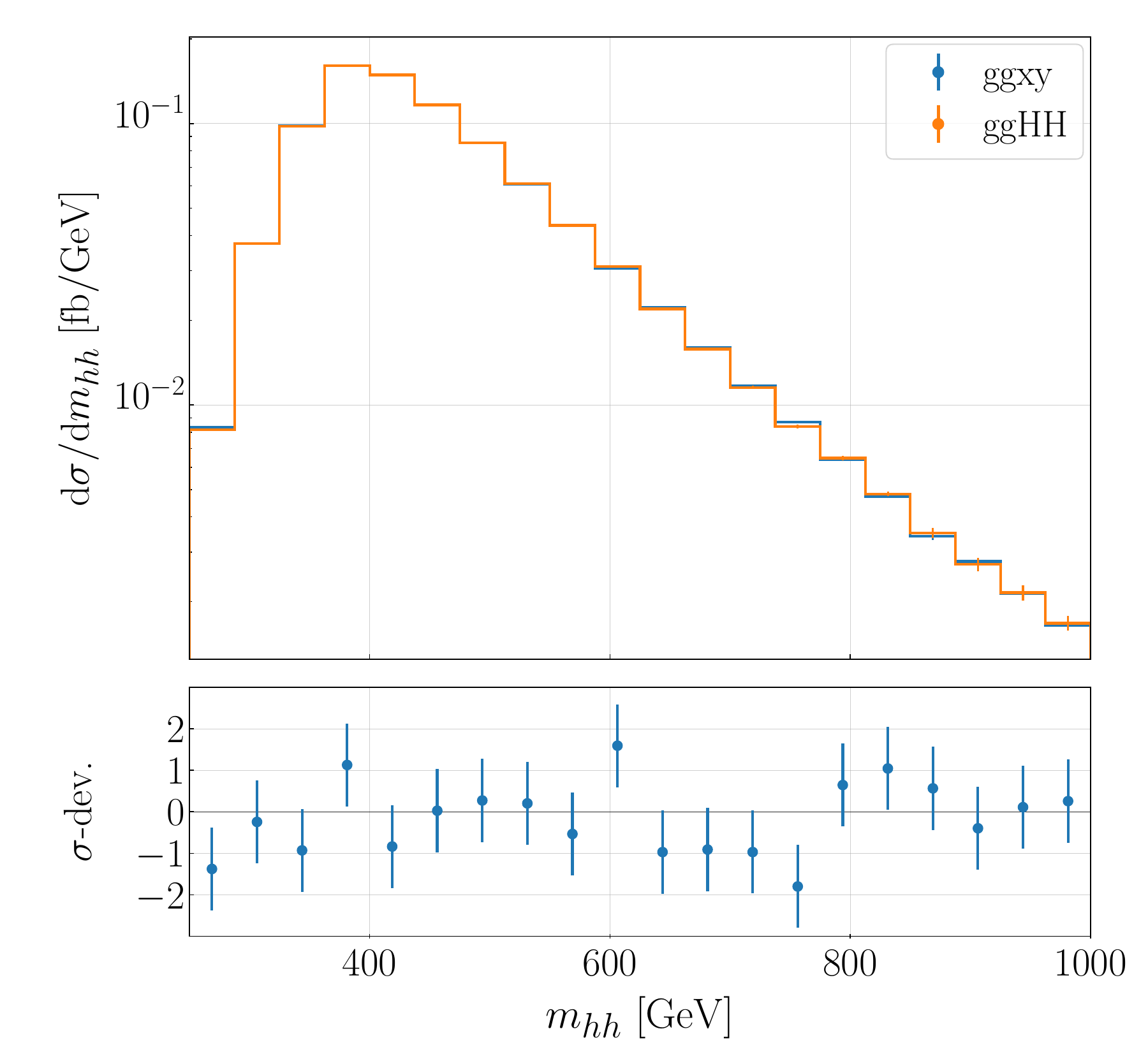}
     \includegraphics[width=0.45\textwidth]{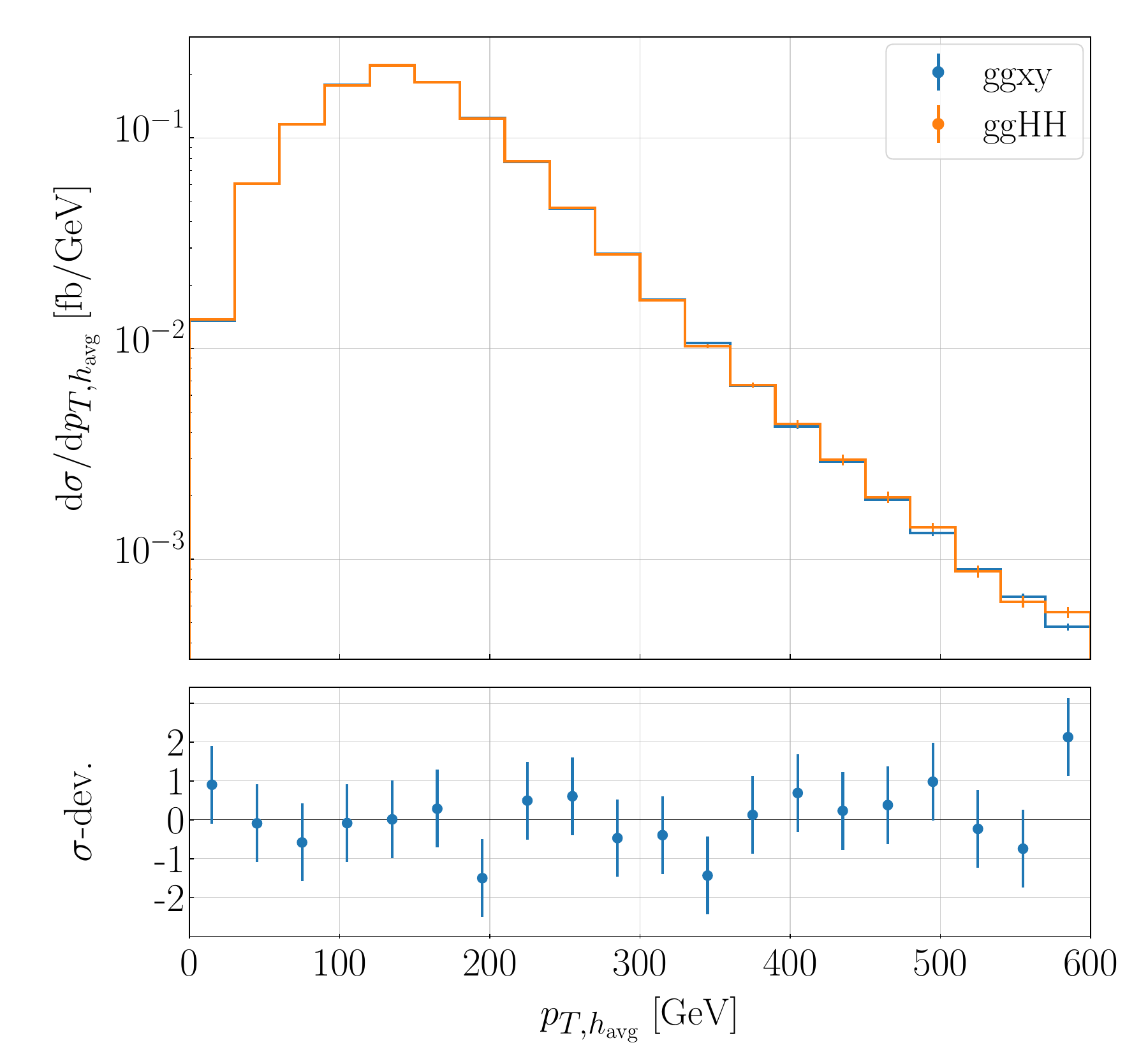}
  \end{tabular}
  \end{center}
  \caption{\label{fig::ggxy_nlo} Comparison of
   NLO predictions generated with
   \texttt{ggxy\_ggHH} and \texttt{ggHH}
   for $m_{hh}$ and $p_{T,h_{\rm avg}}$ distributions for $\sqrt{s}=14~\textup{TeV}$ in the SM. The lower panels display the sigma deviations with respect to the MC uncertainties as explained in the text. Results shown are from Ref.~\cite{Davies:2025qjr}.}
\end{figure}

\texttt{ggxy} has been linked to \texttt{Powheg}~\cite{Alioli:2010xd} (\texttt{ggxy\_ggHH}) to enable the matching to parton showers. Runtime tests have shown that
it is faster by a factor of $4$--$5$ than the previous implementation ({\tt ggHH}),
mainly due to the fast numerical evaluation of the one-loop $2\to 3$ processes for the real corrections with {\tt Recola}~\cite{Actis:2012qn,Actis:2016mpe}. Both \texttt{Powheg} implementations yield identical results for the fixed values of $m_t$ and $m_h$ in {\tt ggHH} as demonstrated in Fig.~\ref{fig::ggxy_nlo}, where we present the Higgs boson pair invariant mass distribution and the average Higgs boson transverse momentum distribution calculated with both tools without parton-shower effects. The lower panels display the total difference of the two implementations,
divided by the squared mean of their MC errors.
It therefore represents the sigma deviations with respect to the statistical
uncertainties coming from the MC integration.
The blue error bars indicate the one-sigma region.

These results will substantially improve the efficiency of future studies that require NLO differential distributions, particularly when studying top quark mass scheme uncertainties and
the high-$p_T$ phase space region.
The results obtained with \texttt{ggxy} agree with those obtained in the SM in
the on-shell (OS) scheme
with the codes presented in Sections~\ref{sec:fullPS} and~\ref{sec:nlo2}.
Furthermore, in the $\overline{\rm MS}$ scheme the output
of \texttt{ggxy} has been compared against the approach of
Section~\ref{sec:expGPL}.
Therefore \texttt{ggxy} constitutes a fast, flexible, and reliable tool to compute NLO QCD
corrections to Higgs boson pair production.

\section{Approximate NNLO QCD corrections} 
\label{sec:nnlo-qcd}

\subsection[The \boldmath NNLO FTapprox combination]{The \boldmath $\textup{NNLO}_{\textup{FTapprox}}$ combination}
\label{sec:NNLO_FTapprox}


\textbf{Massimiliano Grazzini, Gudrun Heinrich, Stephen Jones, Stefan Kallweit, Mat\-thias Kerner, Jonas M. Lindert, Javier Mazzitelli~\cite{Grazzini:2018bsd}.}


\newcommand{\nnloBP}{NNLO$_{\mathrm{B-proj}}$}
\newcommand{\nnloNI}{NNLO$_{\mathrm{NLO-i}}$}
\newcommand{\nnloFT}{NNLO$_{\mathrm{FTapprox}}$}
\newcommand{\nloFT}{$\mathrm{NLO}_{\mathrm{FTapprox}}$}

\noindent QCD corrections to $HH$ production beyond NLO have so far only been computed in approximate form since the relevant two- and three-loop amplitudes for a full NNLO calculation are not yet available. In Ref.~\cite{Grazzini:2018bsd} an approximate NNLO result, dubbed \nnloFT, was presented. In this approach, the exact NLO contribution of Ref.~\cite{Borowka:2016ehy} is incorporated. The approximation of the NNLO correction was constructed starting from the observation that the double-real emission contribution to the NNLO cross section requires only one-loop amplitudes, and can thus be computed exactly with full top quark mass dependence, for example, by using OpenLoops \cite{Cascioli:2011va,Buccioni:2019sur}. However, the approximation of the higher-loop amplitudes entering the NNLO corrections must be done with care, in order not to jeopardize the cancellation of IR singularities.

The approximation is defined as follows. Working in the HTL for each $n$-loop squared amplitude that needs to be computed for a given partonic subprocess, ${\cal A}^{(n)}_{\rm HTL}$, the reweighting factor ${\cal A}_{\rm Full}^{\rm Born}/{\cal A}_{\rm HTL}^{\rm (0)}$ is applied. Here ${\cal A}_{\rm Full}^{\rm Born}$ stands for the exact lowest order (loop-induced) squared amplitude for the corresponding subprocess. The NNLO corrections computed in this way are expected to provide a better estimate of the full result compared to estimates based on a naive reweighting procedure. The impact of NNLO QCD corrections computed in this way ranges from $+11\%$ to $+7\%$ as the collider energy ranges from $13$ to $100$\,TeV \cite{Grazzini:2018bsd}.

Once the employed approximation is defined, the important next question is the size of the remaining uncertainty due to finite top quark mass effects that are still missing compared to an NNLO (3-loop) calculation with full top quark mass dependence.

As a first step to define an uncertainty, we can check the quality of the analogous \nloFT\ approximation, where one can compare to an exact result. At $14$\,TeV the difference between \nloFT\ and the exact NLO cross section is about $4\%$, or, considering only the higher-order corrections, the relative difference between the \nloFT\ corrections and their exact calculation is about $11\%$, which can be translated into a lower bound on the uncertainty of the NNLO cross section. Considering the smaller impact of NNLO with respect to NLO corrections, this translates into a $1.2\%$ effect at NNLO. To be conservative, one can multiply this estimate by a factor of two, thereby obtaining an uncertainty that ranges from $\pm 2.3\%$ at $\sqrt{s}=13$\,TeV to $\pm 3.1\%$ at $\sqrt{s}=100$\,TeV.
However, the study of Ref.~\cite{Grazzini:2018bsd} shows that the difference between the ``best'' prediction, \nnloFT, and the prediction obtained through a simple reweighting of the $HH$ invariant-mass distribution, ${\rm NNLO}_{\rm NLO-I}$ \cite{Borowka:2016ypz}, increases with the energy. Therefore, the uncertainty of the \nnloFT\ prediction is finally defined by taking the half difference of the reference \nnloFT\ result with the ${\rm NNLO}_{\rm NLO-I}$ result. This leads to a final uncertainty due to missing finite top quark mass effects that ranges from $\pm 2.6\%$ at $\sqrt{s}=13$ TeV to $\pm 4.6\%$ at $\sqrt{s}=100$ TeV, and it is therefore slightly more conservative, especially at very high centre-of-mass energies.

Due to the inclusion of the maximal amount of top quark mass dependence in all ingredients, i.e.\ except for those virtual diagrams which are not yet available, the \nnloFT\ prediction had been the recommendation of the LHC Higgs Working Group up to 2026.

\subsection{Matching to parton showers with GENEVA}


\textbf{Simone Alioli, Giulia Marinelli, Davide Napoletano~\cite{Alioli:2025xcu}.}

\noindent Given the absence of exact results for the three-loop NNLO massive contributions,
to achieve NNLO accuracy we can employ several approximations.
In particular, we can exploit the similarity to the production of a single Higgs boson
in gluon fusion, and apply a reweighting procedure to the unknown virtual matrix
elements using the massive Born ones.
One such approximation is dubbed \ftapprox discussed in the previous section~\cite{Frederix:2014hta,Maltoni:2014eza,Grazzini:2018bsd}, where both the real-virtual and double-virtual contributions
are reweighted to the Born matrix elements.
In the case of real-virtual corrections, this implies that
\begin{equation}
  V_{1}\left(\Phi_1\right) = V_{1}\left(\Phi_1,m_t\to\infty\right)
  \frac{B_1\left(\Phi_1,m_t\right)}
  {B_1\left(\Phi_1,m_t\to\infty\right)}\,,
\end{equation}
where $\Phi_1$ represents the double Higgs boson plus one parton phase space, $V_1,\, B_1$
are the relative virtual and Born contributions, respectively.

The case of the double-virtual contribution is instead slightly more subtle.
In the \geneva framework, we include this contribution through the matching
to the resummation of the slicing variable employed to perform the fixed-order
calculation,
which in turn is used to evaluate the contribution to the cross section
below the resolution variable cut~\cite{Alioli:2012fc}.
For this purpose, we employ the zero-jettiness~\cite{Stewart:2010tn} ($\mathcal{T}_0$) as a resolution variable.
When $\mathcal{T}_0 \muchless Q$ (with $Q$ the hard scale,
chosen here as the Higgs boson pair invariant mass $m_{hh}$), the cross section
factorizes according to Soft Collinear Effective Theory (SCET).
In this regime,
large logarithms of $\mathcal{T}_0/Q$ are resummed systematically up to
next-to-next-to-leading logarithmic primed (NNLL$^\prime$) accuracy.
At leading power in $\mathcal{T}_0^{\mathrm{cut}} / m_{hh}$, we can write the differential cross section
as
\begin{equation}\label{eq:convolutionmess}
  \frac{\mathrm{d} \sigma^{\rm SCET}}{\mathrm{d} \Phi_0 \, \mathrm{d} \mathcal{T}_0} =
  H_{{\scriptscriptstyle gg \to HH }}(Q^2,\mu) \int B_g(t_a,x_a,\mu) \,
  B_g(t_b,x_b,\mu) \, S_{gg}\left( \mathcal{T}_0 - \frac{t_a+t_b}{Q}, \mu \right) \,
  \mathrm{d} t_a \, \mathrm{d} t_b \,,
\end{equation}
where $H$ stands for the hard function, while $S$ and $B$ stand for the soft and beam functions, respectively.
The resummed formula is then matched to the appropriate fixed-order calculation,
yielding predictions that are valid across the entire phase space.

In this approach, double-virtual contributions are recovered as they are included in the
two-loop hard function coefficients $H^{(2)}$.
However, in order to ensure the cancellation of the slicing variable dependence, we also need
to reweight the $H^{(1)}$ term which, for $\mathcal{T}_0>\mathcal{T}_0^{\mathrm{cut}}$,
needs to match the real-virtual contribution.
Accordingly, the finite hard coefficients $H^{(2)}_{\mathrm{fin}}$ and
$H^{(1)}_{\mathrm{fin}}$ are rescaled as
\begin{equation}
  \label{eq:twoloophard}
  H^{(i)}_{\mathrm{fin}}(\Phi_0) = H^{(i)}_{\mathrm{fin}}\left(\Phi_0,m_t\to\infty\right)
  \frac{B_0\left(\Phi_0,m_t\right)}
  {B_0\left(\Phi_0,m_t\to\infty\right)}\,, \quad i=1,2\,.
\end{equation}
We evaluate all the remaining contributions with exact top quark mass dependence
either through \openloops~\cite{Cascioli:2011va,Buccioni:2017yxi,Buccioni:2019sur}, where available, or \hhgrid~\cite{Borowka:2016ehy,Borowka:2016ypz,Heinrich:2017kxx,Davies:2019dfy}.
However, as noted in~\cite{Grazzini:2018bsd}, the stability of the one-loop double-real
matrix elements $gg \to HH gg$ near the double infrared limit is not sufficient
to ensure their cancellation against the subtraction matrix elements.
To preserve the cancellation with the subtraction terms, and to enhance
the stability of these matrix elements in the deep infrared region, we employ
an approximation constructed as
\begin{equation}
  {\mathrm{d}\sigma_{gg\to hh
      gg}\left(\Phi_2\right)\Biggr|}_{\min(\alpha_{ij})<\alpha_{\mathrm{cut}}}
  = {\mathrm{d}\sigma_{gg\to hh g}\left(\widetilde{\Phi}_1\right)}
  \frac{\mathrm{d}\sigma_{gg\to hhgg}^{m_t\to\infty}\left(\Phi_2\right)}
  {\mathrm{d}\sigma_{gg\to hh g}^{m_t\to\infty}\left(\widetilde{\Phi}_1\right)} \,,
\end{equation}
where $\widetilde{\Phi}_1$ is obtained by applying the
FKS projection mapping~\cite{Frixione:1995ms} to the original
two-parton phase space point $\Phi_2$.
Similarly to Ref.~\cite{Grazzini:2018bsd}, we adopt $\alpha_{\mathrm{cut}} = 10^{-4}$ as our technical cutoff.
The rationale behind this reweighting choice is that, in
the soft-collinear limit, the ratio 
\begin{equation}
\de\sigma_{gg\to hhgg}^{m_t\to\infty}\left(\Phi_2\right)
/{\de\sigma_{gg\to hh g}^{m_t\to\infty}\left(\widetilde{\Phi}_1\right)}
\end{equation} 
tends towards
the splitting function used in the subtraction, thus reproducing the structure of
the subtraction term.
This ensures that the cancellation between the reweighted $gg\to HHgg$ matrix element and its
counterterms, which are always evaluated with the exact top quark mass dependence, is local and
finite in each singular limit when $\alpha_{\mathrm{cut}} \to 0$.

\begin{figure}[H]
\begin{center}
  \includegraphics[width=0.49\textwidth]{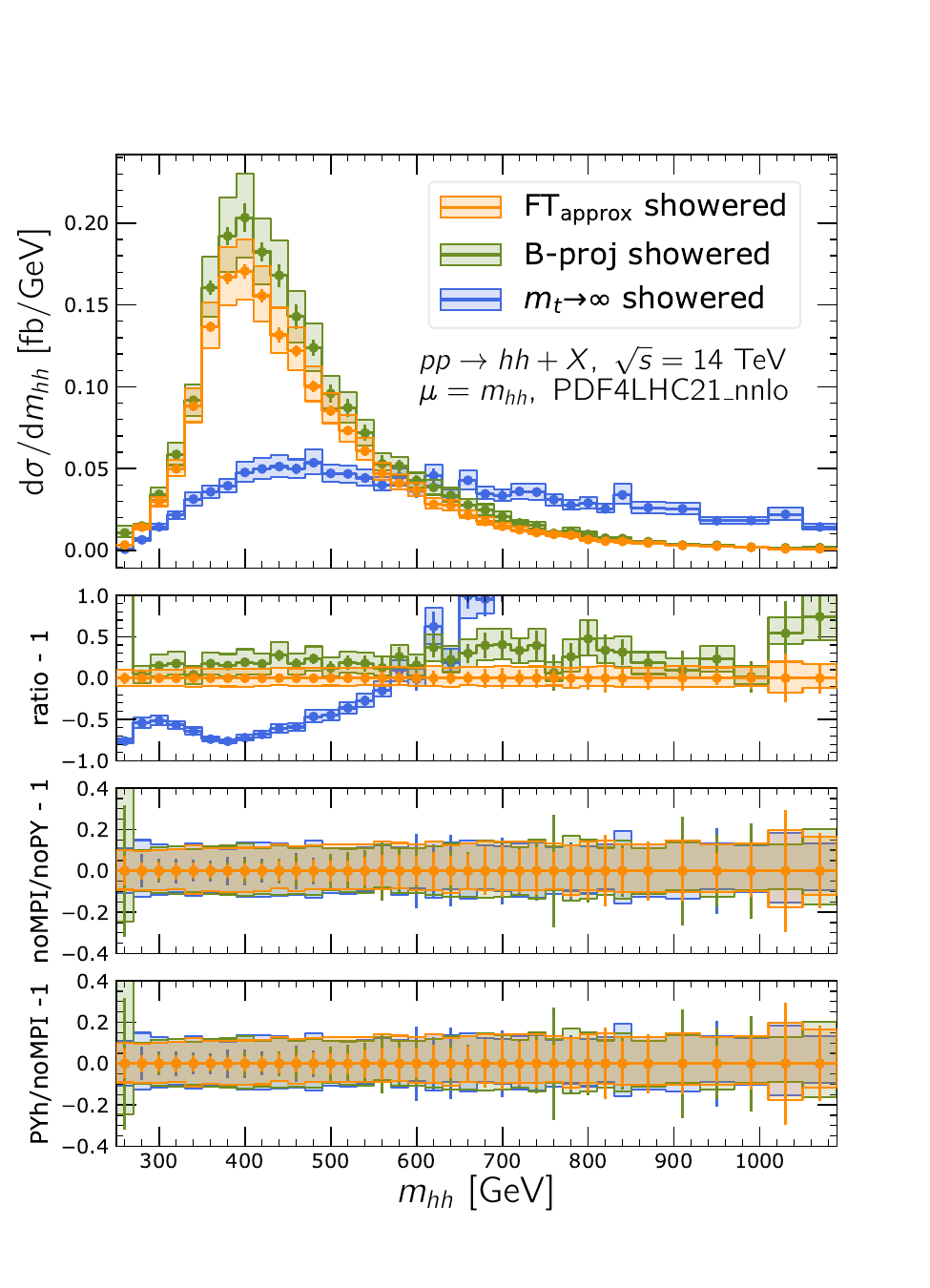} \hfill
  \includegraphics[width=0.49\textwidth]{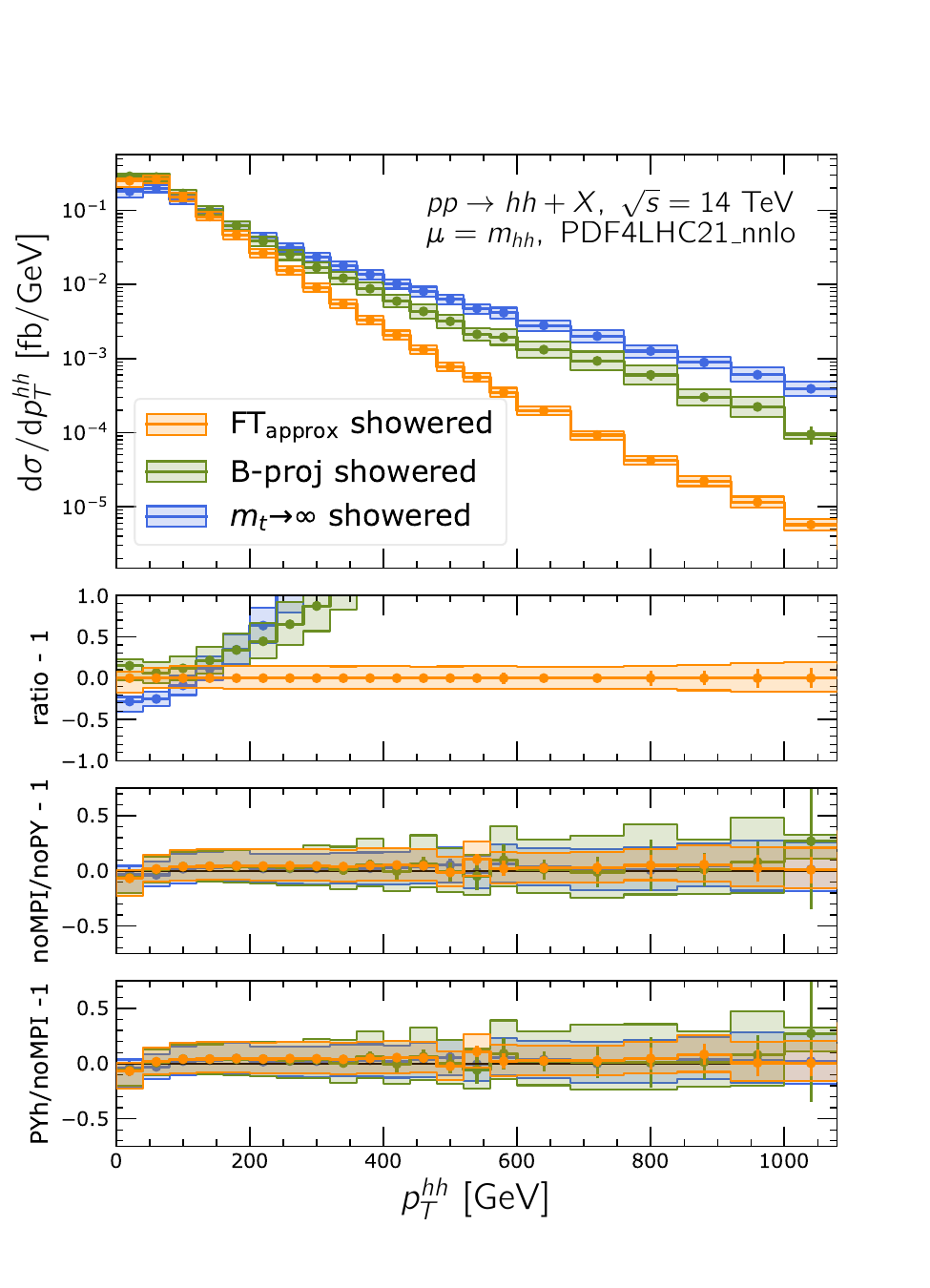}
  \caption{\label{fig:genevaplots}Comparison of showered predictions obtained with the three top quark mass approximations: $m_t \to \infty$
  (blue), B-proj (light green) and \ftapprox (dark orange), shown for various differential distributions in the SM.
  These include the invariant mass of the Higgs boson pair ($m_{hh}$)
  and the transverse momentum of the pair system ($p_T^{hh}$).
  Results are obtained at $\sqrt{s} = 14$ TeV, with central scale $\mu = m_{hh}$,
  using the \texttt{PDF4LHC21\_nnlo} PDF set and 3-point scale variations.
  Results shown are from Ref.~\cite{Alioli:2025xcu}.
}
\end{center}
\end{figure}

Finally, the \geneva results are matched to the \pythiaEight~\cite{Sjostrand:2014zea} shower
following the procedure described in Ref.~\cite{Alioli:2022dkj}.
We present our results for the LHC at 14~TeV centre-of-mass energy in Fig.~\ref{fig:genevaplots}.
We evaluate both factorization and renormalization scales at $\mu=m_{hh}$ and vary them about
their central value by a factor of two simultaneously.
We employ the PDF4LHC21 NNLO~\cite{PDF4LHCWorkingGroup:2022cjn} parton distribution function set, evaluated through
our internal \texttt{LHAPDF6}~\cite{Buckley:2014ana} interface.

We compare our \ftapprox results with other approximations in the literature,
namely the $m_t\to\infty$ limit, corresponding to our previous results of~\cite{Alioli:2022dkj},
and the B-proj approximation described in~\cite{Grazzini:2018bsd}.
The latter corresponds to reweighting all squared matrix elements
-- including those entering the hard function perturbative coefficients -- using
Born-projected matrix elements. Specifically, the reweighting factor is given by
\begin{equation}
  \label{eq:bprojkfact}
  K(\Phi_n) = \frac{B_{gg\to hh}\left(\tilde{\Phi}_0(\Phi_n),m_t\right)}
  {B_{gg\to hh}\left(\tilde{\Phi}_0(\Phi_n),m_t\to\infty\right)}\,.
\end{equation}
This procedure provides the exact top quark mass dependence only at LO.
Starting from NLO and in the presence of additional emissions,
the reweighting factor is evaluated on the projected kinematics $\tilde{\Phi}_0(\Phi_n)$,
defined in Eqs.~(2.2) and~(2.3) of Ref.~\cite{Grazzini:2018bsd} and in Ref.~\cite{Catani:2015vma}.

The structure of the plots in Fig.~\ref{fig:genevaplots} is as follows.
The main panel displays
predictions after the full shower procedure, including hadronization
and multi-parton interaction (MPI) effects, but excluding hadron decays.
The first ratio panel shows the relative difference of each approximation with respect to the
\ftapprox result.
The second ratio panel highlights the impact of the parton shower at parton level (i.e.\ without
hadronization and MPI effects), by displaying the relative difference with respect to the
partonic results before the shower.
The third ratio panel illustrates the impact of hadronization and MPI effects,
shown relative to the parton shower results without these effects.

The inclusion of finite top quark mass effects leads to a significant impact,
much larger than that of the parton shower itself.
For the Higgs boson pair invariant mass this is a consequence of the shower unitarity
with respect to fully inclusive observables and thus follows by construction.
For the transverse momentum of the pair of Higgs bosons, the shower
induces a small effect in the first few bins, of approximately $5\%$ relative to the
partonic predictions.
Finally, by comparing the last two ratio panels, we can observe that the hadronization and MPI
effects are marginal and affect all mass approximations in the same way.

\subsection{Towards NNLO with full top quark mass dependence}


\textbf{Joshua Davies, Kay Schönwald, Matthias Steinhauser, Marco Vitti~\cite{Davies:2023obx,Davies:2024znp,Davies:2025ghl}.}

\noindent In order to increase the precision for the theoretical prediction of double Higgs boson production and tackle the large uncertainties due to the top quark mass renormalization scheme, one needs information about higher-order corrections to the process.
This can be achieved either by resummation of large logarithms in certain kinematic regions, see Section~\ref{sec:scet}, or by calculating the higher-order corrections directly.
There is an ongoing effort to compute the virtual NNLO corrections to double Higgs boson production retaining top quark mass effects beyond the large mass expansion.

\begin{figure}[t]
    \begin{center}
    \includegraphics[width=.24\textwidth]{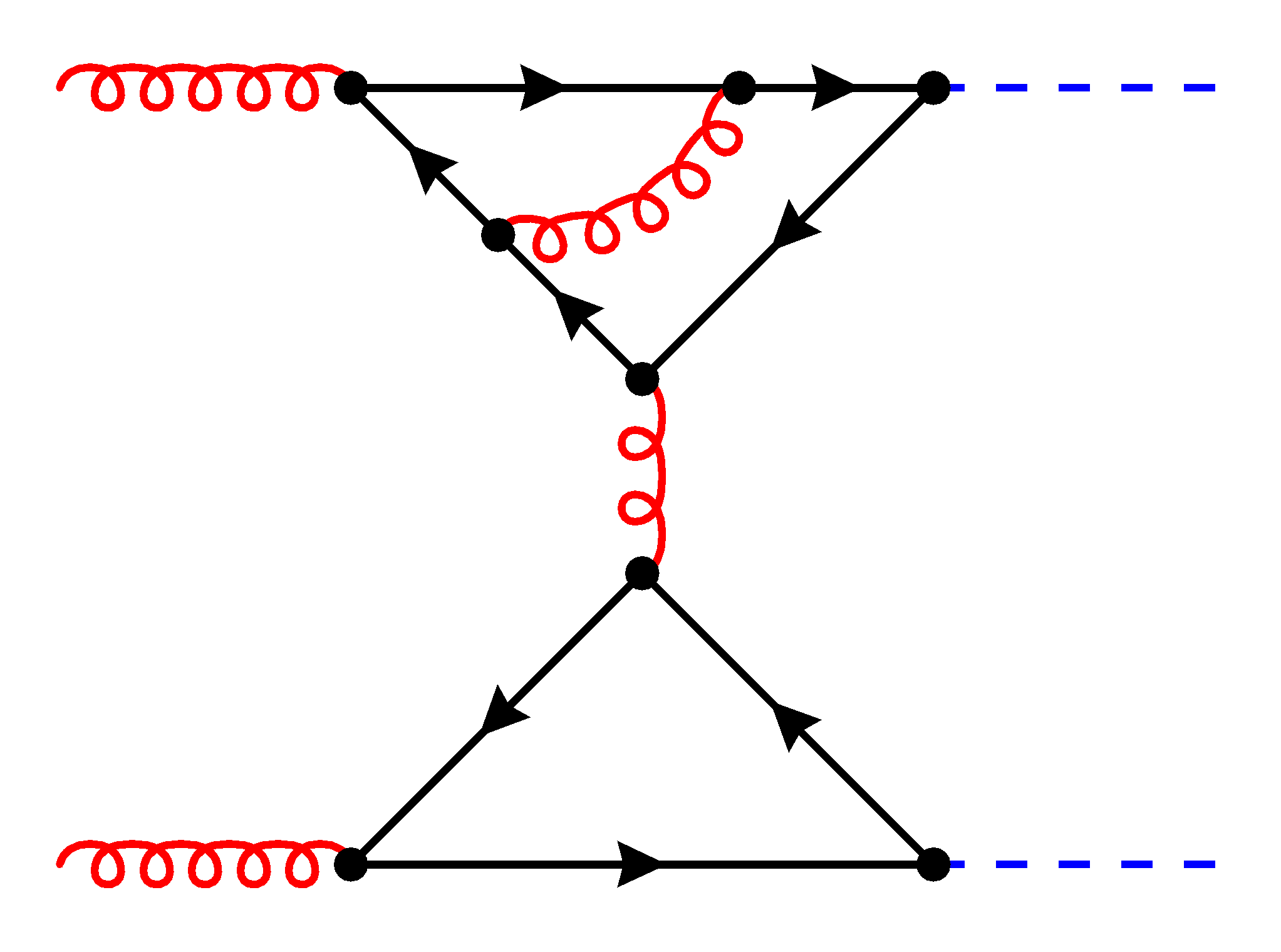}
    \includegraphics[width=.25\textwidth]{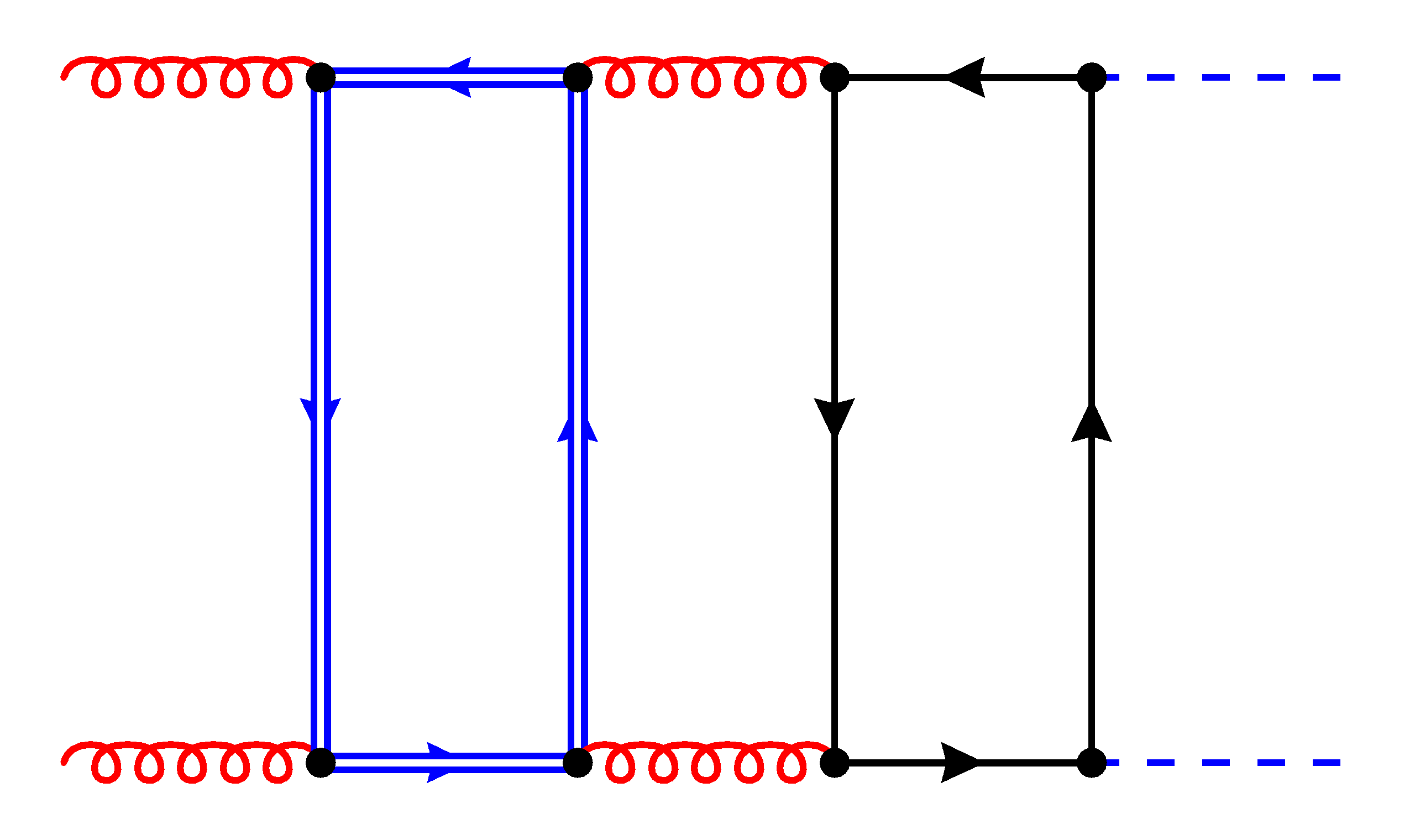}
    \includegraphics[width=.25\textwidth]{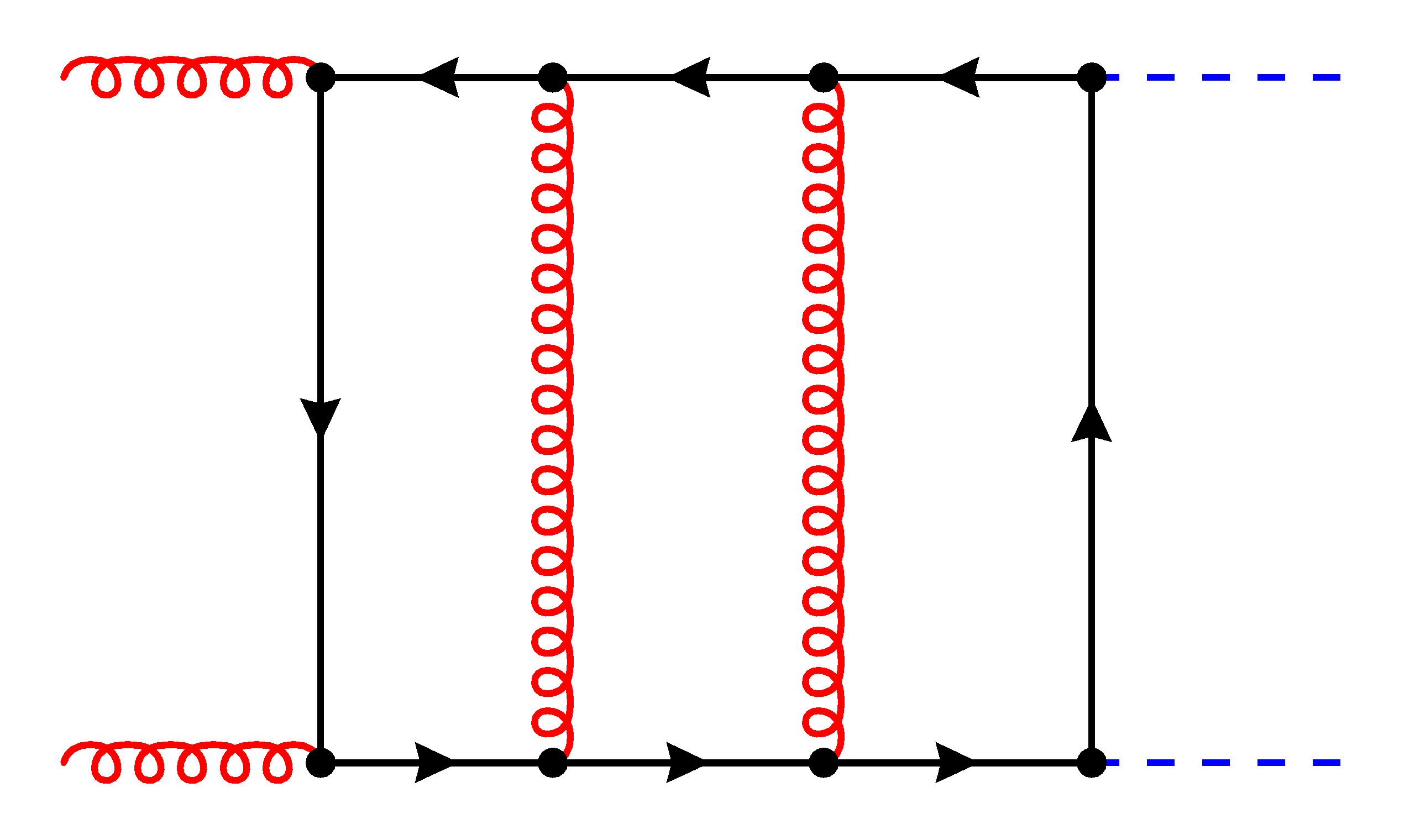}
    \includegraphics[width=.24\textwidth]{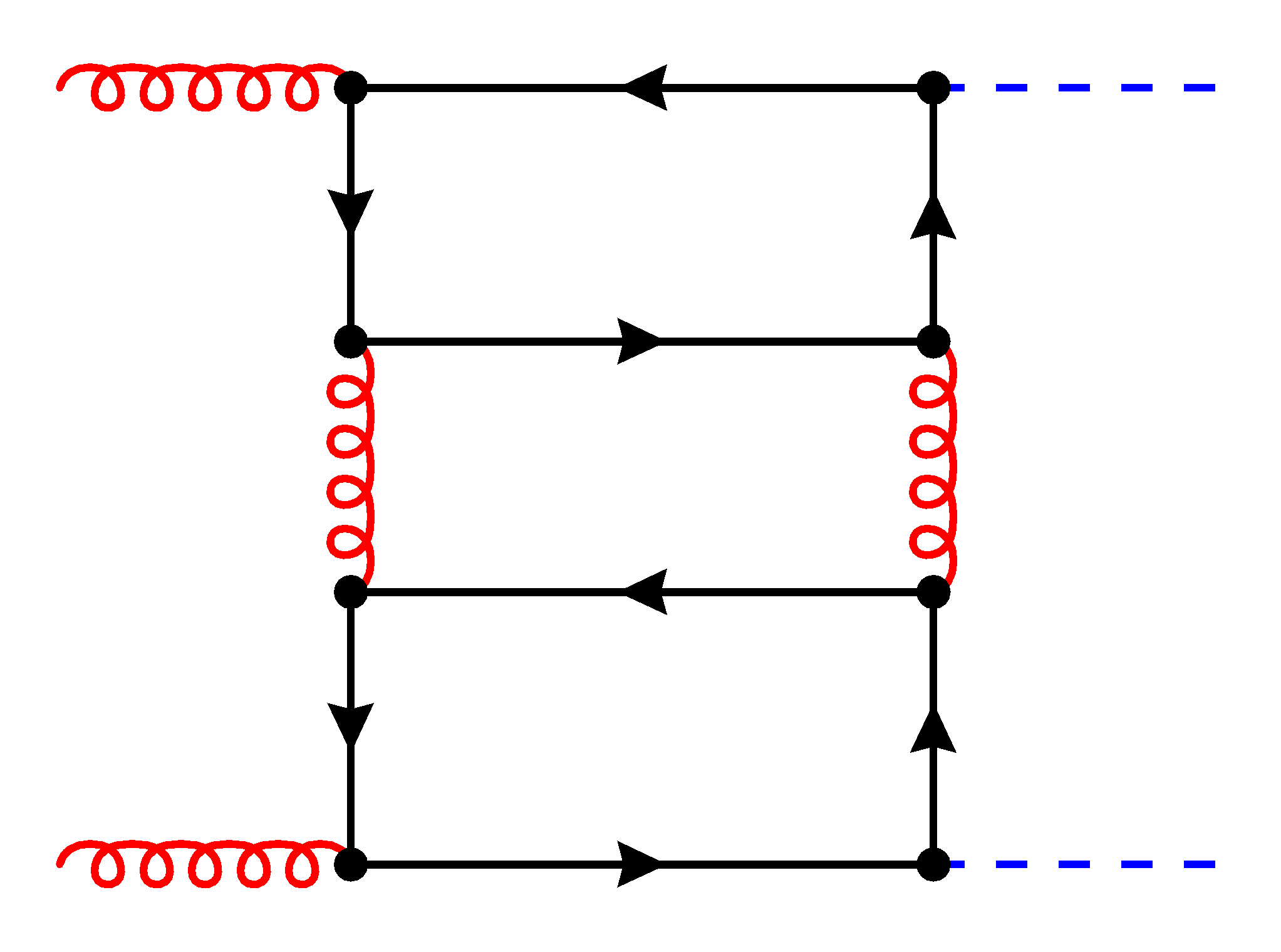}
    (a)
    \hspace{0.2\textwidth}
    (b)
    \hspace{0.2\textwidth}
    (c)
    \hspace{0.2\textwidth}
    (d)
    \end{center}
  \caption{\label{fig::gghh_FDs}Sample Feynman diagrams of contributions to the virtual NNLO corrections.
  The contributions shown represent: (a) reducible contributions,
  (b) contributions with one closed light quark loop,
  (c) contributions with one closed top quark loop,
  (d) contributions with two closed top quark loops.
  Curly (red) lines correspond to gluons,
  single (black) lines to the top quark,
  double (blue) lines to a massless quark
  and dashed (blue) lines to the Higgs boson.
  The diagram classes follow Refs.~\cite{Davies:2023obx,Davies:2024znp,Davies:2025ghl}.
  }
\end{figure}

The computation of the NNLO amplitude can be split into several building blocks of differing complexity, shown in Fig.~\ref{fig::gghh_FDs}.
The one-particle reducible contributions, see Fig.~\ref{fig::gghh_FDs}(a), which are only of one-loop times two-loop complexity, have been calculated in Ref.~\cite{Davies:2024znp} for general kinematics.
The calculation of the genuine three-loop contributions faces technical challenges due to the large number of loops and scales.
They are addressed using an expansion in $p_T$ or, equivalently, $m_h$ and $t$.
The first expansion terms have been obtained for the contributions including a closed light quark loop, see Fig.~\ref{fig::gghh_FDs}(b), and, in the large-$N_c$ limit, for the diagrams involving one closed top quark loop, see Fig.~\ref{fig::gghh_FDs}(c); here $N_c$ stands for the number of colours in QCD.
In these works, amplitude processing, expansion and IBP reduction were performed with {\tt FORM} \cite{Ruijl:2017dtg,Davies:2026cci}, {\tt Feynson} \cite{Maheria:2022dsq}, {\tt Kira} \cite{Lange:2025ofh} and {\tt FIRE} \cite{Smirnov:2025prc}.
The class of diagrams represented in Fig.~\ref{fig::gghh_FDs}(d), which involves two top quark loops, faces the additional challenge of massless cuts in the $t$-channel or isolating an external Higgs boson and therefore requires a more involved asymptotic expansion.
They are not yet known, but remain work in progress.

\begin{figure}[H]
\begin{center}
  \begin{tabular}{cc}
    \includegraphics[width=0.45\textwidth]{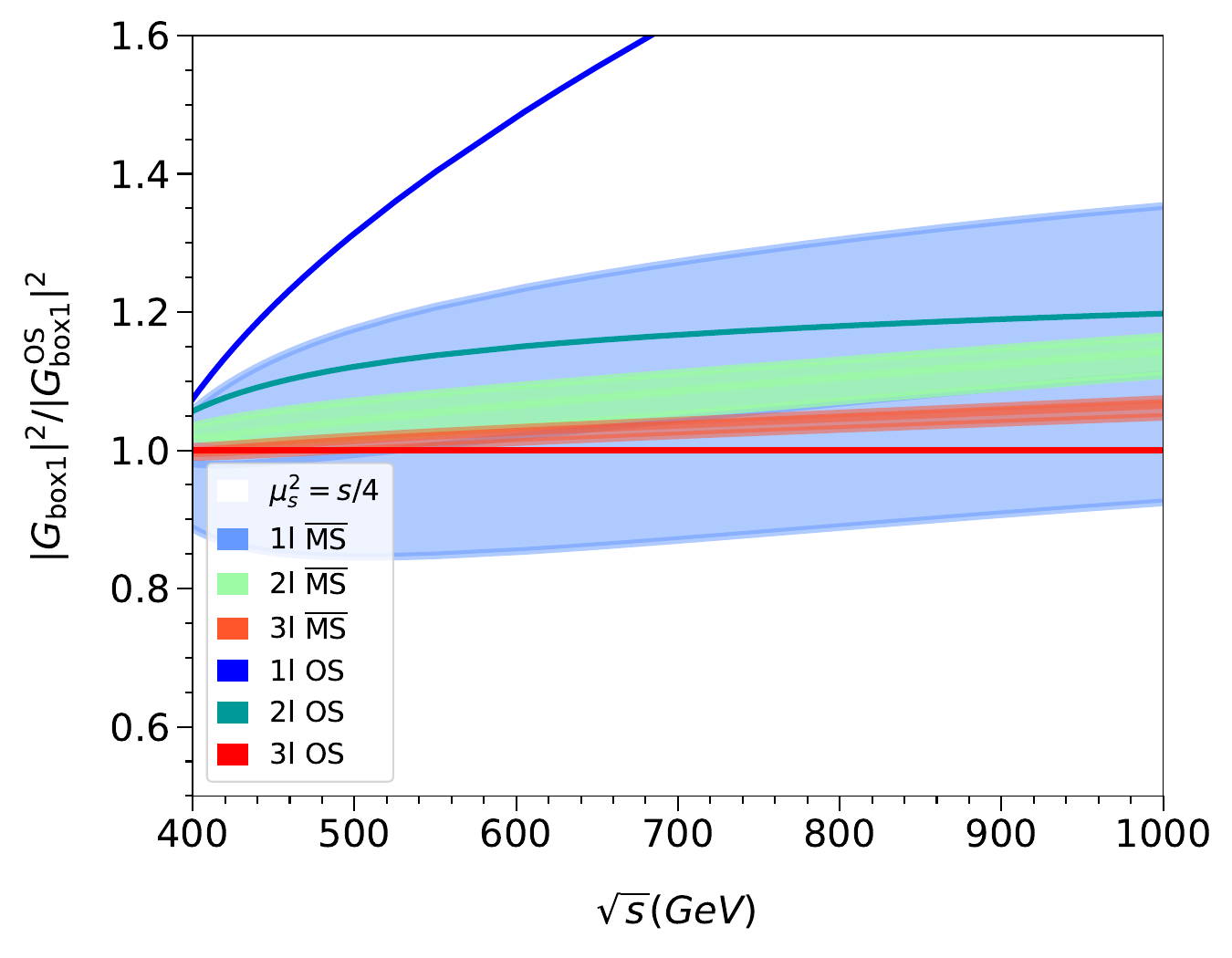}
    &
    \includegraphics[width=0.45\textwidth]{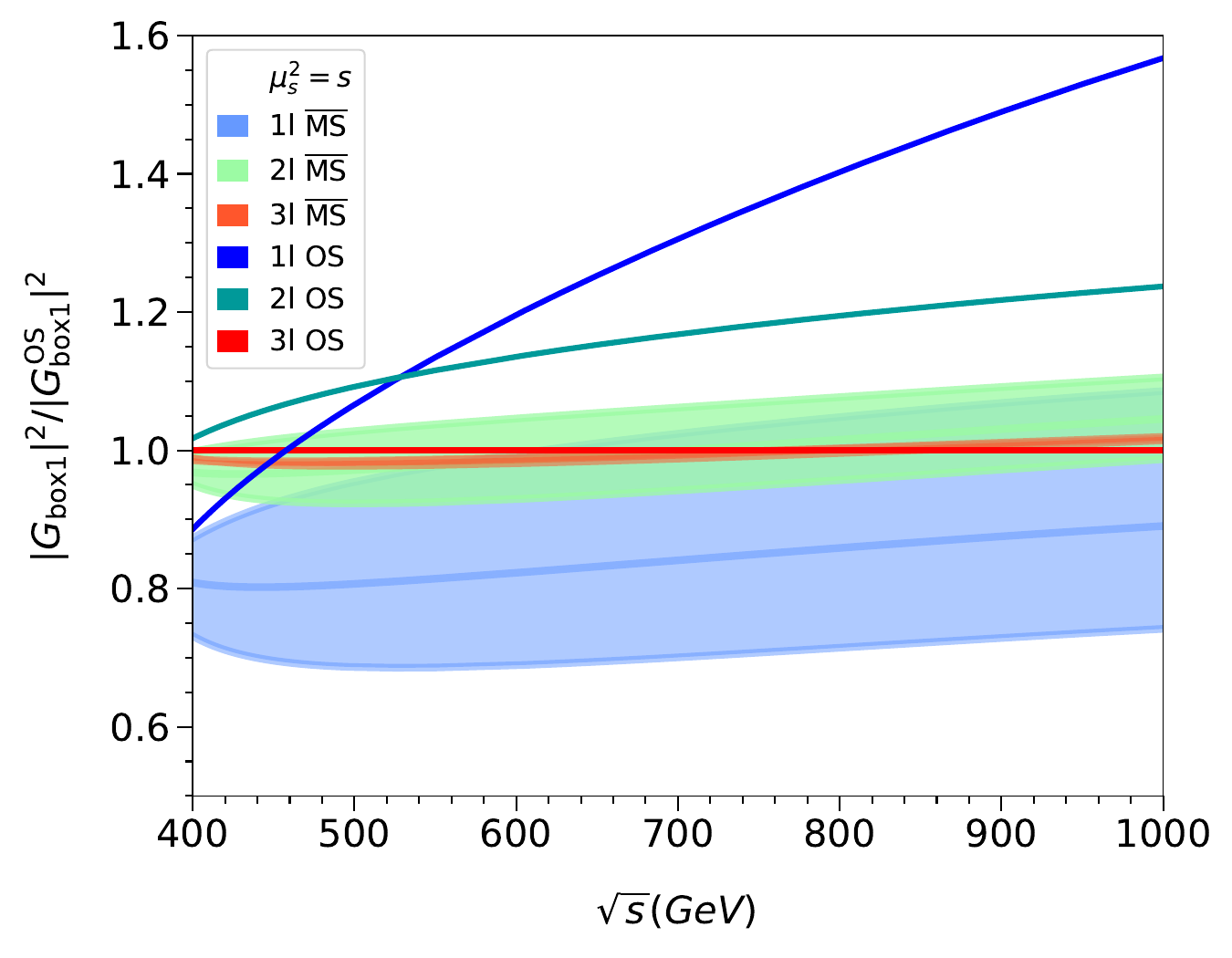}
  \end{tabular}
\end{center}
  \caption{\label{fig::G1_ratio} $G_{\rm box1}$
  computed with the $\overline{\rm MS}$ and on-shell
  definition of the top quark mass for
    $\mu_s^2=s/4$ (left) and $\mu_s^2=s$ (right). $\mu_t^2$ is chosen between $s$ and $s/16$. All curves are normalized to the three-loop on-shell result. Results shown are from Ref.~\cite{Davies:2025ghl}.}
\end{figure}

At LO and NLO the first expansion term in the forward limit provides an approximation with an accuracy at the 20\% level for $p_T < 300 \,$GeV (see, e.g., Ref.~\cite{Davies:2025ghl}).
Therefore, one can use these results to study the dependence on the top quark mass renormalization scheme at NNLO.
In Fig.~\ref{fig::G1_ratio} we show the quantity
$|G_{\rm box1}|^2$, where $G_{\rm box1}$ is the box contribution of the properly normalized form factor $F_1$
(see Ref.~\cite{Davies:2025ghl} for details)
normalized to the three-loop result in the on-shell scheme as a function
of partonic $\sqrt{s}$. The bands for the $\overline{\rm MS}$ results arise from the variation of the renormalization scale
associated with the top quark mass. The two panels correspond to two different choices of $\mu_s$, the renormalization scale of $\alpha_s$.
One observes a good convergence of the perturbative expansion
in both renormalization schemes.
One also observes smaller perturbative corrections when using the $\overline{\mathrm{MS}}$ scheme for the top quark mass and a shrinking of the scale uncertainty bands. Furthermore, the
NNLO predictions are closer together than the corresponding curves at lower orders. All of these are very encouraging features.

A full NNLO analysis also needs the top quark mass dependence of the real radiation amplitudes.
While the one-loop amplitudes with two additional real emissions can be obtained with automated tools, the two-loop amplitudes with one additional real emission are not known yet and constitute another computational challenge.

\section[\texorpdfstring{$\textup{N}^3\textup{LO}$ QCD corrections and resummation}{N3LO QCD corrections and resummation}]{\texorpdfstring{\boldmath$\textup{N}^3\textup{LO}$ QCD corrections and resummation}{N3LO QCD corrections and resummation}}
\label{sec:n3lo-qcd}

This section combines results from several publications. In the HTL, the N$^3$LO QCD corrections to the inclusive total cross section and the Higgs boson pair invariant-mass ($m_{hh}$) distribution were first reported in Refs.~\cite{Chen:2019lzz, Chen:2019fhs}. The N$^3$LL soft-gluon threshold resummation for the same observables was presented in Ref.~\cite{Ajjath:2022kpv}. More recently, fully differential N$^3$LO QCD calculations were presented in Ref.~\cite{Chen:2026zmi}.

In the HTL, the effective Lagrangian for the
coupling of the Higgs field to two gluon field strength tensors
can be written as
\begin{align}\label{eq:effL}
 \mathcal{L}_{\rm eff}= -\frac{1}{4}  G_{\mu\nu}^a G^{a~\mu\nu}
 \left(
  C_{h} \frac{h}{v} - C_{hh} \frac{h^2}{2v^2}
 \right)\,,
\end{align}
where $v$ is the Higgs vacuum expectation value and
$C_h$, $C_{hh}$ are the Wilson coefficients obtained by matching the full Standard Model (SM) onto the effective theory, in which the top quark degree of freedom has been integrated out.
The Wilson coefficients start at $\mathcal{O}(\alpha_s)$ and are known up to $\mathcal{O}(\alpha_s^4)$~\cite{Inami:1982xt, Chetyrkin:1997iv, Chetyrkin:1997un, Schroder:2005hy, Chetyrkin:2005ia, Kniehl:2006bg,Baikov:2016tgj,Spira:2016zna,Gerlach:2018hen}.

In this section, we present our theoretical predictions using the following setup~\cite{Chen:2019lzz,Chen:2019fhs,Ajjath:2022kpv,Chen:2026zmi}. We take the Higgs-boson mass $m_h=125$ GeV, the top quark pole mass $m_t=173$ GeV, and the Higgs vacuum expectation value $v=246.2$ GeV. We employ either the \texttt{PDF4LHC15\ttus nnlo\ttus 30}~\cite{Butterworth:2015oua,Dulat:2015mca,Harland-Lang:2014zoa,NNPDF:2014otw} or the \texttt{PDF4LHC21\ttus 40}~\cite{PDF4LHCWorkingGroup:2022cjn}
(\texttt{NNPDF40\ttus an3lo\ttus as\ttus 01180}~\cite{NNPDF:2024nan}) parton distribution function (PDF) sets for observables without and with fiducial cuts, respectively, together with the corresponding $\alpha_s$ running provided by {\tt LHAPDF6}~\cite{Buckley:2014ana}.\footnote{We note here that the N$^3$LO $k$-factors in the final recommendations of Section~\ref{sec:recs} were produced using the \texttt{PDF4LHC21\ttus 40} set.}
The central renormalization and factorization scales are chosen as $\mu_0 = m_{hh}/2$. Scale uncertainties, which estimate missing higher-order corrections, are obtained from the envelope of the $7$-point variations of
$\mu_F$ and $\mu_R$, defined as $\mu_{R/F} = \xi_{R/F} \mu_0$ with $\xi_{R/F}\in \left\{0.5,1,2\right\}$, excluding the two extreme combinations $(\xi_R, \xi_F)=(0.5, 2)$ and $(2, 0.5)$. We quote only the scale uncertainty, since other parametric uncertainties are largely independent of the perturbative order.
For the fully differential results, we impose the following fiducial cuts~\cite{Chen:2026zmi}:
\begin{equation}
\label{eq:fiducialcuts}
p_{T, h_{1}}>30~\mathrm{GeV}\,, \quad p_{T, h_{2}}>20~\mathrm{GeV}\,, \quad |y_{h}|<2.4\,,
\end{equation}
where $h_1$ and $h_2$ denote the leading and subleading Higgs bosons, ordered by their transverse momenta $p_{T,h_i}$.

\subsection[\texorpdfstring{$\textup{N}^3\textup{LO}$ in the HTL}{N3LO in the HTL}]{\texorpdfstring{\boldmath$\textup{N}^3\textup{LO}$ in the HTL}{N3LO in the HTL}}


\textbf{Long-Bin Chen, Xuan Chen, Yuesheng Dai, Hai Tao Li, Shi-Yuan Li, Hua-Sheng Shao, Jian Wang~\cite{Chen:2019lzz,Chen:2026zmi}.}

\begin{figure}[!hbtp]
    \centering
    \includegraphics[width=\textwidth]{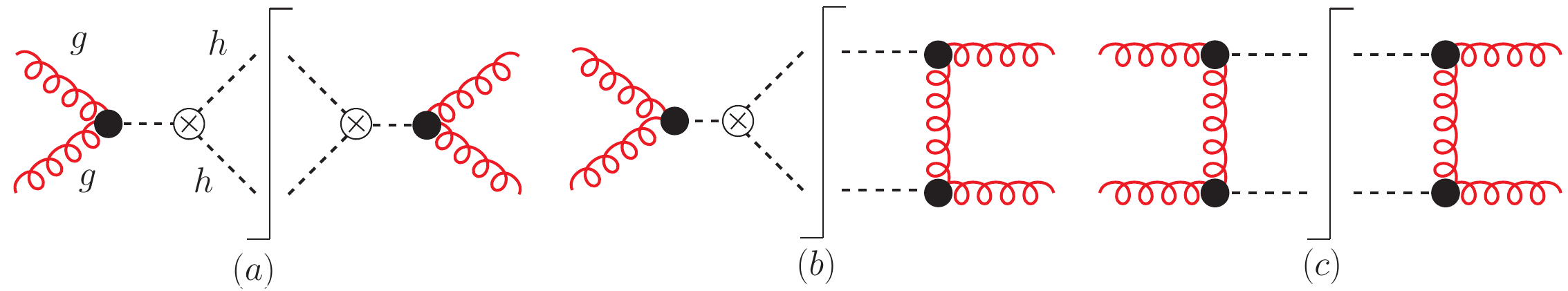}
    \caption{Representative Born-level cut diagrams for Higgs boson pair production via gluon-gluon fusion. Bullets denote the effective vertices described by the Lagrangian in Eq.~\eqref{eq:effL}, while the crossed circle represents the trilinear Higgs boson self-coupling. Figure adapted from Ref.~\cite{Chen:2019lzz}.}
    \label{fig:FeynDia}
\end{figure}

\noindent The fixed-order computation of the gluon-gluon fusion (ggF) Higgs boson pair cross section in the HTL can be organized according to the number of the effective-vertex insertions in the squared amplitude. Three representative Born-level cut diagrams are shown in Fig.~\ref{fig:FeynDia}: from left to right, they involve two, three, and four effective vertices, and are denoted as class-$a$, class-$b$, and class-$c$ contributions, respectively. Accordingly, the phase-space-integrated or differential cross section can be decomposed as
\begin{align}
    d\sigma_{hh} = d\sigma^{a}_{hh} + d\sigma^{b}_{hh} + d\sigma^{c}_{hh} .
\end{align}
Their contributions at different powers of $\alpha_s$ are summarized in Table~\ref{tab:my_label}. To achieve N$^3$LO accuracy in $\alpha_s$, the required inputs are the class-$a$ contribution at $\mathrm{N^3LO}_a$, the class-$b$ contribution at $\mathrm{NNLO}_b$, and the class-$c$ contribution at $\mathrm{NLO}_c$, where the subscripts indicate the perturbative order within the corresponding topology class.

\begin{table}[hbt!]
    \centering
\begin{tabular}{ccccc}
    \toprule
        \multirow{2}{*}{Class}    &  LO  & NLO  & NNLO  & $\textup{N}^3\textup{LO}$  \\
      &  $\mathcal{O}(\alpha_s^2)$  &   $\mathcal{O}(\alpha_s^3)$ &   $\mathcal{O}(\alpha_s^4)$  & $\mathcal{O}(\alpha_s^5)$
     \\
    \midrule
    $a$ &  $\textup{LO}_a$  &   $\textup{NLO}_a$ &   $\textup{NNLO}_a$  & $\textup{N}^3\textup{LO}_a$
     \\
     $b$ &  --- &   $\textup{LO}_b$ &   $\textup{NLO}_b$  &  $\textup{NNLO}_b$
     \\
    $c$ & --- &  --- &   $\textup{LO}_c$  & $\textup{NLO}_c$
     \\
    \bottomrule
    \end{tabular}
            \caption{Perturbative orders in $\alpha_s$ for three topology classes.
    The first and second rows indicate the perturbative expansion orders of the total cross section for Higgs boson pair production,
    while each individual class has its own expansion order, specified by the subscript.}
    \label{tab:my_label}
\end{table}

The class-$a$ contribution can be obtained from the single-Higgs production cross section in the HTL up to N$^3$LO.
For inclusive results, this component is computed using the public program \texttt{iHixs2}~\cite{Dulat:2018rbf}, while the class-$a$ contribution for fully differential results is computed with \texttt{NNLOJet}~\cite{NNLOJET:2025rno} using the $q_T$ phase-space slicing ($q_T$-slicing) method~\cite{Catani:2007vq,Cieri:2018oms}.
The class-$b$ contribution is calculated up to NNLO in $\alpha_s$ using the $q_T$-slicing method~\cite{Catani:2007vq} by combining the \texttt{MadGraph5\_aMC@NLO}~\cite{Alwall:2014hca,Frederix:2018nkq} framework with a private code. The corresponding two-loop amplitude with two effective-vertex insertions was computed in Ref.~\cite{Banerjee:2018lfq}. The double-real and real-virtual contributions in class-$b$, as well as all contributions in class-$c$, are calculated using \texttt{MadGraph5\_aMC@NLO}.

\begin{table}[h]
\begin{center}

\newcolumntype{P}[1]{>{\centering\arraybackslash}p{#1}}
{\renewcommand{\arraystretch}{1.5}
\begin{tabular}{P{2.7cm}P{2.7cm}P{2.7cm}P{2.7cm}}
\toprule
    & $\sigma_{{\rm NLO}}$ $[\textup{fb}]$
    & $\sigma_{{\rm NNLO}}$ $[\textup{fb}]$
    & $\sigma_{{\rm N}^3{\rm LO}}$ $[\textup{fb}]$
    \\
  \midrule
  Inclusive &
  $31.89^{+18\%}_{-15\%}$ &
  $37.55^{+5.2\%}_{-7.6\%}$ &
 $38.65^{+0.5\%}_{-2.7\%}$
 \\
 \midrule
   Fiducial &
  $27.87_{-15 \%}^{+18 \%}$ &
  $32.74_{-7.3 \%}^{+5.2 \%}$ &
 $34.00(4)_{-2.9 \%}^{+1.4 \%}$
 \\
 \bottomrule
\end{tabular}}
\caption{Fixed-order cross sections in the HTL at $\sqrt{s}=14\,\textup{TeV}$. The second line corresponds to the inclusive case, while the third line corresponds to the fiducial case with cuts defined in Eq.~\eqref{eq:fiducialcuts}. The quoted uncertainties are obtained from $7$-point scale variations.}
\label{tab:FON3LOinclusivexs}
\end{center}
\end{table}

\begin{figure*}[t]
\centering
\includegraphics[width=0.49\textwidth]{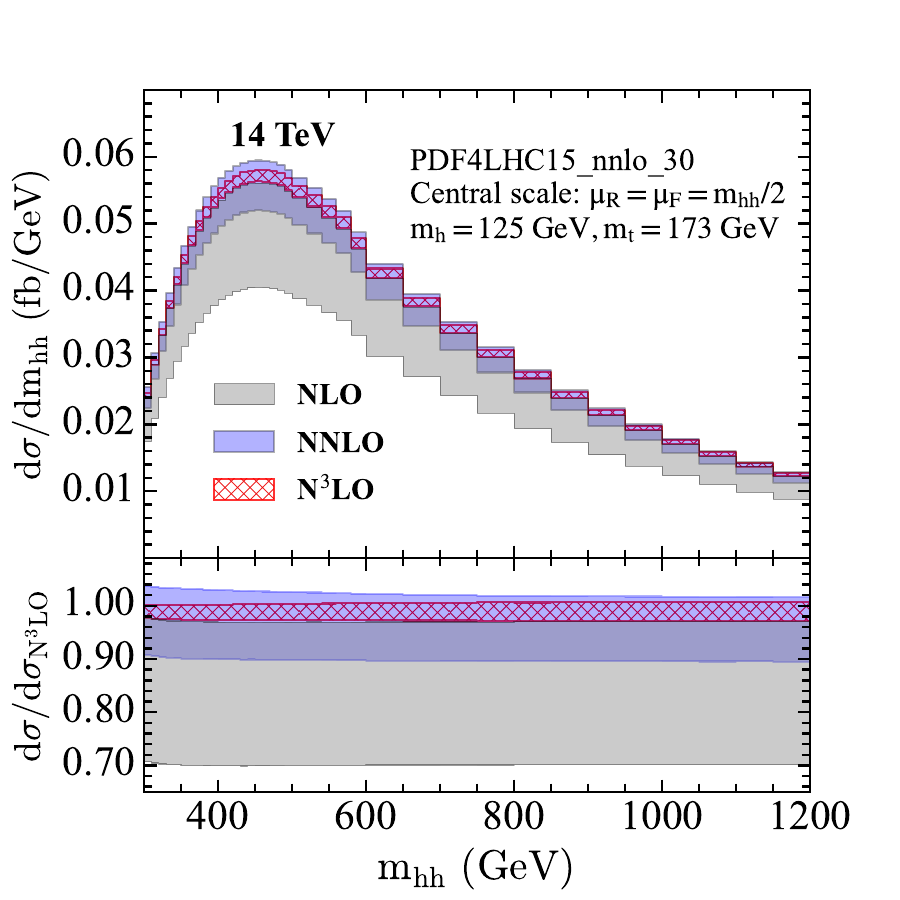}
\includegraphics[width=0.49\textwidth,height=7.52cm]{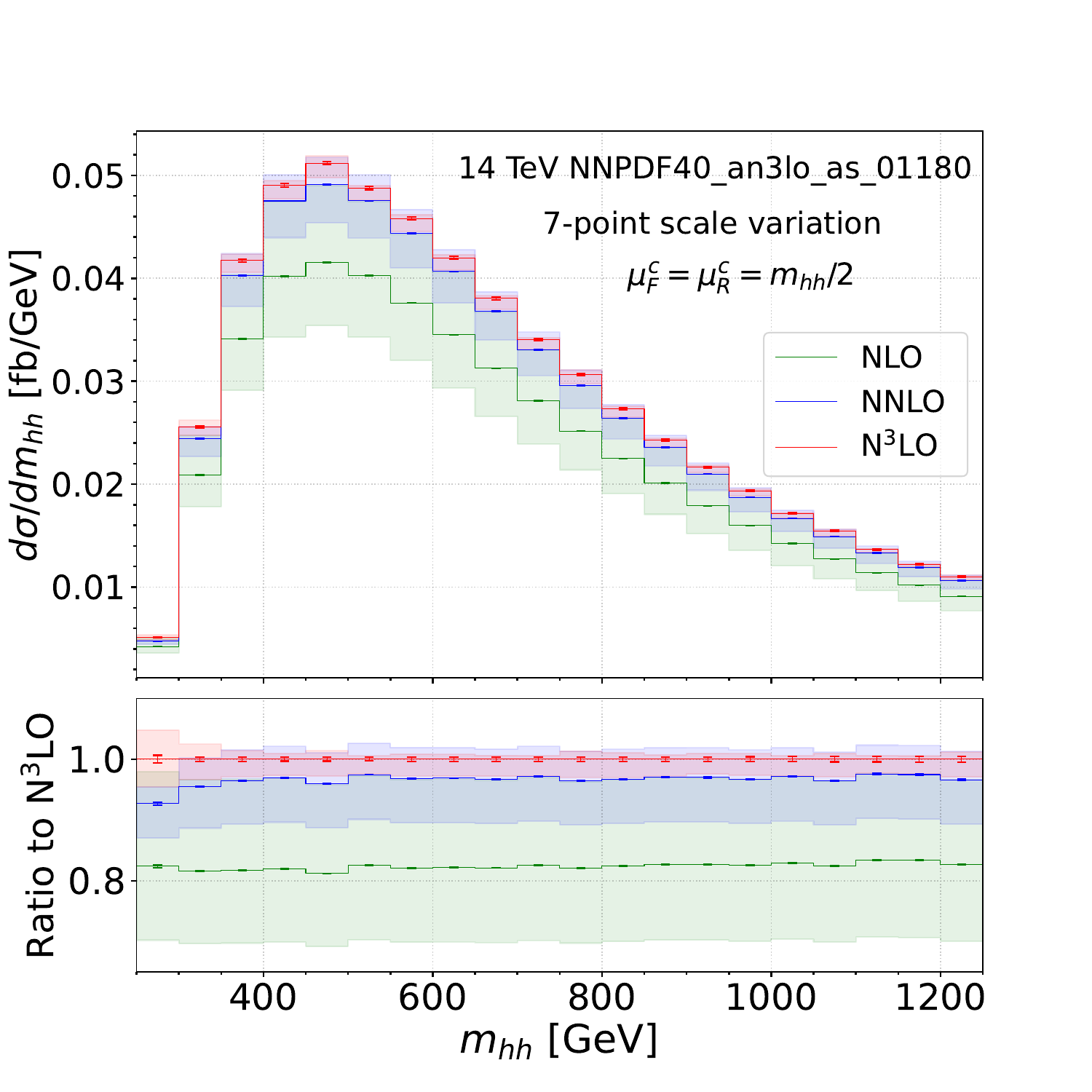}
\caption{Inclusive (left) and fiducial (right) invariant-mass distributions for SM Higgs boson pair production at $\sqrt{s}=14$ TeV in the HTL, from NLO to N$^3$LO. The error bands correspond to $7$-point scale variations. Here $\mu^C$ denotes the central value of the renormalization and factorization scales. Results shown are from Refs.~\cite{Chen:2019lzz,Chen:2019fhs,Chen:2026zmi}.}
\label{fig:xs_vs_mhh_FO_N3LO}
\end{figure*}

\begin{figure*}[ht]
\centering
\includegraphics[width=0.49\textwidth,height=7.52cm]{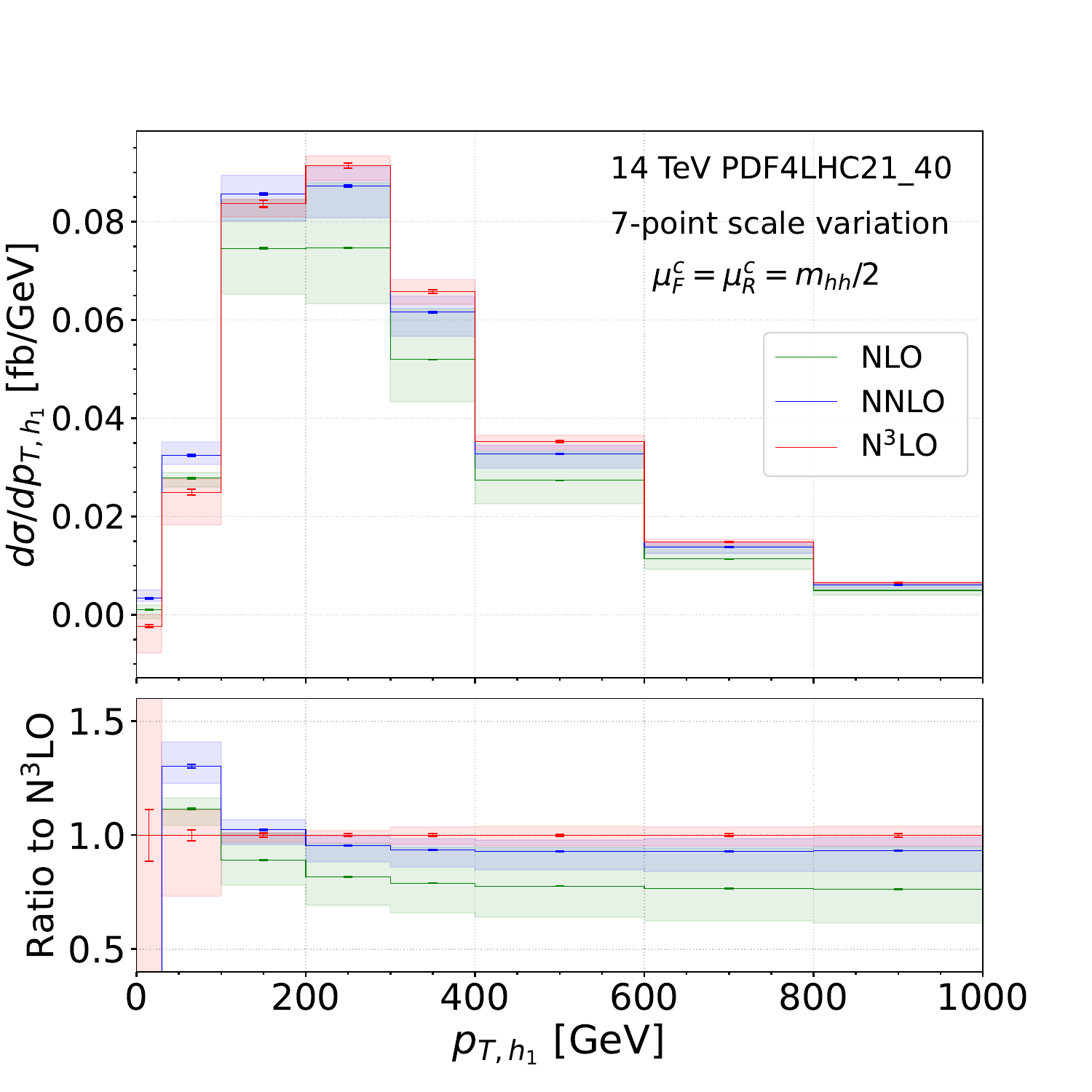}
\includegraphics[width=0.49\textwidth,height=7.52cm]{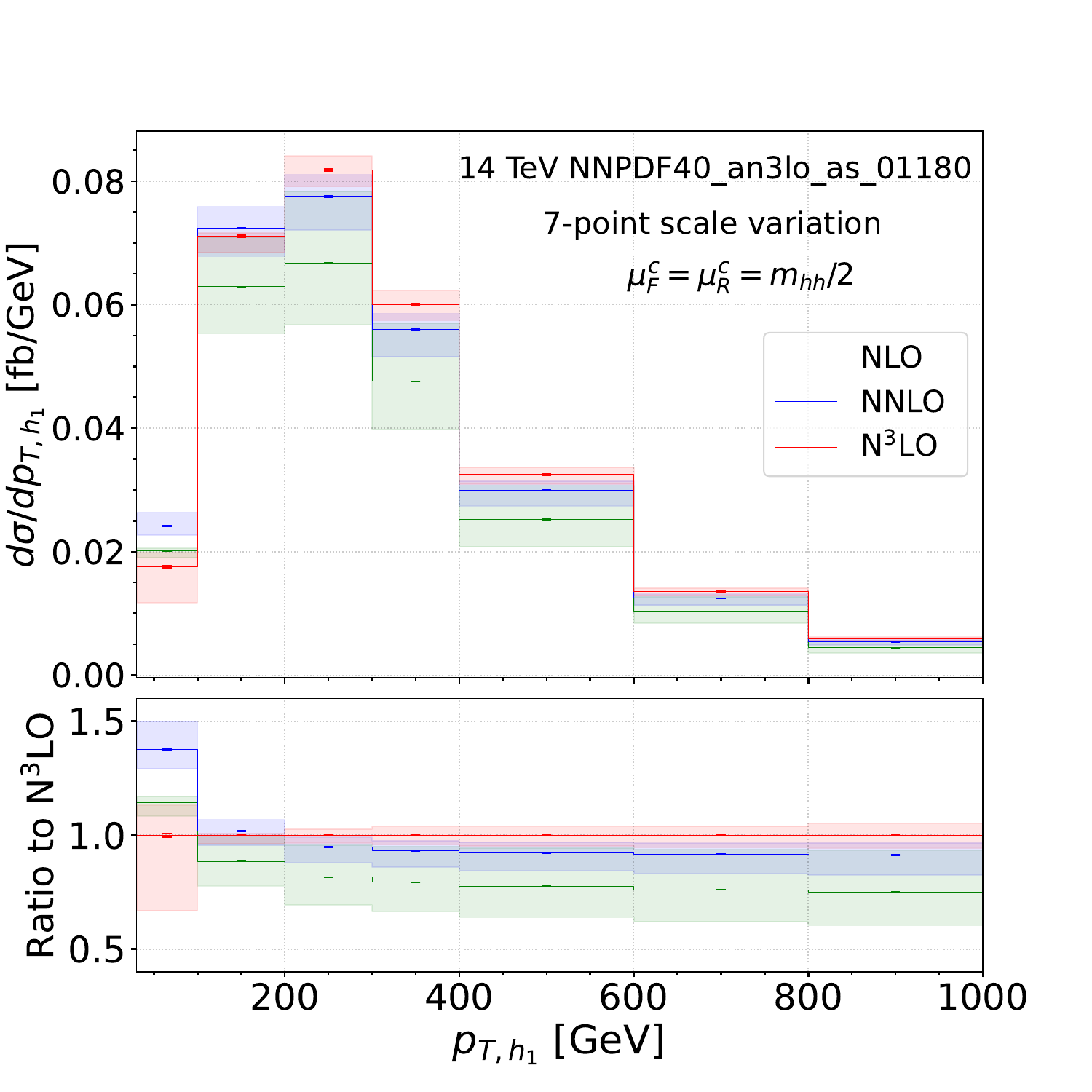}
\caption{Inclusive (left) and fiducial (right) leading-Higgs $p_{T}$ distributions for SM Higgs boson pair production at $\sqrt{s}=14$ TeV in the HTL, from NLO to N$^3$LO. The error bands correspond to $7$-point scale variations. Here $\mu^C$ denotes the central value of the renormalization and factorization scales. Results shown are from Ref.~\cite{Chen:2026zmi}.}
\label{fig:fid_pth1_FO_N3LO}
\end{figure*}

\begin{figure*}[ht]
\centering
\includegraphics[width=0.49\textwidth,height=7.52cm]{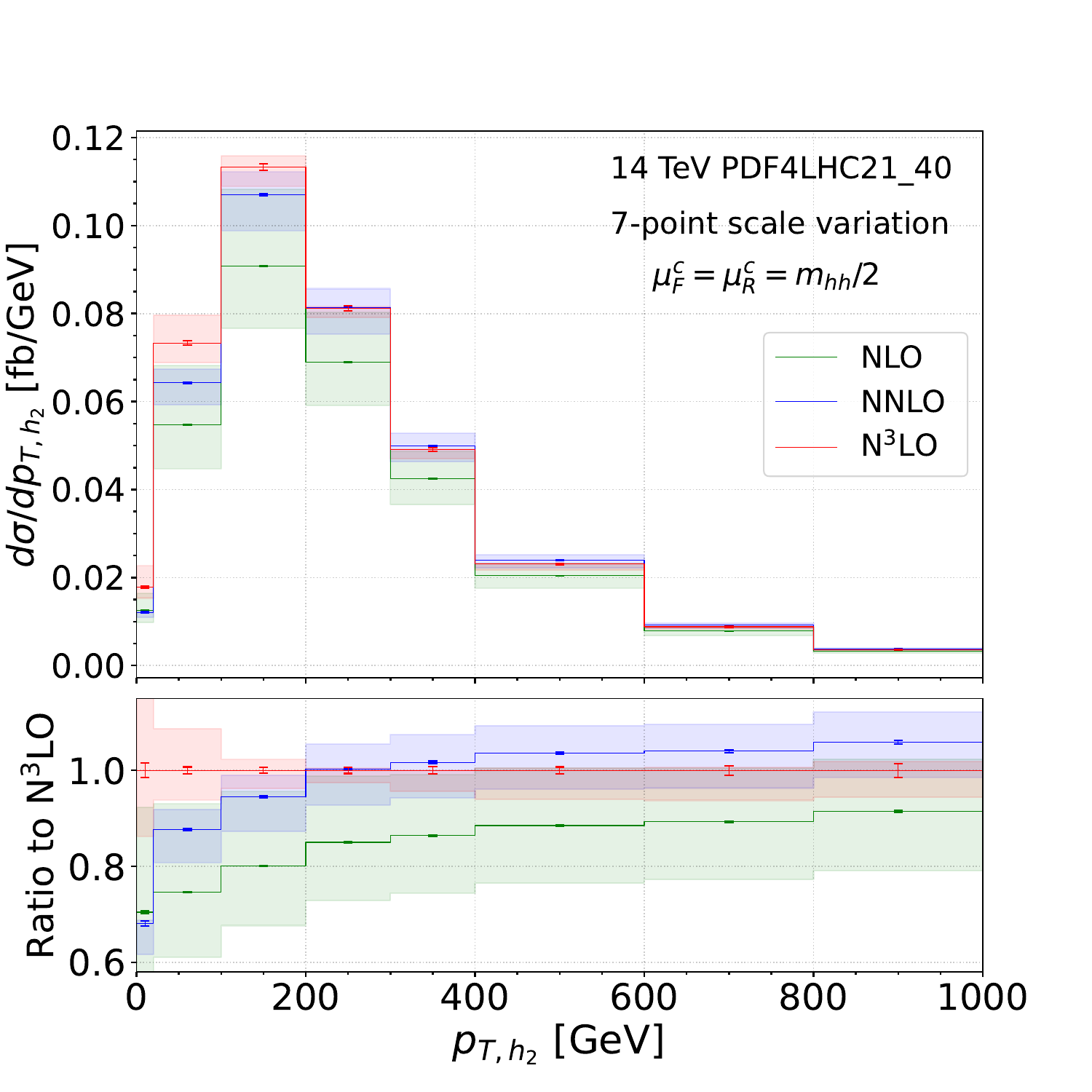}
\includegraphics[width=0.49\textwidth,height=7.52cm]{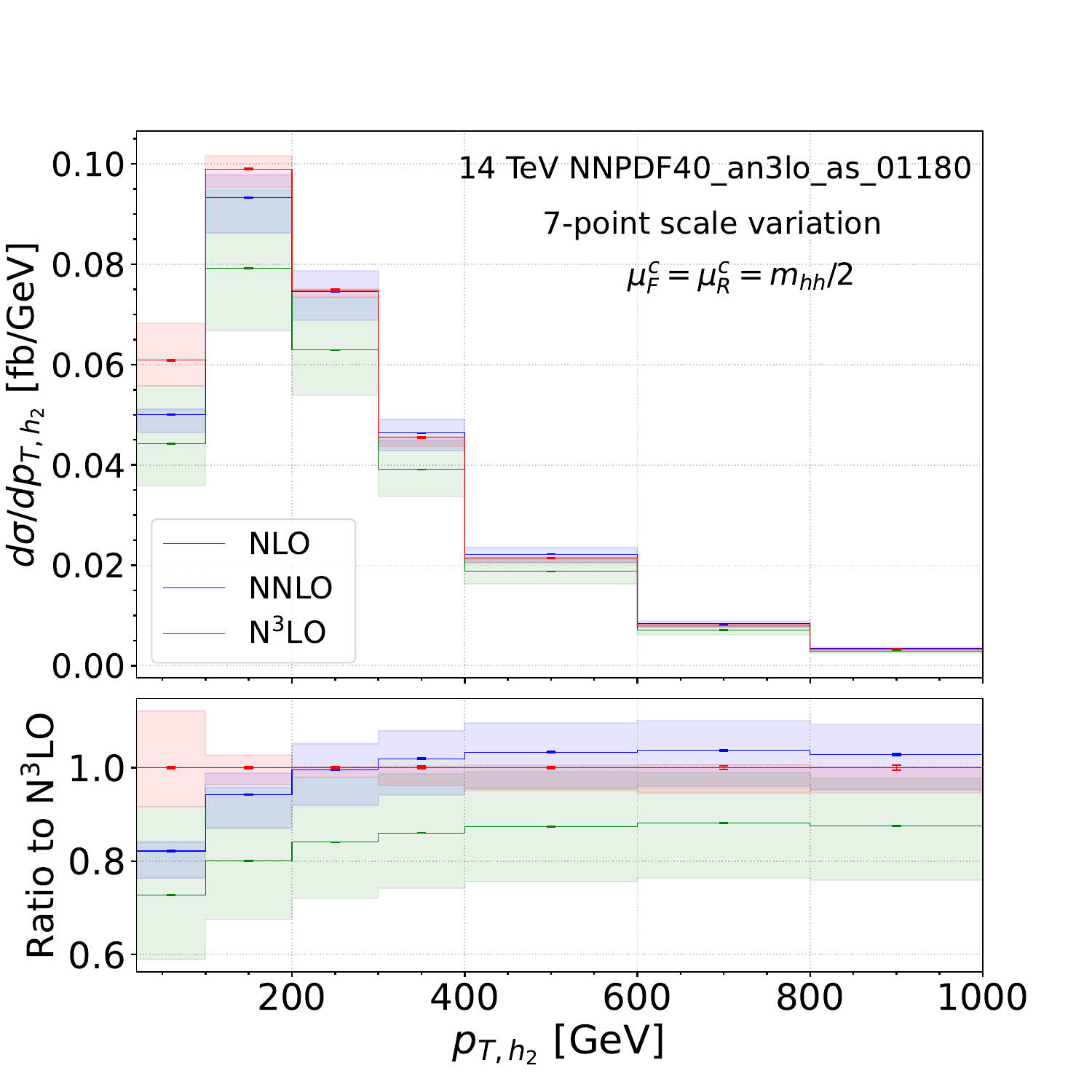}
\caption{Inclusive (left) and fiducial (right) subleading-Higgs $p_{T}$ distributions for SM Higgs boson pair production at $\sqrt{s}=14$ TeV in the HTL, from NLO to N$^3$LO. The error bands correspond to $7$-point scale variations. Here $\mu^C$ denotes the central value of the renormalization and factorization scales. Results shown are from Ref.~\cite{Chen:2026zmi}.
}
\label{fig:fid_pth2_FO_N3LO}
\end{figure*}

The phase-space-integrated cross sections from NLO to N$^3$LO in the HTL at $\sqrt{s}=14$ TeV for proton-proton ($pp$) collisions are listed in Table~\ref{tab:FON3LOinclusivexs}. For the inclusive (fiducial) results, the N$^3$LO QCD corrections increase the NNLO cross section by $3\%$ ($4\%$) and reduce the fractional scale uncertainty by a factor of four (three).
The Higgs boson pair invariant-mass distributions at fixed orders in $\alpha_s$ are shown in Fig.~\ref{fig:xs_vs_mhh_FO_N3LO} for the inclusive (left panel) and fiducial (right panel) cases. In both cases, the higher-order QCD corrections do not shift the peak positions of the distributions. There is, however, a small difference between the two setups. In the inclusive case, the $K$-factor for N$^3$LO over NNLO is nearly flat across a wide range of $m_{hh}$, and the N$^3$LO prediction, with its small scale uncertainty, lies entirely within the NNLO uncertainty band. In contrast, with the fiducial cuts defined in Eq.~\eqref{eq:fiducialcuts}, the N$^3$LO correction in the first bin ($m_{hh}<300$ GeV) amounts to 8$\%$, which is almost a factor of two larger than the corrections in the other bins. Additionally, the scale uncertainty is significantly larger in this bin. The difference in the threshold region mainly stems from linear power corrections in the presence of fiducial cuts. The leading- and subleading-$p_T$ distributions of the two Higgs bosons are shown in Figs.~\ref{fig:fid_pth1_FO_N3LO} and~\ref{fig:fid_pth2_FO_N3LO}. The N$^3$LO QCD corrections modify their shapes, with the largest effects occurring in the small-$p_T$ region, where the scale uncertainties are also most significant. In this region, the fixed-order predictions are not stable, and a resummation calculation is necessary to stabilize the theoretical predictions. Moreover, the fiducial cuts also have a clear impact in this region. The scale uncertainty at N$^3$LO of $p_{T,h_1}$ ($p_{T,h_2}$) in the 20--100 GeV (30--100 GeV) bin changes from 38$\%$ (15$\%$) to 46$\%$ (21$\%$) after imposing the fiducial cuts.
In the intermediate- and large-$p_T$ regions, the fixed-order predictions remain reliable.

\subsection[\texorpdfstring{$\textup{N}^3\textup{LO} + \textup{N}^3\textup{LL}$ in the HTL}{N3LO + N3LL in the HTL}]{\texorpdfstring{\boldmath$\textup{N}^3\textup{LO} + \textup{N}^3\textup{LL}$ in the HTL}{N3LO + N3LL in the HTL}}


\textbf{Ajjath A H, Hua-Sheng Shao~\cite{Ajjath:2022kpv}.}

\noindent Theoretical predictions for the Higgs boson pair production cross section can be further improved by resumming threshold logarithms from soft-gluon emissions. In this subsection, these logarithms are resummed to N$^3$LL accuracy~\cite{Ajjath:2022kpv}, following the formalism outlined briefly below.

Within the QCD-improved parton model, the differential cross section in the invariant mass squared $m_{hh}^2$
of the Higgs boson pair, for hadronic collisions at the centre-of-mass energy $\sqrt{s}$, can be written as a convolution of the partonic flux $\phi_{ab}$ and the perturbative coefficient function $\Delta_{ab\to hh}$
\begin{equation}\label{eq:partonmodel}
    m_{hh}^2 \dfrac{d}{d m_{hh}^2}\sigma_{pp\rightarrow hh}(s,m_{hh}^2) = \tau \sum_{a,b=q,\bar q,g} \int_\tau^1 \frac{dz}{z} \phi_{ab}\left(\frac{\tau}{z},\mu_F^2\right)~ \Delta_{ab\rightarrow hh}  \left(z,m_{hh}^2,\mu_F^2\right)\,,
\end{equation}
where $\tau\equiv m_{hh}^2/s$, $z\equiv m_{hh}^2/\hat{s}$, with $\sqrt{\hat{s}}$
the partonic centre-of-mass energy.

 In the threshold limit, $z\rightarrow1$, the dominant (leading-power, LP) terms arise from both virtual loop corrections and soft, unresolved real-gluon emissions. These contributions take the form of Dirac delta functions, $\delta(1-z)$,  and plus-distributions, such as $ {\cal D}_k(z) \equiv  \left(\frac{\ln^k{(1-z)}}{1-z}\right)_+$, reflecting the incomplete cancellation between virtual and real radiation associated with soft-gluon emissions. These singular threshold contributions can be systematically factorized as
\begin{align}\label{eq:partonicxsec}
    \Delta^{\text{LP}}_{gg\rightarrow hh}(z,m_{hh}^2,\mu_F^2) =& H_{gg\rightarrow hh}(m_{hh}^2,\mu_R^2) ~\delta(1-z)\otimes S_{\Gamma,gg}(z,m_{hh}^2,\mu_F^2,\mu_R^2)\,,
\end{align}
where $H_{gg\rightarrow hh}$ encodes the process-dependent hard virtual corrections after infrared-divergence subtraction, and
 $S_{\Gamma,gg}$
  is the soft-collinear function describing gluon emissions that are soft, collinear, or both with respect to the parent partons. Singular terms are retained in this factorization, while power-suppressed corrections in $(1-z)$ are neglected. The threshold-enhanced logarithms resummed in this approach are universal and can largely be expressed in terms of process-independent anomalous dimensions and splitting functions. Detailed structures of these functions are given in Refs.~\cite{Ravindran:2005vv,Ahmed:2020nci,Ajjath:2022kpv}.

The universal nature of the soft and collinear terms allows the resummation of large logarithms to all orders in $\alpha_s$ using renormalization group methods. This resummation is efficiently performed in Mellin $N$-moment space, where the threshold limit $z\rightarrow 1$ corresponds to $N\rightarrow \infty$:
\begin{align}
    \Delta_{gg\rightarrow hh}^{\rm res}(N,m_{hh}^2,\mu_F^2) =&  \int_0^1 dz ~z^{N-1}\Delta^{\text{LP}}_{gg\rightarrow hh}\left(z,m_{hh}^2,\mu_F^2\right)
    \nonumber \\
    =  g_{0,gg\to hh}(m_{hh}^2,\mu_F^2,\mu_R^2)
    & 
    ~\exp \left(\tilde{C}_{0,gg} (a_s(\mu_R^2)) + g_{1,gg} (\omega)\ln N + \sum_{k=2}{a_s^{k-2}(\mu_R^2)~g_{k,gg} (\omega)}\right)\,,
    \label{DeltaN}
\end{align}
where $a_s\equiv \alpha_s/4\pi$ and $\omega \equiv 2 \beta_0 a_s(\mu_R^2) \ln N$. The factor $g_{0,gg\to hh}$, outside the exponent, arises from the $N$-independent hard function and the Altarelli-Parisi splitting kernels, while the functions $g_{k,gg}$, $k\ge 1$, encode the resummation of logarithmic contributions up to the specified logarithmic accuracy (LL, NLL, NNLL, N$^3$LL, etc.).
Furthermore, the exponent contains an additional $N$-independent coefficient, $\tilde{C}_{0,gg}$, arising from the ${\cal O}(1)$ terms of ${\cal D}_k(z)$ after the Mellin transform; it depends on the Riemann zeta functions $\zeta_n$ and the Euler-Mascheroni constant $\gamma_E$. 

The structure above ensures that the resummation captures all LP threshold logarithms at each order, resulting in a perturbative series with greatly improved convergence and reduced dependence on the renormalization and factorization scales. Matching to fixed-order calculations guarantees the exact reproduction of known results and prevents double counting of terms already included at a given order. At the $\NkLONkLL$ accuracy, the double counting corresponds to $\Delta^{\text{LP}}_{gg\to hh}$ expanded up to N$^k$LO in $\alpha_s$. However, at a given logarithmic accuracy, the right-hand side of the resummation formalism in Eq.~\eqref{DeltaN} is not uniquely defined. The assignment of $N$-independent terms inside or outside the exponent introduces a degree of arbitrariness, referred to as the resummation scheme dependence~\cite{Ajjath:2022kpv}. Four different resummation schemes, denoted ${\rm N_1}$, $\rm{\overline N}_1$, ${\rm N_2}$, and $\rm{\overline N}_2$, are considered in Ref.~\cite{Ajjath:2022kpv} (see their definitions in Section 2.4 of that reference).

\begin{figure*}[ht]
\centering
\includegraphics[width=0.5\textwidth]{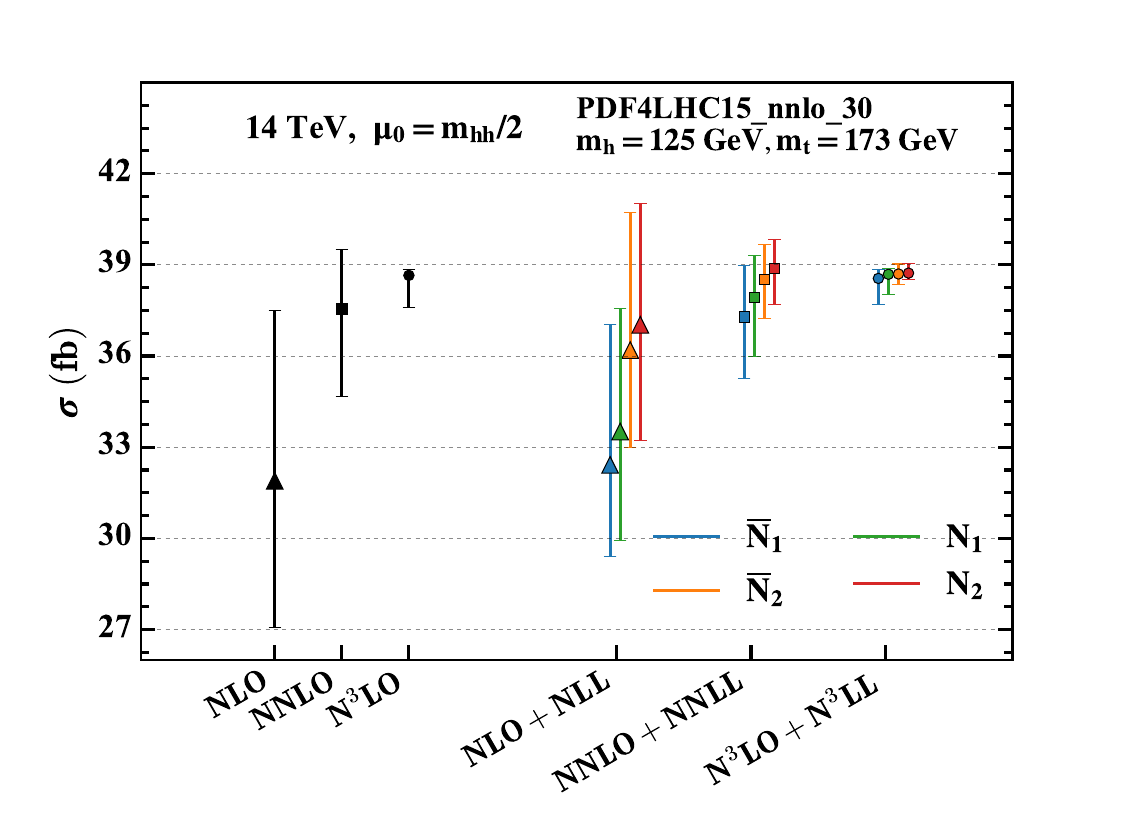}
\caption{Inclusive SM Higgs boson pair cross sections at $\sqrt{s}=14~\textup{TeV}$ in the HTL at several perturbative orders and in different resummation schemes. The error bars indicate scale uncertainties. Results shown are from Ref.~\cite{Ajjath:2022kpv}.}
\label{fig:Verical_prescription}
\end{figure*}

A comparison of inclusive Higgs boson pair cross sections at $\sqrt{s}=14$ TeV in the HTL, at several perturbative orders and in different resummation schemes, is shown in Fig.~\ref{fig:Verical_prescription}. Both the fixed-order N$^k$LO and resummation-improved N$^k$LO+N$^k$LL predictions are presented, with error bars indicating scale uncertainties. We stress a few key points:
\begin{itemize}
    \item Although a significant resummation scheme dependence is observed at NLO+NLL, this dependence decreases as higher orders are included and becomes moderate at N$^3$LO+N$^3$LL.
    \item The results in the $\ResNTwo$ and $\ResNbTwo$ schemes show faster perturbative convergence than those in the other two schemes.
    \item The inclusion of threshold resummation further reduces scale uncertainties. For instance, in the $\ResNbTwo$ scheme, the N$^3$LO+N$^3$LL scale-uncertainty band is reduced by a factor of two compared to the N$^3$LO result, and nearly by a factor of four compared to the NNLO+NNLL result in the same scheme.
    \item The resummation scheme uncertainty at N$^3$LO+N$^3$LL is subdominant compared to the residual scale uncertainty. At $14$ TeV, for example, we have \begin{equation*}\sigma_{\rm N^3LO+N^3LL}=38.70\left(^{+0.85\%}_{-0.87\%}\right)_{\rm scale}\left(^{+0.08\%}_{-0.39\%}\right)_{\rm scheme}~\mathrm{fb}.
    \end{equation*}
\end{itemize}
Therefore, in the following, we consider the $\ResNbTwo$ scheme only. Table~\ref{tab:ResumN3LON3LLinclusivexs} summarizes inclusive cross sections, from NLO+NLL to N$^3$LO+N$^3$LL, in the HTL at $\sqrt{s}=14$ TeV. Inclusion of N$^3$LL resummation corrections increases the N$^3$LO cross section by around $1\%$ and reduces the scale uncertainty by about a factor of two.

\begin{table}[htbp] 
\begin{center}
\newcolumntype{P}[1]{>{\centering\arraybackslash}p{#1}}
{\renewcommand{\arraystretch}{1.5}
\begin{tabular}{P{2.6cm}P{2.6cm}P{2.6cm}}
    \toprule
    $\sigma_{\textup{NLO}+\textup{NLL}}$ $[\textup{fb}]$
    & $\sigma_{\textup{NNLO}+\textup{NNLL}}$ $[\textup{fb}]$
    & $\sigma_{\textup{N}^3\textup{LO}+\textup{N}^3\textup{LL}}$ $[\textup{fb}]$
    \\
    \midrule
  $36.19^{+12.5\%}_{-8.8\%}$ &
  $38.52^{+3.0\%}_{-3.4\%}$ &
 $38.70_{-0.85\%}^{+0.87\%}$
 \\
    \bottomrule
\end{tabular}}
\caption{Soft-gluon resummation–improved inclusive cross sections in the HTL at $\sqrt{s}=14~\textup{TeV}$. The quoted uncertainties correspond to $7$-point scale variations. Only results in the $\ResNbTwo$ scheme are shown.}
\label{tab:ResumN3LON3LLinclusivexs}
\end{center}
\end{table}

\begin{figure*}[!htbp]
\centering
\includegraphics[width=0.5\textwidth]{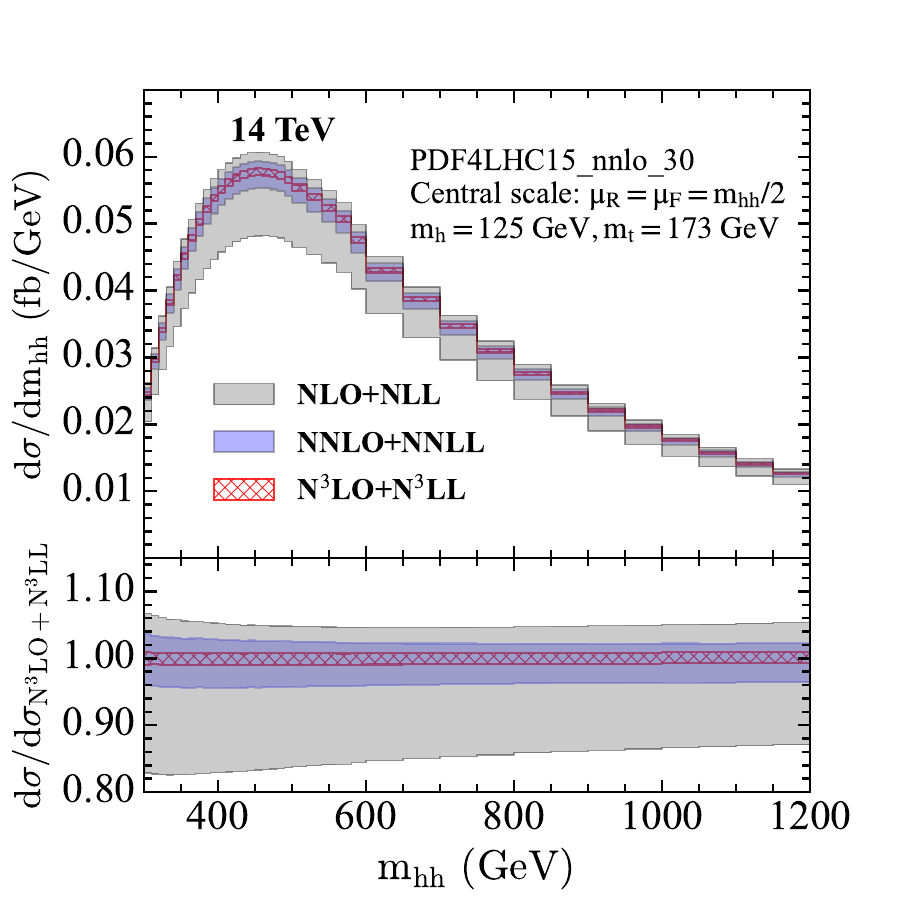}
\\
\caption{Inclusive invariant-mass distributions for SM Higgs boson pair production at $\sqrt{s}=14$ TeV in the HTL, from NLO+NLL to N$^3$LO+N$^3$LL. The error bands correspond to $7$-point scale variations. Results shown are from Ref.~\cite{Ajjath:2022kpv}.}
\label{fig:xs_vs_mhh_Res_N3LON3LL}
\end{figure*}

The differential prediction for the Higgs boson pair invariant mass $m_{hh}$ in Fig.~\ref{fig:xs_vs_mhh_Res_N3LON3LL} further demonstrates the perturbative convergence of the strong-coupling expansion. Soft-gluon resummation induces only a modest shift in the spectrum and further reduces the fractional scale errors.

\subsection{Including partial top quark mass effects}


\textbf{Ajjath A H, Long-Bin Chen, Xuan Chen, Yuesheng Dai, Hai Tao Li, Shi-Yuan Li, Hua-Sheng Shao, Jian Wang~\cite{Chen:2019fhs,Ajjath:2022kpv,Chen:2026zmi}.}

\noindent Since finite top quark mass effects cannot be neglected and are exactly known only at NLO~\cite{Borowka:2016ehy,Borowka:2016ypz,Baglio:2018lrj,Davies:2019dfy}, we improve the NLO QCD computation with full top quark mass dependence~\cite{Heinrich:2017kxx,Heinrich:2019bkc,Davies:2025qjr} (denoted as ``$\NLOmt$") by multiplying it with the higher-order (differential) $K$-factors calculated in the HTL in this subsection. The reweighting is carried out consistently using the on-shell top quark mass scheme in $\NLOmt$ and the on-shell mass in the HTL predictions. The resulting combinations are denoted as $\NkLOtimesNLOmt$ and $\NkLONkLLtimesNLOmt$, where $k$ is a positive integer. Specifically,
\begin{eqnarray}\label{NkLO-fullmt}
d\sigma_{\NkLOtimesNLOmt}&\equiv& d\sigma_{\NLOmt}\frac{d\sigma_{\NkLO}}{d\sigma_{\rm NLO}}\,,\nonumber\\
d\sigma_{\NkLONkLLtimesNLOmt}&\equiv&d\sigma_{\NLOmt}\frac{d\sigma_{\NkLONkLL}}{d\sigma_{\rm NLO}}\,.
\end{eqnarray}
This treatment is justified under certain working assumptions; alternative approaches to incorporating top quark mass effects can be found in Section~3 of Ref.~\cite{Chen:2019fhs}.

\begin{figure*}[!htbp]
\centering
\vspace{-1.5cm}
\includegraphics[width=0.49\textwidth]{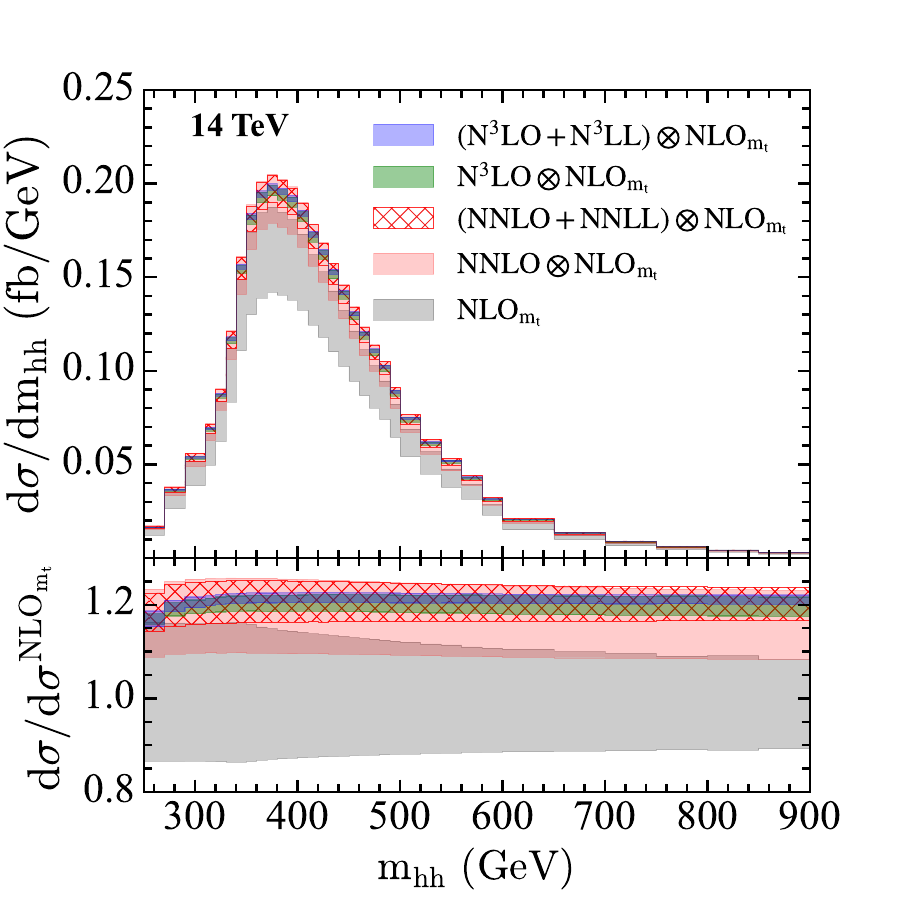}
\includegraphics[width=0.49\textwidth,height=7.52cm]{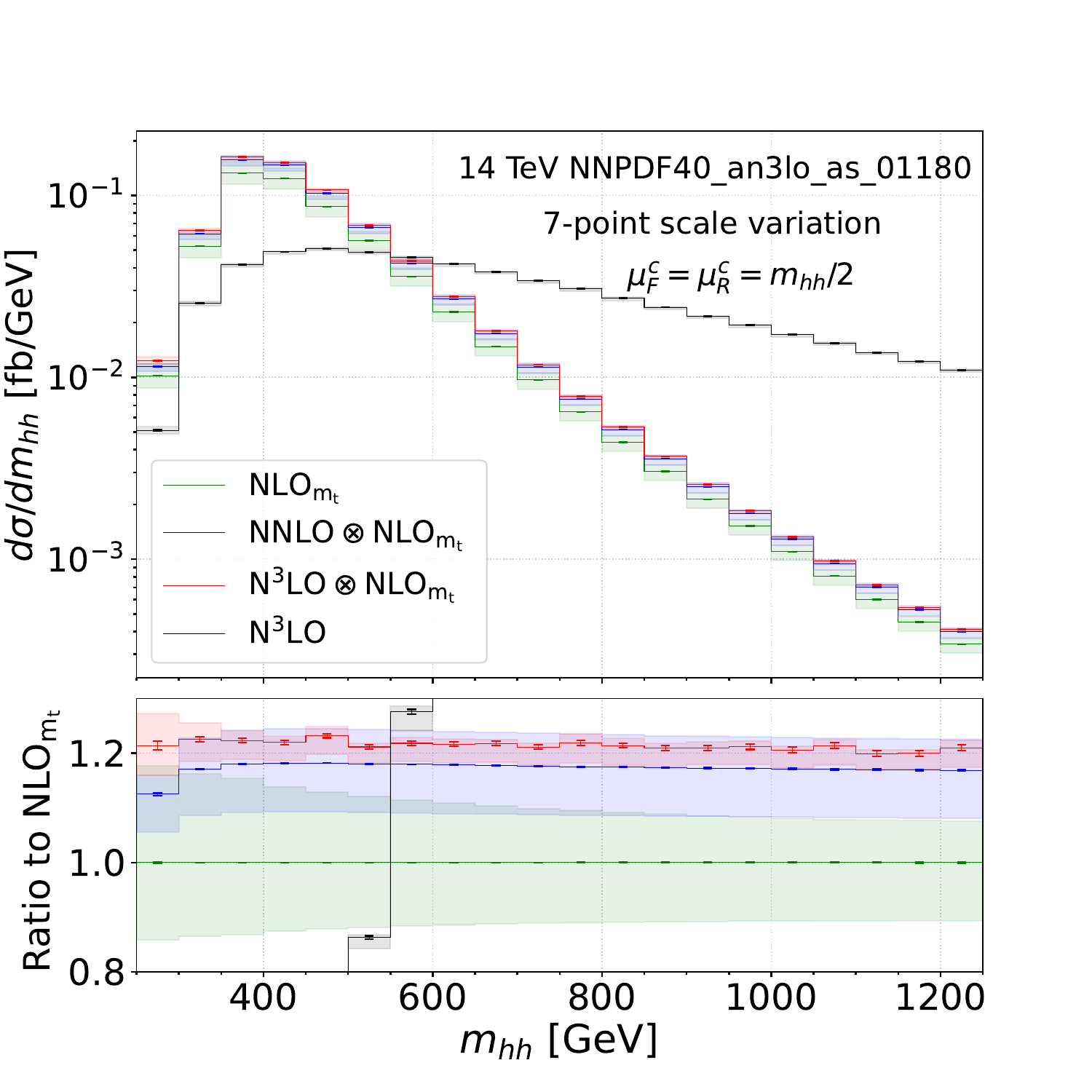}
\\
\caption{Inclusive (left) and fiducial (right) invariant-mass distributions for SM Higgs boson pair production at $\sqrt{s}=14$ TeV, including top quark mass effects. The error bands stem from $7$-point scale variations. Here $\mu^C$ denotes the central value of the renormalization and factorization scales. Results shown are from Refs.~\cite{Chen:2019fhs,Ajjath:2022kpv,Chen:2026zmi}.}
\label{fig:xs_vs_mhh_large_mt1}
\end{figure*}

\begin{sloppypar}
The inclusive phase-space-integrated cross sections are listed in Table~\ref{Tab:fullMt_inc} for LHC $pp$ collisions at $\sqrt{s}=14$ TeV and for five perturbative orders: $\NLOmt$, $\NNLOtimesNLOmt$, $\NNLONNLLtimesNLOmt$, $\NtLOtimesNLOmt$, and $\NtLONtLLtimesNLOmt$. The N$^3$LO QCD corrections enhance the $\NLOmt$ cross section by $21\%$. The N$^3$LL threshold resummation only marginally modifies the $\NtLOtimesNLOmt$ cross section. The scale uncertainty is significantly reduced when higher-order QCD corrections are included: the width of the uncertainty band from the envelope of the $7$-point scale variation in $\NtLOtimesNLOmt$ is only about one quarter of that in the previous fixed-order $\NNLOtimesNLOmt$ result. Inclusion of N$^3$LL threshold resummation further reduces the scale uncertainty to below one percent. A similar pattern is observed in the differential invariant-mass distribution of the Higgs boson pair, as shown in the left panel of Fig.~\ref{fig:xs_vs_mhh_large_mt1}.
\end{sloppypar}

\begin{figure*}[h]
\centering
\vspace{-2cm}
\includegraphics[width=0.49\textwidth,height=7.52cm]{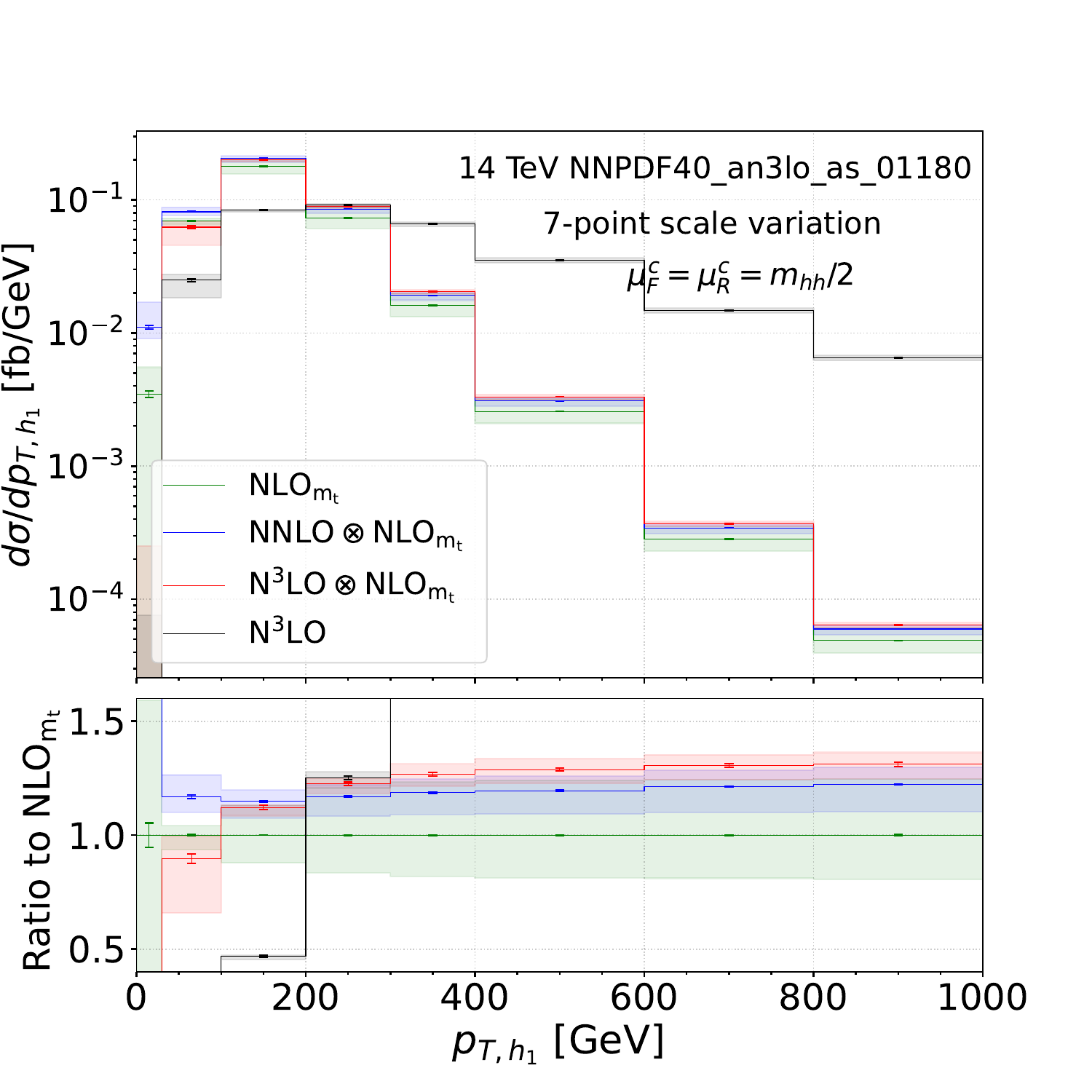}
\includegraphics[width=0.49\textwidth,height=7.52cm]{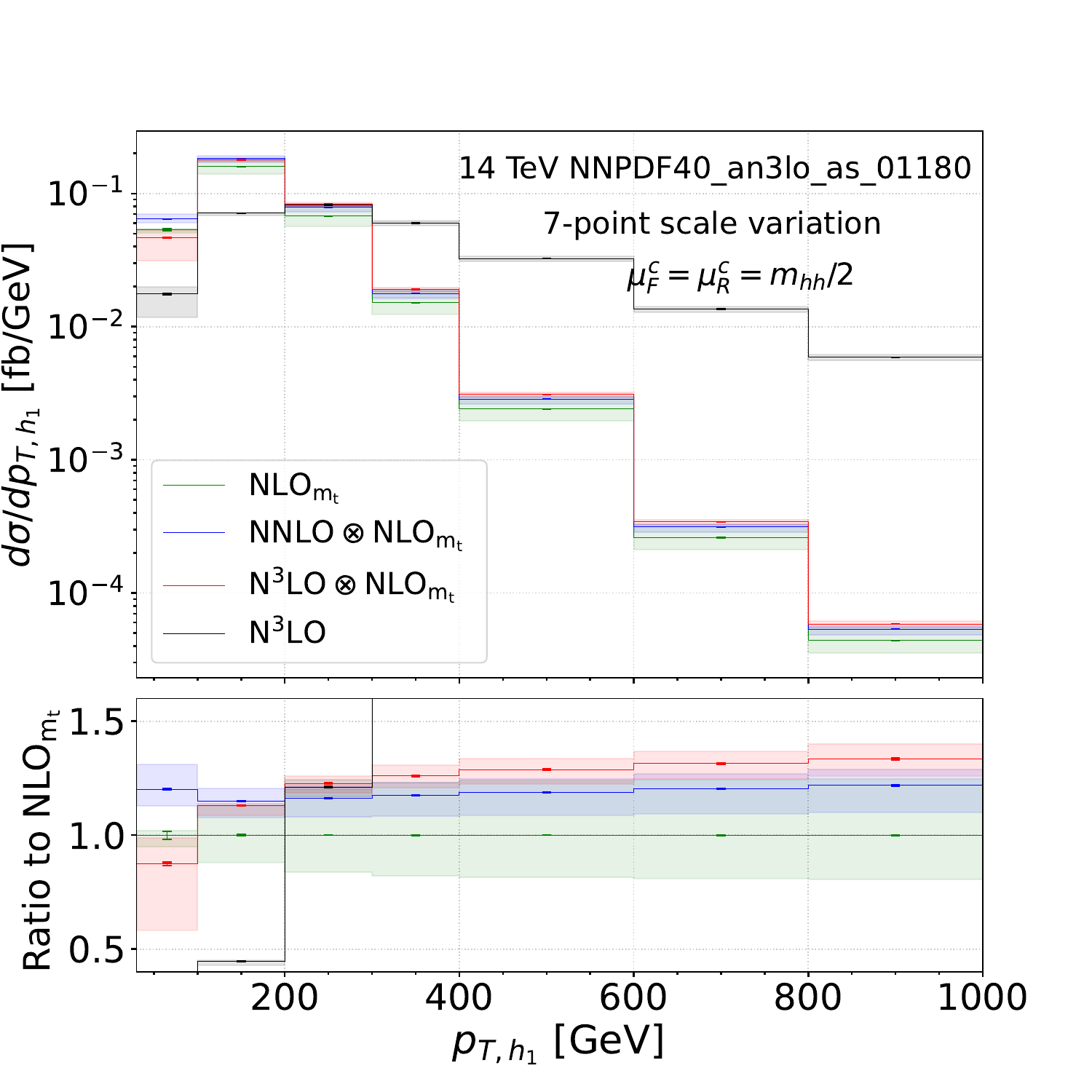}
\caption{Inclusive (left) and fiducial (right) leading-Higgs $p_{T}$ distributions for SM Higgs boson pair production at $\sqrt{s}=14$ TeV, including top quark mass effects. The error bands stem from $7$-point scale variations. Here $\mu^C$ denotes the central value of the renormalization and factorization scales. Results shown are from Ref.~\cite{Chen:2026zmi}.}
\label{fig:fid_pth1_mt}
\end{figure*}

\begin{figure*}[h]
\centering
\includegraphics[width=0.49\textwidth,height=7.52cm]{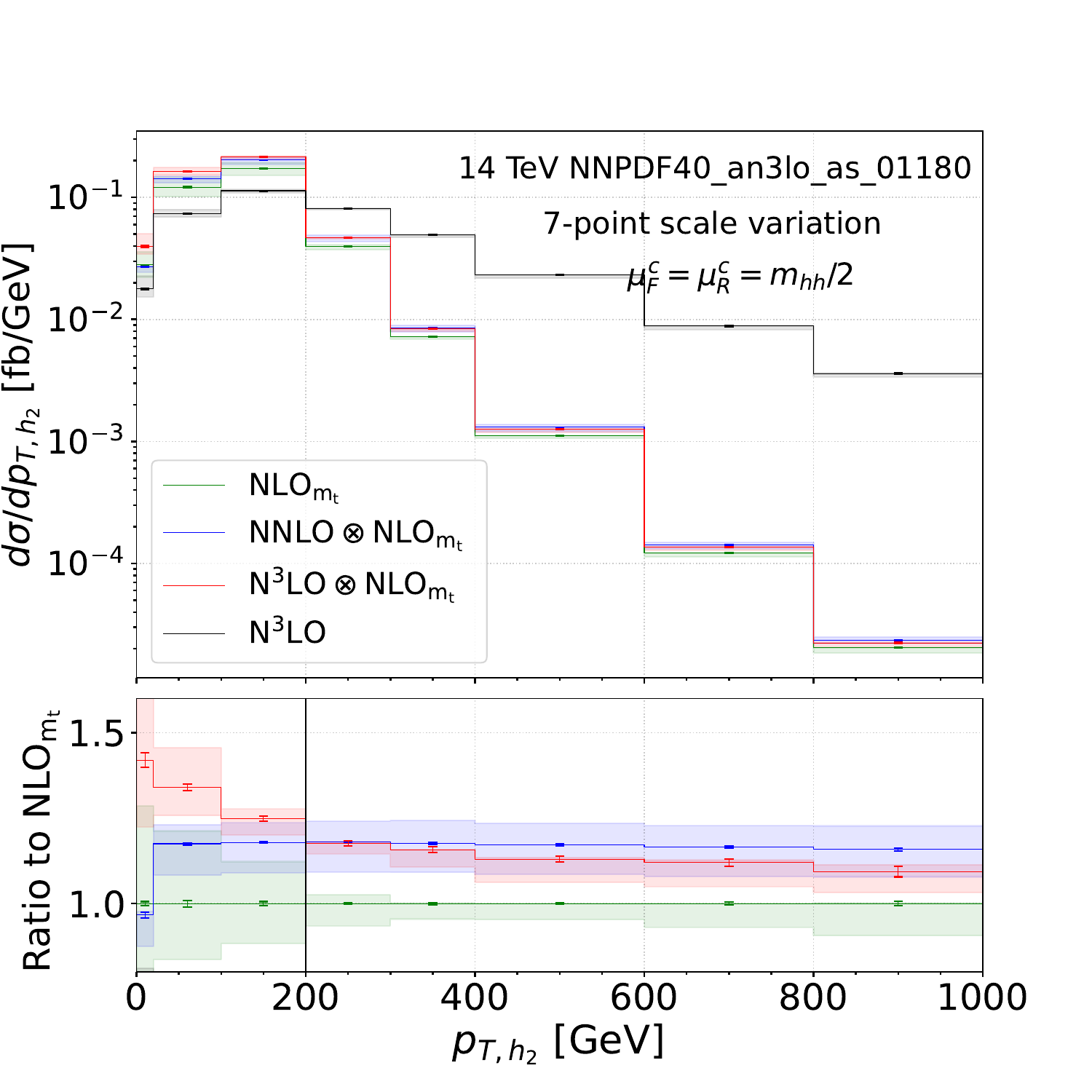}
\includegraphics[width=0.49\textwidth,height=7.52cm]{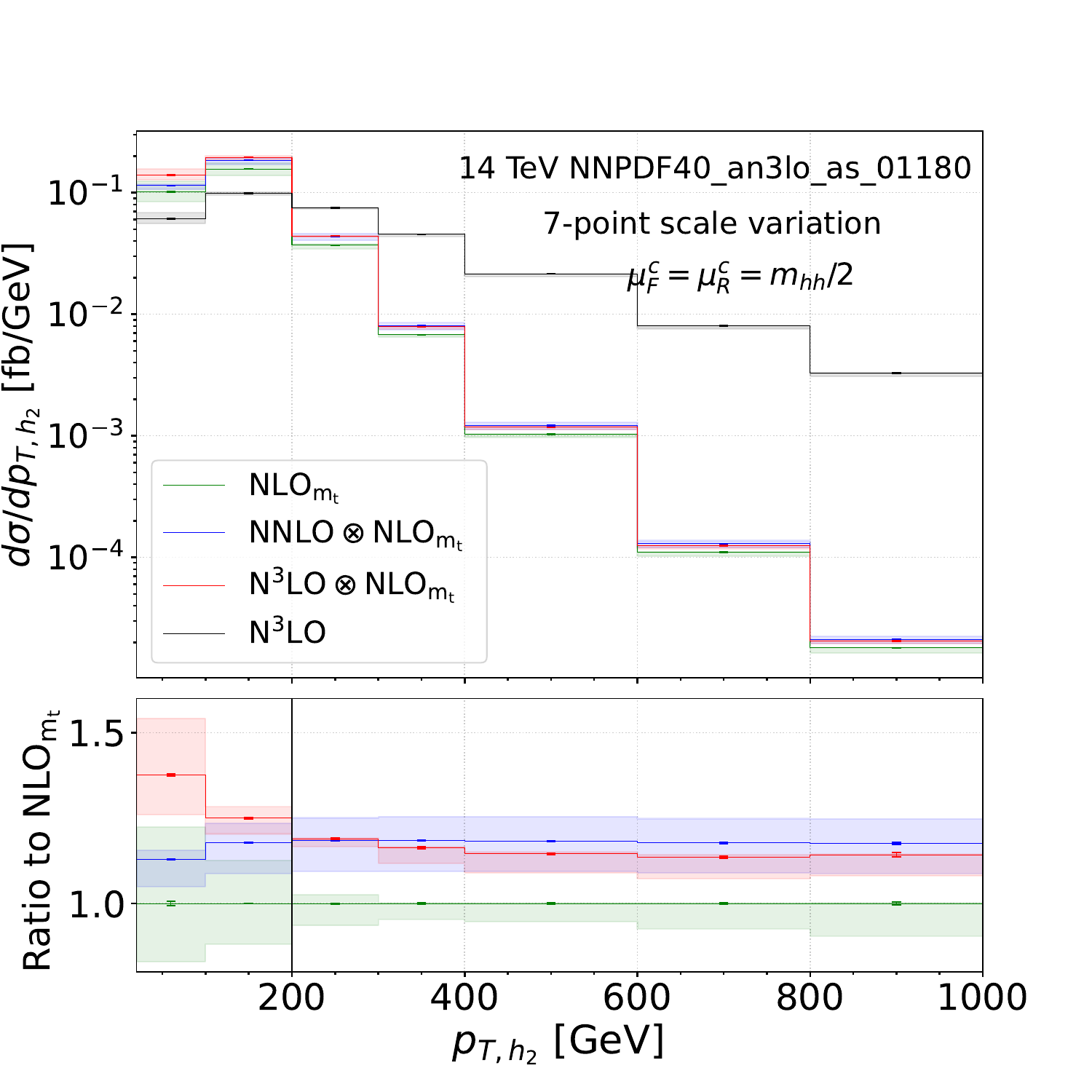}
\caption{Inclusive (left) and fiducial (right) subleading-Higgs $p_{T}$ distributions for SM Higgs boson pair production at $\sqrt{s}=14$ TeV, including top quark mass effects. The error bands stem from $7$-point scale variations. Here $\mu^C$ denotes the central value of the renormalization and factorization scales. Results shown are from Ref.~\cite{Chen:2026zmi}.}
\label{fig:fid_pth2_mt}
\end{figure*}

With the fiducial cuts defined in Eq.~\eqref{eq:fiducialcuts}, only fixed-order predictions are given in Table~\ref{Tab:fullMt_inc}, as fully differential soft-gluon resummation predictions are not available. The N$^3$LO QCD corrections enhance the $\NLOmt$ cross section by $18\%$, which is close to the inclusive case. The $m_{hh}$ distribution with fiducial cuts, shown in the right panel of Fig.~\ref{fig:xs_vs_mhh_large_mt1}, highlights the necessity of including finite top quark mass effects. A similar behaviour is seen in the $p_T$ distributions of the leading and subleading Higgs bosons in Figs.~\ref{fig:fid_pth1_mt} and~\ref{fig:fid_pth2_mt}. For more detailed discussions, we refer the interested reader to Ref.~\cite{Chen:2026zmi}.

\begin{table}
\begin{center}

\newcolumntype{P}[1]{>{\centering\arraybackslash}m{#1}}
{\renewcommand{\arraystretch}{1.5}
\resizebox{0.98\textwidth}{!}{%
\begin{tabular}{P{5cm}P{5cm}P{5cm}}
    \toprule
        Contributions   &Inclusive cross section [$\textup{fb}$]  &Fiducial cross section [$\textup{fb}$] \\
    \midrule
    $\NLOmt$ & $32.64_{-12.5\%}^{+13.5\%}$  &  $28.44_{-12 \%}^{+14 \%}$ \\
    $\NNLOtimesNLOmt$ & $38.42_{-7.6\%}^{+5.2\%}$ & $33.40_{-7.3 \%}^{+5.2 \%}$ \\
   $\NNLONNLLtimesNLOmt$ & $39.42_{-3.4\%}^{+3.0\%}$  & -  \\
    $\NtLOtimesNLOmt$ & $39.56_{-2.7\%}^{+0.50\%}$ & $34.68(4)^{+1.4\%}_{-2.9\%}$ \\  
    $\NtLONtLLtimesNLOmt$ &  $39.60_{-0.87\%}^{+0.85\%}$ & - \\
    \bottomrule
\end{tabular}}}
\caption{Inclusive and fiducial phase-space-integrated cross sections for SM Higgs boson pair production via gluon fusion at $\sqrt{s}=14\,\textup{TeV}$ in $pp$ collisions, including top quark mass effects. The quoted relative uncertainties correspond to $7$-point scale variations.
The fiducial cuts are specified in Eq.~\eqref{eq:fiducialcuts}.}
\label{Tab:fullMt_inc}
\end{center}
\end{table}

\section{Top quark mass scheme and scale uncertainties}
\label{sec:scheme-scale-uncertainties}

\subsection{Top quark mass scheme and scale uncertainties at NLO}
\label{sec:NLO-uncertainties}


\textbf{Julien Baglio, Francisco Campanario, Seraina Glaus, Milada Margarete Mühl\-leitner, Jonathan Ronca, Michael Spira, Juraj Streicher~\cite{Baglio:2018lrj,Baglio:2020ini,Baglio:2020wgt}.}

\noindent The analysis of the top quark mass scheme and scale uncertainties was first performed in Refs.~\cite{Baglio:2018lrj,Baglio:2020wgt,Baglio:2020ini}, including the full NLO QCD calculation (see Section~\ref{sec:nlo2}). The top quark mass scheme and
scale uncertainties drop by roughly a factor of two from LO to NLO. At
LO, we obtain the uncertainties ($Q$ denotes the invariant Higgs boson pair mass $m_{hh}$)
\begin{eqnarray}
\frac{d\sigma(gg\to hh)}{dQ}\Big|_{Q=300~{\rm GeV}} & = &
0.01656^{+62\%}_{-2.4\%}\, \mathrm{fb/GeV},\nonumber\\
\frac{d\sigma(gg\to hh)}{dQ}\Big|_{Q=400~{\rm GeV}} & = &
0.09391^{+0\%}_{-20\%}\, \mathrm{fb/GeV},\nonumber\\
\frac{d\sigma(gg\to hh)}{dQ}\Big|_{Q=600~{\rm GeV}} & = &
0.02132^{+0\%}_{-48\%}\, \mathrm{fb/GeV},\nonumber\\
\frac{d\sigma(gg\to hh)}{dQ}\Big|_{Q=1200~{\rm GeV}} & = &
0.0003223^{+0\%}_{-56\%}\, \mathrm{fb/GeV}
\end{eqnarray}
where the full spread of the cross sections is taken over predictions obtained with the top quark pole mass
(central values) and with the $\overline{\rm MS}$ mass. In the latter case, the scale
$\mu_t$ of $\overline{m}_t(\mu_t)$ is either set to $\overline{m}_t$ or
varied in the range between $Q/4$ and $Q$, as
in the single (off-shell) Higgs case considered earlier. The final NLO
results read \cite{Baglio:2018lrj,Baglio:2020wgt,Baglio:2020ini}
\begin{eqnarray}
\frac{d\sigma(gg\to hh)}{dQ}\Big|_{Q=300~{\rm GeV}} & = &
0.02978(7)^{+6\%}_{-34\%}\, \mathrm{fb/GeV},\nonumber\\
\frac{d\sigma(gg\to hh)}{dQ}\Big|_{Q=400~{\rm GeV}} & = &
0.1609(7)^{+0\%}_{-13\%}\, \mathrm{fb/GeV},\nonumber\\
\frac{d\sigma(gg\to hh)}{dQ}\Big|_{Q=600~{\rm GeV}} & = &
0.03204(9)^{+0\%}_{-30\%}\, \mathrm{fb/GeV},\nonumber\\
\frac{d\sigma(gg\to hh)}{dQ}\Big|_{Q=1200~{\rm GeV}} & = &
0.000435(6)^{+0\%}_{-35\%}\, \mathrm{fb/GeV}
\end{eqnarray}
Since these uncertainties are similar in size to the renormalization
and factorization scale dependencies at NLO, they constitute an important
contribution to the total theoretical uncertainties.

The amplitude may be written as the sum of
two form factors, $F_1$ and $F_2$, describing the scattering of incoming gluons with the
same helicity and opposite helicities, respectively. The contribution of the box
diagrams to the two form factors dominates at high energy.
Expanding the LO and NLO results of Ref.~\cite{Davies:2018qvx} around
large invariant Higgs boson pair mass, $s$, we have, in the on-shell scheme and
in the notation of Ref.~\cite{Davies:2018qvx},
\begin{eqnarray}
F_\mathrm{box,i} & = & F_\mathrm{box,i}^{(0)} +
\frac{\alpha_s(\mu_R)}{\pi} F_\mathrm{box,i}^{(1)} \qquad (i=1,2) \nonumber \\
F_\mathrm{box,i}^{(0)} & = & \frac{m_t^2}{s} c_{0,i} +
\mathcal{O}\left(\frac{1}{s^2}\right)  \nonumber \\
F_\mathrm{box,i}^{(1)} & = & 2 F_\mathrm{box,i}^{(0)} \log\frac{m_t^2}{s}
+ \frac{m_t^2}{s}  c_{1,i} +
\mathcal{O}\left(\frac{1}{s^2}\right)
\label{eq:hhexp}
\end{eqnarray}
where the coefficients $c_{0,i}$ and $c_{1,i}$ do not depend on the
top quark mass. The overall factor of $m_t^2$ for the box contribution
originates entirely from the Yukawa couplings, and examining $F_\mathrm{box,i}^{(0)}$,
we find that the form factors are independent of the propagator top quark mass in the high-energy limit.
Therefore, for a fixed Yukawa coupling, we expect the results in different schemes to asymptote.
Transforming the top quark pole mass $m_t$ into the $\overline{\rm MS}$ mass
$\overline{m}_t(\mu_t)$, the explicit expressions above are modified to
\cite{Baglio:2020ini}
\begin{eqnarray}
F_\mathrm{box,i}^{(0)} & = & \frac{\overline{m}_t^2(\mu_t)}{s} c_{0,i}
+ \mathcal{O}\left(\frac{1}{s^2}\right) \qquad (i=1,2) \nonumber \\
F_\mathrm{box,i}^{(1)} & = & 2 F_\mathrm{box,i}^{(0)}
\left[\log\frac{\mu_t^2}{s} + \frac{4}{3}\right] +
\frac{\overline{m}_t^2(\mu_t)}{s} c_{1,i} +
\mathcal{O}\left(\frac{1}{s^2}\right)
\label{eq:NLO-MSbar}
\end{eqnarray}
This emphasizes that, in order to absorb the large logarithmic terms $\log\frac{m_t^2}{s}$,
the choice $\mu_t=\kappa\sqrt{s}$ is preferred\footnote{A similar dynamical scale choice, e.g.~the transverse mass $\mu_t = \kappa M_T$ with $M_T = tu/s$ in the high-energy limit, would be equally favoured.} when the coefficient $\kappa$ is not too far from unity. However, the central recommended values of the total and differential cross sections use the top quark pole mass, so that this defines the reference prediction for the estimate of the uncertainties.

\subsection{Resummation of top quark mass-dependent logarithms at high energy}
\label{sec:scet}


\textbf{Sebastian Jaskiewicz, Stephen Jones, Robert Szafron, Yannick Ulrich~\cite{Jaskiewicz:2024xkd}.}

\noindent First steps in tackling the large top quark
mass scheme dependence that arises in this process,
as discussed at NLO in Section~\ref{sec:NLO-uncertainties}, were undertaken in~\cite{Jaskiewicz:2024xkd}. 
This study focused on investigating the 
high-energy behaviour of the $gg\to hh$ amplitude, namely the limit $s, |t|, |u| \gg m_t^2 \gg m_h^2$.
In this case, contributions due to the triangle
type diagrams in Fig.~\ref{fg:hhdia}
are power suppressed, and the dominant 
contribution arises from the box type diagrams.

The starting point of~\cite{Jaskiewicz:2024xkd} is
the observation that in the high-energy limit the
result for one- and two-loop box contributions in
the $\mathrm{OS}$ scheme has the structure
given in \eqref{eq:hhexp}.
Interestingly, the leading term proportional to the $C_F$ colour factor at two loops can be understood as originating from the mass renormalization counterterm. Indeed,
converting the top quark mass to the $\overline{{\rm{MS}}}$ scheme yields a logarithm of $\mu_t^2/s$~\cite{Baglio:2020ini},
as can explicitly be seen in~\eqref{eq:NLO-MSbar}.
Since the leading mass-dependent logarithm originates only from the mass renormalization counterterm, the leading behaviour of $F_\mathrm{box,i}^{(1)}$ in the small mass limit can be predicted solely from the leading order $F_\mathrm{box,i}^{(0)}$ result and the mass renormalization counterterm.
The immediate questions arising from this discussion
that were answered in~\cite{Jaskiewicz:2024xkd} are
\begin{enumerate}
    \item Why does the box amplitude have such a simple structure?
    \item Does this simple structure persist at higher orders, or can super-leading/power-enhanced terms enter in the integrals at higher loop orders to spoil this picture?
\end{enumerate}
The analysis carried out in~\cite{Jaskiewicz:2024xkd}
consists of two main parts.
First, a detailed fixed-order study is performed using
the Method of Regions (MoR) approach
\cite{Beneke:1997zp, Smirnov:1990rz, Smirnov:1994tg, Smirnov:2002pj}
investigating the modes and regions that contribute to the relevant
scalar integrals and to the amplitude. Second, an effective
field theory is constructed which formalizes the findings
from the MoR analysis to all orders in perturbation theory.
The relevant framework for the description of the
high-energy structure of the amplitude can be obtained using
Soft-Collinear Effective Field Theory (SCET) \cite{Bauer:2000yr, Bauer:2001yt,Bauer:2002nz, Beneke:2002ph, Beneke:2002ni},
which is suitable for processes with multiple collinear directions.
The upshot of this analysis is that it identifies, for $gg\to hh$ amplitudes at
high energies, the origin of the large logarithms
in $m_t^2/s$, which cause the discrepancy in the description of the
amplitude in the OS and $\overline{{\rm{MS}}}$ schemes.
Resummation of these large logarithms at leading-power
leading-logarithmic level is then shown to reduce
the mass scheme dependence of the amplitude at high energy.

\begin{figure}
     \centering
     \begin{subfigure}[b]{0.48\textwidth}
         \centering
         \includegraphics[width=\textwidth]{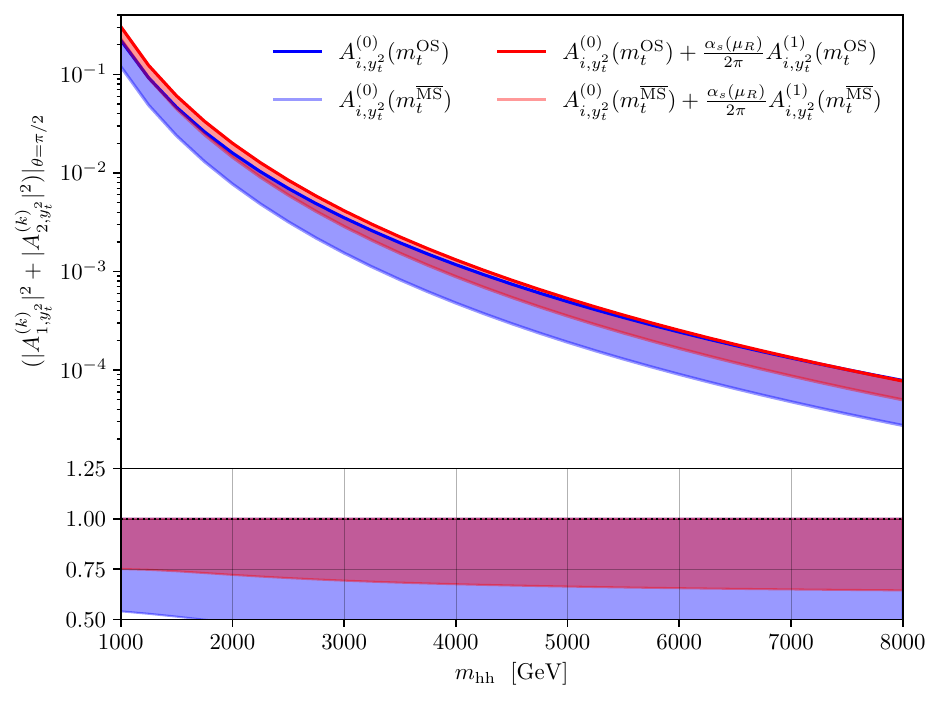}
         \caption{ }
         \label{sfig:all_sq_msbar_vs_os}
     \end{subfigure}
     \hfill
     \begin{subfigure}[b]{0.48\textwidth}
         \centering
         \includegraphics[width=\textwidth]{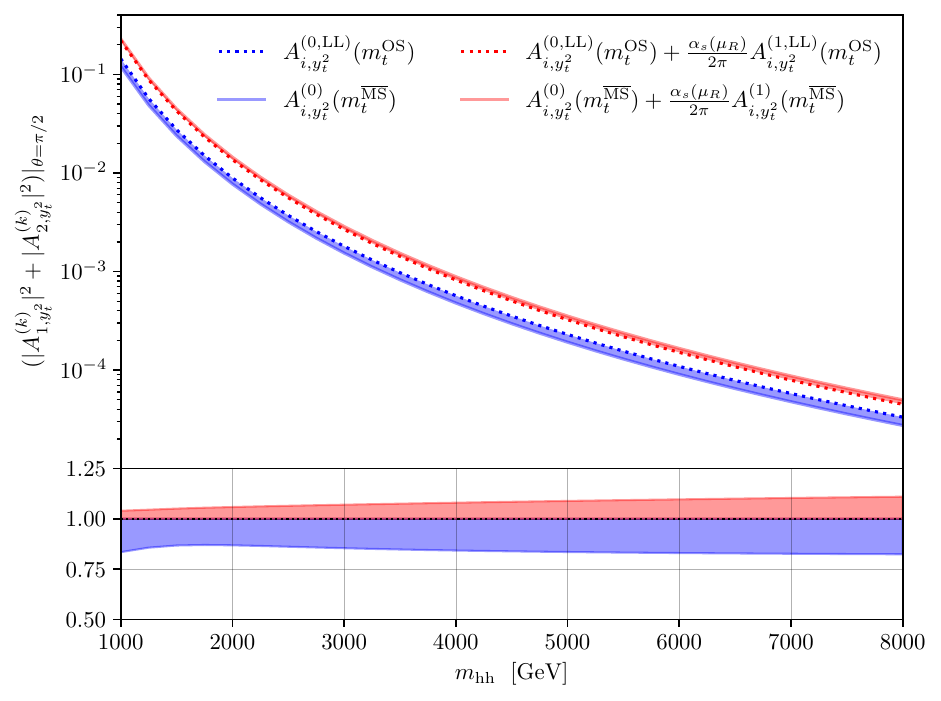}
         \caption{}
         \label{sfig:all_sq_msbar_vs_osll}
     \end{subfigure}
        \caption{Comparison of results for the sum of the squared form factors at one- and
        two-loop order, where the top quark mass is renormalized in the $\overline{\mathrm{MS}}$
        and ${\mathrm{OS}}$ (see panel (a)), and the same quantities with the ${\mathrm{OS}}$
        result supplemented by the resummed tower of leading-power leading logarithms
        (see panel (b)). 
        A significant reduction in the size of the uncertainty due to the choice of the
        top quark mass renormalization scheme is observed. Results shown are from Ref.~\cite{Jaskiewicz:2024xkd}.}
        \label{fig:all_sq_msbar_vs_osll}
\end{figure}

The result of the investigations in Ref.~\cite{Jaskiewicz:2024xkd}
is that the all-order structure of the small top quark mass
power-expanded $y_t^2$ box contribution to the $gg \rightarrow hh$
amplitude can be written as follows in the $\overline{\mathrm{MS}}$ scheme
\begin{align}\label{eq:schem-lo}
& \mathrm{LO}: & &\alpha_s y_t^2 ( \textcolor{blue}{\boldsymbol{c_0}} + m_t \, n_0 ), & \\
& \mathrm{NLO}: & &\alpha_s^2 y_t^2  ( \textcolor{teal}{\boldsymbol{a_1 l_\mu}} + \textcolor{blue}{\boldsymbol{c_1}} + m_t \, n_1  ), & \\
& \mathrm{NNLO}: & &\alpha_s^3 y_t^2  ( \textcolor{teal}{\boldsymbol{a_2 l_\mu^2}} + \textcolor{orange}{\boldsymbol{b_2} \boldsymbol{l_m}} + \textcolor{blue}{\boldsymbol{c_2}} + m_t \, n_2 ), & \\
& \mathrm{N}^3\mathrm{LO}: & &\alpha_s^4 y_t^2 ( \textcolor{teal}{\boldsymbol{a_3 l_\mu^3}} + \textcolor{orange}{\boldsymbol{b_3} \boldsymbol{l_m^2}} + \textcolor{purple}{\boldsymbol{d_3 l_m}} + \textcolor{blue}{\boldsymbol{c_3}} + m_t \, n_3 ), & \\
& \mathrm{N}^i\mathrm{LO}: &
&\alpha_s^{i-1} y_t^2 ( \textcolor{teal}{\boldsymbol{a_i l_\mu^i}} + \textcolor{orange}{\boldsymbol{b_i} \boldsymbol{l_m^{i-1}}}
+   \textcolor{purple}{\boldsymbol{
 d_i l_m^{i-2}}} + \ldots + \textcolor{blue}{\boldsymbol{c_i}} + m_t n_i ),&\label{eq:schem-nklo}
\end{align}
where $l_\mu = \ln(\mu_t^2/s)$ and $l_m$ contains both $\ln(\mu_t^2/s)$ and $\ln(m_t^2/s)$ logarithms. 
The $\textcolor{teal}{\boldsymbol{a_i l_\mu^i}}$ terms are the leading-power LLs, and they are known from the renormalization group running of the top quark mass.
The $\textcolor{orange}{\boldsymbol{b_i} \boldsymbol{l_m^{i-1}}}$ terms are the leading-power NLLs. These terms receive a contribution from the running of the top quark mass and from IR matching (massification)~\cite{Penin:2005eh, Mitov:2006xs, Becher:2007cu, Liu:2017axv, Engel:2018fsb, Wang:2023qbf, Wang:2024pmv}. The leading-power LLs depend on $\mu_t$, which can be set to the order of $\sqrt{s}$. This renders the explicit logarithms $l_\mu$ small. The dependence on the
large ratio of scales, $m_t^2/s$, is taken care of to all orders in the perturbative expansion implicitly through the running of the top quark mass. What drives the discrepancy discussed at NLO in Section~\ref{sec:NLO-uncertainties} is the fact that the conversion between the two schemes is truncated at NLO, so only the first logarithm in $m_t^2/s$ is captured in the OS scheme. In the study of Ref.~\cite{Jaskiewicz:2024xkd}, it was demonstrated that the origin of this
tower of logarithms is in fact known, and can be reinstated in the OS result
through
\begin{align}
A^{(j,\,{\text{LL}})}_{i,y_t^2}(m_t^\mathrm{OS}) = \left(\frac{m^{\text{LL}}(\mu_t)}{m_t^\mathrm{OS}} \right)^2 A^{(j)}_{i,y_t^2}(m_t^\mathrm{OS}), \label{eq:a_ll}
\end{align}
where
\begin{align}\label{eq:schmeme-conv-all-order}
    &m^{\text{LL}}(\mu) = M \exp \left[ a_{\gamma_m}^{\text{LL}}(\mu)\right]\, z_m(M),&
    &a_{\gamma_m}^{\text{LL}}(\mu) = \frac{3 C_F}{2 \beta_0} \ln \left(1 - \frac{\alpha_s(\mu)}{2 \pi} \beta_0 \ln \left(\frac{\mu^2}{M^2} \right) \right).&
\end{align}
The effect of including this tower of leading-power leading logarithms in the OS
result is demonstrated in Fig.~\ref{fig:all_sq_msbar_vs_osll}. As discussed above,
this study provides the first step in taming the top quark mass scheme uncertainty.
Thus far, the improvement can be seen at the level of the $gg\to hh$ amplitude in the
high-energy limit. Future studies are needed to tackle this sizeable uncertainty
in different regions of phase space and at the cross section level.

In future studies, it will be important to extend this analysis
from the amplitude level to physical observables, such as a
cross section, and beyond the high-energy limit. In order to
address the top quark mass scheme uncertainties at the total
cross section level, it is important to develop an understanding
of the uncertainty near the peak of the invariant mass distribution,
i.e.\ the region $300 \mathrm{GeV} < \sqrt{s} < 800 \mathrm{GeV}$,
which is a region that is not covered by the analysis carried out in
\cite{Jaskiewicz:2024xkd}.
One way of tackling this issue is to work out fully the subleading power
contributions, extending the region of validity of the high-energy framework.
For this, the basis has been developed in~\cite{Jaskiewicz:2024xkd}; however,
the factorization structure at subleading powers is much richer than the leading-power
counterpart, requiring, for example, the treatment of endpoint
divergences~\cite{Beneke:2020ibj,Beneke:2022obx,Liu:2019oav,Bell:2022ott}. Hence, it will be
interesting from both a theoretical and phenomenological point of view
to work on extending the framework of~\cite{Jaskiewicz:2024xkd} to next-to-leading power.
Alternatively, it can also be useful, if not necessary, to work on developing the all-order understanding of the structure of top quark mass corrections in a different expansion variable. For instance, it would be interesting to consider the HTL, the threshold expansion at $s \sim 4 m_t^2$, or the small-$p_T$ expansion.

\section{Electroweak corrections in the Standard Model}
\label{sec:ew-sm}

After the inclusion of the higher-order QCD corrections and top quark mass effects discussed above, electroweak effects become a relevant ingredient of a precision prediction for Higgs boson pair production. They probe parts of the Standard Model dynamics that are complementary to QCD, including the top Yukawa interaction, weak gauge-boson contributions, and light-quark initiated channels. In this section we summarize the recent computation of the complete NLO electroweak corrections to the gluon-fusion process, compare it with separately studied subsets of the two-loop contributions, and discuss the small but shape-dependent quark-antiquark channel in the LHC setup relevant for this report.

\subsection{Full SM electroweak corrections}

\label{EW:full}

\textbf{Huan-Yu Bi, Li-Hong Huang, Rui-Jun Huang, Yan-Qing Ma, Huai-Min Yu~\cite{Bi:2023bnq}.}

\noindent Ref.~\cite{Bi:2023bnq} identifies 116 integral families for the complete NLO EW corrections. The loop integrals in each family are reduced to master integrals using the {\tt Blade} package \cite{Blade}, employing {\tt FiniteFlow} \cite{Peraro:2019svx} to solve integration-by-parts (IBP) relations \cite{Chetyrkin:1981qh}, and incorporating a block-triangular form \cite{Liu:2018dmc,Guan:2019bcx} to enhance computational efficiency. Dimensional regularization is realized by choosing $\epsilon=\pm1/1000$, where $D=4-2\epsilon$. This eliminates the need for Laurent expansions in $\epsilon$ for computing the cross sections and thus reduces resource demands \cite{Liu:2022chg,Liu:2022mfb}. Master integrals are evaluated by numerically solving systems of differential equations \cite{Kotikov:1990kg,Remiddi:1997ny,Caffo:2008aw,Czakon:2008zk,Lee:2014ioa,Moriello:2019yhu,Hidding:2020ytt,Armadillo:2022ugh} in $\hat{s}$ and $\hat{t}$,  with boundary conditions computed using the {\tt AMFlow} package \cite{Liu:2022chg} implementing the auxiliary mass flow method \cite{Liu:2022mfb,Liu:2021wks,Liu:2017jxz}. The analytic continuation is performed by introducing an infinitesimal positive imaginary part to $\hat{s}$ to cross physical singularities arising in the master integrals, located at $\sqrt{\hat{s}}=2m_h$, $m_W+m_t$, $2m_t$, $2m_t+m_Z$, and $2m_t+m_h$, corresponding to intermediate particles going on-shell. The on-shell scheme for masses and fields is employed for the renormalization of the bare amplitudes, while the electromagnetic coupling $\alpha$ is renormalized in the $G_{\mu}$ scheme \cite{Denner:2019vbn}.

The cross section is obtained by integrating over the phase space,
\begin{align}\label{eq:Xsection}
\sigma^{\rm LO(NLO)} &= \frac{1}{512\pi}\int_0^1 \mathrm{d}x_{1}\int_0^1 \mathrm{d}x_{2} \int^{\hat{t}_{+}}_{\hat{t}_{-}}\mathrm{d}\hat{t} \,\times \nonumber \\
& \times \frac{1}{\hat{s}^{2}}f_{g/p}(x_{1},\mu)f_{g/p}(x_{2},\mu) M^{\rm LO(NLO)},
\end{align}
where $f_{g/p}(x,\mu)$ denotes the gluon distribution function of the proton, $\mu$ represents the factorization scale, $\hat{t}^{\pm}=m^{2}_{H}-\frac{\hat{s}}{2}(1\mp\sqrt{1-{4m^{2}_{H}}/{\hat{s}}})$, and $M^{\rm LO(NLO)}$ are given by
\begin{align}
M^{\rm LO}&=|F_{1}^{(0)}|^2+|F_{2}^{(0)}|^2 \;,\\
M^{\rm NLO}&=|F_{1}^{(0)}+F_{1}^{(1)}|^2-|F_{1}^{(1)}|^2\nonumber\\
&+|F_{2}^{(0)}+F_{2}^{(1)}|^2  -|F_{2}^{(1)}|^2,
\end{align}
where $F_i^{(0)}$ and $F_i^{(1)}$ correspond to the lowest order and the next-order terms in the $\alpha$ expansion of the form factors. The phase-space integration is carried out by optimized sampling techniques implemented through the {\tt Parni} package \cite{vanHameren:2007pt}.

The masses of the particles are set to
\begin{eqnarray}
\frac{m_h^2}{m_t^2}=\frac{12}{23}
\,,\qquad
\frac{m_Z^2}{m_t^2}=\frac{23}{83}
\,\qquad
\frac{m_W^2}{m_t^2}=\frac{14}{65}
\,,
\end{eqnarray}
with $m_t=172.69\,\textup{GeV}$ \cite{ParticleDataGroup:2022pth}. The electromagnetic coupling $\alpha=1/133.12=7.512\times 10^{-3}$ is derived from the Fermi constant $G_F=1.166378\times 10^{-5}\,\textup{GeV}^{-2}$ via
\begin{eqnarray}
\alpha=\frac{\sqrt{2}}{\pi}G_F m_W^2 \left(1-\frac{m_W^2}{m_Z^2}\right)
.
 \end{eqnarray}
The Cabibbo--Kobayashi--Maskawa (CKM) mixing matrix is set to be the unit matrix. The default PDF set for both LO and NLO calculations is NNPDF3.1 \cite{NNPDF:2017mvq}, in particular \texttt{NNPDF31\ttus nlo\ttus as\ttus 0118}. The running of the strong coupling $\alpha_s$ is taken with two-loop accuracy as provided by the {\tt LHAPDF6} library \cite{Buckley:2014ana}, including five active flavours. The default renormalization and factorization scales are $\mu=m_{hh}/2$, where $m_{hh}$ is the invariant mass of the produced Higgs boson pair.

We generate $1.8\times 10^4$ events at LO and then these events are reweighted to NLO. This enables us to calculate the $\cal{K}$ factors for the total and differential cross sections. We compute an additional 400 reweighted events per bin to reduce the statistical uncertainties for the $m_{hh}$ and $p_T$ distributions.

\begin{table}
\begin{center}
\begin{tabular}{ccccccc}
\toprule
$\mu$  & $m_{hh}/2$  & $\sqrt{p_{T}^2+m_h^2}$ & $m_h$ \\
\midrule
LO   & $19.96(6)$   & $21.11(7)$  & $25.09(8)$ \\
NLO  & $19.12(6)$   & $20.21(6)$  & $23.94(8)$ \\
${\cal K}$-factor  & $0.958(1)$   &     $0.957(1)$   &   $0.954(1)$     \\
\bottomrule
\end{tabular}
\caption{LO and NLO ${\rm EW}$-corrected integrated cross sections (in $\textup{fb}$) with $\sqrt{s}=14\,\textup{TeV}$ based on $1.8\times 10^4$ reweighted events. The uncertainties arise from statistical errors in phase space integration.}
\label{total-cs}
\end{center}
\end{table}

Table~\ref{total-cs} shows LO and NLO EW-corrected total cross sections at $\sqrt{s}=14\,\textup{TeV}$ for varying $\mu$. A difference of about $20\%$ arises for both LO and NLO cross sections from $\alpha_s$ and PDF sensitivity, which can be mitigated by higher-order QCD corrections \cite{Jones:2023uzh}. In contrast, the $\cal{K}$ factor ranges from 0.954 to 0.958 and remains stable, indicating that the EW corrections approximately factorize from the QCD corrections.

To assess the EW uncertainties, we investigate the EW corrections in the $G_{\mu}$, $\alpha_0$, and $\alpha_{m_Z}$ renormalization schemes for the electromagnetic coupling \cite{Denner:2019vbn,Sang:2024vqk}. Table~\ref{total-cs2} presents the results based on these three different schemes. We find that the scale uncertainties (or scheme uncertainties) quantified by
\begin{eqnarray}
\frac{{\rm max}(\sigma_{G_{\mu}},\sigma_{\alpha_0},\sigma_{\alpha_{m_Z}})-
      {\rm min}(\sigma_{G_{\mu}},\sigma_{\alpha_0},\sigma_{\alpha_{m_Z}})}
     {{\rm min}(\sigma_{G_{\mu}},\sigma_{\alpha_0},\sigma_{\alpha_{m_Z}})},
\end{eqnarray}
amount to $13.0\%$ at LO and are further reduced to $0.8\%$ once the EW corrections are included. These results indicate that the dominant scale uncertainties are effectively absorbed at NLO EW.

\begin{table}
\begin{center}

\begin{tabular}{ccccccc}
\toprule
 schemes  & $G_{\mu}$  & $\alpha_0$ & $\alpha_{m_Z}$ \\
\midrule
LO   & $19.96$   & $18.84$  & $21.28$ \\
NLO  & $19.12$   & $19.19$  & $19.03$ \\
${\cal K}$-factor  & $0.958$   &     $1.019$   &   $0.894$     \\
\bottomrule
\end{tabular}
\caption{LO and NLO ${\rm EW}$-corrected integrated cross sections (in $\textup{fb}$) with $\sqrt{s}=14\,\textup{TeV}$ and $\mu=m_{hh}/2$ in the $G_{\mu}$, $\alpha_0$, and $\alpha_{m_Z}$ schemes. The electromagnetic couplings are taken as $\alpha_{G_{\mu}}=1/133.12$, $\alpha_{0}=1/137.035999$, and $\alpha_{m_Z}=1/128.932$, respectively.}
\label{total-cs2}
\end{center}
\end{table}

\begin{figure}
\centering
    \subfloat[]{{\includegraphics[width=0.47\textwidth]{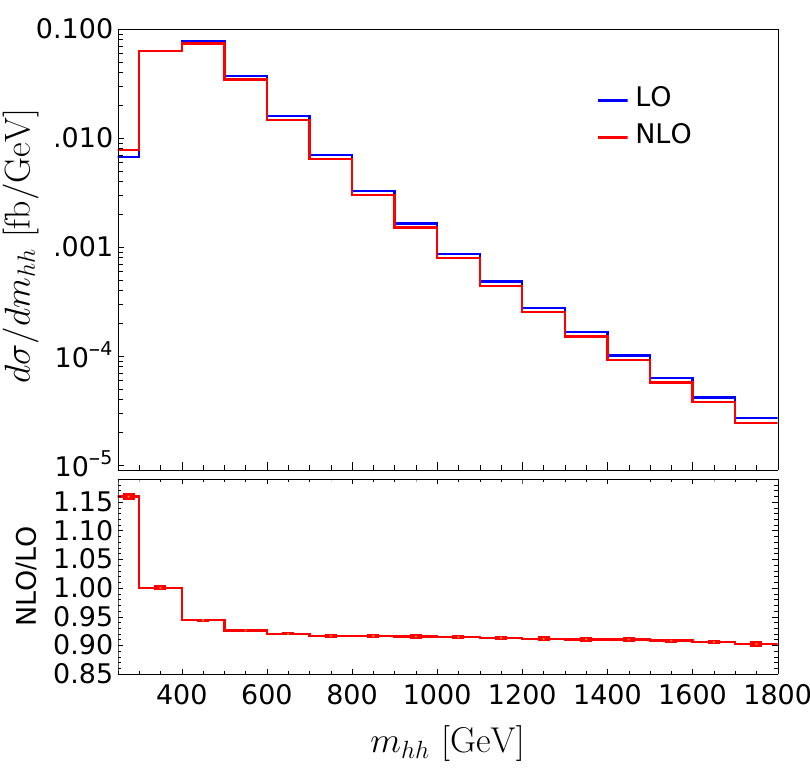}}\label{mhh}}
    \subfloat[]{{\includegraphics[width=0.47\textwidth]{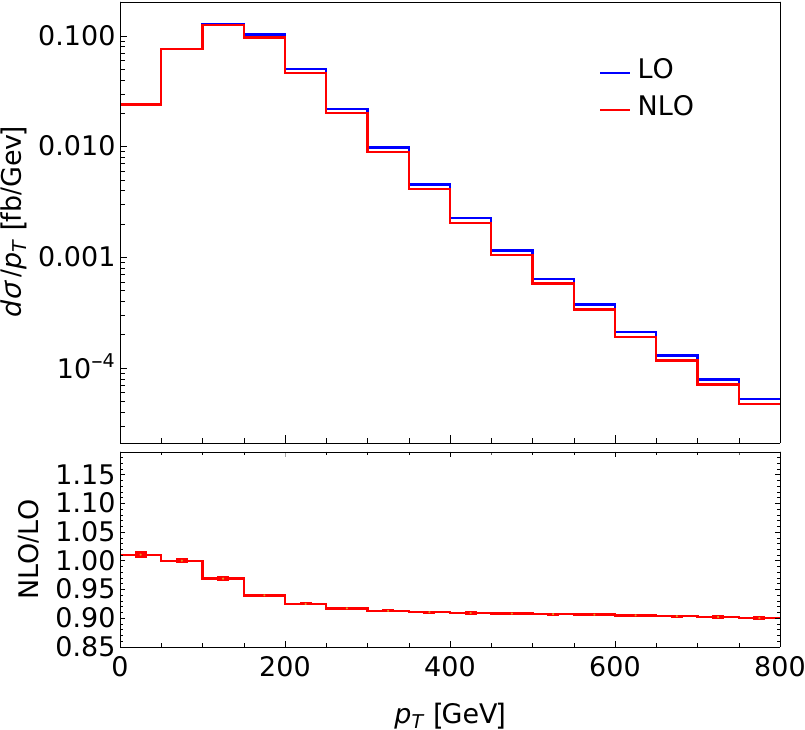}}\label{pth}}
	\\
    \subfloat[]{{\includegraphics[width=0.47\textwidth]{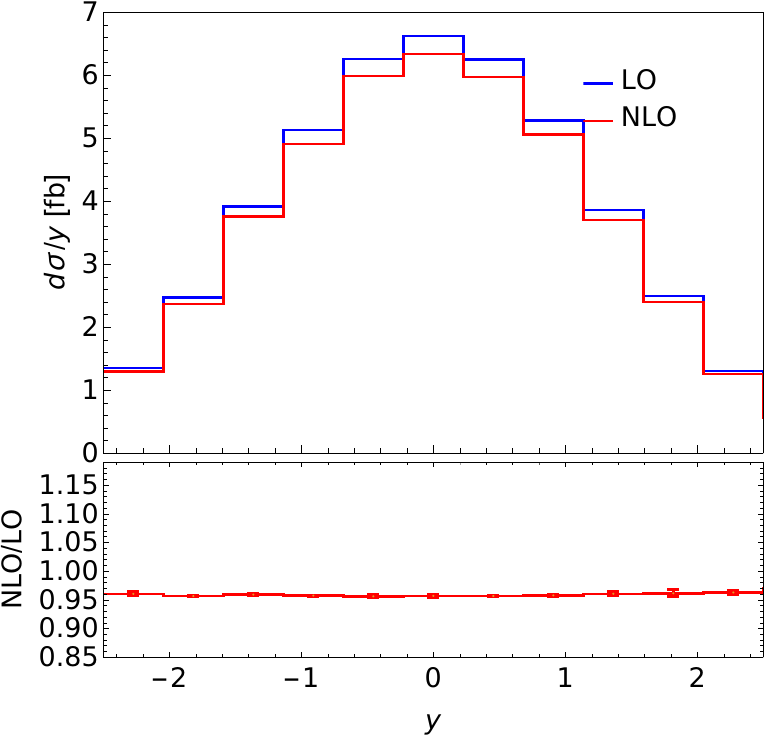}}\label{yh}}
		\caption{Invariant mass distribution of the Higgs boson pair (a), transverse momentum distribution of one of the two Higgs bosons (b), and rapidity distribution of one of the two Higgs bosons (c), all at $\sqrt{s}=14\,\textup{TeV}$ in the SM. In each panel, the upper plot shows absolute predictions and the lower panel displays the differential ${\cal K}$-factor with error bars representing statistical errors. Results shown are from Ref.~\cite{Bi:2023bnq}.}
    \label{fullEWobs}
\end{figure}

The invariant mass distribution of the Higgs boson pair $m_{hh}$ is shown in Fig.~\ref{mhh}. A significant positive correction, of approximately $+15\%$, is observed in the first bin. In fact, we find that the EW correction for phase space points near the $hh$ production threshold can exceed $+70\%$. As $m_{hh}$ increases, the $\mathcal{K}$ factor initially decreases dramatically and then becomes milder farther from the threshold. A similar pattern occurs for the $p_{T}$ distribution in Fig.~\ref{pth}, where the correction is initially positive before turning negative around $100\,\textup{GeV}$. For regions of large $m_{hh}$ or large $p_{T}$, the NLO EW correction is approximately $-10\%$. Note that, while the corrections can reach $-30\%$ for the squared matrix element at $\sqrt{\hat{s}} \approx 14\,\textup{TeV}$, the extreme suppression of the gluon luminosity at high energy makes this impact insignificant.

In Fig.~\ref{yh}, we display the rapidity distribution of one Higgs boson. The $\mathcal{K}$ factor is nearly flat, approximately $0.96$, matching that of the total cross section.

\FloatBarrier
\subsection{Breakdown of individual contributions}

\label{EW:breakdown}

\begin{figure}
\centering
    \includegraphics[width=\textwidth]{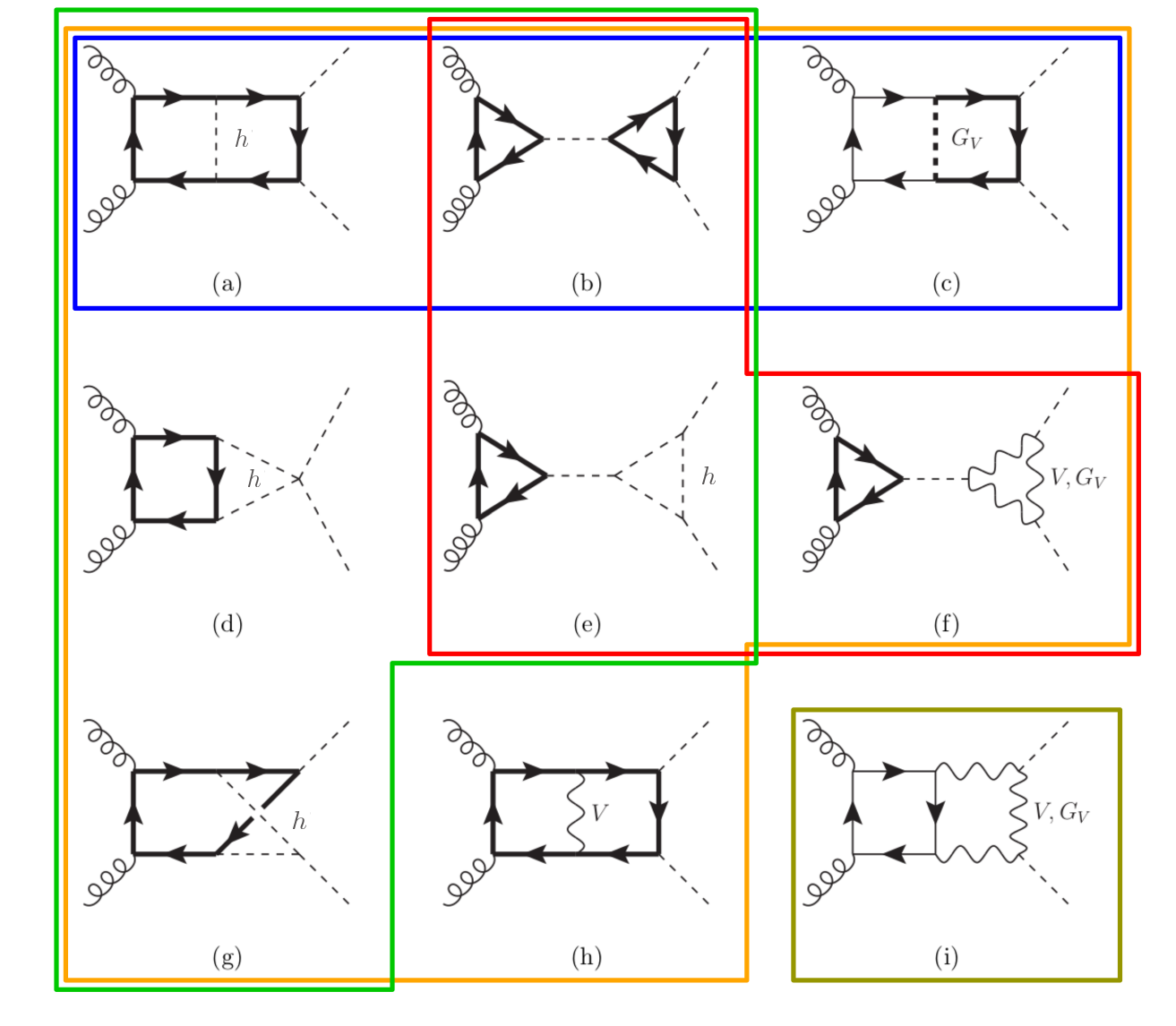}
	\caption{Example diagrams for different contributions. Blue set (a, b, c): top Yukawa-induced electroweak corrections; orange set (a, b, c, d, e, f, g, h): top quark contributions; red set (b, e, f): factorizable contributions; green set (a, b, d, e, g): top Yukawa and Higgs boson self-coupling corrections; olive set (i): light-quark contributions. The results of Section~\ref{EW:full} comprise all the displayed contributions. See Table~\ref{tab:bd}.}
    \label{fig:bd}
\end{figure}

Individual parts of the complete EW corrections to double Higgs boson production in gluon fusion have been computed separately, using a number of different methods.
Fig.~\ref{fig:bd} and Table~\ref{tab:bd} summarize the different contributions.


\begin{table}
\begingroup
\small
\setlength{\tabcolsep}{2pt}
\begin{center}
\begin{tabularx}{\textwidth}{@{}
  >{\raggedright\arraybackslash}p{0.60\textwidth}
  c
  >{\raggedright\arraybackslash}X
@{}}
\toprule
Section & \mbox{Diagram sets} & \mbox{Key references} \\
\midrule
$\imineq{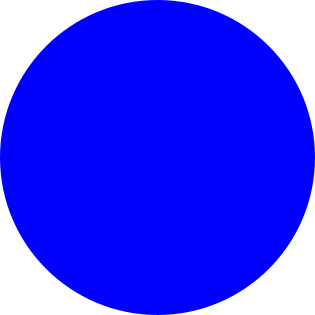}{1}$~Top Yukawa-induced electroweak corrections
  & (a), (b), (c)
  & \cite{Muhlleitner:2022ijf,Bhattacharya_TBA} \\

$\imineq{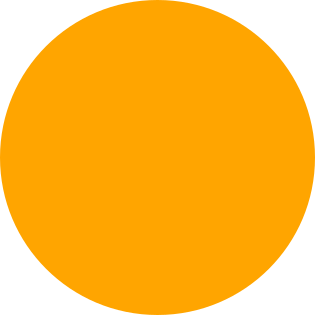}{1}$~Top quark contributions
  & (a) to (h)
  & \cite{Davies:2023npk,Davies:2026wbx} \\

$\imineq{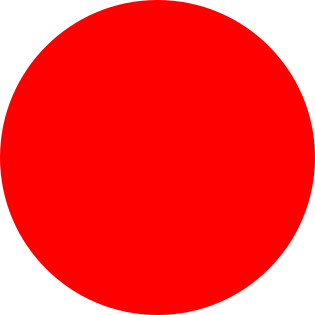}{1}$~Factorizable contributions
  & (b), (e), (f)
  & \cite{Muhlleitner:2022ijf,Zhang:2024rix} \\

$\imineq{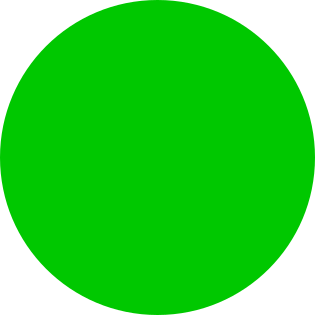}{1}$~Top Yukawa and Higgs boson self-coupling corrections
  & (a), (b), (d), (e), (g)
  & \cite{Davies:2022ram,Davies:2025wke,Heinrich:2024dnz} \\

$\imineq{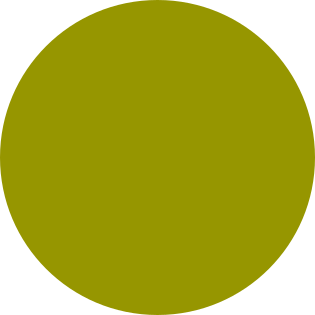}{1}$~Light-quark contributions
  & (i)
  & \cite{Bonetti:2025vfd,Bhattacharya_TBA} \\
\bottomrule
\end{tabularx}
\caption{Breakdown of the individual EW contributions to $gg \to hh$ at two loops; see Fig.~\ref{fig:bd}. The results of Section~\ref{EW:full} correspond to the sum of all terms from (a) to (g).}
\label{tab:bd}
\end{center}
\endgroup
\end{table}

\subsubsection{Top Yukawa-induced electroweak corrections}

\label{EW:Top-Higgs_sector}


\noindent \textbf{Milada Margarete Mühlleitner, Johannes Schlenk, Michael Spira~\cite{Muhlleitner:2022ijf}.}

\begin{figure}[!hbtp]
\centering
\includegraphics[width=0.9\textwidth]{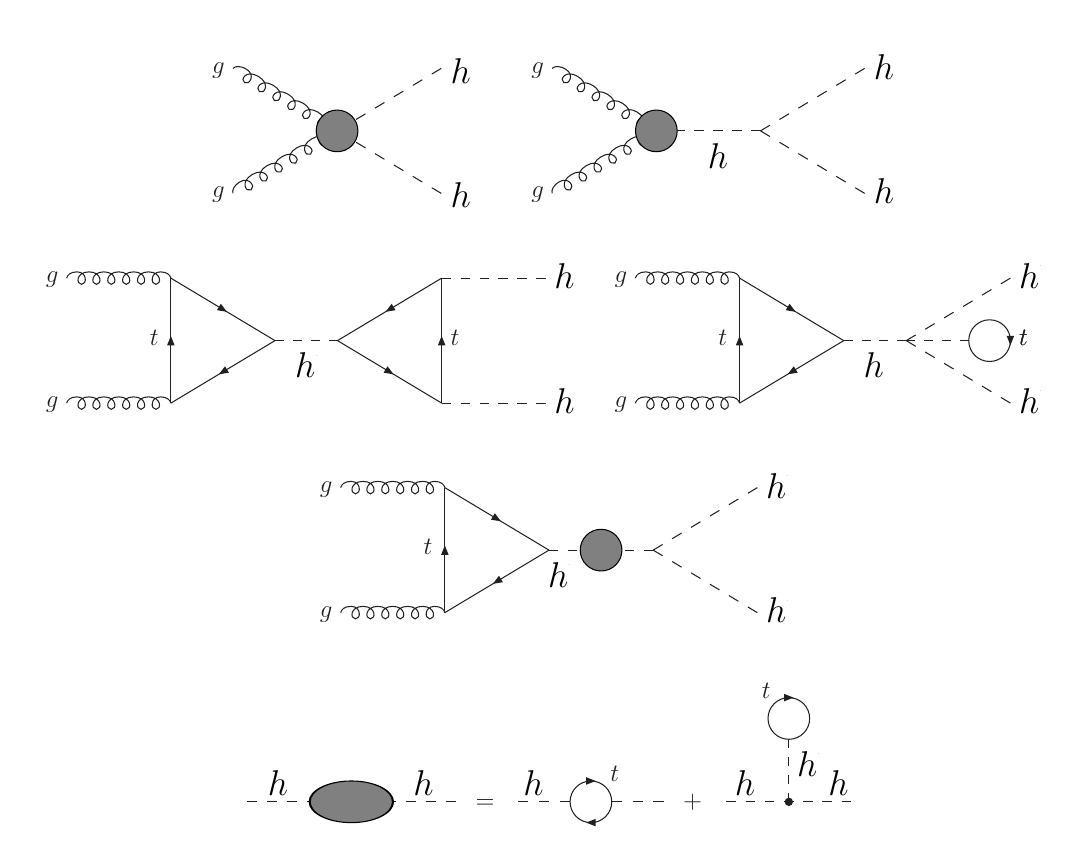}
\caption{\label{fg:dia_htl}Generic diagrams contributing to the top Yukawa-induced electroweak corrections to the $gg\to hh$ process. The first two diagrams have been treated in the HTL. Diagrams adapted from Ref.~\cite{Muhlleitner:2022ijf}.}
\end{figure}

\noindent The first calculation of partial electroweak corrections to Higgs boson pair production via the dominant gluon-fusion process has been performed by investigating the corrections induced by the top Yukawa coupling \cite{Muhlleitner:2022ijf}. In order to set up a complete and consistent framework, these corrections have been obtained in the gaugeless limit, where all electroweak gauge couplings are set to zero, but the (massless) Goldstone contributions are taken into account, see Fig.~\ref{fig:bd}. This approach ensures the preservation of the $SU(2)$ symmetry of the scalar Higgs doublet. While the top-induced corrections to the trilinear Higgs boson self-coupling vertex are treated with the full top quark mass dependence, the triangle and box diagrams involving couplings of one or two Higgs bosons to gluons are treated in the HTL. The corresponding diagrams are displayed in Fig.~\ref{fg:dia_htl}, where the first two represent those that are treated in the HTL. For these two diagrams, we have used the effective Lagrangian \cite{Djouadi:1994ge,Muhlleitner:2022ijf}
\begin{equation}
{\cal L}_{eff} = \frac{\alpha_s}{12\pi} G^{a\mu\nu} G^a_{\mu\nu} \left\{
(1+\delta_1) \frac{H}{v} + ( 1 + \eta_1) \frac{H^2}{2v^2} + {\cal
O}(H^3) \right\}
\label{eq:leff}
\end{equation}
where
\begin{eqnarray}
\delta_1 & = & \frac{x_t}{2} + {\cal O}(x_t^2) \,, \qquad\qquad
\eta_1   = 4 x_t + {\cal O}(x_t^2) \,,
\label{eq:leff_coeff}
\end{eqnarray}
Here $x_t = G_F m_t^2/(8\sqrt{2}\pi^2)$, $G_F$ denotes the Fermi constant, $v$ is the vacuum expectation value (vev) of the Higgs field, and $m_t$ is the top quark mass. This effective Lagrangian describes the electroweak corrections induced by $x_t$ to the $hgg$ and $hhgg$ vertices in the HTL. We would like to point out explicitly that the square root of the wave-function counterterm of the external Higgs boson(s) is already taken into account in this effective Lagrangian.
In Fig.~\ref{fg:dia_htl} the tadpole diagrams are displayed explicitly. We have used both the Fleischer--Jegerlehner scheme and the conventional scheme, in which the tadpole diagrams are omitted but retained in the corresponding counterterm of the trilinear Higgs boson self-coupling. We have obtained identical results in the final sum of the radiative corrections, once the vev $v$ is expressed in terms of the Fermi constant $G_F$.

Because the full top quark mass dependence of the one-loop times one-loop diagrams describing the radiatively corrected trilinear Higgs boson self-coupling vertex (all diagrams of Fig.~\ref{fg:dia_htl} except the first two) has been kept, the use of the effective trilinear Higgs boson self-coupling (derived from the effective Coleman--Weinberg potential \cite{Coleman:1973jx,Weinberg:1973ua,Jackiw:1974cv})
\begin{equation}
\lambda_{hhh} = 3 \frac{m_h^2}{v} \, , \qquad\qquad \lambda_{hhh}^{\rm eff} = \lambda_{hhh} - \frac{3 m_t^4}{\pi^2 v^3} \approx 0.91 \times \lambda_{hhh}
\end{equation}
to accommodate the dominant part of the radiative corrections can be tested. The result is that process-dependent, finite-momentum-dependent corrections are of the same size as the corrections above, growing with $m_t^4$,
\begin{eqnarray}
\sigma & = & K_{elw} \times \sigma_{LO} \nonumber \\
K_{elw} & \approx & 1.002 \hspace*{1.0cm} \mbox{($\lambda_{hhh}$)} \nonumber \\
K^{eff}_{elw} & \approx & 0.938 \hspace*{1cm} \mbox{($\lambda^{\rm eff}_{hhh}$)}.
\end{eqnarray}
Since the effective trilinear Higgs boson self-coupling generates about $6\%$ electroweak corrections to the total cross section artificially and larger corrections to the distributions, using the LO-like trilinear Higgs boson self-coupling $\lambda_{hhh}$ is favoured \cite{Muhlleitner:2022ijf}. The top Yukawa-induced electroweak corrections are small except in the region close to the production threshold due to the complete cancellation of the leading top quark mass terms in the LO matrix element, to which the corrections are normalized.

~

\noindent \textbf{Arunima Bhattacharya, Francisco Campanario, Sauro Carlotti, Jamie Chang, Javier Mazzitelli, Milada Margarete Mühlleitner, Jonathan Ronca, Michael Spira ~\cite{Bhattacharya_TBA}.}

\noindent The analysis of the top Yukawa-induced electroweak corrections discussed above has been extended to include the full top quark mass dependence. Specifically, the first two diagrams of Fig.~\ref{fg:dia_htl}, previously evaluated in the HTL, have been resolved into the corresponding two-loop diagrams involving top, bottom, Higgs, and Goldstone propagators. To preserve the SU(2) symmetry of the Higgs sector, this work has been performed in the gaugeless limit, enabling a gauge-invariant definition of the contributions associated with the top Yukawa coupling. In this limit, the Goldstone bosons are massless but do not generate infrared singularities. The method used for this calculation involves applying the projectors to the two form factors (in $D=4-2\epsilon$ dimensions), followed by a diagram-by-diagram Feynman parametrization. Since no tensor reduction has been used, the resulting Feynman integrands remain quite compact. The singularities were extracted by appropriate end-point subtractions. Moreover, this allowed us to keep, alongside the top quark mass, the bottom and the Higgs masses as free symbolic parameters. The virtual thresholds have been treated by introducing complex masses of the virtual propagators,
\begin{equation}
m_{t/b}^2 \to m_{t/b}^2 (1-i\bar\epsilon) \, , \qquad m_h^2 \to m_h^2 (1-i\bar\epsilon)
\end{equation}
with the regulator $\bar\epsilon$ to be sent to 0. The latter was achieved by applying a Richardson extrapolation \cite{Richardson} to a series of different values of $\bar\epsilon$. The minimal value of $\bar\epsilon$ used for numerical integrations varied between $10^{-8}$ and 0.1 depending on the diagram and the value of the invariant Higgs boson pair mass $Q=m_{hh}$. The top quark mass has been renormalized on-shell, and the vev has been renormalized according to the proper definition of the Fermi constant in muon decay.
\begin{figure}[!hbtp]
    \centering
    \includegraphics[width=0.8\textwidth]{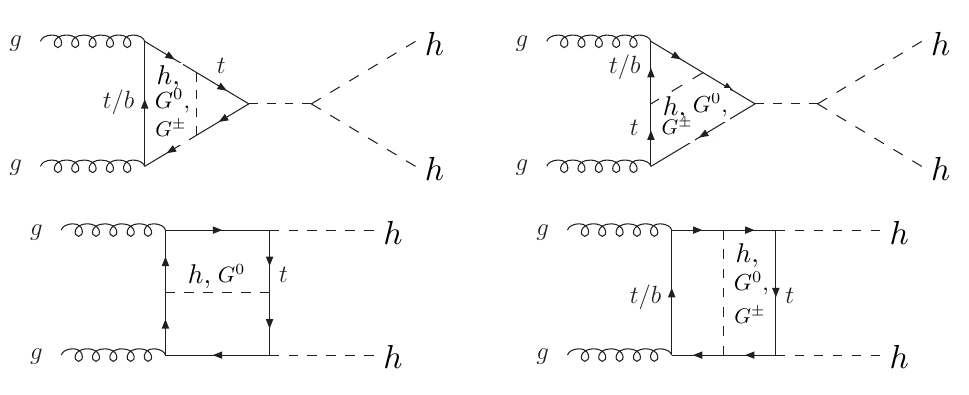}
    \caption{\label{fg:dia_top}Sample two-loop triangle and box diagrams of the top Yukawa-induced electroweak corrections to Higgs boson pair production involving Higgs $h$ and Goldstone $G^0,G^\pm$ exchanges. The bottom propagators only contribute to the diagrams with charged Goldstone exchange. Diagrams from Ref.~\cite{Bhattacharya_TBA}.}
\end{figure}
\begin{figure}[!hbtp]
    \centering
    \includegraphics[width=0.7\textwidth]{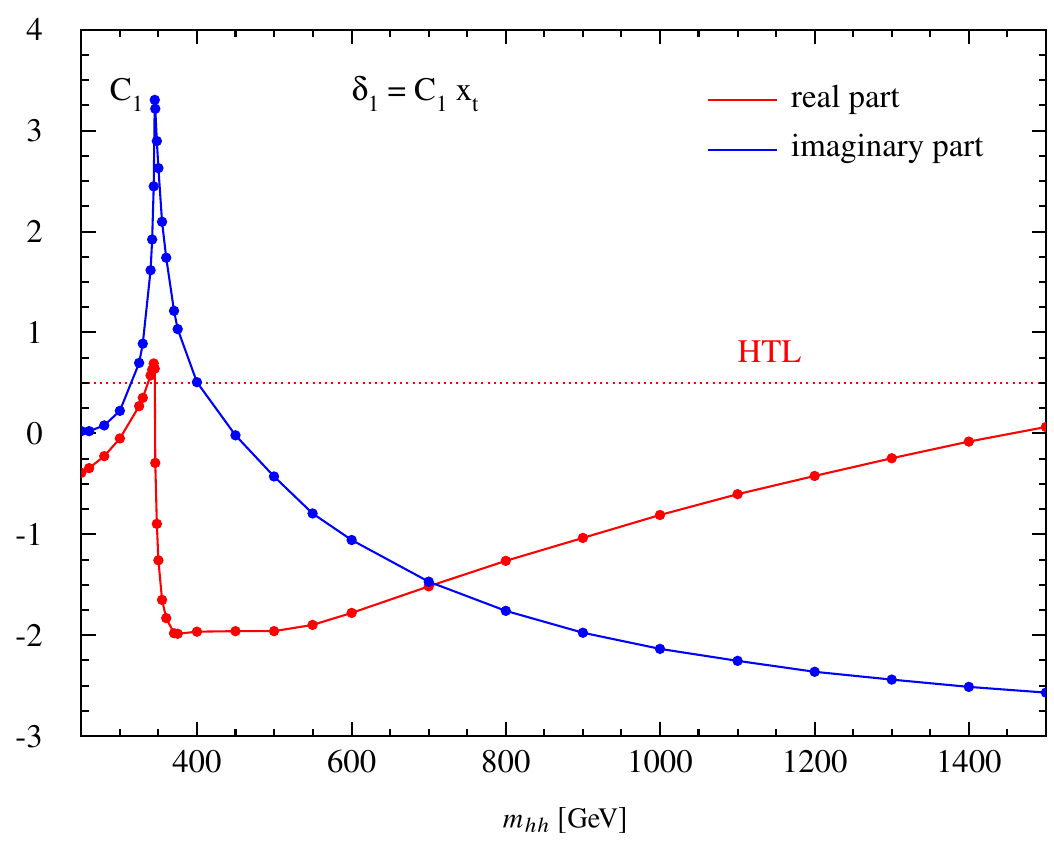}
    \caption{\label{fg:delta_1}The full top quark mass dependence of the coefficient $\delta_1$ of Eq.~\eqref{eq:leff_coeff} as a function of the invariant Higgs boson pair mass. The dotted line shows the purely real HTL result. The sizeable imaginary part arises from the diagrams with charged Goldstone exchange that include diagrams with $b\bar b$ thresholds. The points on the curve indicate the $m_{hh}$ values for which $\delta_1$ has been computed. The numerical errors are negligible and thus not shown. Results shown are from Ref.~\cite{Bhattacharya_TBA}.}
\end{figure}

The result of the two-loop triangle diagrams, see Fig.~\ref{fg:dia_top}, which can be directly compared to the coefficient $\delta_1$ in Eq.~\eqref{eq:leff_coeff}, is shown in Fig.~\ref{fg:delta_1}. As a cross-check, the HTL result of Eq.~\eqref{eq:leff_coeff} has been reproduced within uncertainties by artificially increasing the value of the top quark mass. The results indicate differences between the HTL and the full result at the (sub)percent level for the distribution in the Higgs boson pair mass.

\begin{figure}[!hbtp]
    \centering
\includegraphics[width=0.49\textwidth]{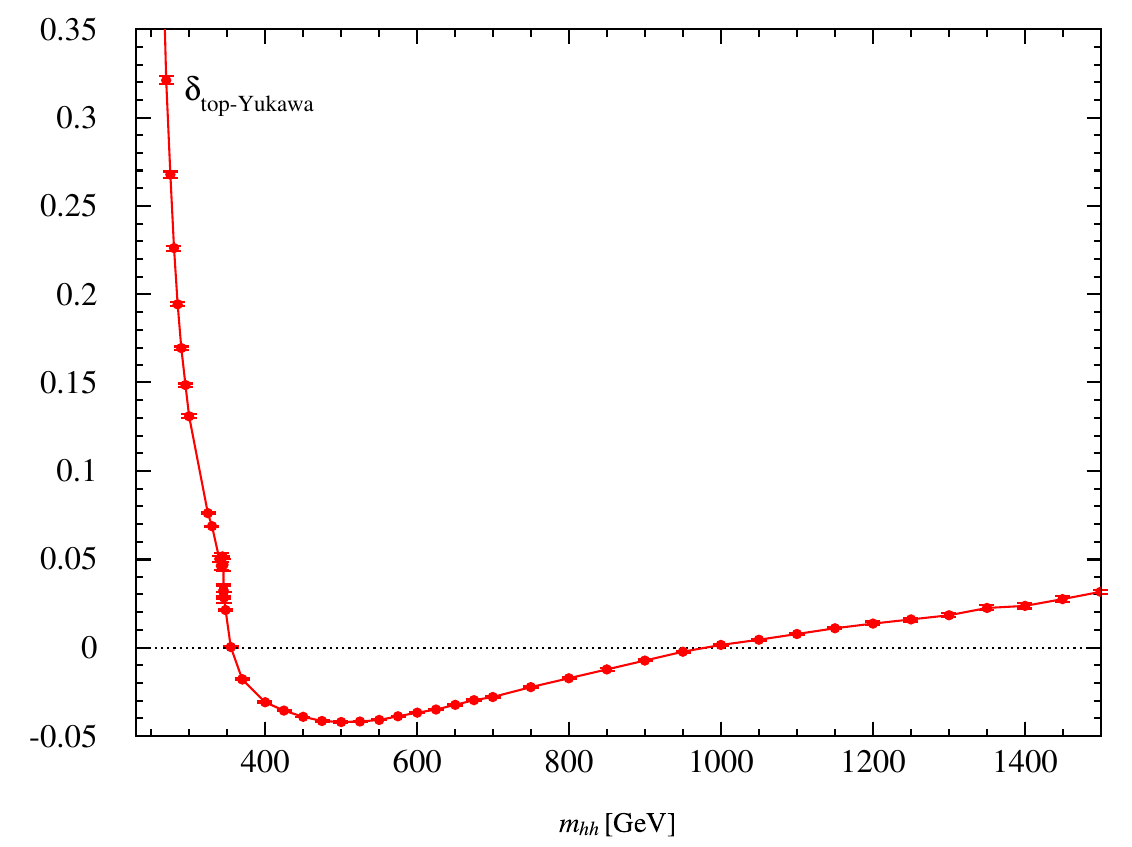}
\includegraphics[width=0.49\textwidth]{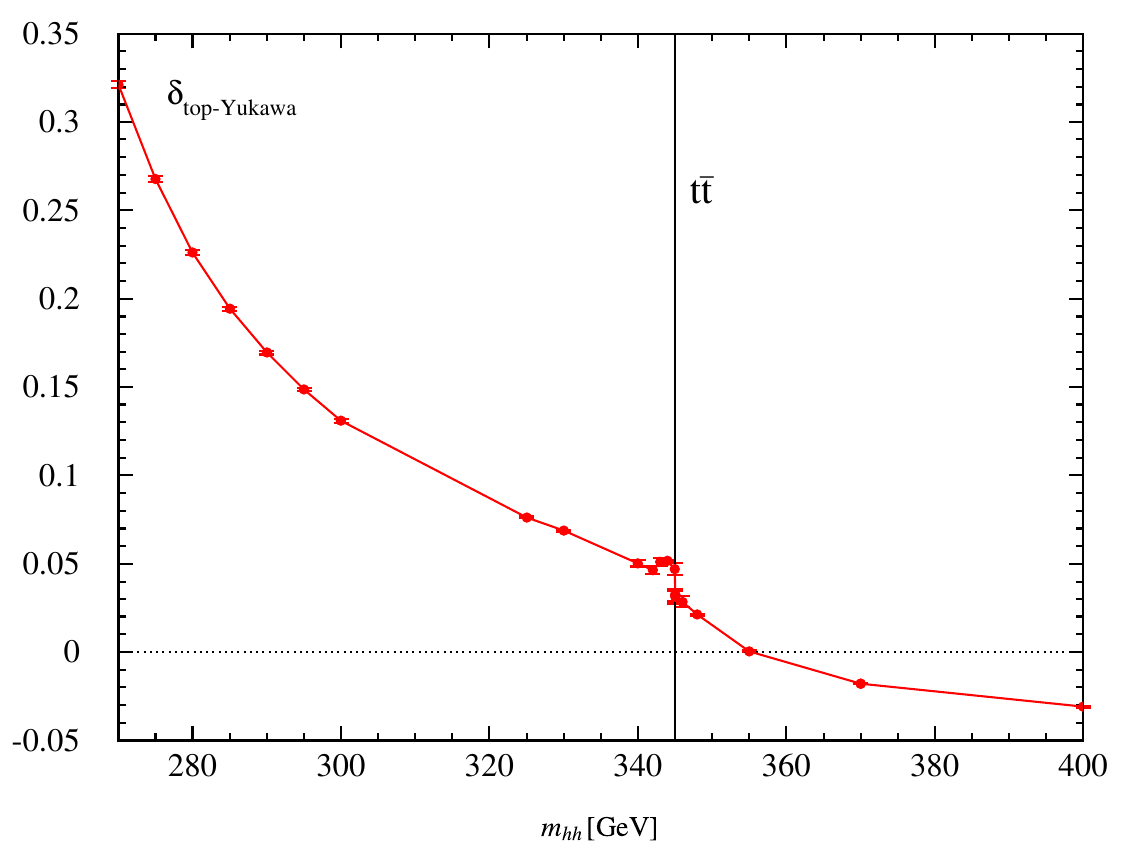}
    \caption{The full relative top Yukawa-induced electroweak corrections to the $gg \to hh$ process. The points on the curve indicate the $m_{hh}$ values for which the corrections have been computed. The right panel shows a magnified view of the structure of the $t\bar t$ threshold. Numerical error bars are shown for each point. The vertical line indicates the position of the $t\bar t$ threshold. Results shown are from Ref.~\cite{Bhattacharya_TBA}.}
    \label{fg:del_top}
\end{figure}

The calculation of the two-loop box diagrams, see Fig.~\ref{fg:dia_top}, is more involved, since the additional phase space dependence does not allow for an immediate comparison with the HTL. Applying the same method as for the two-loop triangle diagrams and including the phase space integration in the numerical analysis, the result of the two-loop renormalized box diagrams plus the two-loop triangle diagrams and the previous one-loop times one-loop diagrams related to the radiative corrections to the trilinear Higgs boson self-coupling vertex is shown in Fig.~\ref{fg:del_top}, where the relative corrections are defined as
\begin{equation}
\sigma (gg\to hh) = \sigma_{LO} (gg\to hh)~( 1 + \delta_{t\mathrm{-Yukawa}} + \delta_{\mathrm{light\,quarks}} )
\end{equation}
with the light-quark contributions $\delta_{\mathrm{light\,quarks}}$ discussed in Section~\ref{EW:light-quark}. The total corrections are about $-5\%$ for intermediate values of $m_{hh}$ and are smaller for large values, while they are large close to the production threshold due to the large cancellation in the LO matrix element, to which the corrections are normalized. At large values of $m_{hh}$, the corrections develop a visible slope. The right panel of Fig.~\ref{fg:del_top} shows details of the $t\bar t$ threshold, which develops a sign-changing interference behaviour with the LO matrix element, with a maximum below the $t\bar t$ threshold.

\subsubsection{Top quark contributions}

\label{EW:top}

\noindent \textbf{Joshua Davies, Kay Schönwald, Matthias Steinhauser, Hantian Zhang~\cite{Davies:2023npk,Davies:2026wbx}.}

\noindent The full top quark contributions, including all sectors of the SM, have been computed analytically in
Refs.~\cite{Davies:2023npk,Davies:2026wbx} in the inverse top quark mass expansion and the high-energy expansion.

In the large-$m_t$ limit~\cite{Davies:2023npk}, expansions up to order $1/m_t^{10}$ have been computed in the Feynman gauge in the limit
\begin{align}
m_t^2 \, \gg \, s, \, |t|, \, m_h^2, \, m_W^2 , \, m_Z^2 \,,
\end{align}
where no additional hierarchy is assumed among the scales on the right-hand side.
The calculation has been analytically validated in the general $R_\xi$-gauge by assuming the hierarchy
\begin{align}
    m_t^2 \, \gg \, \xi_W \, m_W^2, \, \xi_Z \, m_Z^2  \, \gg \, s, \, |t|, \, m_h^2, \, m_W^2 , \, m_Z^2 \,,
\end{align}
where $\xi_W, \xi_Z$ are gauge parameters for the $W$ and $Z$ bosons.
These parameters appear in combination with gauge boson masses in the gauge boson and Goldstone propagators.
It is a valuable and non-trivial analytic validation that $\xi_W$ and $\xi_Z$ drop out in the renormalized amplitudes. 
Note that the gauge parameter for the photon, $\xi_\gamma$, drops out after summing all bare two-loop diagrams.

In the high-energy limit~\cite{Davies:2026wbx},
the electroweak calculation is much more challenging compared to the QCD case and the large-$m_t$ EW case. In this limit, we perform nested expansions in several small parameters in the following hierarchies in the Feynman gauge:
\begin{align}
    s, \, |t| \,  \gg \, m_t^2, \, m_W^2, \, m_Z^2 \,, (m_h^{\rm int})^2 \, \gg \,  (m_h^{\rm ext})^2\,,
\end{align}
where $m_h^{\rm int}$ denotes the mass of Higgs propagators inside the loop and $m_h^{\rm ext}$ is the mass of the final-state Higgs boson.
By identifying these hierarchies, we can first perform an external Higgs mass expansion, followed by expansions in the mass differences among several internal mass parameters as $\delta_X = 1 - m_X/ m_t $ for $ X=W,Z,H$.
Finally and most importantly, we perform a deep high-energy expansion with more than 100 expansion terms around the small-$m_t$ limit.
Schematically, the generic structure of the high-energy expansion of the form factors can be written as
\begin{align}
  F &= \sum_{n=-4}^{108} \sum_{i=0}^{2} \sum_{k_W=0}^{{4}}\sum_{k_Z=0}^{{4}}\sum_{k_H=0}^{{4}} 
        c_{ni}^{k_W k_Z k_H}  \: m_t^{n} \:
        (m^{\rm ext}_H)^{2i} \:
        \delta_W^{k_W} \: \delta_Z^{k_Z} \: \delta_H^{k_H} 
        \,,
\end{align}
where $c_{ni}^{k_W k_Z k_H}$ are coefficient functions of $s,t$ and $\log(m_t^2)$, featuring polylogarithmic and elliptic structures.
These analytic expressions are numerically evaluated through a Pad\'{e} procedure to increase the radius of convergence of the high-energy expansion across a larger phase space region.
The high-energy computation is technically challenging at both the form factor and Feynman integral level, due to drastically increased complexity when the full electroweak sector is taken into account.
We have successfully tackled both complexities and achieved a deep high-energy expansion up to the orders $m_t^{108}$, $(m_{H}^{\rm ext})^4$ and $\delta_X^4$. 
We have systematically analysed the convergence properties of our expansion, and we estimate the truncation uncertainty, dominated by the mass-difference expansion, to be about $\pm 1\%$ of the two-loop corrections.
For technical details of master integral calculations for the electroweak topologies, we refer to Refs.~\cite{Davies:2025wke,Davies:2022ram,Zhang:2024fcu}.

The electroweak renormalization follows the standard procedure as outlined in Refs.~\cite{Denner:1991kt,Denner:2019vbn}. 
The one-loop form factors are expressed in terms of parameters $\{e,m_W,m_Z,m_t,m_h\}$, and the parameter renormalization is performed with corresponding renormalization constants in the on-shell scheme.
The wave functions of the external Higgs bosons are also renormalized in the on-shell scheme.
Throughout the entire calculation, all tadpole contributions are consistently included such that all parameter renormalization constants are gauge-parameter independent, while the gauge-parameter dependence only appears in the wave function renormalization constant for the external Higgs boson.
This prescription is equivalent to the so-called \textit{Fleischer--Jegerlehner tadpole scheme}~\cite{Fleischer:1980ub}.
%

The large-$m_t$ result~\cite{Davies:2023npk} supports the findings of Ref.~\cite{Muhlleitner:2022ijf},
where the top Yukawa-induced effects have been studied, namely that the electroweak corrections near the Higgs boson pair production threshold can lead to a few tens of percent effects with respect to the leading order contribution.
This interesting phenomenon is due to the fact that the electroweak corrections lift the destructive interference near the production threshold, where the matrix element is suppressed by the cancellation between triangle and box form factors at leading order.
Similarly, large top Yukawa-induced corrections near the production threshold have been found in Ref.~\cite{Heinrich:2024dnz}.
Although the radius of convergence is
limited, this result also serves as an important benchmark for other calculations. 
For example, it has been compared with Ref.~\cite{Bi:2023bnq} in a large-$m_t$ limit and good agreement was observed.

\begin{figure}[t]
\centering
    \includegraphics[width=.65\textwidth]{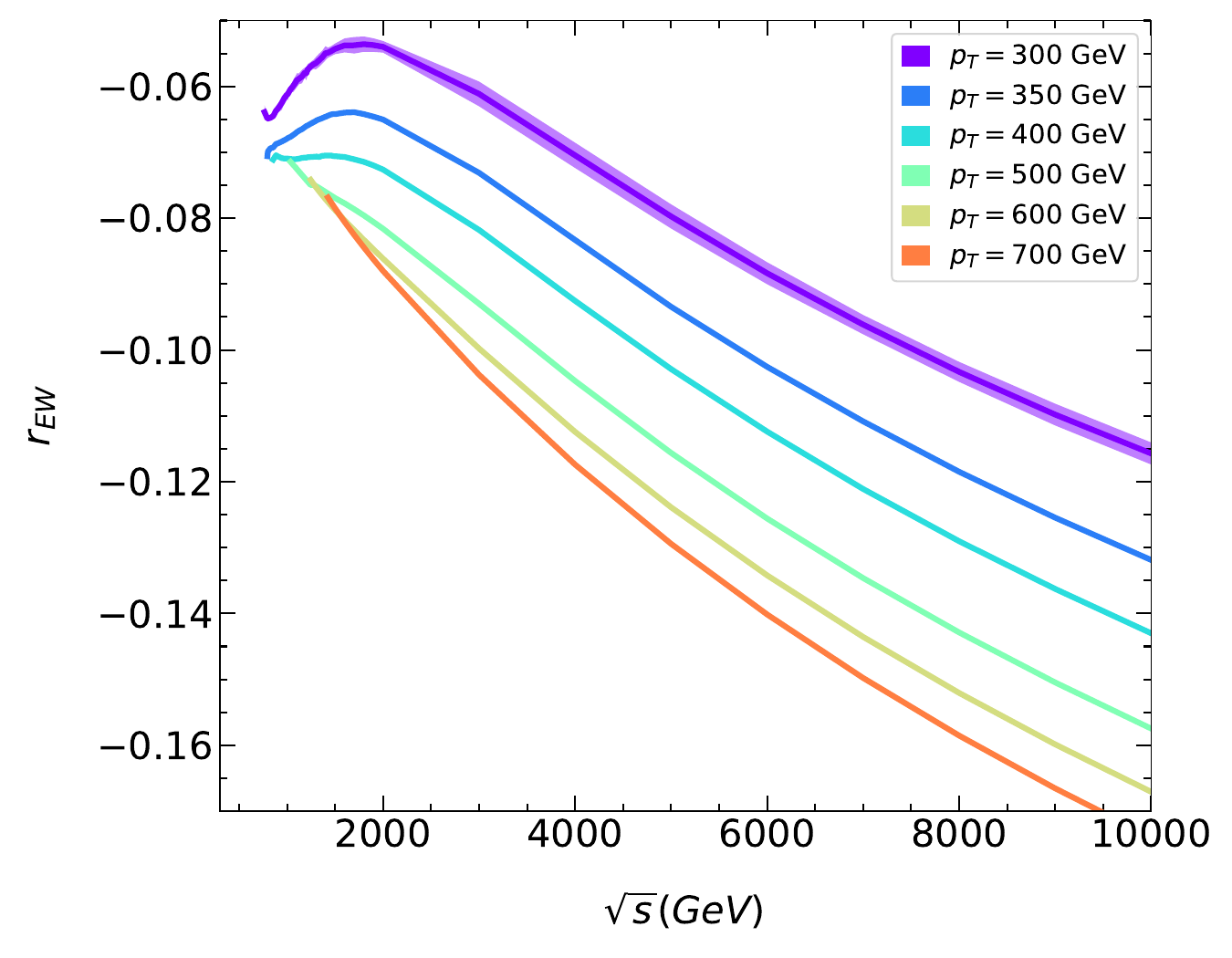}
  \caption{\label{fig::rEW_HE}
   $r_{\rm EW}$ for various values of $p_T$ as a function of $\sqrt{s}$.
   The plots show the same data for different ranges of $\sqrt{s}$ on the $x$ axis. The uncertainty band (only visible for $p_T = 300$~GeV) is obtained from the Pad\'{e} approximation, and the truncation uncertainty in the $\delta_X$ and $m_h^{\rm ext}$ expansions is not shown but is estimated to be $\pm1\%$.
   Results shown are from Ref.~\cite{Davies:2026wbx}.
   }
\end{figure}

The high-energy result~\cite{Davies:2026wbx} has several interesting features.
In the limiting case of $m_h = 0$ and $m_t \to 0$, the leading high-energy expansion starts at $(m_t^2/s)^0$ at two loops, while it only starts at $(m_t^2/s)^1$ at one loop.
At NLO EW in this limit, we observe a quadratic and cubic logarithm ($\alpha \log^2(m_t^2/s)$ and $\alpha \log^3(m_t^2/s)$) as the leading logarithmic contribution for the box contributions in form factors $F_1$ and $F_2$, respectively. Note that the triangle contributions are suppressed at high energies due to the $s$-channel Higgs propagator.
In Fig.~\ref{fig::rEW_HE}, we present our result in terms of a ``partonic $K$-factor'' $r_{EW} = \frac{\alpha}{\pi} \big( \, \mathcal{U}^{(0,1)} / \mathcal{U}^{(0)} \big)$ up to $\sqrt{s} = 10$~TeV, where we assume the perturbative NLO QCD and EW expansion as $|\mathcal{M}|^2 = \bar{X}_0 \big( \, \mathcal{U}^{(0)} + \frac{\alpha_s}{\pi}\, \mathcal{U}^{(1,0)} + \frac{\alpha}{\pi} \, \mathcal{U}^{(0,1)} \big)$ with an overall prefactor $\bar{X}_0$.
Our result covers a large phase space ranging from the high-energy limit to fairly low values of Higgs transverse momentum $p_T \gtrsim 350$~GeV.
It shows that these electroweak corrections lead to an effect of order $-10\%$, supporting the observation from Ref.~\cite{Bi:2023bnq} for the hadronic invariant mass or $p_T$ distribution.

\subsubsection{Factorizable contributions}

\label{EW:1loopX1loop}


Factorizable contributions consist of diagrams containing two separated one-loop corrections, see Fig.~\ref{fig:bd}, cases (b), (e), and (f).

\noindent \textbf{Milada Margarete Mühlleitner, Johannes Schlenk, Michael Spira~\cite{Muhlleitner:2022ijf}.}

\noindent The subset of factorizable contributions containing Yukawa interactions between the Higgs and the top quark (case (b)) has been evaluated analytically retaining full dependence on masses and kinematics. Analytical results have been presented in Ref.~\cite{Muhlleitner:2022ijf}.

\noindent \textbf{Joshua Davies, Kay Schönwald, Matthias Steinhauser, Hantian Zhang~\cite{Zhang:2024rix}.}

\noindent Exact analytic results for the complete set of factorizable contributions in the SM have been obtained (cases (b), (e), and (f)). No approximation has been applied. Computer-readable results for the form factors are provided, expressed in terms of standard $B_0$ and $C_0$ functions.

\subsubsection{Top Yukawa and Higgs boson self-coupling corrections}

\label{EW:Top-Higgs_self}


\noindent \textbf{Joshua Davies, Kay Schönwald, Matthias Steinhauser, Hantian Zhang~\cite{Davies:2022ram,Davies:2025wke}.}

\noindent Analytic high-energy results for the subset of Feynman diagrams involving top Yukawa and Higgs boson self-energy couplings have been computed in Refs.~\cite{Davies:2022ram,Davies:2025wke}.  Such diagrams involve as dimensionful quantities the Mandelstam variables $s$ and $t$, the top quark mass and the Higgs boson mass.  In the calculation the final-state Higgs mass and the one in the propagators inside the loop diagrams are treated separately. This allows us to perform as a first step an expansion in the final-state mass and subsequently an expansion in $\delta = 1-m_h^{\rm int}/m_t$ with $m_h^{\rm int}$ being the internal propagator mass. Both expansions are simple Taylor expansions.
Afterwards, we realize the high-energy limit through an expansion for small $m_t$ which requires the application of a nontrivial asymptotic expansion.
Technical details of the master integral calculations in the high-energy limit are given in Refs.~\cite{Zhang:2024fcu,Davies:2022ram,Davies:2025wke}, together with a publicly available tool \texttt{AsyInt}~\cite{Zhang:2024fcu}.
We typically compute about $100$ expansion terms in $m_t$ and subsequently apply a Pad\'e procedure that provides, for each phase-space point, a central value and an uncertainty. It has been shown in Refs.~\cite{Davies:2022ram, Davies:2023vmj} that this approach leads to precise results down to fairly low values of the Higgs transverse momentum $p_T$ in the case of QCD and leading Yukawa corrections.

\begin{figure}
\centering
    \subfloat[]{{\includegraphics[width=0.48\textwidth]{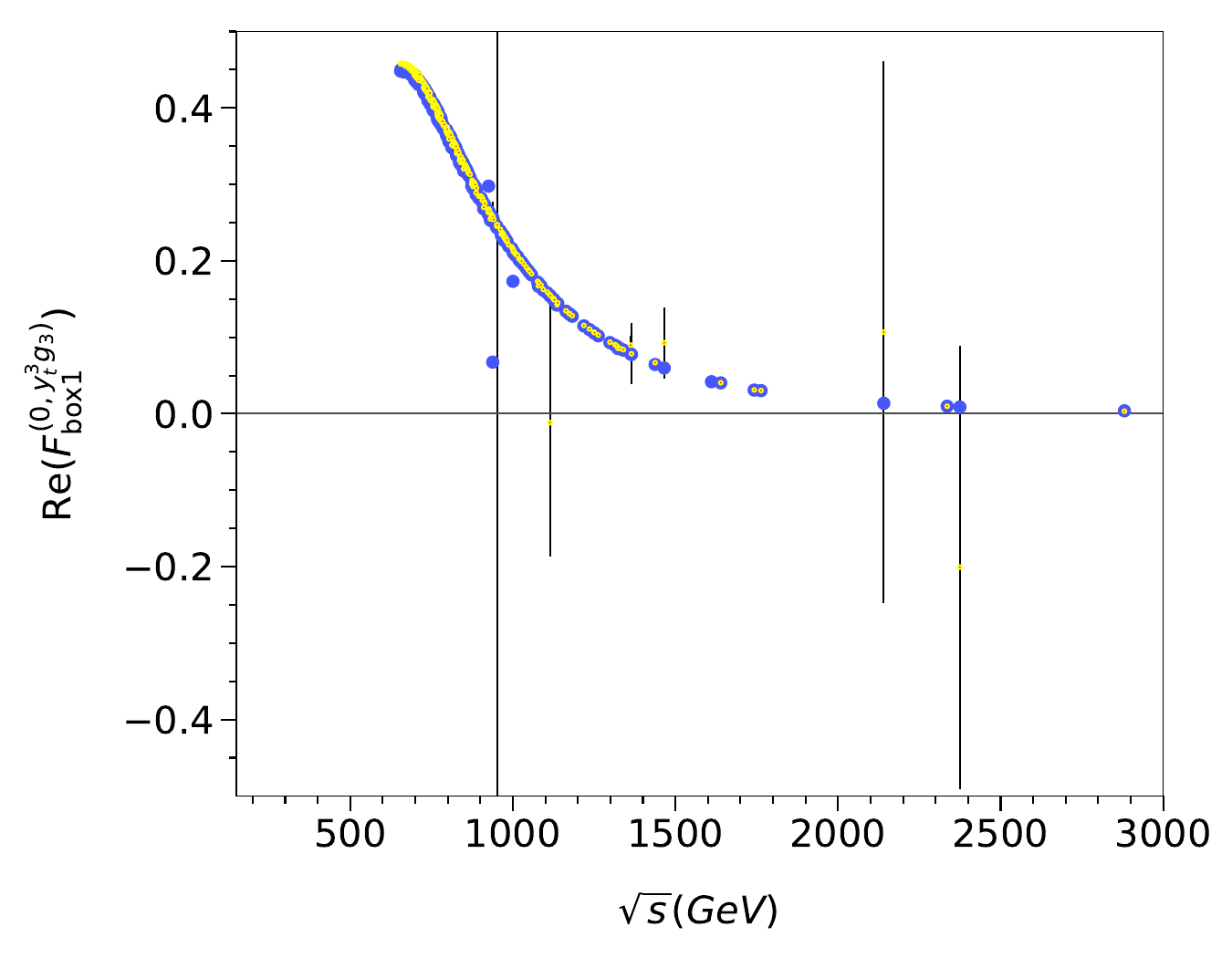}}}
	\\
    \subfloat[]{{\includegraphics[width=0.48\textwidth]{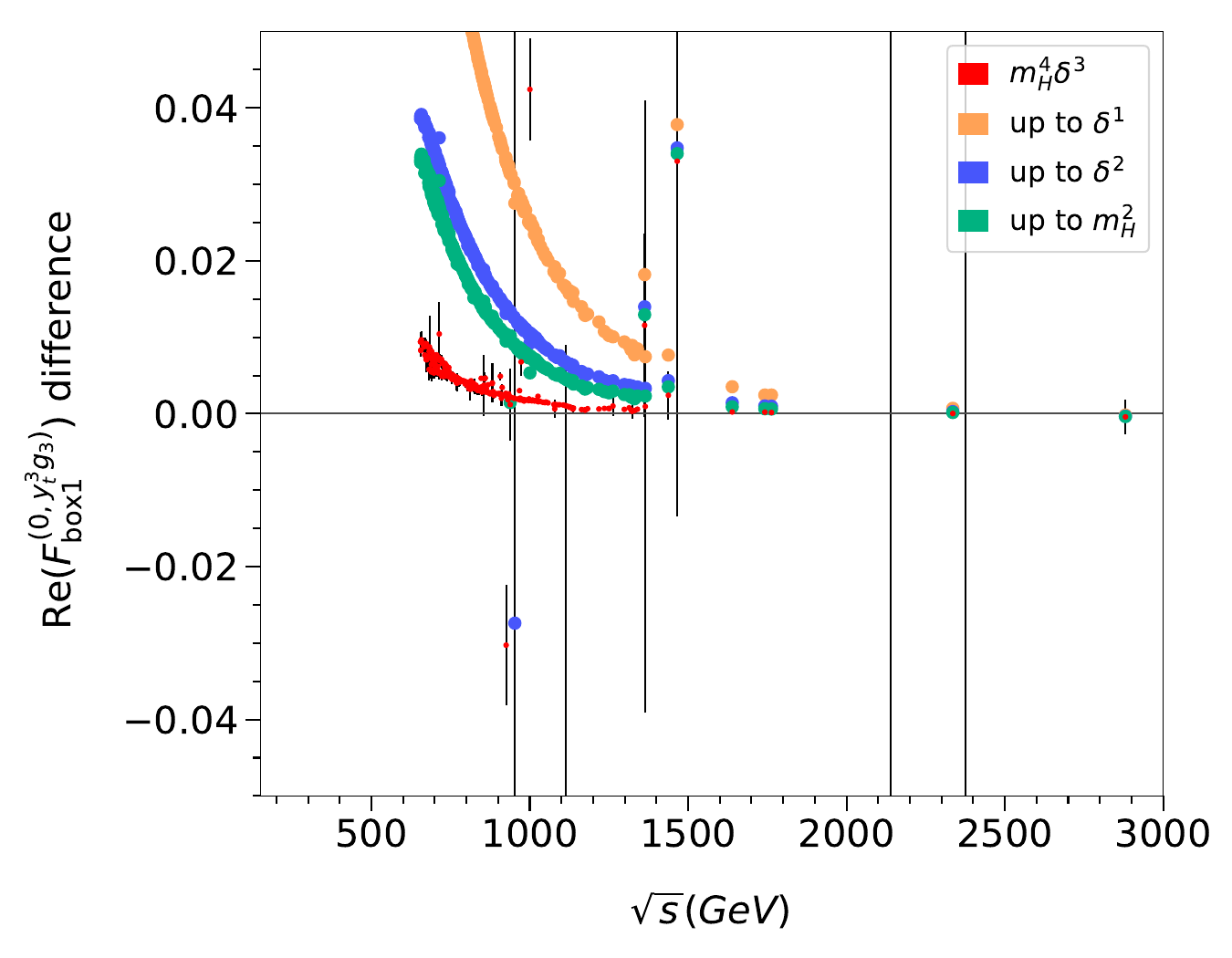}}}
	\quad
    \subfloat[]{{\includegraphics[width=0.48\textwidth]{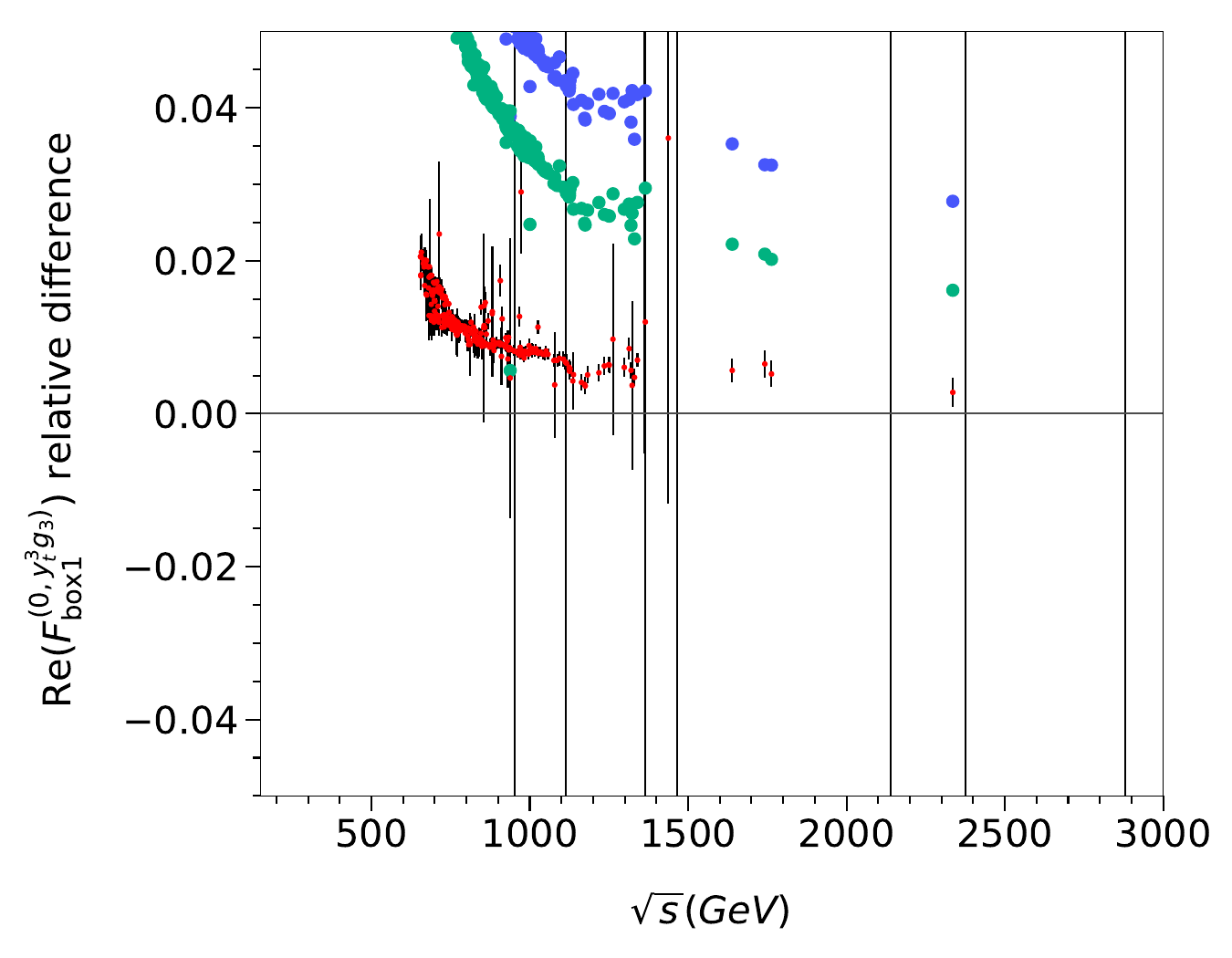}}}
	\caption{Real part of the form factor (a) from Feynman diagrams like those shown in Fig.~\ref{fig::diags}. (b) and (c) show the absolute and relative difference compared to Ref.~\cite{Heinrich:2024dnz}. Analytic results shown are from Ref.~\cite{Davies:2025wke}.}
    \label{fig::diff_to_SD_yt3lam1}
\end{figure}

\begin{figure}
\centering
    \subfloat[]{{\includegraphics[width=0.3\textwidth]{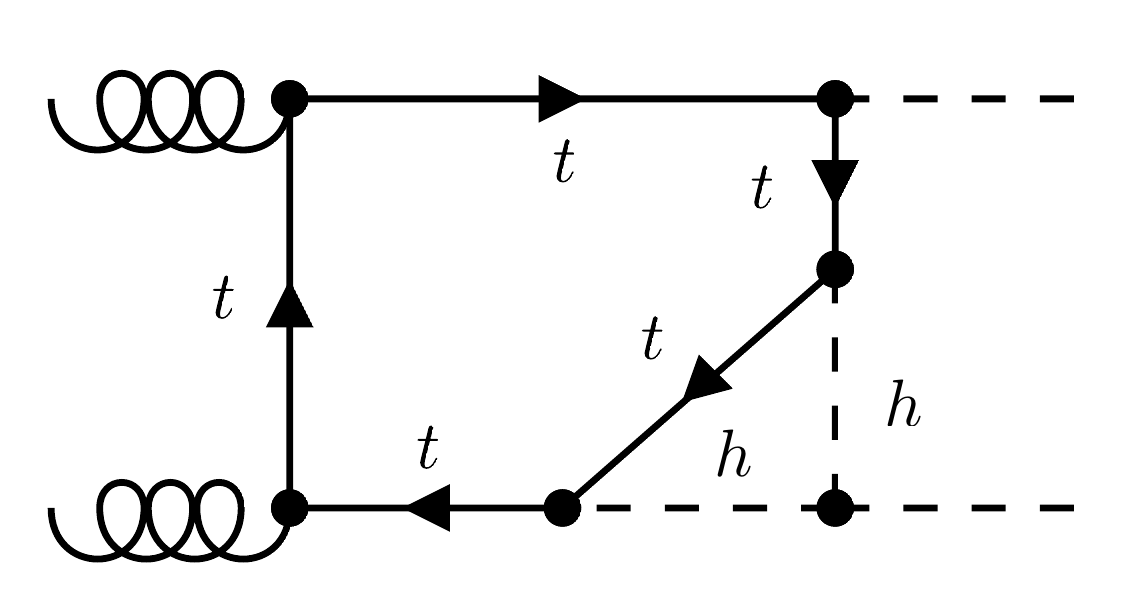}}}
	\quad
    \subfloat[]{{\includegraphics[width=0.3\textwidth]{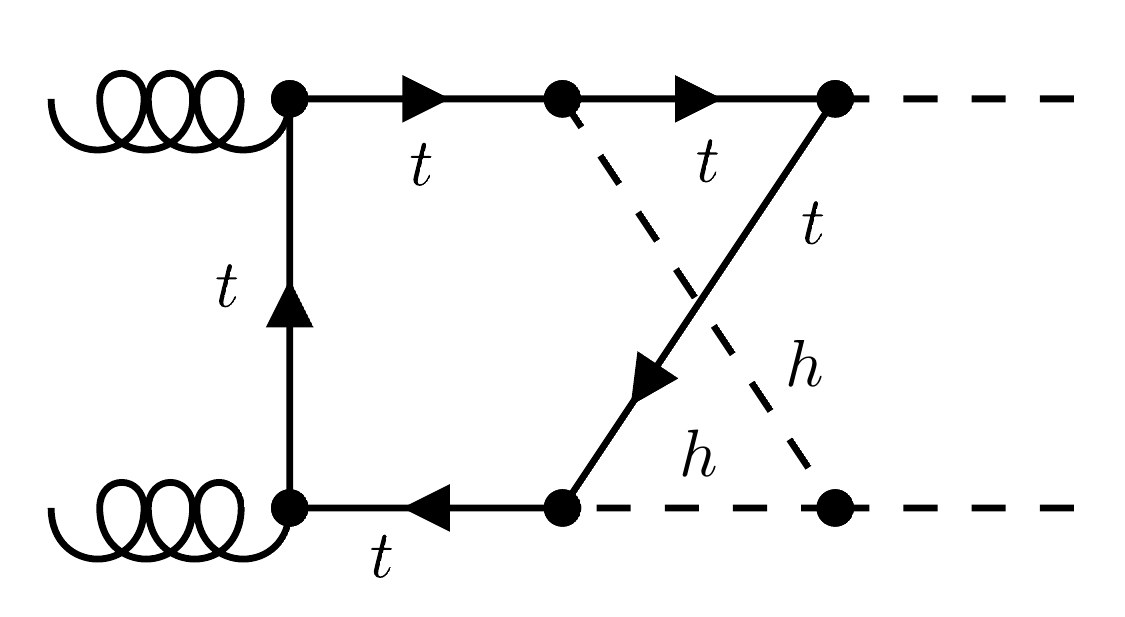}}}
	\caption{Sample Feynman diagrams. Solid, dashed and curly lines represent top quarks, Higgs bosons and gluons, respectively. Diagrams from Ref.~\cite{Davies:2025wke}.}
    \label{fig::diags}
\end{figure}

As an illustration we show in Fig.~\ref{fig::diff_to_SD_yt3lam1} the comparison of the real part of the box contribution to form factor $F_1$ originating from diagrams with three top quark Yukawa couplings and one trilinear Higgs boson self-coupling (see Fig.~\ref{fig::diags} for sample diagrams) to the numerical results from Ref.~\cite{Heinrich:2024dnz}. After including expansion terms up to order $m_h^4\delta^3$ one observes agreement at the percent level.

Note that the analytic results provide full flexibility concerning the change of renormalization schemes and the variation of input parameters.

~

\noindent \textbf{Gudrun Heinrich, Stephen Jones, Matthias Kerner, Thomas Stone, Augustin Vestner~\cite{Heinrich:2024dnz}.}

\noindent In Ref.~\cite{Heinrich:2024dnz}, fully differential results for the Yukawa- and Higgs boson self-coupling-type electroweak corrections to Higgs boson pair production in gluon fusion are presented.
The results retain the full dependence on the mass of the top quark and the Higgs boson.
At the amplitude level, the results are further separated into the individual coupling structures, allowing for the bare amplitude to be computed with arbitrary top Yukawa coupling, trilinear Higgs boson self-coupling, and quartic Higgs boson self-coupling.

\begin{figure}
\centering
    \subfloat[]{{\includegraphics[width=0.3\textwidth]{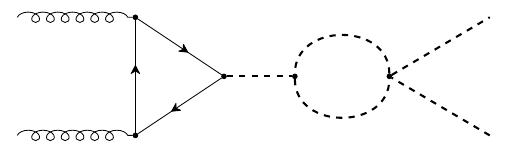}}}
	\qquad
    \subfloat[]{{\includegraphics[width=0.3\textwidth]{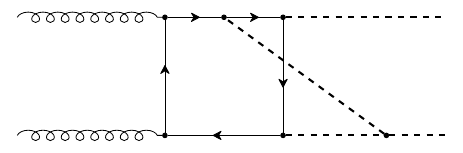}}}
	\qquad
    \subfloat[]{{\includegraphics[width=0.3\textwidth]{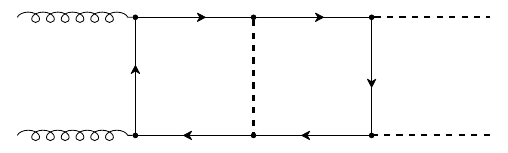}}}
		\caption{Example diagrams contributing to the different coupling structures into which the bare two-loop amplitude is separated. Diagrams from Ref.~\cite{Heinrich:2024dnz}.}
    \label{fig:240704653_examplediags}
\end{figure}

The precise class of corrections which are computed is defined via a Yukawa model with only an up-type quark (the top quark) and a scalar field (the Higgs boson). Example Feynman diagrams are shown in Fig.~\ref{fig:240704653_examplediags}.
This model selects a well-defined subset of diagrams which can be obtained from considering the SM in the unitary gauge and then setting $(g, g^\prime) \rightarrow (0,0)$, which removes the electroweak gauge bosons (and their associated ghost fields).
The electroweak input-parameter scheme $\left\{ M_Z=0, M_W=0, G_F\right\}$ + $\left\{m_t,m_h\right\}$ is used, where all masses are specified in the on-shell scheme.
Results are presented with the Higgs vev fixed in the $G_\mu$ scheme (with $M_Z=0$, $M_W=0$). However, the intermediate results are sufficiently general to allow another renormalization scheme to be adopted.
Tadpole contributions are treated within the Fleischer--Jegerlehner tadpole scheme (FJTS)~\cite{Fleischer:1980ub}.

The calculation is performed by projecting the bare amplitude onto the two form factors described above in $D=4-2\epsilon$ dimensions.
The unreduced amplitude is obtained using two separate calculations based on either \texttt{alibrary}~\cite{alibrary}, a \texttt{Mathematica} and \texttt{Form}~\cite{Kuipers:2013pba} package for computing multi-loop amplitudes, or \texttt{Reduze 2}~\cite{vonManteuffel:2012np}, allowing for a detailed cross-check.
The form factors are then reduced fully symbolically, retaining the dependence on the Mandelstam invariants ($s$, $t$), the masses ($m_h$, $m_t$) and the space-time dimension $D$, to a basis of 494 finite master integrals using the public programs \texttt{Kira}~\cite{Klappert:2020nbg}, \texttt{Ratracer}~\cite{Magerya:2022hvj}, and \texttt{Firefly}~\cite{Klappert:2019emp,Klappert:2020aqs}.
The master integrals appearing in the reduced form factors are then evaluated numerically for each phase-space point using \texttt{pySecDec}~\cite{Borowka:2017idc,Borowka:2018goh,Heinrich:2021dbf,Heinrich:2023til}. The evaluation of the master integrals is additionally cross-checked at several physical points using \texttt{DiffExp}~\cite{Hidding:2020ytt,Moriello:2019yhu}.

\begin{figure}
\centering
    \subfloat[]{{\includegraphics[width=0.48\textwidth]{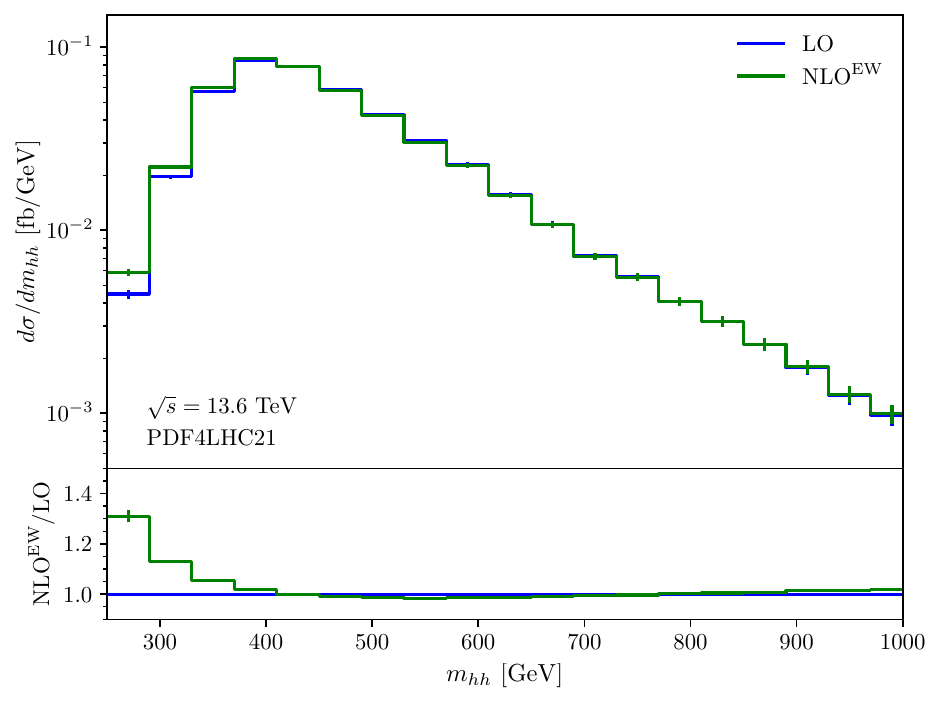}}}
	\quad
    \subfloat[]{{\includegraphics[width=0.48\textwidth]{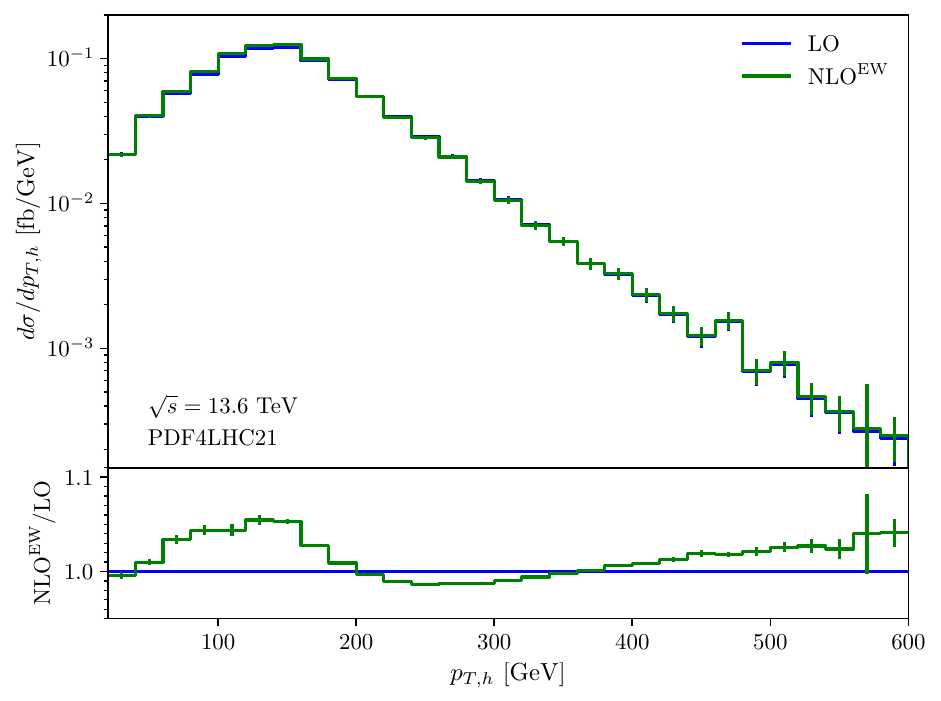}}}
    \caption{Invariant mass (a) and transverse momentum (b) distributions for SM Higgs boson pair production at LO and $\mathrm{NLO}^\mathrm{EW}$ including only the Yukawa and Higgs boson self-coupling-type corrections. Results shown are from Ref.~\cite{Heinrich:2024dnz}.}
    \label{fig:240704653_dists}
\end{figure}

In Fig.~\ref{fig:240704653_dists}, results for the Higgs boson pair invariant mass and $p_{T,H}$~distribution are shown.
The results presented here are obtained using the \texttt{PDF4LHC21\_40}~\cite{PDF4LHCWorkingGroup:2022cjn} parton distribution functions interfaced via \texttt{LHAPDF}~\cite{Buckley:2014ana} and setting the factorization and renormalization scale to $\mu_r = \mu_f = m_{hh}/2$. The masses of the Higgs boson and top quark are set to $m_h = 125~\mathrm{GeV}$, $m_t = \sqrt{23/12}\,m_h = 173.055$ GeV, respectively, and $G_F = 1.1663787 \cdot 10^{-5}\,\textup{GeV}^{-2}$, corresponding to $v=246.22~\mathrm{GeV}$.
For the subset of the electroweak correction considered, very large shape distortions (around $30\%$ for the binning selected) are observed close to threshold for the invariant mass distribution, with much smaller corrections above $400\,\textup{GeV}$ (up to $1\,\textup{TeV}$).
The corrections in the $p_{T,H}$ distribution are less localized, with distortions of up to $5\%$ present across the spectrum.
Overall, the top Yukawa and Higgs boson self-coupling corrections amount to a $1\%$ enhancement of the total cross section.

An important observation concerns the renormalizability of models in which the top Yukawa coupling, the trilinear Higgs boson self-coupling, and the quartic Higgs boson self-coupling are varied from their SM values.
In particular, it is observed that the $1/\epsilon$ pole structure of the vev counterterm obtained by demanding the finiteness of the Yukawa or Higgs boson self-coupling vertices differs unless they are each set to their SM value.
This implies that the $\kappa$-framework used to extract the trilinear Higgs boson self-coupling modifier, $\kappa_\lambda \equiv \lambda_{hhh}/\lambda_{hhh}^{\rm SM}$, cannot directly be used when including the electroweak corrections to this process.
Furthermore, even though the quartic Higgs boson self-coupling first enters at $\mathrm{NLO}_\mathrm{EW}$, the subset of diagrams in which it appears is divergent prior to UV renormalization, implying that the $\kappa_4$ coupling also cannot naively be varied.

\subsubsection{Light-quark contributions}

\label{EW:light-quark}


\begin{figure}
\centering
    \subfloat[$VVh$]{{\includegraphics[height=0.10\textheight]{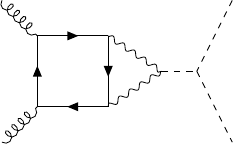}}\label{fig:VVHV}}
	\qquad\qquad
    \subfloat[$VVhh$]{{\includegraphics[height=0.10\textheight]{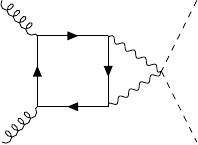}}\label{fig:HHVV}}
	\qquad\qquad
    \subfloat[$VVV$]{{\includegraphics[height=0.10\textheight]{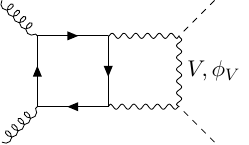}}\label{fig:VVV}}
	\caption{Representative diagrams for the light-quark contributions from Ref.~\cite{Bonetti:2025vfd}.}
    \label{fig:amplgg}
\end{figure}

Double Higgs boson production in gluon fusion mediated by light quarks appears for the first time at two loops: the gluons couple to a light-quark box, to which either two $W$ or two $Z$ bosons are attached, subsequently generating the Higgs boson(s). Three blocks of contributions can be identified: one-particle reducible diagrams containing one vector-vector-Higgs vertex and one trilinear Higgs boson self-coupling vertex (dubbed $VVH$, see Fig.~\ref{fig:VVHV}); diagrams containing a vector-vector-Higgs-Higgs vertex ($VVHH$, Fig.~\ref{fig:HHVV}); and diagrams containing two vector-vector-Higgs vertices ($VVV$, Fig.~\ref{fig:VVV}).\footnote{The $VVV$ type of diagrams may also contain a Goldstone boson connecting the two Higgs bosons.} Each one of these blocks is further divided into diagrams containing either $W$ or $Z$ bosons only. Each one of the aforementioned terms is explicitly gauge-invariant and UV- and IR-finite. Four light flavours are considered in diagrams containing $W$ bosons, while five light flavours have been taken into account in diagrams containing $Z$ bosons.

~

\noindent \textbf{Marco Bonetti, Philipp Rendler, William Torres Bobadilla~\cite{Bonetti:2025vfd}.}

\noindent In Ref.~\cite{Bonetti:2025vfd} fully symbolic expressions for each of the blocks described above have been calculated. The amplitude has been decomposed as a linear combination of the standard tensors $T_1^{\mu\nu}$ and $T_2^{\mu\nu}$ of Section~\ref{intro_ampli}.\footnote{No axial terms are generated by the fermion loop thanks to charge-parity conservation.}

The two form factors are extracted, and the set of scalar two-loop Feynman integrals appearing there is reduced to master integrals. Differential equations are derived for specific linear combinations of the master integrals with respect to the Mandelstam variables and the masses, such that their solution can be written in terms of iterated integrals over logarithmic kernels. This choice of basis is further improved by applying a series of rotations to eliminate redundant integration kernels. Integration constants are fixed by matching the expressions of the master integrals to their large-$m_{W,Z}$ expansion.

The results are employed to obtain analytic expressions for the form factors, further identifying independent functions therein for which dedicated differential equations are built for optimized numerical evaluation in terms of generalized series expansions evolved from the large-mass limit to arbitrary points in the physical region of the phase space. A code implementation for the numerical evaluation of the functions is provided, based on the \texttt{Mathematica} package \texttt{DiffExp}~\cite{Hidding:2020ytt}. Grids for the light-quark amplitude, as well as for its interference with the LO amplitude, have been implemented in the POWHEG-BOX framework or are available upon request. The interference between the LO amplitude and the first form factor of the light-quark amplitude represents the main source of contributions except for the high-energy tail, where the second form factor becomes relevant, although very small overall. Large cancellations occur when considering the interference between LO triangle and box diagrams with each of the $VVh$, $VVhh$, and $VVV$ blocks, suggesting that the light-quark terms are highly sensitive to variations of the trilinear Higgs boson self-coupling.

\begin{figure}
\centering
    \subfloat[]{{\includegraphics[width=0.48\textwidth]{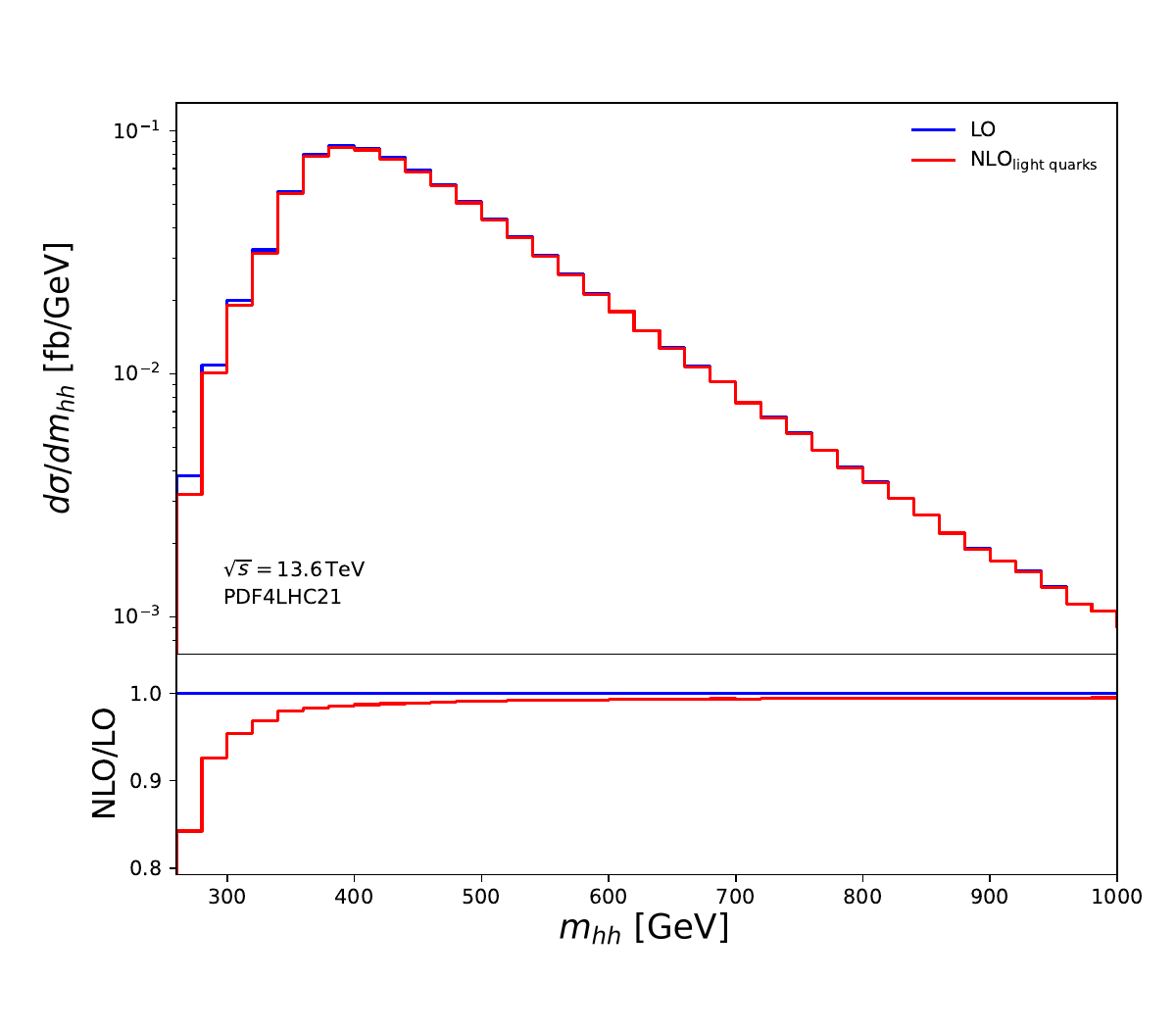}}\label{fig:lq_mHH}}
	\quad
    \subfloat[]{{\includegraphics[width=0.48\textwidth]{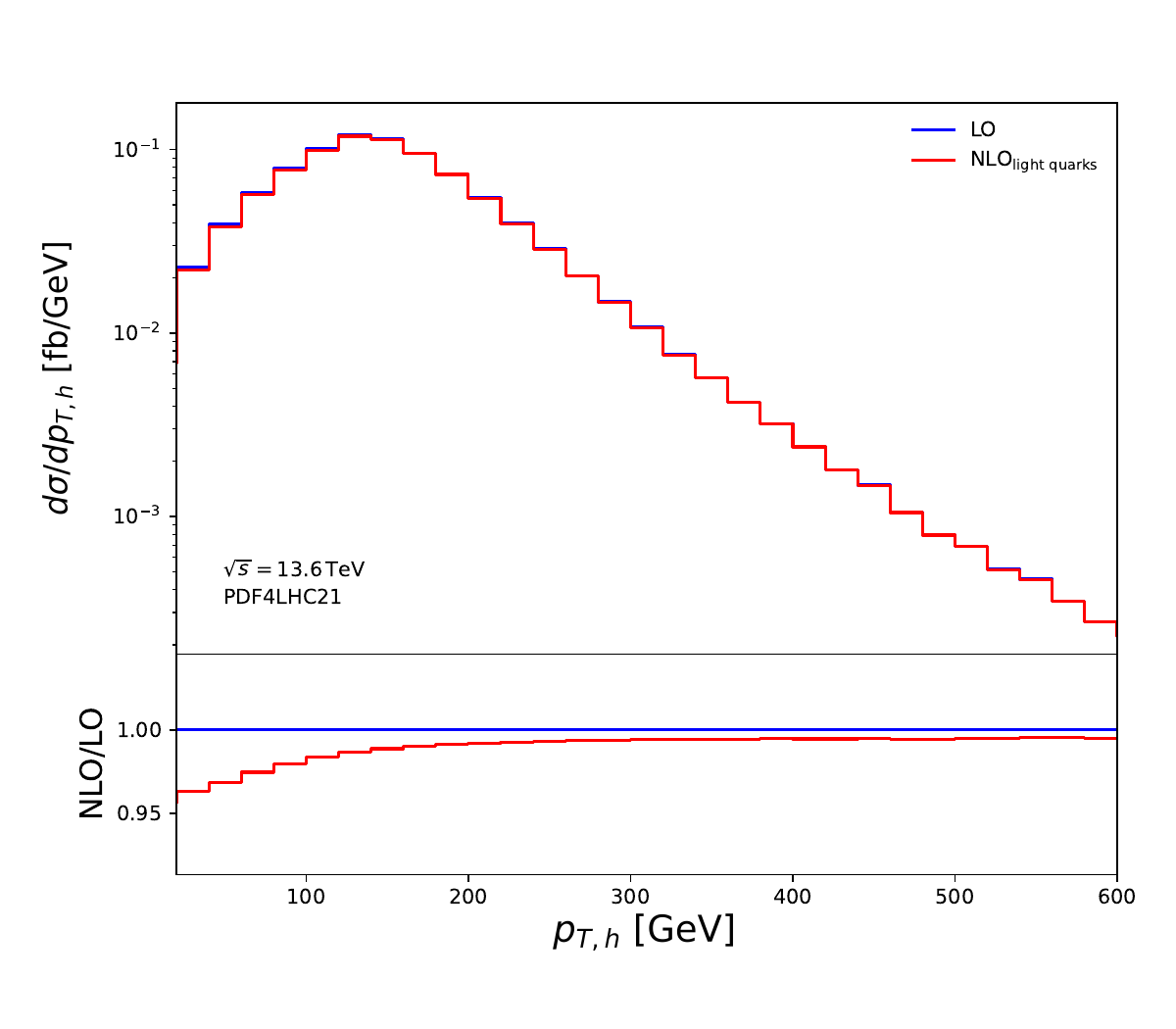}}\label{fig:lq_pTH}}
	\caption{Light-quark and Yukawa contributions to the Higgs boson pair invariant mass (a) and Higgs transverse momentum distributions (b).}
    \label{fig:lq_obs}
\end{figure}

The impact of the light-quark corrections to the Higgs boson pair invariant mass distribution and the transverse momentum distribution is depicted in Figs.~\ref{fig:lq_mHH} and~\ref{fig:lq_pTH}, respectively. The light-quark corrections suppress both the invariant mass and the transverse momentum distribution at all centre-of-mass energies, with particular relevance for the region of the phase space close to the production threshold, where they have a negative effect of more than $-15\%$ on the $m_{hh}$ distribution and around $-4\%$ on the $p_{T,h}$ distribution. In both cases, the corrections quickly approach zero as the centre-of-mass energy grows.

~

\noindent \textbf{Arunima Bhattacharya, Francisco Campanario, Sauro Carlotti, Jamie Chang, Javier Mazzitelli, Milada Margarete Mühlleitner, Jonathan Ronca, Michael Spira ~\cite{Bhattacharya_TBA}.}

\noindent Ref.~\cite{Bhattacharya_TBA} provides an independent evaluation of the light-quark contributions to $gg \to hh$. The bottleneck of this calculation is the evaluation of the genuine two-loop box diagrams $VVV$ (see Fig.~\ref{fig:VVV}). The numerical stability of these diagrams is quite demanding, since strong cancellations emerge between all diagrams, which require the form factors of the individual diagrams to be numerically determined with very significant precision. The numerical method applied to these diagrams is the same as in Section~\ref{EW:Top-Higgs_sector}, i.e.\ projection on the two form factors of Eq.~\eqref{eq:FFdeco} in $D=4-2\epsilon$ dimensions, no tensor reduction, Feynman parametrization and analytical continuation of all propagator masses including the $W$ and $Z$ boson masses, $M_{W/Z}^2 \to M_{W/Z}^2 (1-i\bar\epsilon)$. To isolate the singularities, multiple end-point subtractions need to be performed. In order to reach the narrow-width limit $\bar\epsilon\to 0$, a Richardson extrapolation was performed with a minimal regulator down to $\bar\epsilon = 0.025$ depending on $m_{hh}$.

The two-loop triangle diagrams $VVh$ (see Fig.~\ref{fig:VVHV}) can be adopted from the single-Higgs calculation \cite{Aglietti:2004nj,Aglietti:2006yd} and translated to the two-loop box diagrams with a 4-point vertex $VVhh$ (see Fig.~\ref{fig:HHVV}).

\begin{figure}[!hbtp]
    \centering
\includegraphics[width=0.48\textwidth]{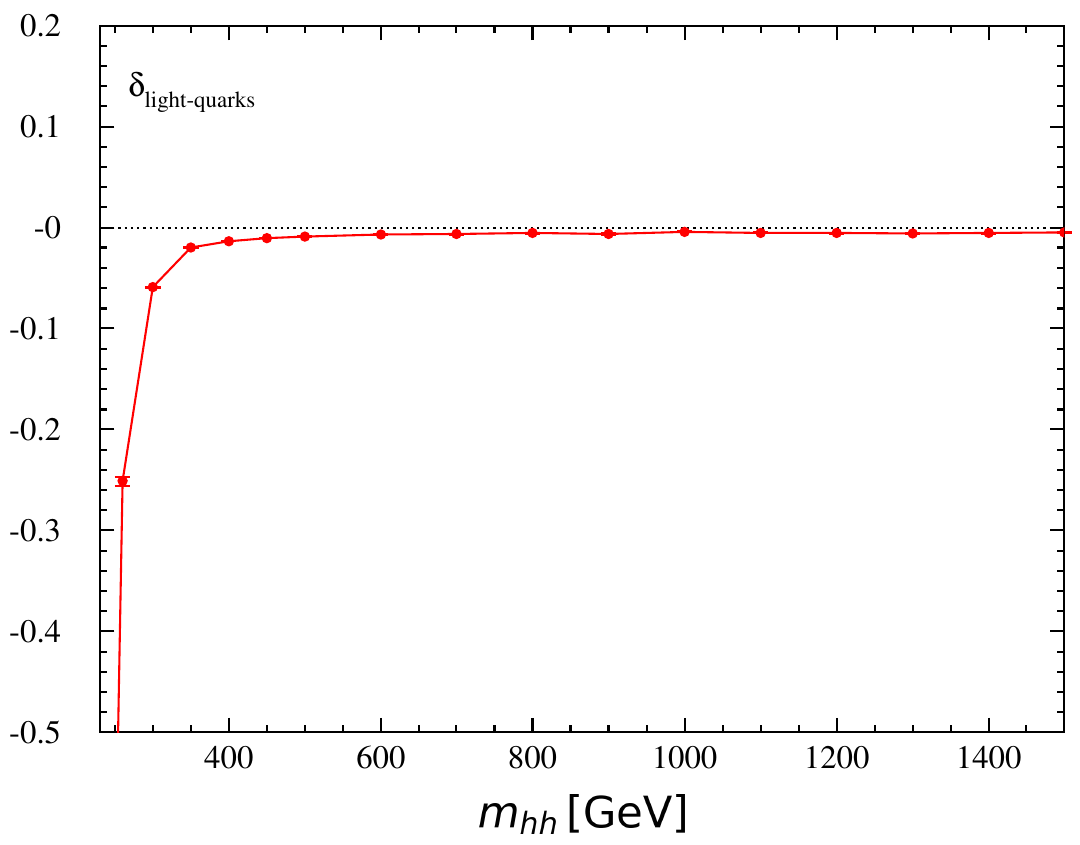}
\caption{\label{fg:delta_lq}The full relative light-quark loop-induced electroweak corrections to the $gg \to hh$ process. The points on the curve together with their numerical error bars indicate the $m_{hh}$ values for which the corrections have been computed. Results shown are from Ref.~\cite{Bhattacharya_TBA}.}
\end{figure}

The total sum of all light-quark loop diagrams is finite, and no renormalization is required. The result is a small correction at the sub-percent level for large values of $m_{hh}$, see Fig.~\ref{fg:delta_lq}, while the corrections are larger close to the production threshold due to the cancellation of the leading top quark mass terms of the LO matrix element, to which the results are normalized. In total, the light-quark loops do not play a dominant role in Higgs boson pair production via gluon fusion in contrast to single-Higgs production \cite{Actis:2008ts,Actis:2008ug}.

~

\paragraph{Comparison.}

The two independent evaluations of the light-quark contributions mentioned above have been found to agree within the numerical uncertainties.

\subsection{Quark-antiquark-initiated contributions}

\label{EW:qqbar}

A Higgs boson pair can also be produced in the quark-antiquark channel. The Higgs bosons can be generated either through direct Yukawa couplings to massive bottom quarks or through intermediate EW vector bosons. Direct Yukawa production already occurs at tree level but is suppressed by the small Yukawa couplings, while EW-boson-mediated production is loop-induced.

\noindent \textbf{Marco Bonetti, Gudrun Heinrich, Philipp Rendler, William  Torres~Bobadilla~\cite{Bonetti:2026cih}.}

\begin{figure}
\centering
    \subfloat[LO.]{{\includegraphics[height=0.10\textheight]{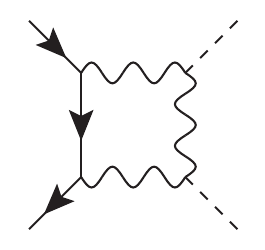}}\label{fig:qqbarLO}}
	\qquad\qquad
    \subfloat[NLO virtual.]{{\includegraphics[height=0.10\textheight]{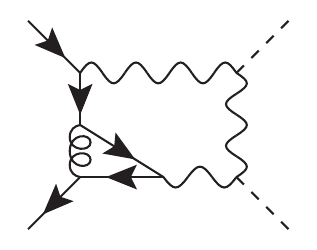}}\label{fig:qqbarV}}
	\qquad\qquad
    \subfloat[NLO real.]{{\includegraphics[height=0.10\textheight]{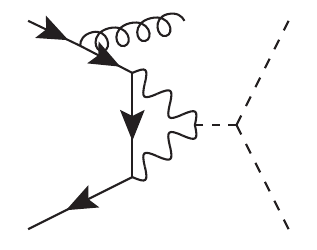}}\label{fig:qqbarR}}
	\qquad\quad

	\caption{Representative diagrams for light-quark-antiquark-initiated contributions from Ref.~\cite{Bonetti:2026cih}.}
    \label{fig:amplqqbar}
\end{figure}

\noindent The process $q\overline{q} \to hh$, mediated by weak bosons in the loop, has been calculated in Ref.~\cite{Bonetti:2026cih}, including NLO QCD corrections and considering only light quarks. Fig.~\ref{fig:amplqqbar} shows illustrative diagrams at LO and at NLO QCD. Only the first two generations of quarks are considered as initial states.

The $q\overline{q}hh$ two-loop diagrams and the $q\overline{q}hhg$ real emission diagrams contributing at NLO can be decomposed into subsets according to the number of $VVH$ or $VVhh$ vertices and vector boson propagators they contain, analogous to the light-quark contributions to $gg \to hh$ described in Section~\ref{EW:light-quark} and calculated in Ref.~\cite{Bonetti:2025vfd}.

For the two-loop virtual diagrams, only the $VVV$ subset (exemplified in Fig.~\ref{fig:amplqqbar}b) is non-zero, due to angular momentum conservation. Consequently, real emissions belonging to the $VVh$ and $VVhh$ blocks do not develop IR poles when integrating over the phase space of the unresolved extra gluon.

The one-loop amplitudes have been evaluated via the computer code \texttt{GoSam-3.0}~\cite{Braun:2025afl}, while the two-loop virtual contributions have been calculated analytically, retaining full dependence on the Mandelstam invariants and masses. The two-loop Feynman integrals have been reduced to master integrals, which have been expressed in terms of Chen iterated integrals over logarithmic kernels. Differential equations tailored to the independent transcendental functions at the level of the form factors have been constructed to produce precise numerical evaluations through the \texttt{Mathematica} package \texttt{DiffExp}~\cite{Hidding:2020ytt}.

The full set of NLO amplitudes has been implemented in the form of numerical grids in the {\tt POWHEG-BOX-V2} framework. NLO QCD corrections increase the total cross section with respect to LO $q\overline{q} \to hh$ by about $+60\%$.

\begin{figure}
\centering
\includegraphics[width=0.49\textwidth]{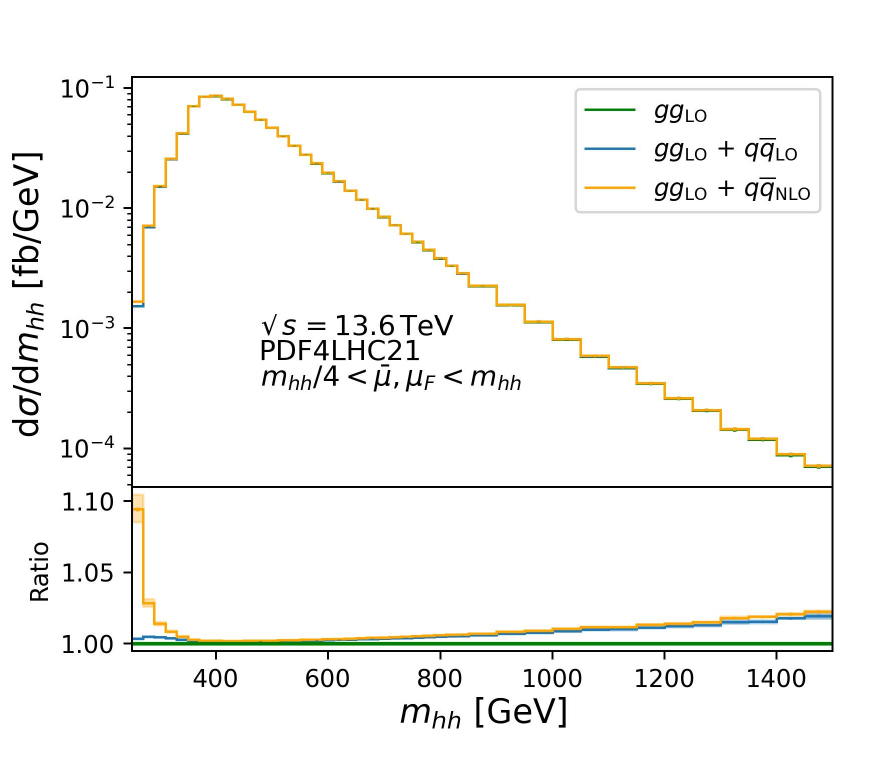}
\includegraphics[width=0.49\textwidth]{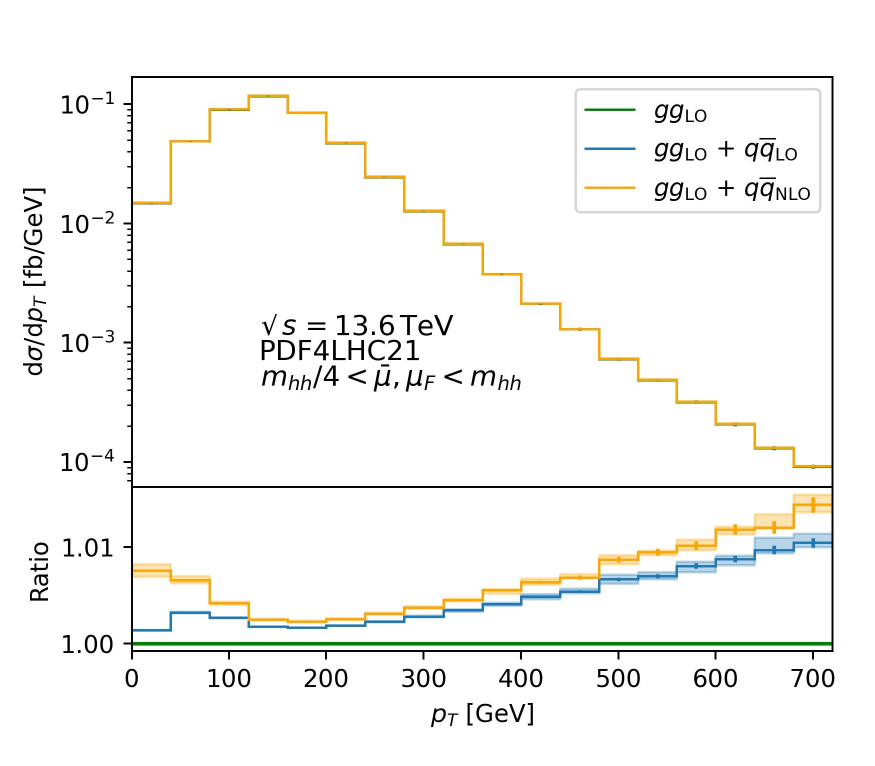}
\caption{Invariant mass distribution of the Higgs boson pair (left) and transverse momentum distribution (right) of a single Higgs boson at a proton-proton centre-of-mass energy of $\sqrt{s} = 13.6\,\textup{TeV}$ in the SM. The error bands represent the scale uncertainties of the quark-antiquark channel and the error bars indicate the statistical uncertainties from the Monte Carlo integration. Figures from Ref.~\cite{Bonetti:2026cih}.
}
\label{fig:invariantMass}
\end{figure}

On the one hand, the contribution to the total cross section coming from the quark-antiquark channel is negligible when compared to the gluon channel, amounting to about $+0.35\%$. On the other hand, the NLO QCD corrections to $q\overline{q} \to hh$ have a large shape-distorting effect at low energies in the $m_{hh}$ differential distribution, increasing the leftmost bin by about $10\%$ (cf.\ Fig.~\ref{fig:invariantMass}).\footnote{Bins with a $20\,\textup{GeV}$ width are considered.} As the low $m_{hh}$ region is also very sensitive to deviations of the trilinear Higgs boson self-coupling from the SM value, these contributions should not be neglected in high-precision predictions for Higgs boson pair production.

\section{Cross-section recommendations}\label{sec:recs}

The previous sections summarize the perturbative ingredients entering the current state-of-the-art prediction for Higgs boson pair production via gluon fusion. In this section we combine these results into the recommended reference cross sections, uncertainties, and invariant-mass distribution. Unless a dedicated alternative setup is required, the numbers reported here should be used for phenomenological studies and experimental interpretations.

\subsection[\texorpdfstring{NNLO$_{\textup{FTapprox}}$}{NNLO FTapprox}]{\texorpdfstring{\boldmath$\textup{NNLO}_{\textup{FTapprox}}$}{NNLO FTapprox}}

\begin{table}[htbp]
\begin{center}
\begin{tabular}{lccc}
\toprule
$\sqrt{s}$ [TeV] & \textbf{13} & \textbf{13.6} & \textbf{14} \\
\midrule
$\sigma$ [fb] & 30.75 & 34.01 & 36.27 \\\hline
$\pm$ PDF unc. [\%] & 1.96 & 1.92 & 1.89 \\
$\pm \alpha_s$ unc. [\%] & 1.51 & 1.49 & 1.47 \\
{\bfseries\boldmath $\pm$ PDF $+$ $\alpha_s$ unc. [\%]} & \textbf{2.47} & \textbf{2.43} & \textbf{2.39} \\\hline
QCD scale unc. [\%] & $^{+2.11}_{-4.98}$ & $^{+2.07}_{-4.85}$ & $^{+2.01}_{-4.78}$ \\
$\pm$ TH$_\mathrm{num}$ [\%] & 0.14 & 0.19 & 0.21 \\
\bottomrule
\end{tabular}
\caption{Total cross sections in $\textup{NNLO}_{\textup{FTapprox}}$~\cite{Grazzini:2018bsd}, at $13$, $13.6$ and $14\,\textup{TeV}$ proton centre-of-mass energies, with the Higgs boson mass set to $m_h = 125\,\textup{GeV}$. The central renormalization and factorization scale is taken to be $\mu_0  = m_{hh}/2$, and the scale uncertainties were obtained through a 7-point variation around $\mu_0$. The top quark mass was set to $m_t=173.0$~GeV and the bottom quark mass was set to zero, employing a 5-flavour scheme. The theory uncertainty (TH$_\mathrm{num}$) covers numerical imprecision in the final prediction stemming from MC integration and the extrapolation of the technical cut $r_\mathrm{cut} \to 0$. The PDF set \texttt{PDF4LHC21\ttus 40} was used. The PDF $+$ $\alpha_s$ uncertainty is obtained by adding the two contributions in quadrature.}
\label{tab:nnlo_ftapprox_xs_125}
\end{center}
\end{table}

The total cross sections in the $\textup{NNLO}_{\textup{FTapprox}}$ approximation~\cite{Grazzini:2018bsd} are shown in Table~\ref{tab:nnlo_ftapprox_xs_125} for proton centre-of-mass energies of 13, 13.6 and 14~TeV, with the Higgs boson mass set to $m_h = 125$~GeV. The central renormalization and factorization scale is taken to be $\mu_0  = m_{hh}/2$, and the scale uncertainties were obtained through a 7-point variation around $\mu_0$.
The top quark mass was set to $m_t=173.0$~GeV and the bottom quark mass was set to zero, employing a five-flavour scheme (5FS). The numerical uncertainty (TH$_\mathrm{num}$) combines MC integration errors with
a systematic uncertainty on the technical cut $r_\mathrm{cut}$ employed in the $\textup{NNLO}_{\textup{FTapprox}}$ calculation, where $r_\mathrm{cut}\rightarrow 0$ is extrapolated numerically.
This uncertainty is negligible, but we quote it for completeness. The \texttt{PDF4LHC21\ttus 40} PDF set was used.

\subsection[\texorpdfstring{$\textup{N}^3\textup{LO}+\textup{N}^3\textup{LL}$ $K$-factors}{N3LO+N3LL K-factors}]{\texorpdfstring{\boldmath$\textup{N}^3\textup{LO}+\textup{N}^3\textup{LL}$ $K$-factors}{N3LO+N3LL K-factors}}

The N$^3$LO+N$^3$LL QCD $K$-factor~\cite{Chen:2019lzz,Ajjath:2022kpv} is defined as
\begin{equation}\label{eq:n3lokfac}
K_3 = \left(\frac{\mathrm{N^3LO+N^3LL~HTL} }{\mathrm{NNLO~HTL}}\right)\;.
\end{equation}

\begin{sloppypar}
The vacuum expectation value was set to $v=246.2197$~GeV. The corresponding Fermi constant was $G_F = 1.166378 \times 10^{-5}$~GeV$^{-2}$. The top quark mass was set to $m_t=173.0$~GeV. The central renormalization and factorization scale was taken to be $\mu_0  = m_{hh}/2$.
\end{sloppypar}

The uncertainty due to the use of an NNLO PDF instead of an N$^3$LO PDF is estimated via: 
\begin{equation}
 \Delta ^{\mathrm{MHOU}}_\mathrm{NNLO} =  \left| \frac { \sigma^{\mathrm{N}^3\mathrm{LO}}_{\mathrm{N}^3\mathrm{LO-PDF}} - \sigma^{\mathrm{N}^3\mathrm{LO}}_{\mathrm{NNLO-PDF}} } { \sigma^{\mathrm{N}^3\mathrm{LO}}_{\mathrm{N}^3\mathrm{LO-PDF}}  }  \right|\;,
 \label{eq:deltan3lopdf}
\end{equation}
where the N$^3$LO and NNLO PDFs used were the (approximate) \texttt{MSHT20xNNPDF40\ttus aN3LO\ttus qed} and \texttt{MSHT20xNNPDF40\ttus NNLO\ttus qed}, respectively.

\begin{table}[htbp]
\begin{center}
\begin{tabular}{lccc}
\toprule
$\sqrt{s}$ [TeV] & \textbf{13} & \textbf{13.6} & \textbf{14} \\
\midrule
$K_3$ & 1.030913 & 1.030812 & 1.030736 \\
$\Delta K_{3\rm down}/K_3$ [\%] & 9.3460 & 9.2396 & 9.1721 \\
$\Delta K_{3\rm up}/K_3$ [\%] & 4.9088 & 4.8756 & 4.8537 \\
$\Delta^{\mathrm{MHOU}}_{\mathrm{NNLO}}$ [\%] & 2.1207 & 2.1700 & 2.1799 \\
\bottomrule
\end{tabular}
\caption{$K$-factors $\mathrm{N^3LO{+}N^3LL}/\mathrm{NNLO}$ for the total cross section (relative uncertainties).}
\label{tab:kfactor_n3lo_n3ll_over_nnlo}
\end{center}
\end{table}

The total cross section results are shown in Table~\ref{tab:kfactor_n3lo_n3ll_over_nnlo}, including the upper and lower variations. The last row estimates the effect of missing higher-order corrections associated with using NNLO PDFs in the N$^3$LO calculation. This uncertainty is quoted for completeness, but is not combined in the final recommendation; the corresponding final uncertainty is inherited from the $\textup{NNLO}_{\textup{FTapprox}}$ calculation.

\subsection[\texorpdfstring{NLO electroweak $K$-factors}{NLO electroweak K-factors}]{\texorpdfstring{NLO electroweak \boldmath$K$-factors}{NLO electroweak K-factors}}

The $K$-factor due to NLO electroweak corrections is defined as
\begin{equation}\label{eq:NLOEWkfac}
K_\mathrm{EW} = \frac{\sigma_\mathrm{NLO~EW}}{\sigma_\mathrm{LO}} \;. 
\end{equation}

The input parameters were taken to be as in Ref.~\cite{Bi:2023bnq}:
\begin{itemize}
\item $m_t = 172.69$ GeV,
\item $m_h^2 / m_t^2 = 12 / 23 \Rightarrow m_h = 124.74$~GeV,
\item $m_Z^2 / m_t^2 = 23 / 83 \Rightarrow m_Z = 90.91$~GeV,
\item $m_W^2 / m_t^2 = 14 / 65 \Rightarrow m_W = 80.14$~GeV,
\item $G_F = 1.166378 \times 10^{-5}$~GeV$^{-2}$.
\end{itemize}
We note that the top and Higgs boson masses differ from those used in the $\textup{NNLO}_{\textup{FTapprox}}$ and N$^3$LO calculations. However, the $K$-factors are not particularly sensitive to these.

\begin{table}[htbp]
\begin{center}
\begin{tabular}{lccc}
\toprule
$\sqrt{s}$ [TeV] & \textbf{13} & \textbf{13.6} & \textbf{14} \\
\midrule
$K_\mathrm{EW}$ & 0.959 & 0.959 & 0.958 \\
\bottomrule
\end{tabular}
\caption{NLO electroweak $K$-factors for the total cross section, defined in Eq.~\eqref{eq:NLOEWkfac}, at $13$, $13.6$ and $14\,\textup{TeV}$.}
\label{tab:nlo_ew_kfac}
\end{center}
\end{table}

The NLO electroweak $K$-factors for the total cross section at 13, 13.6 and 14~TeV are shown in Table~\ref{tab:nlo_ew_kfac}. The scale variation uncertainties on these values are at the per-mille level or better, and therefore we do not include them here. We also note that at present, no electroweak effects are included in the PDF sets employed. 

\subsection{Combination}\label{subsec:combination}

The final recommended cross section is obtained by multiplying the $\textup{NNLO}_{\textup{FTapprox}}$ prediction by the N$^3$LO+N$^3$LL QCD $K$-factor and the NLO electroweak $K$-factor,
\begin{equation}\label{eq:xs_reco}
\sigma_\mathrm{rec.}(\sqrt{s}) \;=\; \sigma_{\textup{NNLO}_{\textup{FTapprox}}}(\sqrt{s}) \times K_3(\sqrt{s}) \times K_\mathrm{EW}(\sqrt{s}) \;,
\end{equation}
where $K_3$ and $K_\mathrm{EW}$ are given in Eqs.~\eqref{eq:n3lokfac} and~\eqref{eq:NLOEWkfac}, respectively. The central values are computed using the
central entries of Tables~\ref{tab:nnlo_ftapprox_xs_125},~\ref{tab:kfactor_n3lo_n3ll_over_nnlo} and~\ref{tab:nlo_ew_kfac}.

\paragraph{Scale uncertainties.}
The final prediction combines information from the $\textup{NNLO}_{\textup{FTapprox}}$ calculation and from the N$^3$LO+N$^3$LL QCD $K$-factor. The scale variations of these two ingredients should therefore not be interpreted as independent uncertainties and combined again. For the final recommendation, we take the QCD scale uncertainty directly from the $\textup{NNLO}_{\textup{FTapprox}}$ prediction,
\begin{equation}
\delta_\mathrm{scale}^\mathrm{final}(\sqrt{s}) \;=\; \delta_{\mathrm{scale},\,\textup{NNLO}_{\textup{FTapprox}}}(\sqrt{s}) \;,
\end{equation}
as quoted in Table~\ref{tab:nnlo_ftapprox_xs_125}. The scale variation of the N$^3$LO+N$^3$LL $K$-factor is shown separately in Table~\ref{tab:kfactor_n3lo_n3ll_over_nnlo} and is not combined with the scale uncertainty of the final recommendation. No additional scale uncertainty is assigned to $K_\mathrm{EW}$ in this prescription. The TH$_\mathrm{num}$ assigned to the final recommendation, which accounts for the MC integration errors and $r_\mathrm{cut}\to0$ extrapolation uncertainties in the $\textup{NNLO}_{\textup{FTapprox}}$ calculation, is likewise taken from Table~\ref{tab:nnlo_ftapprox_xs_125}.

\paragraph{PDF and $\alpha_s$ uncertainties.}
The quantity $\Delta^{\mathrm{MHOU}}_\mathrm{NNLO}$ in Eq.~\eqref{eq:deltan3lopdf} estimates the possible effect of using NNLO PDFs in the N$^3$LO component, rather than a fully consistent N$^3$LO PDF set. We use this comparison only to assess the possible size of the PDF-order effect and do not add it to the final uncertainty budget. The quoted PDF and $\alpha_s$ uncertainties of the final recommendation are therefore inherited from the $\textup{NNLO}_{\textup{FTapprox}}$ calculation, as given in Table~\ref{tab:nnlo_ftapprox_xs_125}. Their combined contribution is obtained by adding the PDF and $\alpha_s$ uncertainties in quadrature.

\paragraph{Top quark mass scheme uncertainty.}
An additional uncertainty is assigned to the definition of the virtual top quark mass entering the top Yukawa coupling and the propagators, following the prescription of Ref.~\cite{Baglio:2020wgt} and the discussion in Section~\ref{sec:scheme-scale-uncertainties}. In this approach the NLO prediction obtained with the top quark pole mass is compared with predictions in the $\overline{\mathrm{MS}}$ scheme.

We obtain the corresponding NLO predictions with \texttt{ggxy}~\cite{Davies:2025qjr}, which provides the required total cross sections and differential distributions for the different top quark mass schemes. Specifically, we consider four predictions: with the top quark pole mass $m_t$, and with $\overline{\mathrm{MS}}$ mass $\overline{m}_t(\mu_t)$ with $\mu_t = \overline{m}_t$, $\mu_t = m_{hh}$, and $\mu_t = m_{hh}/4$ (i.e.~the latter two being a factor of two around the central renormalization and factorization scale $\mu_R=\mu_F=m_{hh}/2$). In \texttt{ggxy} we run at the highest available order, i.e.~at five-loop for the strong coupling $\alpha_s$ and the $\overline{\mathrm{MS}}$ mass, and four-loop for the conversion between the pole mass and the $\overline{\mathrm{MS}}$ mass. We then construct, in the $m_{hh}$ distribution, the bin-by-bin maximum and minimum across the four top quark mass schemes; the corresponding inclusive uncertainty is obtained by integrating the bin-wise extrema. We investigated several methods for the numerical integration: similarly to~Ref.~\cite{Baglio:2020ini}, with piecewise integration (Boole's rule~\cite{boole1880,Abramowitz1964} for $m_{hh} < 300$ GeV and the trapezoidal method for $m_{hh}>300$ GeV), as well as the trapezoidal method with finer bins and a continuous fit. The final uncertainties change by about 1\% depending on the method and binning used.



\paragraph{Final recommendations.}
The resulting central values and uncertainties are summarized in Table~\ref{tab:final_xs_reco_125} for $\sqrt{s}=13, 13.6, 14$~TeV, including the top quark mass scheme uncertainty and its linear combination with the scale and TH$_\mathrm{num}$ uncertainties. For the additional choices of the Higgs boson mass shown in Table~\ref{tab:final_xs_reco_125_mh}, we use \texttt{ggxy}~\cite{Davies:2025qjr} to compute the NLO cross section at the target value of $m_h$ and at $m_h=125$~GeV with the same setup. We then form the ratio $\sigma_{\mathrm{NLO}}(m_h)/\sigma_{\mathrm{NLO}}(125~\mathrm{GeV})$ at each centre-of-mass energy and multiply the corresponding $m_h=125$~GeV final recommendation by this factor.

\begin{table}[htbp]
\begin{center}
\begin{tabular}{lccc}
\toprule
$\sqrt{s}$ [TeV] & \textbf{13} & \textbf{13.6} & \textbf{14} \\
\midrule
$\sigma_\mathrm{rec.}$ [fb] & 30.40 & 33.62 & 35.81 \\\hline
$\pm$ PDF unc. [\%] & 1.96 & 1.92 & 1.89 \\
$\pm \alpha_s$ unc. [\%] & 1.51 & 1.49 & 1.47 \\
{\bfseries\boldmath $\pm$ PDF $+$ $\alpha_s$ unc. [\%]} & \textbf{2.47} & \textbf{2.43} & \textbf{2.39} \\\hline
QCD scale unc. [\%] & $^{+2.11}_{-4.98}$ & $^{+2.07}_{-4.85}$ & $^{+2.01}_{-4.78}$ \\
$m_t$ scheme unc. [\%] & $^{+5}_{-16}$ & $^{+5}_{-16}$ & $^{+5}_{-16}$ \\
$\pm$ TH$_\mathrm{num}$ [\%] & 0.14 & 0.19 & 0.21 \\
{\bfseries\boldmath Scale $+$ TH$_\mathrm{num}+m_t$ scheme unc. [\%]} & {\bfseries\boldmath $^{+7.25}_{-21.12}$} & {\bfseries\boldmath $^{+7.26}_{-21.04}$} & {\bfseries\boldmath $^{+7.22}_{-20.99}$} \\
\bottomrule
\end{tabular}
\caption{Final recommended inclusive total cross sections for Higgs boson pair production via gluon fusion obtained using Eq.~\eqref{eq:xs_reco}.
The QCD scale and TH$_\mathrm{num}$ uncertainties, as well as the $\alpha_s$ and PDF uncertainties, are taken directly from the $\textup{NNLO}_{\textup{FTapprox}}$ prediction (Table~\ref{tab:nnlo_ftapprox_xs_125}). The PDF $+$ $\alpha_s$ uncertainty is obtained by adding the two contributions \textit{in quadrature}. The final combined scale, TH$_\mathrm{num}$ and $m_t$ scheme uncertainty is obtained by adding the corresponding components \textit{linearly}.}
\label{tab:final_xs_reco_125}
\end{center}
\end{table}

\begin{table}[htbp]
\centering
\begin{tabular}{@{}lccc@{}}
\toprule
& \multicolumn{3}{c}{$\sigma_{\mathrm{rec.}}$ [fb]} \\
\cmidrule(lr){2-4}
$m_h$ [GeV] & $\sqrt{s}=13$~TeV
& $\sqrt{s}=13.6$~TeV
& $\sqrt{s}=14$~TeV \\
\midrule
124.8 & 30.48 & 33.68 & 35.90 \\
125 & 30.40 & 33.62 & 35.81 \\
125.2 & 30.29 & 33.54 & 35.68 \\
\bottomrule
\end{tabular}
\caption{Inclusive cross sections for Higgs boson pair production via gluon fusion
at different values of $m_h$, obtained from those at $m_h = 125$~GeV after rescaling
with the ratio $\sigma_{\mathrm{NLO}}(m_h)/\sigma_{\mathrm{NLO}}(125~\mathrm{GeV})$
for centre-of-mass energies of 13, 13.6, and 14~TeV.}
\label{tab:final_xs_reco_125_mh}
\end{table}

\subsection{Higgs boson pair invariant-mass distribution}\label{subsec:mhh_distribution}
The differential distribution in the Higgs boson pair invariant mass, $m_{hh}$, is constructed bin by bin following the multiplicative prescription of Eq.~\eqref{eq:xs_reco}. The $\textup{NNLO}_{\textup{FTapprox}}$ spectrum is multiplied by the differential QCD factor $K_3$ and the electroweak factor $K_\mathrm{EW}$, while the uncertainty band is inherited from the $7$-point scale variation of the $\textup{NNLO}_{\textup{FTapprox}}$ calculation. The top quark mass scheme uncertainty quoted in Table~\ref{tab:final_xs_reco_125} is not included in this differential uncertainty band.

\begin{figure*}[htbp]
\centering
\includegraphics[width=0.9\textwidth]{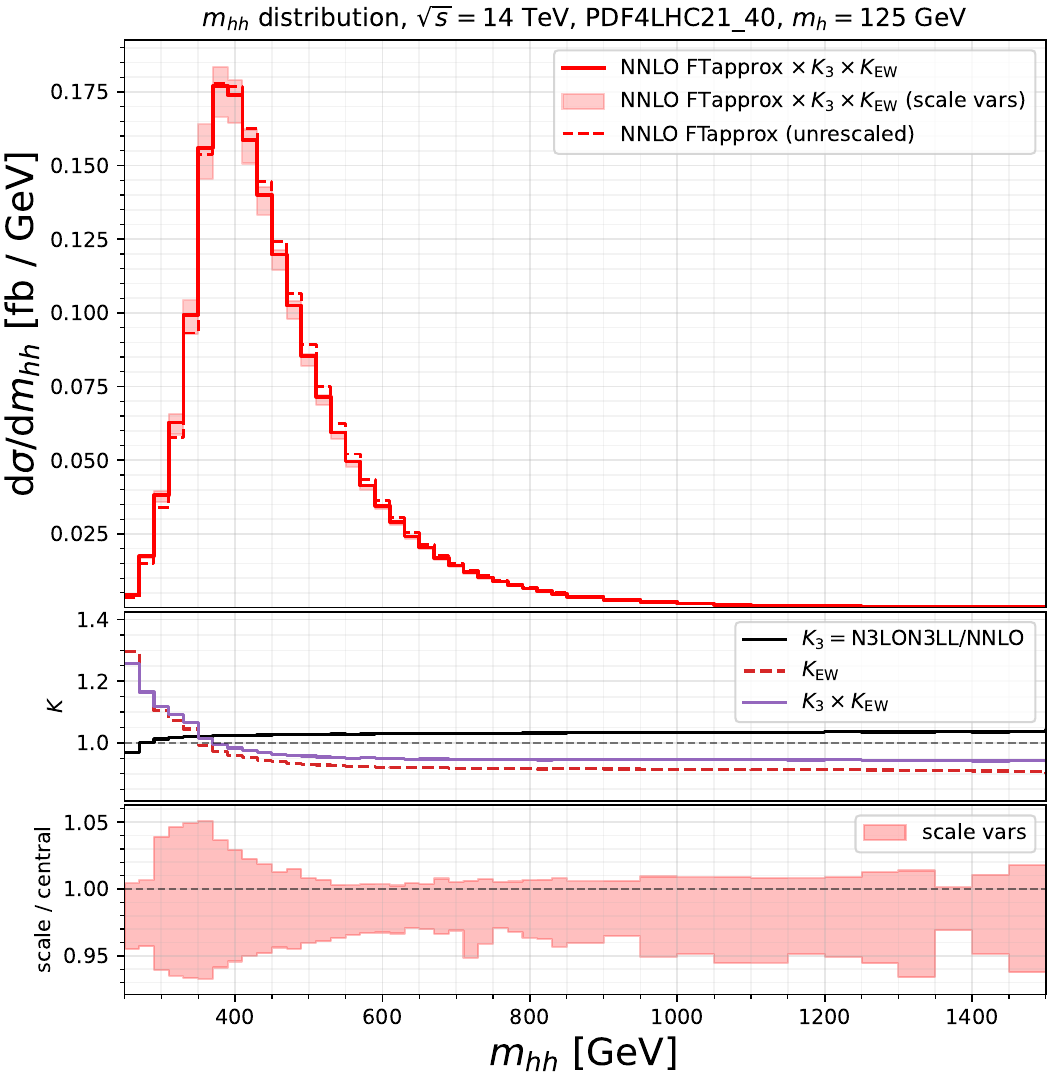}
\\
\caption{SM Higgs boson pair invariant-mass distribution at $\sqrt{s}=14$~TeV, constructed bin by bin following the multiplicative prescription of Eq.~\eqref{eq:xs_reco}. The solid red curve in the main panel is the final spectrum obtained by multiplying the $\textup{NNLO}_{\textup{FTapprox}}$ distribution by the differential (N$^3$LO+N$^3$LL)/NNLO QCD and NLO EW $K$-factors. The dashed red curve shows the unrescaled $\textup{NNLO}_{\textup{FTapprox}}$ input, and the red band gives the inherited $7$-point scale variation of the $\textup{NNLO}_{\textup{FTapprox}}$ calculation. The upper ratio inset shows the QCD factor $K_3=(\textup{N}^3\textup{LO}+\textup{N}^3\textup{LL})/\textup{NNLO}$, the electroweak factor $K_\mathrm{EW}$, and their product, and the lower ratio inset shows the uncertainty due to scale variations.}
\label{fig:xs_vs_mhh_combination_14}
\end{figure*}

The upper ratio inset in Fig.~\ref{fig:xs_vs_mhh_combination_14} shows $K_3$, $K_\mathrm{EW}$ and their product, making explicit the bin-wise rescaling applied to the $\textup{NNLO}_{\textup{FTapprox}}$ spectrum. The lower ratio inset shows the uncertainty due to scale variations. 

The shape of the spectrum and of the correction factors changes only mildly between the 13, 13.6 and 14~TeV centre-of-mass energies considered here. We therefore show the 14~TeV result as a representative case in Fig.~\ref{fig:xs_vs_mhh_combination_14}. The corresponding binned data files and the 13 and 13.6~TeV distributions are available from the LHC Higgs WG4 wiki~\cite{wg4wiki}, along with the differential $K$-factors shown in the figure, which can be used to rescale predictions obtained at lower order.


\FloatBarrier

\section{Conclusions}
\label{sec:conclusions}

This report has summarized the current status of precision predictions for
Standard Model Higgs boson pair production via gluon fusion. The discussion
brings together NLO QCD calculations with full top quark mass dependence,
approximate NNLO QCD predictions, N$^3$LO QCD corrections and soft-gluon
resummation, NLO electroweak corrections, and details the main sources of theoretical
uncertainty entering the theoretical prediction.

The final recommendations provide state-of-the-art SM reference predictions for phenomenological
studies and LHC analyses. They combine higher-order QCD (exact NLO, approximate NNLO and N$^3$LO + N$^3$LL) and EW (NLO) corrections, together with a full uncertainty budget. The combined inclusive cross sections, including the dominant uncertainty associated with the top-quark mass scheme, are collected in Table~\ref{tab:final_xs_reco_125}, while their dependence on the Higgs-boson mass is given in Table~\ref{tab:final_xs_reco_125_mh}. Additionally we provide differential distributions in $m_{hh}$ (Section~\ref{subsec:mhh_distribution}),
along with corresponding $K$-factors from the higher-order calculations.
These numbers should be used as the definitive
predictions of this report, superseding the intermediate results shown in the
preceding sections where different input parameters or PDF choices are used.
It is worth noting that while the present work does not reduce the overall
uncertainty with respect to the previous recommendation, its central prediction includes N$^3$LO+N$^3$LL QCD corrections in the HTL, NLO
electroweak effects and updated PDF sets, and should therefore provide a more
accurate reference value.

Further improvements in the SM prediction will come from reducing uncertainties associated with finite top quark mass effects and
mass-scheme choice, extending fully differential predictions with
consistently combined higher-order QCD and electroweak effects, and updating the
recommendations as parton distributions and input parameters evolve.

\section*{Acknowledgements}

\begin{sloppypar}
Ajjath A H and S.~Jones acknowledge that this research was supported in part by the UK Science and Technology Facilities Council under contracts ST/X000745/1 and ST/X003167/1. S.~Jones and Ajjath A H are supported by a Royal Society University Research Fellowship (URF/R1/201268, URF/R/251034).

A.~Bhattacharya and F.~Campanario are supported by the Generalitat Valenciana, the Spanish government, and ERDF funds from the European Commission ``NextGenerationEU/PRTR'' (CNS2022-136165, PID2023-151418NB-I00, MCIN/AEI/\allowbreak 10.13039/\allowbreak 501100011033). J.~Chang is supported by the Swiss National Science Foundation (SNSF). J.~Ronca acknowledges support from INFN. A.~Bhattacharya, F.~Campanario, S.~Carlotti, J.~Chang, J.~Mazzitelli, M.~M\"uhlleitner, J.~Ronca and M.~Spira acknowledge support from the state of Baden-W\"urttemberg through bwHPC and from the German Research Foundation (DFG) through grant no.\ INST~39/963-1~FUGG (bwForCluster NEMO).
\end{sloppypar}

M.~Bonetti, S.~Carlotti, G.~Heinrich, M.~M\"uhlleitner, P.~Rendler and A.~Vestner acknowledge support from the \textit{Deutsche Forschungsgemeinschaft} (DFG, German Research Foundation) under grant 396021762 -- TRR 257.

L.-B.~Chen is supported by the National Natural Science Foundation of China (NSFC) under Grant No.~12175048.

X.~Chen is supported by the National Science Foundation of China (NSFC) with grants No.~12475085 and No.~12321005.

J.~Davies is supported by the STFC Consolidated Grant ST/X000699/1.

\begin{sloppypar}
P.~P.~Giardino is supported by the Ram\'on y Cajal grant RYC2022-038517-I funded by MCIN/AEI/\allowbreak 10.13039/\allowbreak 501100011033 and by FSE+, and by the Spanish Research Agency (Agencia Estatal de Investigaci\'on) through the grant IFT Centro de Excelencia Severo Ochoa No.~CEX2020-001007-S.
\end{sloppypar}

R.~Gr\"ober acknowledges support from a STARS@UniPD grant under the acronym HiggsPairs.

S.~Jaskiewicz is supported by the Swiss National Science Foundation Ambizione grant PZ00P2\_223524.

\begin{sloppypar}
H.~T.~Li is supported by the National Science Foundation of China under Grant Nos.~12275156 and 12321005.
\end{sloppypar}

G.~Marinelli acknowledges funding from the European Research Council (ERC) under the European Union's Horizon 2020 research and innovation programme (Grant agreements Nos.~714788 REINVENT and 101002090 COLORFREE) and support from the Deutsche Forschungsgemeinschaft (DFG) under Germany's Excellence Strategy -- EXC 2121 ``Quantum Universe'' -- 390833306.

A.~Papaefstathiou acknowledges support from the U.S. Department of Energy, Office of Science, Office of Nuclear Physics under Award Number DE-SC0025728 and by the U.S. National Science Foundation under Grant No.\ PHY 2210161.

K.~Sch\"onwald and H.~Zhang are supported by the European Union under the Marie Sk{\l}odowska-Curie Actions (MSCA) Grants No.~101204018 and No.~101202083.

L.~Scyboz is supported by the Australian Research Council through a
Discovery Early Career Researcher Award (project number DE230100867).

H.-S.~Shao acknowledges support by grants from the ERC (grant 101041109 ``BOSON'') and the French ANR (grant ANR-20-CE31-0015 ``PrecisOnium''). Views and opinions expressed are however those of the authors only and do not necessarily reflect those of the European Union or the European Research Council Executive Agency. Neither the European Union nor the granting authority can be held responsible for them.

R.~Szafron is supported by the U.S. Department of Energy under Grant Contract DE-SC0012704.

W.~J.~Torres Bobadilla is supported by the Leverhulme Trust, LIP-2021-014.

J.~Wang is supported in part by the National Natural Science Foundation of China under Grant Nos.~12321005 and 12375076.

\bibliography{HHCitations}

\begin{thebibliography}{100}
\providecommand{\url}[1]{\texttt{#1}}
\providecommand{\urlprefix}{URL }
\expandafter\ifx\csname urlstyle\endcsname\relax
  \providecommand{\doi}[1]{doi:\discretionary{}{}{}#1}\else
  \providecommand{\doi}{doi:\discretionary{}{}{}\begingroup
  \urlstyle{rm}\Url}\fi
\providecommand{\eprint}[2][]{\url{#2}}

\bibitem{Aad:2012tfa}
ATLAS Collaboration, G.~Aad \emph{et~al.},
\newblock \emph{{Observation of a new particle in the search for the Standard
  Model Higgs boson with the ATLAS detector at the LHC}},
\newblock Phys. Lett. B \textbf{716}, 1 (2012),
\newblock \doi{10.1016/j.physletb.2012.08.020},
\newblock \eprint{1207.7214}.

\bibitem{Chatrchyan:2012xdj}
CMS Collaboration, S.~Chatrchyan \emph{et~al.},
\newblock \emph{{Observation of a New Boson at a Mass of 125 GeV with the CMS
  Experiment at the LHC}},
\newblock Phys. Lett. B \textbf{716}, 30 (2012),
\newblock \doi{10.1016/j.physletb.2012.08.021},
\newblock \eprint{1207.7235}.

\bibitem{Higgs:1964ia}
P.~W. Higgs,
\newblock \emph{{Broken symmetries, massless particles and gauge fields}},
\newblock Phys. Lett. \textbf{12}, 132 (1964),
\newblock \doi{10.1016/0031-9163(64)91136-9}.

\bibitem{Higgs:1964pj}
P.~W. Higgs,
\newblock \emph{{Broken Symmetries and the Masses of Gauge Bosons}},
\newblock Phys. Rev. Lett. \textbf{13}, 508 (1964),
\newblock \doi{10.1103/PhysRevLett.13.508}.

\bibitem{Englert:1964et}
F.~Englert and R.~Brout,
\newblock \emph{{Broken Symmetry and the Mass of Gauge Vector Mesons}},
\newblock Phys. Rev. Lett. \textbf{13}, 321 (1964),
\newblock \doi{10.1103/PhysRevLett.13.321}.

\bibitem{Guralnik:1964eu}
G.~S. Guralnik, C.~R. Hagen and T.~W.~B. Kibble,
\newblock \emph{{Global Conservation Laws and Massless Particles}},
\newblock Phys. Rev. Lett. \textbf{13}, 585 (1964),
\newblock \doi{10.1103/PhysRevLett.13.585}.

\bibitem{Higgs:1966ev}
P.~W. Higgs,
\newblock \emph{{Spontaneous Symmetry Breakdown without Massless Bosons}},
\newblock Phys. Rev. \textbf{145}, 1156 (1966),
\newblock \doi{10.1103/PhysRev.145.1156}.

\bibitem{Kibble:1967sv}
T.~W.~B. Kibble,
\newblock \emph{{Symmetry breaking in nonAbelian gauge theories}},
\newblock Phys. Rev. \textbf{155}, 1554 (1967),
\newblock \doi{10.1103/PhysRev.155.1554}.

\bibitem{Khachatryan:2016vau}
ATLAS, CMS Collaboration, G.~Aad \emph{et~al.},
\newblock \emph{{Measurements of the Higgs boson production and decay rates and
  constraints on its couplings from a combined ATLAS and CMS analysis of the
  LHC pp collision data at $ \sqrt{s}=7 $ and 8 TeV}},
\newblock JHEP \textbf{08}, 045 (2016),
\newblock \doi{10.1007/JHEP08(2016)045},
\newblock \eprint{1606.02266}.

\bibitem{ATLAS:2019nkf}
ATLAS Collaboration, G.~Aad \emph{et~al.},
\newblock \emph{{Combined measurements of Higgs boson production and decay
  using up to $80$ fb$^{-1}$ of proton-proton collision data at $\sqrt{s}=$ 13
  TeV collected with the ATLAS experiment}},
\newblock Phys. Rev. D \textbf{101}(1), 012002 (2020),
\newblock \doi{10.1103/PhysRevD.101.012002},
\newblock \eprint{1909.02845}.

\bibitem{CMS:2020xwi}
CMS Collaboration, A.~M. Sirunyan \emph{et~al.},
\newblock \emph{{Evidence for Higgs boson decay to a pair of muons}},
\newblock JHEP \textbf{01}, 148 (2021),
\newblock \doi{10.1007/JHEP01(2021)148},
\newblock \eprint{2009.04363}.

\bibitem{Glover:1987nx}
E.~W.~N. Glover and J.~J. van~der Bij,
\newblock \emph{{HIGGS BOSON PAIR PRODUCTION VIA GLUON FUSION}},
\newblock Nucl. Phys. B \textbf{309}, 282 (1988),
\newblock \doi{10.1016/0550-3213(88)90083-1}.

\bibitem{Plehn:1996wb}
T.~Plehn, M.~Spira and P.~M. Zerwas,
\newblock \emph{{Pair production of neutral Higgs particles in gluon-gluon
  collisions}},
\newblock Nucl. Phys. B \textbf{479}, 46 (1996),
\newblock \doi{10.1016/0550-3213(96)00418-X},
\newblock [Erratum: Nucl.Phys.B 531, 655--655 (1998)],
\newblock \eprint{hep-ph/9603205}.

\bibitem{Dawson:1998py}
S.~Dawson, S.~Dittmaier and M.~Spira,
\newblock \emph{{Neutral Higgs boson pair production at hadron colliders: QCD
  corrections}},
\newblock Phys. Rev. D \textbf{58}, 115012 (1998),
\newblock \doi{10.1103/PhysRevD.58.115012},
\newblock \eprint{hep-ph/9805244}.

\bibitem{Borowka:2016ehy}
S.~Borowka, N.~Greiner, G.~Heinrich, S.~P. Jones, M.~Kerner, J.~Schlenk,
  U.~Schubert and T.~Zirke,
\newblock \emph{{Higgs Boson Pair Production in Gluon Fusion at Next-to-Leading
  Order with Full Top-Quark Mass Dependence}},
\newblock Phys. Rev. Lett. \textbf{117}(1), 012001 (2016),
\newblock \doi{10.1103/PhysRevLett.117.079901},
\newblock [Erratum: Phys.Rev.Lett. 117, 079901 (2016)],
\newblock \eprint{1604.06447}.

\bibitem{Borowka:2016ypz}
S.~Borowka, N.~Greiner, G.~Heinrich, S.~P. Jones, M.~Kerner, J.~Schlenk and
  T.~Zirke,
\newblock \emph{{Full top quark mass dependence in Higgs boson pair production
  at NLO}},
\newblock JHEP \textbf{10}, 107 (2016),
\newblock \doi{10.1007/JHEP10(2016)107},
\newblock \eprint{1608.04798}.

\bibitem{Baglio:2018lrj}
J.~Baglio, F.~Campanario, S.~Glaus, M.~M{\"u}hlleitner, M.~Spira and
  J.~Streicher,
\newblock \emph{{Gluon fusion into Higgs pairs at NLO QCD and the top mass
  scheme}},
\newblock Eur. Phys. J. C \textbf{79}(6), 459 (2019),
\newblock \doi{10.1140/epjc/s10052-019-6973-3},
\newblock \eprint{1811.05692}.

\bibitem{Baglio:2020ini}
J.~Baglio, F.~Campanario, S.~Glaus, M.~M{\"u}hlleitner, J.~Ronca, M.~Spira and
  J.~Streicher,
\newblock \emph{{Higgs-Pair Production via Gluon Fusion at Hadron Colliders:
  NLO QCD Corrections}},
\newblock JHEP \textbf{04}, 181 (2020),
\newblock \doi{10.1007/JHEP04(2020)181},
\newblock \eprint{2003.03227}.

\bibitem{Baglio:2020wgt}
J.~Baglio, F.~Campanario, S.~Glaus, M.~M{\"u}hlleitner, J.~Ronca and M.~Spira,
\newblock \emph{{$gg\to HH$ : Combined uncertainties}},
\newblock Phys. Rev. D \textbf{103}(5), 056002 (2021),
\newblock \doi{10.1103/PhysRevD.103.056002},
\newblock \eprint{2008.11626}.

\bibitem{deFlorian:2013uza}
D.~de~Florian and J.~Mazzitelli,
\newblock \emph{{Two-loop virtual corrections to Higgs pair production}},
\newblock Phys. Lett. B \textbf{724}, 306 (2013),
\newblock \doi{10.1016/j.physletb.2013.06.046},
\newblock \eprint{1305.5206}.

\bibitem{deFlorian:2013jea}
D.~de~Florian and J.~Mazzitelli,
\newblock \emph{{Higgs Boson Pair Production at Next-to-Next-to-Leading Order
  in QCD}},
\newblock Phys. Rev. Lett. \textbf{111}, 201801 (2013),
\newblock \doi{10.1103/PhysRevLett.111.201801},
\newblock \eprint{1309.6594}.

\bibitem{Grigo:2014jma}
J.~Grigo, K.~Melnikov and M.~Steinhauser,
\newblock \emph{{Virtual corrections to Higgs boson pair production in the
  large top quark mass limit}},
\newblock Nucl. Phys. B \textbf{888}, 17 (2014),
\newblock \doi{10.1016/j.nuclphysb.2014.09.003},
\newblock \eprint{1408.2422}.

\bibitem{Banerjee:2018lfq}
P.~Banerjee, S.~Borowka, P.~K. Dhani, T.~Gehrmann and V.~Ravindran,
\newblock \emph{{Two-loop massless QCD corrections to the $g + g \to H + H$
  four-point amplitude}},
\newblock JHEP \textbf{11}, 130 (2018),
\newblock \doi{10.1007/JHEP11(2018)130},
\newblock \eprint{1809.05388}.

\bibitem{Chen:2019lzz}
L.-B. Chen, H.~T. Li, H.-S. Shao and J.~Wang,
\newblock \emph{{Higgs boson pair production via gluon fusion at N$^3$LO in
  QCD}},
\newblock Phys. Lett. B \textbf{803}, 135292 (2020),
\newblock \doi{10.1016/j.physletb.2020.135292},
\newblock \eprint{1909.06808}.

\bibitem{Chen:2019fhs}
L.-B. Chen, H.~T. Li, H.-S. Shao and J.~Wang,
\newblock \emph{{The gluon-fusion production of Higgs boson pair: N$^3$LO QCD
  corrections and top-quark mass effects}},
\newblock JHEP \textbf{03}, 072 (2020),
\newblock \doi{10.1007/JHEP03(2020)072},
\newblock \eprint{1912.13001}.

\bibitem{Chen:2026zmi}
X.~Chen, Y.~Dai, H.~T. Li, S.-Y. Li, H.-S. Shao and J.~Wang,
\newblock \emph{{Fully differential Higgs boson pair production at N$^{3}$LO
  with top quark mass effects}},
\newblock JHEP \textbf{06}, 005 (2026),
\newblock \doi{10.1007/JHEP06(2026)005},
\newblock \eprint{2601.19990}.

\bibitem{Ajjath:2022kpv}
A.~A~H and H.-S. Shao,
\newblock \emph{{N$^{3}$LO+N$^{3}$LL QCD improved Higgs pair cross sections}},
\newblock JHEP \textbf{02}, 067 (2023),
\newblock \doi{10.1007/JHEP02(2023)067},
\newblock \eprint{2209.03914}.

\bibitem{Heinrich:2017kxx}
G.~Heinrich, S.~P. Jones, M.~Kerner, G.~Luisoni and E.~Vryonidou,
\newblock \emph{{NLO predictions for Higgs boson pair production with full top
  quark mass dependence matched to parton showers}},
\newblock JHEP \textbf{08}, 088 (2017),
\newblock \doi{10.1007/JHEP08(2017)088},
\newblock \eprint{1703.09252}.

\bibitem{Jones:2017giv}
S.~Jones and S.~Kuttimalai,
\newblock \emph{{Parton Shower and NLO-Matching uncertainties in Higgs Boson
  Pair Production}},
\newblock JHEP \textbf{02}, 176 (2018),
\newblock \doi{10.1007/JHEP02(2018)176},
\newblock \eprint{1711.03319}.

\bibitem{Bagnaschi:2023rbx}
E.~Bagnaschi, G.~Degrassi and R.~Gr{\"o}ber,
\newblock \emph{{Higgs boson pair production at NLO in the POWHEG approach and
  the top quark mass uncertainties}},
\newblock Eur. Phys. J. C \textbf{83}(11), 1054 (2023),
\newblock \doi{10.1140/epjc/s10052-023-12238-8},
\newblock \eprint{2309.10525}.

\bibitem{Grazzini:2018bsd}
M.~Grazzini, G.~Heinrich, S.~Jones, S.~Kallweit, M.~Kerner, J.~M. Lindert and
  J.~Mazzitelli,
\newblock \emph{{Higgs boson pair production at NNLO with top quark mass
  effects}},
\newblock JHEP \textbf{05}, 059 (2018),
\newblock \doi{10.1007/JHEP05(2018)059},
\newblock \eprint{1803.02463}.

\bibitem{Bi:2023bnq}
H.-Y. Bi, L.-H. Huang, R.-J. Huang, Y.-Q. Ma and H.-M. Yu,
\newblock \emph{{Electroweak Corrections to Double Higgs Production at the
  LHC}},
\newblock Phys. Rev. Lett. \textbf{132}(23), 231802 (2024),
\newblock \doi{10.22323/1.478.0120},
\newblock \eprint{2311.16963}.

\bibitem{Heinrich:2019bkc}
G.~Heinrich, S.~P. Jones, M.~Kerner, G.~Luisoni and L.~Scyboz,
\newblock \emph{{Probing the trilinear Higgs boson coupling in di-Higgs
  production at NLO QCD including parton shower effects}},
\newblock JHEP \textbf{06}, 066 (2019),
\newblock \doi{10.1007/JHEP06(2019)066},
\newblock \eprint{1903.08137}.

\bibitem{Binoth:2000ps}
T.~Binoth and G.~Heinrich,
\newblock \emph{{An automatized algorithm to compute infrared divergent
  multi-loop integrals}},
\newblock Nucl. Phys. B \textbf{585}, 741 (2000),
\newblock \doi{10.1016/S0550-3213(00)00429-6},
\newblock \eprint{hep-ph/0004013}.

\bibitem{Borowka:2015mxa}
S.~Borowka, G.~Heinrich, S.~P. Jones, M.~Kerner, J.~Schlenk and T.~Zirke,
\newblock \emph{{SecDec-3.0: Numerical evaluation of multi-scale integrals
  beyond one loop}},
\newblock Comput. Phys. Commun. \textbf{196}, 470 (2015),
\newblock \doi{10.1016/j.cpc.2015.05.022},
\newblock \eprint{1502.06595}.

\bibitem{GoSam:2014yla}
GoSam Collaboration, G.~Cullen \emph{et~al.},
\newblock \emph{{G$\scriptsize{O}$S$\scriptsize{AM}$-2.0: a tool for automated
  one-loop calculations within the Standard Model and beyond}},
\newblock Eur. Phys. J. C \textbf{74}(8), 3001 (2014),
\newblock \doi{10.1140/epjc/s10052-014-3001-5},
\newblock \eprint{1404.7096}.

\bibitem{Braun:2025afl}
J.~Braun, B.~Campillo~Aveleira, G.~Heinrich, M.~H{\"o}fer, S.~P. Jones,
  M.~Kerner, J.~Lang and V.~Magerya,
\newblock \emph{{One-loop calculations in effective field theories with
  GoSam-3.0}},
\newblock SciPost Phys. Codeb. \textbf{62}, 1 (2026),
\newblock \doi{10.21468/SciPostPhysCodeb.62},
\newblock \eprint{2507.23549}.

\bibitem{vonManteuffel:2012np}
A.~von Manteuffel and C.~Studerus,
\newblock \emph{{Reduze 2 - Distributed Feynman Integral Reduction}}  (2012),
\newblock \eprint{1201.4330}.

\bibitem{Borowka:2017idc}
S.~Borowka, G.~Heinrich, S.~Jahn, S.~P. Jones, M.~Kerner, J.~Schlenk and
  T.~Zirke,
\newblock \emph{{pySecDec: A toolbox for the numerical evaluation of
  multi-scale integrals}},
\newblock Comput. Phys. Commun. \textbf{222}, 313 (2018),
\newblock \doi{10.1016/j.cpc.2017.09.015},
\newblock \eprint{1703.09692}.

\bibitem{Heinrich:2023til}
G.~Heinrich, S.~P. Jones, M.~Kerner, V.~Magerya, A.~Olsson and J.~Schlenk,
\newblock \emph{{Numerical scattering amplitudes with pySecDec}},
\newblock Comput. Phys. Commun. \textbf{295}, 108956 (2024),
\newblock \doi{10.1016/j.cpc.2023.108956},
\newblock \eprint{2305.19768}.

\bibitem{Catani:1996vz}
S.~Catani and M.~H. Seymour,
\newblock \emph{{A General algorithm for calculating jet cross-sections in NLO
  QCD}},
\newblock Nucl. Phys. B \textbf{485}, 291 (1997),
\newblock \doi{10.1016/S0550-3213(96)00589-5},
\newblock [Erratum: Nucl.Phys.B 510, 503--504 (1998)],
\newblock \eprint{hep-ph/9605323}.

\bibitem{Maltoni:2014eza}
F.~Maltoni, E.~Vryonidou and M.~Zaro,
\newblock \emph{{Top-quark mass effects in double and triple Higgs production
  in gluon-gluon fusion at NLO}},
\newblock JHEP \textbf{11}, 079 (2014),
\newblock \doi{10.1007/JHEP11(2014)079},
\newblock \eprint{1408.6542}.

\bibitem{Alioli:2010xd}
S.~Alioli, P.~Nason, C.~Oleari and E.~Re,
\newblock \emph{{A general framework for implementing NLO calculations in
  shower Monte Carlo programs: the POWHEG BOX}},
\newblock JHEP \textbf{06}, 043 (2010),
\newblock \doi{10.1007/JHEP06(2010)043},
\newblock \eprint{1002.2581}.

\bibitem{Alwall:2014hca}
J.~Alwall, R.~Frederix, S.~Frixione, V.~Hirschi, F.~Maltoni, O.~Mattelaer,
  H.~S. Shao, T.~Stelzer, P.~Torrielli and M.~Zaro,
\newblock \emph{{The automated computation of tree-level and next-to-leading
  order differential cross sections, and their matching to parton shower
  simulations}},
\newblock JHEP \textbf{07}, 079 (2014),
\newblock \doi{10.1007/JHEP07(2014)079},
\newblock \eprint{1405.0301}.

\bibitem{Gleisberg:2008ta}
T.~Gleisberg, S.~Hoeche, F.~Krauss, M.~Schonherr, S.~Schumann, F.~Siegert and
  J.~Winter,
\newblock \emph{{Event generation with SHERPA 1.1}},
\newblock JHEP \textbf{02}, 007 (2009),
\newblock \doi{10.1088/1126-6708/2009/02/007},
\newblock \eprint{0811.4622}.

\bibitem{Buchalla:2018yce}
G.~Buchalla, M.~Capozi, A.~Celis, G.~Heinrich and L.~Scyboz,
\newblock \emph{{Higgs boson pair production in non-linear Effective Field
  Theory with full $m_t$-dependence at NLO QCD}},
\newblock JHEP \textbf{09}, 057 (2018),
\newblock \doi{10.1007/JHEP09(2018)057},
\newblock [Erratum: JHEP 06, 094 (2025)],
\newblock \eprint{1806.05162}.

\bibitem{Heinrich:2020ckp}
G.~Heinrich, S.~P. Jones, M.~Kerner and L.~Scyboz,
\newblock \emph{{A non-linear EFT description of $gg\to HH$ at NLO interfaced
  to POWHEG}},
\newblock JHEP \textbf{10}, 021 (2020),
\newblock \doi{10.1007/JHEP10(2020)021},
\newblock \eprint{2006.16877}.

\bibitem{Heinrich:2022idm}
G.~Heinrich, J.~Lang and L.~Scyboz,
\newblock \emph{{SMEFT predictions for gg {\textrightarrow} hh at full NLO QCD
  and truncation uncertainties}},
\newblock JHEP \textbf{08}, 079 (2022),
\newblock \doi{10.1007/JHEP08(2022)079},
\newblock [Erratum: JHEP 10, 086 (2023)],
\newblock \eprint{2204.13045}.

\bibitem{Heinrich:2023rsd}
G.~Heinrich and J.~Lang,
\newblock \emph{{Combining chromomagnetic and four-fermion operators with
  leading SMEFT operators for gg {\textrightarrow} hh at NLO QCD}},
\newblock JHEP \textbf{05}, 121 (2024),
\newblock \doi{10.1007/JHEP05(2024)121},
\newblock \eprint{2311.15004}.

\bibitem{Heinrich:2024rtg}
G.~Heinrich and J.~Lang,
\newblock \emph{{Renormalisation group effects in SMEFT for di-Higgs
  production}},
\newblock SciPost Phys. \textbf{18}(3), 113 (2025),
\newblock \doi{10.21468/SciPostPhys.18.3.113},
\newblock \eprint{2409.19578}.

\bibitem{Alasfar:2023xpc}
L.~Alasfar \emph{et~al.},
\newblock \emph{{Effective Field Theory descriptions of Higgs boson pair
  production}},
\newblock SciPost Phys. Comm. Rep. \textbf{2024}, 2 (2024),
\newblock \doi{10.21468/SciPostPhysCommRep.2},
\newblock \eprint{2304.01968}.

\bibitem{Sjostrand:2014zea}
T.~Sj{\"o}strand, S.~Ask, J.~R. Christiansen, R.~Corke, N.~Desai, P.~Ilten,
  S.~Mrenna, S.~Prestel, C.~O. Rasmussen and P.~Z. Skands,
\newblock \emph{{An introduction to PYTHIA 8.2}},
\newblock Comput. Phys. Commun. \textbf{191}, 159 (2015),
\newblock \doi{10.1016/j.cpc.2015.01.024},
\newblock \eprint{1410.3012}.

\bibitem{Bellm:2017bvx}
J.~Bellm \emph{et~al.},
\newblock \emph{{Herwig 7.1 Release Note}}  (2017),
\newblock \eprint{1705.06919}.

\bibitem{Bellm:2025pcw}
J.~Bellm \emph{et~al.},
\newblock \emph{{The Physics of Herwig 7}}  (2025),
\newblock \eprint{2512.16645}.

\bibitem{Richardson}
L.~F. Richardson,
\newblock \emph{{The approximate arithmetical solution by finite differences of
  physical problems including differential equations, with an application to
  the stresses in a masonry dam}},
\newblock Philosophical Transactions of the Royal Society \textbf{A210}, 307
  (1911),
\newblock \doi{10.1098/rsta.1911.0009}.

\bibitem{Hahn:2000kx}
T.~Hahn,
\newblock \emph{{Generating Feynman diagrams and amplitudes with FeynArts 3}},
\newblock Comput. Phys. Commun. \textbf{140}, 418 (2001),
\newblock \doi{10.1016/S0010-4655(01)00290-9},
\newblock \eprint{hep-ph/0012260}.

\bibitem{Hahn:1998yk}
T.~Hahn and M.~Perez-Victoria,
\newblock \emph{{Automatized one loop calculations in four-dimensions and
  D-dimensions}},
\newblock Comput. Phys. Commun. \textbf{118}, 153 (1999),
\newblock \doi{10.1016/S0010-4655(98)00173-8},
\newblock \eprint{hep-ph/9807565}.

\bibitem{Denner:2016kdg}
A.~Denner, S.~Dittmaier and L.~Hofer,
\newblock \emph{{Collier: a fortran-based Complex One-Loop LIbrary in Extended
  Regularizations}},
\newblock Comput. Phys. Commun. \textbf{212}, 220 (2017),
\newblock \doi{10.1016/j.cpc.2016.10.013},
\newblock \eprint{1604.06792}.

\bibitem{Graudenz:1992pv}
D.~Graudenz, M.~Spira and P.~M. Zerwas,
\newblock \emph{{QCD corrections to Higgs boson production at proton proton
  colliders}},
\newblock Phys. Rev. Lett. \textbf{70}, 1372 (1993),
\newblock \doi{10.1103/PhysRevLett.70.1372}.

\bibitem{Spira:1995rr}
M.~Spira, A.~Djouadi, D.~Graudenz and P.~M. Zerwas,
\newblock \emph{{Higgs boson production at the LHC}},
\newblock Nucl. Phys. B \textbf{453}, 17 (1995),
\newblock \doi{10.1016/0550-3213(95)00379-7},
\newblock \eprint{hep-ph/9504378}.

\bibitem{Anastasiou:2016cez}
C.~Anastasiou, C.~Duhr, F.~Dulat, E.~Furlan, T.~Gehrmann, F.~Herzog,
  A.~Lazopoulos and B.~Mistlberger,
\newblock \emph{{High precision determination of the gluon fusion Higgs boson
  cross-section at the LHC}},
\newblock JHEP \textbf{05}, 058 (2016),
\newblock \doi{10.1007/JHEP05(2016)058},
\newblock \eprint{1602.00695}.

\bibitem{Bonciani:2018omm}
R.~Bonciani, G.~Degrassi, P.~P. Giardino and R.~Gr{\"o}ber,
\newblock \emph{{Analytical Method for Next-to-Leading-Order QCD Corrections to
  Double-Higgs Production}},
\newblock Phys. Rev. Lett. \textbf{121}(16), 162003 (2018),
\newblock \doi{10.1103/PhysRevLett.121.162003},
\newblock \eprint{1806.11564}.

\bibitem{Bonciani:2018uvv}
R.~Bonciani, G.~Degrassi, P.~P. Giardino and R.~Gr{\"o}ber,
\newblock \emph{{A Numerical Routine for the Crossed Vertex Diagram with a
  Massive-Particle Loop}},
\newblock Comput. Phys. Commun. \textbf{241}, 122 (2019),
\newblock \doi{10.1016/j.cpc.2019.03.014},
\newblock \eprint{1812.02698}.

\bibitem{Pozzorini:2005ff}
S.~Pozzorini and E.~Remiddi,
\newblock \emph{{Precise numerical evaluation of the two loop sunrise graph
  master integrals in the equal mass case}},
\newblock Comput. Phys. Commun. \textbf{175}, 381 (2006),
\newblock \doi{10.1016/j.cpc.2006.05.005},
\newblock \eprint{hep-ph/0505041}.

\bibitem{Aglietti:2007as}
U.~Aglietti, R.~Bonciani, L.~Grassi and E.~Remiddi,
\newblock \emph{{The Two loop crossed ladder vertex diagram with two massive
  exchanges}},
\newblock Nucl. Phys. B \textbf{789}, 45 (2008),
\newblock \doi{10.1016/j.nuclphysb.2007.07.019},
\newblock \eprint{0705.2616}.

\bibitem{Davies:2019dfy}
J.~Davies, G.~Heinrich, S.~P. Jones, M.~Kerner, G.~Mishima, M.~Steinhauser and
  D.~Wellmann,
\newblock \emph{{Double Higgs boson production at NLO: combining the exact
  numerical result and high-energy expansion}},
\newblock JHEP \textbf{11}, 024 (2019),
\newblock \doi{10.1007/JHEP11(2019)024},
\newblock \eprint{1907.06408}.

\bibitem{Bellafronte:2022jmo}
L.~Bellafronte, G.~Degrassi, P.~P. Giardino, R.~Gr{\"o}ber and M.~Vitti,
\newblock \emph{{Gluon fusion production at NLO: merging the transverse
  momentum and the high-energy expansions}},
\newblock JHEP \textbf{07}, 069 (2022),
\newblock \doi{10.1007/JHEP07(2022)069},
\newblock \eprint{2202.12157}.

\bibitem{Davies:2018qvx}
J.~Davies, G.~Mishima, M.~Steinhauser and D.~Wellmann,
\newblock \emph{{Double Higgs boson production at NLO in the high-energy limit:
  complete analytic results}},
\newblock JHEP \textbf{01}, 176 (2019),
\newblock \doi{10.1007/JHEP01(2019)176},
\newblock \eprint{1811.05489}.

\bibitem{Davies:2018ood}
J.~Davies, G.~Mishima, M.~Steinhauser and D.~Wellmann,
\newblock \emph{{Double-Higgs boson production in the high-energy limit: planar
  master integrals}},
\newblock JHEP \textbf{03}, 048 (2018),
\newblock \doi{10.1007/JHEP03(2018)048},
\newblock \eprint{1801.09696}.

\bibitem{Hirschi:2011pa}
V.~Hirschi, R.~Frederix, S.~Frixione, M.~V. Garzelli, F.~Maltoni and R.~Pittau,
\newblock \emph{{Automation of one-loop QCD corrections}},
\newblock JHEP \textbf{05}, 044 (2011),
\newblock \doi{10.1007/JHEP05(2011)044},
\newblock \eprint{1103.0621}.

\bibitem{Davies:2025qjr}
J.~Davies, K.~Sch{\"o}nwald, M.~Steinhauser and D.~Stremmer,
\newblock \emph{{ggxy: A flexible library to compute gluon-induced cross
  sections}},
\newblock Comput. Phys. Commun. \textbf{320}, 109933 (2026),
\newblock \doi{10.1016/j.cpc.2025.109933},
\newblock \eprint{2506.04323}.

\bibitem{Davies:2023vmj}
J.~Davies, G.~Mishima, K.~Sch{\"o}nwald and M.~Steinhauser,
\newblock \emph{{Analytic approximations of 2 {\textrightarrow} 2 processes
  with massive internal particles}},
\newblock JHEP \textbf{06}, 063 (2023),
\newblock \doi{10.1007/JHEP06(2023)063},
\newblock \eprint{2302.01356}.

\bibitem{Grober:2017uho}
R.~Gr{\"o}ber, A.~Maier and T.~Rauh,
\newblock \emph{{Reconstruction of top-quark mass effects in Higgs pair
  production and other gluon-fusion processes}},
\newblock JHEP \textbf{03}, 020 (2018),
\newblock \doi{10.1007/JHEP03(2018)020},
\newblock \eprint{1709.07799}.

\bibitem{Xu:2018eos}
X.~Xu and L.~L. Yang,
\newblock \emph{{Towards a new approximation for pair-production and
  associated-production of the Higgs boson}},
\newblock JHEP \textbf{01}, 211 (2019),
\newblock \doi{10.1007/JHEP01(2019)211},
\newblock \eprint{1810.12002}.

\bibitem{Jaskiewicz:2024xkd}
S.~Jaskiewicz, S.~Jones, R.~Szafron and Y.~Ulrich,
\newblock \emph{{The structure of quark mass corrections in the gg
  {\textrightarrow} HH amplitude at high-energy}},
\newblock JHEP \textbf{09}, 015 (2025),
\newblock \doi{10.1007/JHEP09(2025)015},
\newblock \eprint{2501.00587}.

\bibitem{NNPDF:2017mvq}
NNPDF Collaboration, R.~D. Ball \emph{et~al.},
\newblock \emph{{Parton distributions from high-precision collider data}},
\newblock Eur. Phys. J. C \textbf{77}(10), 663 (2017),
\newblock \doi{10.1140/epjc/s10052-017-5199-5},
\newblock \eprint{1706.00428}.

\bibitem{Actis:2012qn}
S.~Actis, A.~Denner, L.~Hofer, A.~Scharf and S.~Uccirati,
\newblock \emph{{Recursive generation of one-loop amplitudes in the Standard
  Model}},
\newblock JHEP \textbf{04}, 037 (2013),
\newblock \doi{10.1007/JHEP04(2013)037},
\newblock \eprint{1211.6316}.

\bibitem{Actis:2016mpe}
S.~Actis, A.~Denner, L.~Hofer, J.-N. Lang, A.~Scharf and S.~Uccirati,
\newblock \emph{{RECOLA: REcursive Computation of One-Loop Amplitudes}},
\newblock Comput. Phys. Commun. \textbf{214}, 140 (2017),
\newblock \doi{10.1016/j.cpc.2017.01.004},
\newblock \eprint{1605.01090}.

\bibitem{Cascioli:2011va}
F.~Cascioli, P.~Maierhofer and S.~Pozzorini,
\newblock \emph{{Scattering Amplitudes with Open Loops}},
\newblock Phys. Rev. Lett. \textbf{108}, 111601 (2012),
\newblock \doi{10.1103/PhysRevLett.108.111601},
\newblock \eprint{1111.5206}.

\bibitem{Buccioni:2019sur}
F.~Buccioni, J.-N. Lang, J.~M. Lindert, P.~Maierh{\"o}fer, S.~Pozzorini,
  H.~Zhang and M.~F. Zoller,
\newblock \emph{{OpenLoops 2}},
\newblock Eur. Phys. J. C \textbf{79}(10), 866 (2019),
\newblock \doi{10.1140/epjc/s10052-019-7306-2},
\newblock \eprint{1907.13071}.

\bibitem{Alioli:2025xcu}
S.~Alioli, G.~Marinelli and D.~Napoletano,
\newblock \emph{{NNLO+PS double Higgs boson production with top-quark mass
  corrections in GENEVA}},
\newblock JHEP \textbf{09}, 206 (2025),
\newblock \doi{10.1007/JHEP09(2025)206},
\newblock \eprint{2507.08558}.

\bibitem{Frederix:2014hta}
R.~Frederix, S.~Frixione, V.~Hirschi, F.~Maltoni, O.~Mattelaer, P.~Torrielli,
  E.~Vryonidou and M.~Zaro,
\newblock \emph{{Higgs pair production at the LHC with NLO and parton-shower
  effects}},
\newblock Phys. Lett. B \textbf{732}, 142 (2014),
\newblock \doi{10.1016/j.physletb.2014.03.026},
\newblock \eprint{1401.7340}.

\bibitem{Alioli:2012fc}
S.~Alioli, C.~W. Bauer, C.~J. Berggren, A.~Hornig, F.~J. Tackmann, C.~K.
  Vermilion, J.~R. Walsh and S.~Zuberi,
\newblock \emph{{Combining Higher-Order Resummation with Multiple NLO
  Calculations and Parton Showers in GENEVA}},
\newblock JHEP \textbf{09}, 120 (2013),
\newblock \doi{10.1007/JHEP09(2013)120},
\newblock \eprint{1211.7049}.

\bibitem{Stewart:2010tn}
I.~W. Stewart, F.~J. Tackmann and W.~J. Waalewijn,
\newblock \emph{{N-Jettiness: An Inclusive Event Shape to Veto Jets}},
\newblock Phys. Rev. Lett. \textbf{105}, 092002 (2010),
\newblock \doi{10.1103/PhysRevLett.105.092002},
\newblock \eprint{1004.2489}.

\bibitem{Buccioni:2017yxi}
F.~Buccioni, S.~Pozzorini and M.~Zoller,
\newblock \emph{{On-the-fly reduction of open loops}},
\newblock Eur. Phys. J. C \textbf{78}(1), 70 (2018),
\newblock \doi{10.1140/epjc/s10052-018-5562-1},
\newblock \eprint{1710.11452}.

\bibitem{Frixione:1995ms}
S.~Frixione, Z.~Kunszt and A.~Signer,
\newblock \emph{{Three jet cross-sections to next-to-leading order}},
\newblock Nucl. Phys. B \textbf{467}, 399 (1996),
\newblock \doi{10.1016/0550-3213(96)00110-1},
\newblock \eprint{hep-ph/9512328}.

\bibitem{Alioli:2022dkj}
S.~Alioli, G.~Billis, A.~Broggio, A.~Gavardi, S.~Kallweit, M.~A. Lim,
  G.~Marinelli, R.~Nagar and D.~Napoletano,
\newblock \emph{{Double Higgs production at NNLO interfaced to parton showers
  in GENEVA}},
\newblock JHEP \textbf{06}, 205 (2023),
\newblock \doi{10.1007/JHEP06(2023)205},
\newblock \eprint{2212.10489}.

\bibitem{PDF4LHCWorkingGroup:2022cjn}
PDF4LHC Working Group Collaboration, R.~D. Ball \emph{et~al.},
\newblock \emph{{The PDF4LHC21 combination of global PDF fits for the LHC Run
  III}},
\newblock J. Phys. G \textbf{49}(8), 080501 (2022),
\newblock \doi{10.1088/1361-6471/ac7216},
\newblock \eprint{2203.05506}.

\bibitem{Buckley:2014ana}
A.~Buckley, J.~Ferrando, S.~Lloyd, K.~Nordstr{\"o}m, B.~Page, M.~R{\"u}fenacht,
  M.~Sch{\"o}nherr and G.~Watt,
\newblock \emph{{LHAPDF6: parton density access in the LHC precision era}},
\newblock Eur. Phys. J. C \textbf{75}, 132 (2015),
\newblock \doi{10.1140/epjc/s10052-015-3318-8},
\newblock \eprint{1412.7420}.

\bibitem{Catani:2015vma}
S.~Catani, D.~de~Florian, G.~Ferrera and M.~Grazzini,
\newblock \emph{{Vector boson production at hadron colliders:
  transverse-momentum resummation and leptonic decay}},
\newblock JHEP \textbf{12}, 047 (2015),
\newblock \doi{10.1007/JHEP12(2015)047},
\newblock \eprint{1507.06937}.

\bibitem{Davies:2023obx}
J.~Davies, K.~Sch{\"o}nwald and M.~Steinhauser,
\newblock \emph{{Towards~$gg\to HH$ at next-to-next-to-leading order:
  Light-fermionic three-loop corrections}},
\newblock Phys. Lett. B \textbf{845}, 138146 (2023),
\newblock \doi{10.1016/j.physletb.2023.138146},
\newblock \eprint{2307.04796}.

\bibitem{Davies:2024znp}
J.~Davies, K.~Sch{\"o}nwald, M.~Steinhauser and M.~Vitti,
\newblock \emph{{Three-loop corrections to Higgs boson pair production:
  reducible contribution}},
\newblock JHEP \textbf{08}, 096 (2024),
\newblock \doi{10.1007/JHEP08(2024)096},
\newblock \eprint{2405.20372}.

\bibitem{Davies:2025ghl}
J.~Davies, K.~Sch{\"o}nwald and M.~Steinhauser,
\newblock \emph{{Three-loop large-N$_{c}$ virtual corrections to gg
  {\textrightarrow} HH in the forward limit}},
\newblock JHEP \textbf{08}, 192 (2025),
\newblock \doi{10.1007/JHEP08(2025)192},
\newblock \eprint{2503.17449}.

\bibitem{Ruijl:2017dtg}
B.~Ruijl, T.~Ueda and J.~Vermaseren,
\newblock \emph{{FORM version 4.2}}  (2017),
\newblock \eprint{1707.06453}.

\bibitem{Davies:2026cci}
J.~Davies, T.~Kaneko, C.~Marinissen, T.~Ueda and J.~A.~M. Vermaseren,
\newblock \emph{{FORM Version 5.0}}  (2026),
\newblock \eprint{2601.19982}.

\bibitem{Maheria:2022dsq}
V.~Maheria,
\newblock \emph{{Semi- and Fully-Inclusive Phase-Space Integrals at Four
  Loops}},
\newblock Ph.D. thesis, Hamburg U. (2022).

\bibitem{Lange:2025ofh}
F.~Lange, J.~Usovitsch and Z.~Wu,
\newblock \emph{{Kira 3: integral reduction with efficient seeding and
  optimized equation selection}},
\newblock Comput. Phys. Commun. \textbf{322}, 109999 (2026),
\newblock \doi{10.1016/j.cpc.2025.109999},
\newblock \eprint{2505.20197}.

\bibitem{Smirnov:2025prc}
A.~V. Smirnov and M.~Zeng,
\newblock \emph{{FIRE 7: Automatic Reduction with Modular Approach}}  (2025),
\newblock \eprint{2510.07150}.

\bibitem{Inami:1982xt}
T.~Inami, T.~Kubota and Y.~Okada,
\newblock \emph{{Effective Gauge Theory and the Effect of Heavy Quarks in Higgs
  Boson Decays}},
\newblock Z. Phys. C \textbf{18}, 69 (1983),
\newblock \doi{10.1007/BF01571710}.

\bibitem{Chetyrkin:1997iv}
K.~G. Chetyrkin, B.~A. Kniehl and M.~Steinhauser,
\newblock \emph{{Hadronic Higgs decay to order $\alpha_s^4$}},
\newblock Phys. Rev. Lett. \textbf{79}, 353 (1997),
\newblock \doi{10.1103/PhysRevLett.79.353},
\newblock \eprint{hep-ph/9705240}.

\bibitem{Chetyrkin:1997un}
K.~G. Chetyrkin, B.~A. Kniehl and M.~Steinhauser,
\newblock \emph{{Decoupling relations to $\mathcal{O}(\alpha_s^3)$ and their
  connection to low-energy theorems}},
\newblock Nucl. Phys. B \textbf{510}, 61 (1998),
\newblock \doi{10.1016/S0550-3213(97)00649-4},
\newblock \eprint{hep-ph/9708255}.

\bibitem{Schroder:2005hy}
Y.~Schroder and M.~Steinhauser,
\newblock \emph{{Four-loop decoupling relations for the strong coupling}},
\newblock JHEP \textbf{01}, 051 (2006),
\newblock \doi{10.1088/1126-6708/2006/01/051},
\newblock \eprint{hep-ph/0512058}.

\bibitem{Chetyrkin:2005ia}
K.~G. Chetyrkin, J.~H. Kuhn and C.~Sturm,
\newblock \emph{{QCD decoupling at four loops}},
\newblock Nucl. Phys. B \textbf{744}, 121 (2006),
\newblock \doi{10.1016/j.nuclphysb.2006.03.020},
\newblock \eprint{hep-ph/0512060}.

\bibitem{Kniehl:2006bg}
B.~A. Kniehl, A.~V. Kotikov, A.~I. Onishchenko and O.~L. Veretin,
\newblock \emph{{Strong-coupling constant with flavor thresholds at five loops
  in the anti-MS scheme}},
\newblock Phys. Rev. Lett. \textbf{97}, 042001 (2006),
\newblock \doi{10.1103/PhysRevLett.97.042001},
\newblock \eprint{hep-ph/0607202}.

\bibitem{Baikov:2016tgj}
P.~A. Baikov, K.~G. Chetyrkin and J.~H. K{\"u}hn,
\newblock \emph{{Five-Loop Running of the QCD Coupling Constant}},
\newblock Phys. Rev. Lett. \textbf{118}(8), 082002 (2017),
\newblock \doi{10.1103/PhysRevLett.118.082002},
\newblock \eprint{1606.08659}.

\bibitem{Spira:2016zna}
M.~Spira,
\newblock \emph{{Effective Multi-Higgs Couplings to Gluons}},
\newblock JHEP \textbf{10}, 026 (2016),
\newblock \doi{10.1007/JHEP10(2016)026},
\newblock \eprint{1607.05548}.

\bibitem{Gerlach:2018hen}
M.~Gerlach, F.~Herren and M.~Steinhauser,
\newblock \emph{{Wilson coefficients for Higgs boson production and decoupling
  relations to $ \mathcal{O}\left({\alpha}_s^4\right) $}},
\newblock JHEP \textbf{11}, 141 (2018),
\newblock \doi{10.1007/JHEP11(2018)141},
\newblock \eprint{1809.06787}.

\bibitem{Butterworth:2015oua}
J.~Butterworth \emph{et~al.},
\newblock \emph{{PDF4LHC recommendations for LHC Run II}},
\newblock J. Phys. G \textbf{43}, 023001 (2016),
\newblock \doi{10.1088/0954-3899/43/2/023001},
\newblock \eprint{1510.03865}.

\bibitem{Dulat:2015mca}
S.~Dulat, T.-J. Hou, J.~Gao, M.~Guzzi, J.~Huston, P.~Nadolsky, J.~Pumplin,
  C.~Schmidt, D.~Stump and C.~P. Yuan,
\newblock \emph{{New parton distribution functions from a global analysis of
  quantum chromodynamics}},
\newblock Phys. Rev. D \textbf{93}(3), 033006 (2016),
\newblock \doi{10.1103/PhysRevD.93.033006},
\newblock \eprint{1506.07443}.

\bibitem{Harland-Lang:2014zoa}
L.~A. Harland-Lang, A.~D. Martin, P.~Motylinski and R.~S. Thorne,
\newblock \emph{{Parton distributions in the LHC era: MMHT 2014 PDFs}},
\newblock Eur. Phys. J. C \textbf{75}(5), 204 (2015),
\newblock \doi{10.1140/epjc/s10052-015-3397-6},
\newblock \eprint{1412.3989}.

\bibitem{NNPDF:2014otw}
NNPDF Collaboration, R.~D. Ball \emph{et~al.},
\newblock \emph{{Parton distributions for the LHC Run II}},
\newblock JHEP \textbf{04}, 040 (2015),
\newblock \doi{10.1007/JHEP04(2015)040},
\newblock \eprint{1410.8849}.

\bibitem{NNPDF:2024nan}
NNPDF Collaboration, R.~D. Ball \emph{et~al.},
\newblock \emph{{The path to $\hbox {N}^3\hbox {LO}$ parton distributions}},
\newblock Eur. Phys. J. C \textbf{84}(7), 659 (2024),
\newblock \doi{10.1140/epjc/s10052-024-12891-7},
\newblock \eprint{2402.18635}.

\bibitem{Dulat:2018rbf}
F.~Dulat, A.~Lazopoulos and B.~Mistlberger,
\newblock \emph{{iHixs 2 {\textemdash} Inclusive Higgs cross sections}},
\newblock Comput. Phys. Commun. \textbf{233}, 243 (2018),
\newblock \doi{10.1016/j.cpc.2018.06.025},
\newblock \eprint{1802.00827}.

\bibitem{NNLOJET:2025rno}
NNLOJET Collaboration, A.~Huss \emph{et~al.},
\newblock \emph{{NNLOJET: A parton-level event generator for jet cross sections
  at NNLO QCD accuracy}},
\newblock SciPost Phys. Codeb. \textbf{69}, 1 (2026),
\newblock \doi{10.21468/SciPostPhysCodeb.69},
\newblock \eprint{2503.22804}.

\bibitem{Catani:2007vq}
S.~Catani and M.~Grazzini,
\newblock \emph{{An NNLO subtraction formalism in hadron collisions and its
  application to Higgs boson production at the LHC}},
\newblock Phys. Rev. Lett. \textbf{98}, 222002 (2007),
\newblock \doi{10.1103/PhysRevLett.98.222002},
\newblock \eprint{hep-ph/0703012}.

\bibitem{Cieri:2018oms}
L.~Cieri, X.~Chen, T.~Gehrmann, E.~W.~N. Glover and A.~Huss,
\newblock \emph{{Higgs boson production at the LHC using the $q_T$ subtraction
  formalism at N$^3$LO QCD}},
\newblock JHEP \textbf{02}, 096 (2019),
\newblock \doi{10.1007/JHEP02(2019)096},
\newblock \eprint{1807.11501}.

\bibitem{Frederix:2018nkq}
R.~Frederix, S.~Frixione, V.~Hirschi, D.~Pagani, H.~S. Shao and M.~Zaro,
\newblock \emph{{The automation of next-to-leading order electroweak
  calculations}},
\newblock JHEP \textbf{07}, 185 (2018),
\newblock \doi{10.1007/JHEP11(2021)085},
\newblock [Erratum: JHEP 11, 085 (2021)],
\newblock \eprint{1804.10017}.

\bibitem{Ravindran:2005vv}
V.~Ravindran,
\newblock \emph{{On Sudakov and soft resummations in QCD}},
\newblock Nucl. Phys. B \textbf{746}, 58 (2006),
\newblock \doi{10.1016/j.nuclphysb.2006.04.008},
\newblock \eprint{hep-ph/0512249}.

\bibitem{Ahmed:2020nci}
T.~Ahmed, A.~A~H, G.~Das, P.~Mukherjee, V.~Ravindran and S.~Tiwari,
\newblock \emph{{Soft-virtual correction and threshold resummation for
  $n$-colorless particles to fourth order in QCD: Part I}}  (2020),
\newblock \eprint{2010.02979}.

\bibitem{Beneke:1997zp}
M.~Beneke and V.~A. Smirnov,
\newblock \emph{{Asymptotic expansion of Feynman integrals near threshold}},
\newblock Nucl. Phys. B \textbf{522}, 321 (1998),
\newblock \doi{10.1016/S0550-3213(98)00138-2},
\newblock \eprint{hep-ph/9711391}.

\bibitem{Smirnov:1990rz}
V.~A. Smirnov,
\newblock \emph{{Asymptotic expansions in limits of large momenta and masses}},
\newblock Commun. Math. Phys. \textbf{134}, 109 (1990),
\newblock \doi{10.1007/BF02102092}.

\bibitem{Smirnov:1994tg}
V.~A. Smirnov,
\newblock \emph{{Asymptotic expansions in momenta and masses and calculation of
  Feynman diagrams}},
\newblock Mod. Phys. Lett. A \textbf{10}, 1485 (1995),
\newblock \doi{10.1142/S0217732395001617},
\newblock \eprint{hep-th/9412063}.

\bibitem{Smirnov:2002pj}
V.~A. Smirnov,
\newblock \emph{{Applied asymptotic expansions in momenta and masses}},
\newblock Springer Tracts Mod. Phys. \textbf{177}, 1 (2002).

\bibitem{Bauer:2000yr}
C.~W. Bauer, S.~Fleming, D.~Pirjol and I.~W. Stewart,
\newblock \emph{{An Effective field theory for collinear and soft gluons: Heavy
  to light decays}},
\newblock Phys. Rev. D \textbf{63}, 114020 (2001),
\newblock \doi{10.1103/PhysRevD.63.114020},
\newblock \eprint{hep-ph/0011336}.

\bibitem{Bauer:2001yt}
C.~W. Bauer, D.~Pirjol and I.~W. Stewart,
\newblock \emph{{Soft collinear factorization in effective field theory}},
\newblock Phys. Rev. D \textbf{65}, 054022 (2002),
\newblock \doi{10.1103/PhysRevD.65.054022},
\newblock \eprint{hep-ph/0109045}.

\bibitem{Bauer:2002nz}
C.~W. Bauer, S.~Fleming, D.~Pirjol, I.~Z. Rothstein and I.~W. Stewart,
\newblock \emph{{Hard scattering factorization from effective field theory}},
\newblock Phys. Rev. D \textbf{66}, 014017 (2002),
\newblock \doi{10.1103/PhysRevD.66.014017},
\newblock \eprint{hep-ph/0202088}.

\bibitem{Beneke:2002ph}
M.~Beneke, A.~P. Chapovsky, M.~Diehl and T.~Feldmann,
\newblock \emph{{Soft collinear effective theory and heavy to light currents
  beyond leading power}},
\newblock Nucl. Phys. B \textbf{643}, 431 (2002),
\newblock \doi{10.1016/S0550-3213(02)00687-9},
\newblock \eprint{hep-ph/0206152}.

\bibitem{Beneke:2002ni}
M.~Beneke and T.~Feldmann,
\newblock \emph{{Multipole expanded soft collinear effective theory with
  nonAbelian gauge symmetry}},
\newblock Phys. Lett. B \textbf{553}, 267 (2003),
\newblock \doi{10.1016/S0370-2693(02)03204-5},
\newblock \eprint{hep-ph/0211358}.

\bibitem{Penin:2005eh}
A.~A. Penin,
\newblock \emph{{Two-loop photonic corrections to massive Bhabha scattering}},
\newblock Nucl. Phys. B \textbf{734}, 185 (2006),
\newblock \doi{10.1016/j.nuclphysb.2005.11.016},
\newblock \eprint{hep-ph/0508127}.

\bibitem{Mitov:2006xs}
A.~Mitov and S.~Moch,
\newblock \emph{{The Singular behavior of massive QCD amplitudes}},
\newblock JHEP \textbf{05}, 001 (2007),
\newblock \doi{10.1088/1126-6708/2007/05/001},
\newblock \eprint{hep-ph/0612149}.

\bibitem{Becher:2007cu}
T.~Becher and K.~Melnikov,
\newblock \emph{{Two-loop QED corrections to Bhabha scattering}},
\newblock JHEP \textbf{06}, 084 (2007),
\newblock \doi{10.1088/1126-6708/2007/06/084},
\newblock \eprint{0704.3582}.

\bibitem{Liu:2017axv}
T.~Liu, A.~A. Penin and N.~Zerf,
\newblock \emph{{Three-loop quark form factor at high energy: the leading mass
  corrections}},
\newblock Phys. Lett. B \textbf{771}, 492 (2017),
\newblock \doi{10.1016/j.physletb.2017.06.002},
\newblock \eprint{1705.07910}.

\bibitem{Engel:2018fsb}
T.~Engel, C.~Gnendiger, A.~Signer and Y.~Ulrich,
\newblock \emph{{Small-mass effects in heavy-to-light form factors}},
\newblock JHEP \textbf{02}, 118 (2019),
\newblock \doi{10.1007/JHEP02(2019)118},
\newblock \eprint{1811.06461}.

\bibitem{Wang:2023qbf}
G.~Wang, T.~Xia, L.~L. Yang and X.~Ye,
\newblock \emph{{On the high-energy behavior of massive QCD amplitudes}},
\newblock JHEP \textbf{05}, 082 (2024),
\newblock \doi{10.1007/JHEP05(2024)082},
\newblock \eprint{2312.12242}.

\bibitem{Wang:2024pmv}
G.~Wang, T.~Xia, L.~L. Yang and X.~Ye,
\newblock \emph{{Two-loop QCD amplitudes for $ t\overline{t}H $ production from
  boosted limit}},
\newblock JHEP \textbf{07}, 121 (2024),
\newblock \doi{10.1007/JHEP07(2024)121},
\newblock \eprint{2402.00431}.

\bibitem{Beneke:2020ibj}
M.~Beneke, M.~Garny, S.~Jaskiewicz, R.~Szafron, L.~Vernazza and J.~Wang,
\newblock \emph{{Large-x resummation of off-diagonal deep-inelastic parton
  scattering from d-dimensional refactorization}},
\newblock JHEP \textbf{10}, 196 (2020),
\newblock \doi{10.1007/JHEP10(2020)196},
\newblock \eprint{2008.04943}.

\bibitem{Beneke:2022obx}
M.~Beneke, M.~Garny, S.~Jaskiewicz, J.~Strohm, R.~Szafron, L.~Vernazza and
  J.~Wang,
\newblock \emph{{Next-to-leading power endpoint factorization and resummation
  for off-diagonal {\textquotedblleft}gluon{\textquotedblright} thrust}},
\newblock JHEP \textbf{07}, 144 (2022),
\newblock \doi{10.1007/JHEP07(2022)144},
\newblock \eprint{2205.04479}.

\bibitem{Liu:2019oav}
Z.~L. Liu and M.~Neubert,
\newblock \emph{{Factorization at subleading power and endpoint-divergent
  convolutions in $h\to\gamma\gamma$ decay}},
\newblock JHEP \textbf{04}, 033 (2020),
\newblock \doi{10.1007/JHEP04(2020)033},
\newblock \eprint{1912.08818}.

\bibitem{Bell:2022ott}
G.~Bell, P.~B{\"o}er and T.~Feldmann,
\newblock \emph{{Muon-electron backward scattering: a prime example for
  endpoint singularities in SCET}},
\newblock JHEP \textbf{09}, 183 (2022),
\newblock \doi{10.1007/JHEP09(2022)183},
\newblock \eprint{2205.06021}.

\bibitem{Blade}
X.~Guan, X.~Liu, Y.-Q. Ma and W.-H. Wu,
\newblock \emph{{Blade: A package for block-triangular form improved Feynman
  integrals decomposition}},
\newblock Comput. Phys. Commun. \textbf{310}, 109538 (2025),
\newblock \doi{10.1016/j.cpc.2025.109538},
\newblock \eprint{2405.14621}.

\bibitem{Peraro:2019svx}
T.~Peraro,
\newblock \emph{{$\text{FiniteFlow}$: multivariate functional reconstruction
  using finite fields and dataflow graphs}},
\newblock JHEP \textbf{07}, 031 (2019),
\newblock \doi{10.1007/JHEP07(2019)031},
\newblock \eprint{1905.08019}.

\bibitem{Chetyrkin:1981qh}
K.~G. Chetyrkin and F.~V. Tkachov,
\newblock \emph{{Integration by parts: The algorithm to calculate
  $\beta$-functions in 4 loops}},
\newblock Nucl. Phys. B \textbf{192}, 159 (1981),
\newblock \doi{10.1016/0550-3213(81)90199-1}.

\bibitem{Liu:2018dmc}
X.~Liu and Y.-Q. Ma,
\newblock \emph{{Determining arbitrary Feynman integrals by vacuum integrals}},
\newblock Phys. Rev. D \textbf{99}(7), 071501 (2019),
\newblock \doi{10.1103/PhysRevD.99.071501},
\newblock \eprint{1801.10523}.

\bibitem{Guan:2019bcx}
X.~Guan, X.~Liu and Y.-Q. Ma,
\newblock \emph{{Complete reduction of integrals in two-loop five-light-parton
  scattering amplitudes}},
\newblock Chin. Phys. C \textbf{44}(9), 093106 (2020),
\newblock \doi{10.1088/1674-1137/44/9/093106},
\newblock \eprint{1912.09294}.

\bibitem{Liu:2022chg}
X.~Liu and Y.-Q. Ma,
\newblock \emph{{AMFlow: A Mathematica package for Feynman integrals
  computation via auxiliary mass flow}},
\newblock Comput. Phys. Commun. \textbf{283}, 108565 (2023),
\newblock \doi{10.1016/j.cpc.2022.108565},
\newblock \eprint{2201.11669}.

\bibitem{Liu:2022mfb}
Z.-F. Liu and Y.-Q. Ma,
\newblock \emph{{Determining Feynman Integrals with Only Input from Linear
  Algebra}},
\newblock Phys. Rev. Lett. \textbf{129}(22), 222001 (2022),
\newblock \doi{10.1103/PhysRevLett.129.222001},
\newblock \eprint{2201.11637}.

\bibitem{Kotikov:1990kg}
A.~V. Kotikov,
\newblock \emph{{Differential equations method: New technique for massive
  Feynman diagrams calculation}},
\newblock Phys. Lett. B \textbf{254}, 158 (1991),
\newblock \doi{10.1016/0370-2693(91)90413-K}.

\bibitem{Remiddi:1997ny}
E.~Remiddi,
\newblock \emph{{Differential equations for Feynman graph amplitudes}},
\newblock Nuovo Cim. A \textbf{110}, 1435 (1997),
\newblock \doi{10.1007/BF03185566},
\newblock \eprint{hep-th/9711188}.

\bibitem{Caffo:2008aw}
M.~Caffo, H.~Czyz, M.~Gunia and E.~Remiddi,
\newblock \emph{{BOKASUN: A Fast and precise numerical program to calculate the
  Master Integrals of the two-loop sunrise diagrams}},
\newblock Comput. Phys. Commun. \textbf{180}, 427 (2009),
\newblock \doi{10.1016/j.cpc.2008.10.011},
\newblock \eprint{0807.1959}.

\bibitem{Czakon:2008zk}
M.~Czakon,
\newblock \emph{{Tops from Light Quarks: Full Mass Dependence at Two-Loops in
  QCD}},
\newblock Phys. Lett. B \textbf{664}, 307 (2008),
\newblock \doi{10.1016/j.physletb.2008.05.028},
\newblock \eprint{0803.1400}.

\bibitem{Lee:2014ioa}
R.~N. Lee,
\newblock \emph{{Reducing differential equations for multiloop master
  integrals}},
\newblock JHEP \textbf{04}, 108 (2015),
\newblock \doi{10.1007/JHEP04(2015)108},
\newblock \eprint{1411.0911}.

\bibitem{Moriello:2019yhu}
F.~Moriello,
\newblock \emph{{Generalised power series expansions for the elliptic planar
  families of Higgs + jet production at two loops}},
\newblock JHEP \textbf{01}, 150 (2020),
\newblock \doi{10.1007/JHEP01(2020)150},
\newblock \eprint{1907.13234}.

\bibitem{Hidding:2020ytt}
M.~Hidding,
\newblock \emph{{DiffExp, a Mathematica package for computing Feynman integrals
  in terms of one-dimensional series expansions}},
\newblock Comput. Phys. Commun. \textbf{269}, 108125 (2021),
\newblock \doi{10.1016/j.cpc.2021.108125},
\newblock \eprint{2006.05510}.

\bibitem{Armadillo:2022ugh}
T.~Armadillo, R.~Bonciani, S.~Devoto, N.~Rana and A.~Vicini,
\newblock \emph{{Evaluation of Feynman integrals with arbitrary complex masses
  via series expansions}},
\newblock Comput. Phys. Commun. \textbf{282}, 108545 (2023),
\newblock \doi{10.1016/j.cpc.2022.108545},
\newblock \eprint{2205.03345}.

\bibitem{Liu:2021wks}
X.~Liu and Y.-Q. Ma,
\newblock \emph{{Multiloop corrections for collider processes using auxiliary
  mass flow}},
\newblock Phys. Rev. D \textbf{105}(5), L051503 (2022),
\newblock \doi{10.1103/PhysRevD.105.L051503},
\newblock \eprint{2107.01864}.

\bibitem{Liu:2017jxz}
X.~Liu, Y.-Q. Ma and C.-Y. Wang,
\newblock \emph{{A Systematic and Efficient Method to Compute Multi-loop Master
  Integrals}},
\newblock Phys. Lett. B \textbf{779}, 353 (2018),
\newblock \doi{10.1016/j.physletb.2018.02.026},
\newblock \eprint{1711.09572}.

\bibitem{Denner:2019vbn}
A.~Denner and S.~Dittmaier,
\newblock \emph{{Electroweak Radiative Corrections for Collider Physics}},
\newblock Phys. Rept. \textbf{864}, 1 (2020),
\newblock \doi{10.1016/j.physrep.2020.04.001},
\newblock \eprint{1912.06823}.

\bibitem{vanHameren:2007pt}
A.~van Hameren,
\newblock \emph{{PARNI for importance sampling and density estimation}},
\newblock Acta Phys. Polon. B \textbf{40}, 259 (2009),
\newblock \eprint{0710.2448}.

\bibitem{ParticleDataGroup:2022pth}
Particle Data Group Collaboration, R.~L. Workman \emph{et~al.},
\newblock \emph{{Review of Particle Physics}},
\newblock PTEP \textbf{2022}, 083C01 (2022),
\newblock \doi{10.1093/ptep/ptac097}.

\bibitem{Jones:2023uzh}
S.~P. Jones,
\newblock \emph{{An Overview of Standard Model Calculations for Higgs Boson
  Production {\&} Decay}},
\newblock LHEP \textbf{2023}, 442 (2023),
\newblock \doi{10.31526/lhep.2023.442}.

\bibitem{Sang:2024vqk}
W.-L. Sang, F.~Feng and Y.~Jia,
\newblock \emph{{Next-to-leading-order electroweak correction to
  H{\textrightarrow}Z0{\ensuremath{\gamma}}}},
\newblock Phys. Rev. D \textbf{110}(5), L051302 (2024),
\newblock \doi{10.1103/PhysRevD.110.L051302},
\newblock \eprint{2405.03464}.

\bibitem{Muhlleitner:2022ijf}
M.~M{\"u}hlleitner, J.~Schlenk and M.~Spira,
\newblock \emph{{Top-Yukawa-induced corrections to Higgs pair production}},
\newblock JHEP \textbf{10}, 185 (2022),
\newblock \doi{10.1007/JHEP10(2022)185},
\newblock \eprint{2207.02524}.

\bibitem{Bhattacharya_TBA}
A.~Bhattacharya, F.~Campanario, S.~Carlotti, J.~Chang, J.~Mazzitelli,
  M.~M{\"u}hlleitner, J.~Ronca and M.~Spira,
\newblock \emph{{Higgs-Pair Production via Gluon Fusion: Top-Yukawa- and
  light-quark-induced electroweak Corrections}}  (2025),
\newblock \eprint{2512.14823}.

\bibitem{Davies:2023npk}
J.~Davies, K.~Sch{\"o}nwald, M.~Steinhauser and H.~Zhang,
\newblock \emph{{Next-to-leading order electroweak corrections to $gg \to HH$
  and $gg \to gH$ in the large-$m_t$ limit}},
\newblock JHEP \textbf{10}, 033 (2023),
\newblock \doi{10.1007/JHEP10(2023)033},
\newblock \eprint{2308.01355}.

\bibitem{Davies:2026wbx}
J.~Davies, K.~Sch{\"o}nwald, M.~Steinhauser and H.~Zhang,
\newblock \emph{{Analytic next-to-leading order electroweak corrections to
  Higgs boson pair production at high energies}}  (2026),
\newblock \eprint{2603.08789}.

\bibitem{Zhang:2024rix}
H.~Zhang, K.~Sch{\"o}nwald, M.~Steinhauser and J.~Davies,
\newblock \emph{{Electroweak corrections to gg -{\ensuremath{>}} HH:
  Factorizable contributions}},
\newblock PoS \textbf{LL2024}, 014 (2024),
\newblock \doi{10.22323/1.467.0014},
\newblock \eprint{2407.05787}.

\bibitem{Davies:2022ram}
J.~Davies, G.~Mishima, K.~Sch{\"o}nwald, M.~Steinhauser and H.~Zhang,
\newblock \emph{{Higgs boson contribution to the leading two-loop Yukawa
  corrections to gg {\textrightarrow} HH}},
\newblock JHEP \textbf{08}, 259 (2022),
\newblock \doi{10.1007/JHEP08(2022)259},
\newblock \eprint{2207.02587}.

\bibitem{Davies:2025wke}
J.~Davies, K.~Sch{\"o}nwald, M.~Steinhauser and H.~Zhang,
\newblock \emph{{Analytic next-to-leading order Yukawa and Higgs boson
  self-coupling corrections to gg {\textrightarrow} HH at high energies}},
\newblock JHEP \textbf{04}, 193 (2025),
\newblock \doi{10.1007/JHEP04(2025)193},
\newblock \eprint{2501.17920}.

\bibitem{Heinrich:2024dnz}
G.~Heinrich, S.~Jones, M.~Kerner, T.~Stone and A.~Vestner,
\newblock \emph{{Electroweak corrections to Higgs boson pair production: the
  top-Yukawa and self-coupling contributions}},
\newblock JHEP \textbf{11}, 040 (2024),
\newblock \doi{10.1007/JHEP11(2024)040},
\newblock \eprint{2407.04653}.

\bibitem{Bonetti:2025vfd}
M.~Bonetti, P.~Rendler and W.~J. Torres~Bobadilla,
\newblock \emph{{Two-loop light-quark Electroweak corrections to Higgs boson
  pair production in gluon fusion}},
\newblock JHEP \textbf{07}, 024 (2025),
\newblock \doi{10.1007/JHEP07(2025)024},
\newblock \eprint{2503.16620}.

\bibitem{Djouadi:1994ge}
A.~Djouadi and P.~Gambino,
\newblock \emph{{Leading electroweak correction to Higgs boson production at
  proton colliders}},
\newblock Phys. Rev. Lett. \textbf{73}, 2528 (1994),
\newblock \doi{10.1103/PhysRevLett.73.2528},
\newblock \eprint{hep-ph/9406432}.

\bibitem{Coleman:1973jx}
S.~R. Coleman and E.~J. Weinberg,
\newblock \emph{{Radiative Corrections as the Origin of Spontaneous Symmetry
  Breaking}},
\newblock Phys. Rev. D \textbf{7}, 1888 (1973),
\newblock \doi{10.1103/PhysRevD.7.1888}.

\bibitem{Weinberg:1973ua}
S.~Weinberg,
\newblock \emph{{Perturbative Calculations of Symmetry Breaking}},
\newblock Phys. Rev. D \textbf{7}, 2887 (1973),
\newblock \doi{10.1103/PhysRevD.7.2887}.

\bibitem{Jackiw:1974cv}
R.~Jackiw,
\newblock \emph{{Functional evaluation of the effective potential}},
\newblock Phys. Rev. D \textbf{9}, 1686 (1974),
\newblock \doi{10.1103/PhysRevD.9.1686}.

\bibitem{Zhang:2024fcu}
H.~Zhang,
\newblock \emph{{Massive two-loop four-point Feynman integrals at high energies
  with AsyInt}},
\newblock JHEP \textbf{09}, 069 (2024),
\newblock \doi{10.1007/JHEP09(2024)069},
\newblock \eprint{2407.12107}.

\bibitem{Denner:1991kt}
A.~Denner,
\newblock \emph{{Techniques for calculation of electroweak radiative
  corrections at the one loop level and results for W physics at LEP-200}},
\newblock Fortsch. Phys. \textbf{41}, 307 (1993),
\newblock \doi{10.1002/prop.2190410402},
\newblock \eprint{0709.1075}.

\bibitem{Fleischer:1980ub}
J.~Fleischer and F.~Jegerlehner,
\newblock \emph{{Radiative Corrections to Higgs Decays in the Extended
  Weinberg-Salam Model}},
\newblock Phys. Rev. D \textbf{23}, 2001 (1981),
\newblock \doi{10.1103/PhysRevD.23.2001}.

\bibitem{alibrary}
V.~Magerya,
\newblock \emph{{Alibrary}},
\newblock \url{https://github.com/magv/alibrary}.

\bibitem{Kuipers:2013pba}
J.~Kuipers, T.~Ueda and J.~A.~M. Vermaseren,
\newblock \emph{{Code Optimization in FORM}},
\newblock Comput. Phys. Commun. \textbf{189}, 1 (2015),
\newblock \doi{10.1016/j.cpc.2014.08.008},
\newblock \eprint{1310.7007}.

\bibitem{Klappert:2020nbg}
J.~Klappert, F.~Lange, P.~Maierh{\"o}fer and J.~Usovitsch,
\newblock \emph{{Integral reduction with Kira 2.0 and finite field methods}},
\newblock Comput. Phys. Commun. \textbf{266}, 108024 (2021),
\newblock \doi{10.1016/j.cpc.2021.108024},
\newblock \eprint{2008.06494}.

\bibitem{Magerya:2022hvj}
V.~Magerya,
\newblock \emph{{Rational Tracer: a Tool for Faster Rational Function
  Reconstruction}}  (2022),
\newblock \eprint{2211.03572}.

\bibitem{Klappert:2019emp}
J.~Klappert and F.~Lange,
\newblock \emph{{Reconstructing rational functions with FireFly}},
\newblock Comput. Phys. Commun. \textbf{247}, 106951 (2020),
\newblock \doi{10.1016/j.cpc.2019.106951},
\newblock \eprint{1904.00009}.

\bibitem{Klappert:2020aqs}
J.~Klappert, S.~Y. Klein and F.~Lange,
\newblock \emph{{Interpolation of dense and sparse rational functions and other
  improvements in FireFly}},
\newblock Comput. Phys. Commun. \textbf{264}, 107968 (2021),
\newblock \doi{10.1016/j.cpc.2021.107968},
\newblock \eprint{2004.01463}.

\bibitem{Borowka:2018goh}
S.~Borowka, G.~Heinrich, S.~Jahn, S.~P. Jones, M.~Kerner and J.~Schlenk,
\newblock \emph{{A GPU compatible quasi-Monte Carlo integrator interfaced to
  pySecDec}},
\newblock Comput. Phys. Commun. \textbf{240}, 120 (2019),
\newblock \doi{10.1016/j.cpc.2019.02.015},
\newblock \eprint{1811.11720}.

\bibitem{Heinrich:2021dbf}
G.~Heinrich, S.~Jahn, S.~P. Jones, M.~Kerner, F.~Langer, V.~Magerya,
  A.~P{\"o}ldaru, J.~Schlenk and E.~Villa,
\newblock \emph{{Expansion by regions with pySecDec}},
\newblock Comput. Phys. Commun. \textbf{273}, 108267 (2022),
\newblock \doi{10.1016/j.cpc.2021.108267},
\newblock \eprint{2108.10807}.

\bibitem{Aglietti:2004nj}
U.~Aglietti, R.~Bonciani, G.~Degrassi and A.~Vicini,
\newblock \emph{{Two loop light fermion contribution to Higgs production and
  decays}},
\newblock Phys. Lett. B \textbf{595}, 432 (2004),
\newblock \doi{10.1016/j.physletb.2004.06.063},
\newblock \eprint{hep-ph/0404071}.

\bibitem{Aglietti:2006yd}
U.~Aglietti, R.~Bonciani, G.~Degrassi and A.~Vicini,
\newblock \emph{{Two-loop electroweak corrections to Higgs production in
  proton-proton collisions}},
\newblock In \emph{{TeV4LHC Workshop: 2nd Meeting}} (2006),
  \eprint{hep-ph/0610033}.

\bibitem{Actis:2008ts}
S.~Actis, G.~Passarino, C.~Sturm and S.~Uccirati,
\newblock \emph{{NNLO Computational Techniques: The Cases $H \to \gamma \gamma$
  and $H\to g g$}},
\newblock Nucl. Phys. B \textbf{811}, 182 (2009),
\newblock \doi{10.1016/j.nuclphysb.2008.11.024},
\newblock \eprint{0809.3667}.

\bibitem{Actis:2008ug}
S.~Actis, G.~Passarino, C.~Sturm and S.~Uccirati,
\newblock \emph{{NLO Electroweak Corrections to Higgs Boson Production at
  Hadron Colliders}},
\newblock Phys. Lett. B \textbf{670}, 12 (2008),
\newblock \doi{10.1016/j.physletb.2008.10.018},
\newblock \eprint{0809.1301}.

\bibitem{Bonetti:2026cih}
M.~Bonetti, G.~Heinrich, P.~Rendler and W.~J. Torres~Bobadilla,
\newblock \emph{{NLO QCD corrections to the electroweak production of a Higgs
  boson pair in the quark-antiquark channel}},
\newblock JHEP \textbf{04}, 131 (2026),
\newblock \doi{10.1007/JHEP04(2026)131},
\newblock \eprint{2601.16924}.

\bibitem{boole1880}
G.~Boole,
\newblock \emph{A Treatise on the Calculus of Finite Differences},
\newblock Macmillan and Company, London, 3rd edn. (1880).

\bibitem{Abramowitz1964}
M.~Abramowitz and I.~A. Stegun,
\newblock \emph{Handbook of Mathematical Functions with Formulas, Graphs, and
  Mathematical Tables}, vol.~55 of \emph{Applied Mathematics Series},
\newblock Dover Publications, New York (1964).

\bibitem{wg4wiki}
{LHC Higgs Working Group},
\newblock \emph{{LHC Higgs WG4 group (formerly LHC-HH sub group)}},
\newblock \url{https://twiki.cern.ch/twiki/bin/view/LHCPhysics/LHCHWGHH},
\newblock Accessed: 28 April 2026 (2026).

\end{thebibliography}

\end{document}